\documentclass[preprint]{aastex63}
\usepackage{rotating}

\def\l{$\lambda$}

\def\mbh{$M_{\rm BH}$\/}

\def\lledd{$L/L_{\rm Edd}$}

\def\rfe{$R_{\rm FeII}$}

\def\feiiq{\rm Fe{\sc ii}$\lambda$4570\/}
\def\msol{M$_\odot$\/}

\def\ltsima{$\; \buildrel < \over \sim \;$}
\def\ltsim{\lower.5ex\hbox{\ltsima}}  
\def\gtsima{$\; \buildrel > \over \sim \;$}

\def\gtsim{\lower.5ex\hbox{\gtsima}} 
\def\h{$H_{0}$}
\def\lya{{ Ly}$\alpha$}
\def\civ{{\sc{Civ}}$\lambda$1549\/}

\def\cm3{cm$^{-3}$\/}
\def\hb{{\sc{H}}$\beta$\/}

\def\hbbc{{\sc{H}}$\beta_{\rm BC}$\/}

\def\hbnc{{\sc{H}}$\beta_{\rm NC}$\/}
\def\mgii{{Mg\sc{ii}}$\lambda$2800\/}

\def\oiiiopt{{\sc{[Oiii]}}$\lambda\lambda$4959,5007\/}
\def\oii{{\sc{[Oii]}}$\lambda$3727\/}

\def\siiiifull{Si{\sc iii]}$\lambda$1892\/}
\def\aliiifull{Al{\sc iii}$\lambda$1860\/}
\def\ciiifull{C{\sc iii]}$\lambda$1909\/}
\def\siiii{Si{\sc iii]}\/}
\def\aliii{Al{\sc iii}\/}
\def\ciii{C{\sc iii]}\/}

\def\feiiuv{{{Fe\sc{ii}}}$_{\rm UV}$\/}

\def\feii{{Fe\sc{ii}}\/}
\def\siii{{Si\sc{ii}}$\lambda$1816\/}
\def\feiii{{Fe\sc{iii}}\/}
\def\fe{{\sc{Fe}}\/}

\def\heiiopt{He{\sc{ii}}$\lambda$4686\/}
\def\fe76087{{\sc [Fe vii]}$\lambda$6087\/}

\def\b{$\beta$}
\def\kms{km~s$^{-1}$}
\def\rk{$R_\mathrm{K}$\/}
\def\ergss{erg\, s$^{-1}$\/}

\def\heiiopt{{\sc H}e{\sc ii}$\lambda$4686\/}

\usepackage{enumerate}
\definecolor{darkorange}{rgb}{1,0.612,0}
\definecolor{aquamarine}{rgb}{0.498,1,0.8314}

\received{}
\revised{}
\accepted{}
\submitjournal{ApJS}

\shorttitle{\aliii\ and \ciii\ line broadening}
\shortauthors{{\em Marziani et al.}}

\defcitealias{sulenticetal07}{S07}
\defcitealias{sulenticetal17}{S17}
\defcitealias{marzianietal19}{M19}
\defcitealias{marzianietal13a}{M13a}
\defcitealias{marzianietal13}{M13}
\defcitealias{marzianietal03a}{M03}
\defcitealias{marzianisulentic14}{MS14}
\begin{document}
\title{The intermediate-ionization lines  as   virial broadening estimators for Population A quasars   \footnote{Based in part  on observations made with ESO Telescopes at the Paranal Observatory under programme 082.B-0572(A), and 083.B-0273(A).}}

\correspondingauthor{Paola Marziani}
\email{paola.marziani@inaf.it}
\author[0000-0002-6058-4912]{Paola Marziani}
\affiliation{National Institute for Astrophysics (INAF), Os\-servatorio Astronomico di Padova, vicolo dell' Osservatorio 5, IT 35122, Padova, Italy}
\author[0000-0003-2055-9816]{Ascensi\'on del Olmo}
\affiliation{Instituto de Astrofis\'{\i}ca de Andaluc\'{\i}a, IAA-CSIC, Glorieta  de la Astronomia s/n 18008 Granada, Spain}
\author[0000-0002-1656-827X]{C. Alenka Negrete}
\affiliation{Universidad Nacional Auton\'oma de M\'exico Instituto de Astronom\'ia: Ciudad de Mexico, Distrito Federal, MX 04510, Mexico}
\author[0000-0001-5756-8842]{Deborah Dultzin}
\affiliation{Universidad Nacional Auton\'oma de M\'exico Instituto de Astronom\'ia: Ciudad de Mexico, Distrito Federal, MX 04510, Mexico}
\author[0000-0001-9095-2782]{Enrico Piconcelli}
\affiliation{National Institute for Astrophysics (INAF), Os\-servatorio Astronomico di Roma, Via Frascati 33, 00040, Monte Porzio Catone (RM), Italy}
\author[0000-0001-9155-8875]{Giustina Vietri}
\affiliation{National Institute for Astrophysics (INAF), Os\-servatorio Astronomico di Roma, Via Frascati 33, 00040, Monte Porzio Catone (RM), Italy}
\affiliation{National Institute for Astrophysics (INAF),  Istituto di Astrofisica Spaziale e Fisica Cosmica, Via A. Corti 12, 20133 Milano, Italy}
\author[0000-0002-7843-7689]{Mary Loli Mart\'inez-Aldama}
\affiliation{Center for Theoretical Physics (PAS), Al. Lotnik{\'o}w 32/46, 02-668 Warsaw, Poland}
\author[0000-0001-6441-9044]{Mauro D'Onofrio}
\affiliation{Dipartimento di Fisica e Astronomia,  Universit\`a di Padova, Vicolo dell' Osservatorio 3, Padova, Italy}
\author[0000-0002-0465-8112]{Edi Bon}\author[0000-0002-3462-4888]{Natasa Bon}
\affiliation{Belgrade Astronomical Observatory, Belgrade, Serbia}
\author[0000-0002-7630-0777]{Alice Deconto Machado}
\affiliation{Instituto de Astrofis\'{\i}ca de Andaluc\'{\i}a, IAA-CSIC, Glorieta  de la Astronomia s/n 18008 Granada, Spain}
\author[0000-0002-3702-8731]{Giovanna M. Stirpe}
\affiliation{National Institute for Astrophysics (INAF), Osservatorio di Astrofisica e Scienza dello Spazio, Italy,  via Gobetti 93/3,
40129 Bologna, Italy}
\author{Tania Mayte Buendia Rios}
\affiliation{Universidad Nacional Auton\'oma de M\'exico Instituto de Astronom\'ia: Ciudad de Mexico, Distrito Federal, MX 04510, Mexico}






\begin{abstract}
The identification of a virial broadening estimator  in the quasar UV rest frame suitable for black hole mass computation at high redshift has become an important issue. We compare the HI Balmer  \hb\  line width to the ones of two intermediate ionization lines: the  \aliiifull\ doublet and the \ciiifull\ line, over a wide interval of redshift and luminosity ($0 \lesssim z \lesssim 3.5$;  $ 43 \lesssim \log L \lesssim 48.5$\ [\ergss]), for 48 sources belonging to the quasar population characterized by mid-to-high values of the Eddington ratio (Population A).  The present analysis indicates that the line width  of \aliiifull\ { and \hb\ are highly correlated, and can be considered equivalent for most Population A quasars over five orders of magnitude in luminosity; } for \ciiifull, multiplication by a constant correction factor $\xi \approx 1.25$\ is sufficient to bring the FWHM of \ciii\ in agreement with the one of \hb.  The statistical concordance between  low-ionization and intermediate-ionization lines  suggests that they  predominantly arise from the same virialized part of the broad line region.  However, blueshifts of modest amplitude (few hundred \kms) with respect to the quasar rest frame  and an excess ($\lesssim 1.1$) \aliii\ broadening with respect to \hb\ are found  { in a fraction of our sample}.   
{Scaling laws to estimate \mbh\ of high redshift quasar using the \aliii\ and the \ciii\ line widths  have  rms scatter  $\approx$ 0.3  dex. { The \aliii\ scaling law takes the form   $\log M_\mathrm{BH} \approx 
  0.58 \log L_{1700,44} +
 2 \log \mathrm{FWHM} + 0.49$ [\msol]. } 
 }
 \end{abstract}


\thispagestyle{empty}


\section{Introduction} \label{sec:intro}


The energetics of all accretion-related   phenomena occurring in active galactic nuclei (AGN) can be tied down to the mass of the central black hole.  The mass (\mbh) of the  black holes at the origin of the AGN phenomenon is now reputed a key parameter in the evolution of galaxies and in cosmology as well \citep[e.g.,][]{kormendyho13,Vogelsbergeretal2014,Heckman&Best2014}, and its estimation   has become an important branch of extragalactic research. Black hole mass estimates on large type-1 AGN samples  are carried out employing a deceptively-simple formulation of the virial theorem, under the assumption that all the mass of the system is concentrated in the center of gravity provided by the black hole \citep[see e.g.,][for reviews]{marzianisulentic12,shen13,peterson14}: \mbh = $f_\mathrm{S} r_\mathrm{BLR} (\delta v)^{2} / G$, where     $f_\mathrm{S}$\ is a structure factor (a.k.a. virial or form factor) dependent on the emitting region geometry and dynamics, the radius  $r_\mathrm{BLR} $\ the distance of the line emitting region from the continuum source, and $\delta v$\ a suitable measure of the line broadening (e.g., FWHM  or dispersion $\sigma$,    { \citealt{vestergaardpeterson06, petersonetal04}}). The main underlying assumptions are that the broadening is due to Doppler effect because of the line emitting gas motion, and that the velocity field  is such that the emitting gas remains gravitationally bound to the black hole.   

Early UV and optical inter-line shift analysis provided evidence that not all the line emitting gas is bound to the black hole \citep[e.g., ][]{gaskell82,tytlerfan92,brothertonetal94,marzianietal96,leighlymoore04}.  The scenario emerging from more recent studies is that outflows are ubiquitous in active galactic nuclei.  They occur under a wide range of physical conditions, and are detected in almost every band of the electromagnetic spectrum and on a wide range of spatial scales, from a few  gravitational radii to tens of kpc \citep[e.g.,][]{capettietal96,colbertetal98,everett07,carnianietal15,bischettietal17,komossaetal18,kakkadetal20,vietrietal20,laurentietal21}. 
At high luminosity, massive outflows provide feedback effects to the host galaxy \citep[e.g.,][]{fabian12,kingpounds15,kingmuldrew16,baraietal18}, and are  invoked to account for the \mbh-bulge velocity dispersion correlation \citep[e.g.,][and references therein]{kormendyho13}. For $z \gtrsim 4$, \mbh\ estimates rely on the \civ\ high-ionization line, and  the highest-$z$\ sources appear to be almost always high-accretors \citep{banadosetal18,nardinietal19}. Two studies pointed out  20 years ago the similarity between X-ray and UV properties of high-$z$ quasars and local quasars accreting at high rates  \citep[e.g.,][]{mathur00,sulenticetal00a}. The source of concerns is that  high-ionization lines such as \civ\ are subject to a considerable broadening and blueshifts  associated with outflow motions already at low redshift \citep[][see \citealt{marinelloetal20b} for  a detailed study of the prototypical source PHL 1092]{coatmanetal16,sulenticetal17,marinelloetal20a}. Overestimates of the virial broadening by a factor as large as  { $\approx$}5 \citep{netzeretal07,sulenticetal07,mejia-restrepoetal16,mejia-restrepoetal18}   for supermassive black holes at high $z$ may even pose a spurious challenge to concordance cosmology \citep[e.g.,][]{trakhtenbrotetal15} and lead to erroneous inferences on the properties of the seed black holes believed to be fledgling precursors of massive black holes.

This paper is focused on the measurement of the line width of the UV intermediate-ionization lines  at $\approx 1900$\ \AA\ and on their use for black hole mass measurements for large quasar samples, over a wide interval of luminosity and of redshift.  The blend at $\lambda1900$ \AA\ is due, at least in part, to the \aliiifull\ doublet and to the \siiiifull\ and \ciiifull\ lines.  \aliii\ is a resonant doublet ($^{2}P^{o}_{\frac{3}{2},\frac{1}{2}} \rightarrow ^{2}S_{\frac{1}{2}}$) while \siiii\ and   \ciii\ are due to inter-combination transitions ($^{3}P^{o}_{1} \rightarrow ^{1}S_{0}$) with widely different critical densities ($\approx 2 \cdot 10^{11}$ \cm3\ and $\approx 3 \cdot 10^{9}$ \cm3, respectively; \citealt{zheng88,negreteetal12}). The parent ionic species imply ionization potentials  $15 \lesssim \chi_\mathrm{i} \lesssim 30$\ eV, intermediate between the ones of low-ionization lines (LILs), and the ones of high-ionization lines (HILs; $\chi_\mathrm{i} \gtrsim 40-50$ eV).   The intermediate-ionization lines (IILs) at 1900 \AA\ are well-placed to provide a high redshift estimator; they can be observed with optical spectrometers up to $z \sim 4$. Observations can be extended in the NIR (13,500 \AA) up to $z \sim 5.7$ without solution of continuity, thereby sampling  a redshift domain that is crucial for understanding  the primordial growth of massive black holes and galaxy formation. In principle,  observations could be extended to the H band to cover the as yet mostly uncharged range $6.5 \lesssim z \lesssim 8$, a feat that may well become possible with the advent of James Webb Space Telescope \citep{gardneretal06}, of the ESO Extremely Large Telescope \citep{gilmozzispyromilio07}, and of the next-generation  large-aperture telescopes \citep[see e.g.,][for a review of foreseeable technological developments]{donofriomarziani18}. 

The  quasar main sequence provides much needed discerning abilities for the exploitation of the IILs  \citep[e.g.,][]{sulenticetal00a,bachevetal04,marzianietal01,shenho14,pandaetal18}, as line profiles and intensities of individual sources are not considered as isolated entities, but interpreted as part of consolidated trends in the main sequence context. Broad line measurements involving \hb\ line width and  \feii\ strength are not randomly distributed but instead  define a sequence that has become known as the quasar  ``main sequence'' (MS; e.g., \citealt{sulenticetal00a,shenho14,pandaetal19,wildyetal19}). The \feii\ strength is parameterized by the intensity ratio involving  the \feii\ blue blend at 4570 \AA\ and broad \hb\ i.e., \rfe = I(\feiiq)/I(\hb), and the Hydrogen \hb\ line width by   its FWHM.  MS sources  with higher \rfe\   show  narrower broad \hb\  (Population A, FWHM(\hb)$\lesssim 4000$ \kms),  and  sources with broader \hb\ profiles tendentially show low \rfe\  (Pop. B with FWHM(\hb)$\gtrsim 4000$ \kms, \citealt{sulenticetal00a,sulenticetal11}). It is also known that   optical and UV observational  properties  are correlated \citep[e.g.,][]{sulenticetal00b,bachevetal04,sulenticetal07,duetal16a,sniegowskaetal18}. In this paper, the attention is restricted to sources radiating at relatively high Eddington ratio (\lledd\ $\gtrsim 0.1-0.2$) i.e., to Population A that accounts for the large majority of sources discovered at high $z$. In the course of our analysis we realised that sources radiating at lower Eddington ratio ($0.01\ \lesssim $\lledd\ $\lesssim 0.1-0.2$)\ show a different behaviour of the 1900 blend  and will be considered elsewhere.

The coverage of the \hb\ spectral range greatly easies the determination of the redshift as well as the positional classification of sources along the MS. In addition, FWHM \hb\ has been employed as a virial broadening estimator of \mbh\ since the earliest single-epoch observations of large samples of quasars, and in more recent times as well \citep[e.g.,][]{mclurejarvis02,mcluredunlop04,vestergaardpeterson06,assefetal11,trakhtenbrotnetzer12,shenliu12}. The \hb\ line is likely to be still the most widely  used line for \mbh\ computations for low redshift quasars ($z \lesssim 1$).  Our analysis relies on the availability of {\rm both} \hb\ and the 1900 blend lines, as we will consider FWHM \hb\ as the reference ``virial broadening estimator” (VBE). 

Section \ref{sample} introduces the samples used in the present work, covering a wide range in redshift and luminosity, $0 \lesssim z \lesssim 3.5$;  $ 44 \lesssim \log L \lesssim 48.5$\ [\ergss].  The data were  obtained with instruments operating in  widely different spectral ranges (UV, optical, IR); as a consequence, S/N ratio values vary widely and the uncertainty assessment requires a dedicated approach (Section \ref{dataan}, and Appendix \ref{app:unc}).     The   Section \ref{results} introduces paired fits to \hb\ and the 1900 blend (an atlas is provided in Appendix \ref{app:fits}), along with several line width measures, and the relation between \hb\ and \aliii\ measurements. A scaling law for \mbh\ determination equivalent to the one based on \hb\ but based on the IIL broadening is discussed in Section \ref{disc}. 

\section{Sample}
\label{sample}

\paragraph{Low-luminosity 1900 and \hb\ data (FOS$^{\star}$ sample)} We considered a Faint Object Spectrograph (FOS) sample from  \citet[][hereafter \citetalias{sulenticetal07}]{sulenticetal07} as a low-$L$ and low-$z$  sample. For the sake of the present paper, we restrict the \citetalias{sulenticetal07} sample to 28  sources covering the 1900 blend spectral range and with previous measurements for the \hb\ profile and \rfe\ (FOS$^{\star}$\ sample). The spectra covering the \hb\ spectral range come from \citet[][hereafter \citetalias{marzianietal03a}]{marzianietal03a}, as well as   from the SDSS  \citep{yorketal00}  and the 6dF (\citealt{jonesetal04}; Table \ref{tab:sample} provides information on  the provenience of of individual spectra). The FOS high-resolution grisms yielded an inverse resolution $\lambda/\delta \lambda \sim 1000$, equivalent to typical resolution of the data of \citetalias{marzianietal03a} and of the SDSS. The S/N is above $\gtrsim 20$\ for both the  optical and UV  low-redshift data. The FOS$^{\star}$\ sample has a typical bolometric luminosity $\log L \sim 45.6$\ [\ergss] and a redshift $z \lesssim  0.5$.

\paragraph{High-luminosity VLT and TNG data for Hamburg-ESO quasars (HE sample)} The sample of high-$L$\ quasars  includes 10 sources  identified in the Hamburg-ESO survey \citep[][]{wisotzkietal00} in the redshift range $1.4 \lesssim   z  \lesssim 2.6$.  All HE quasars satisfy the condition on the bolometric luminosity $\log L \gtrsim 10^{47.5}$ \ergss\ and are discussed in detail by \citet[][hereafter \citetalias{sulenticetal17}]{sulenticetal17}, where \civ\ and \hb\ were analyzed. The sample used in this paper is restricted to the 9  Population A sources with VLT/FORS1 spectra and 1 TNG/DOLORES (HE1347-2457) spectrum that cover the 1900 \AA\ blend.   The spectral resolutions  at FWHM  are $\lesssim 300$ \kms\ and $\lesssim$\ 600 \kms\ for the spectrographs FORS1 and DOLORES, respectively. The resolution of the ISAAC spectra covering H$\beta$\  is $\approx$ 300 \kms\ \citep{sulenticetal04}. Typical S/N values are $\gtrsim 50$. 

\paragraph{Additional high-luminosity sources (ISAAC sample)} Additional ISAAC spectra were obtained under programme 083.B-0273(A), for three targets \object{SDSS\- J005700.18\-+143737.7}, \object{SDSS\- J132012.33+142037.1},  \object{SDSS\- J161458.33\-+144836.9}.  They have been reduced following the same procedures employed for the HE quasars. The data will be presented in a forthcoming paper (Deconto Machado et al., in preparation).  Matching rest-frame UV spectra were collected from the SDSS and BOSS \citep{smeeetal13}, with a resolving power $R = \lambda$/FWHM $\sim$ 2000. 


\paragraph{High-luminosity sources from the WISSH (WISSH sample)} We included  near-infrared (NIR) spectroscopic observations of 7 WISSH Population A quasars QSOs \citep{vietrietal18}, obtained with LUCI at the Large Binocular Telescope  and in one case with SINFONI at VLT. Basic information on this sample is provided in Table 1 of \citet{vietrietal18}. The matching rest-frame UV spectra are from the SDSS.  The higher resolution implies a somewhat lower S/N with respect to the ISAAC spectra; we restrict our analysis to the spectra above a minimum S/N $\approx\, 15$. Redshifts measured for this paper agree very well with the values reported by \citet{vietrietal18} if the \hb\ profile is sharp; they are lower by 300-400 \kms\ in four cases with relatively shallow profiles due to the different fitting techniques.  



\paragraph{Joint  sample}  Table \ref{tab:sample}  lists in the following order  source identification, redshift, specific rest-frame flux in the UV at 1700 \AA\ $f_{\lambda,1700}$, S/N at 1700 \AA\, reference to the origin of the spectrum,  specific flux in the optical at 5100 \AA\ ($f_{\lambda,5100}$),  S/N at 5100 \AA, and reference to the origin of the optical spectrum. Table footnotes list references to the flux scale origin, in case the spectrum had uncertain of no absolute spectrophotometric flux calibration.  Notes include the radio loudness classification \citep{zamfiretal08,gancietal19}: radio-loud (RL), radio-intermediate (RI), and radio-quiet (RQ). Only two sources (\object{HE 0043-2300} and \object{3C 57}) are ``jetted'' in the sense of having a powerful relativistic jet \citep{padovani17}. \object{HE 0043-2300} is listed as a flat-spectrum radio quasar  with dominant blazar characteristics in the Roma-BZCAT \citep{massaroetal09}, and \object{3C 57} is a compact-steep source (CSS; \citealt{odea98}, \citealt{sulenticetal15}).  
Two other sources qualify as radio-intermediate (\object{HE 0132-4313} and \object{HE0248-3628}), and are briefly discussed  
in Appendix \ref{app:notes}.  


\begin{figure}
 \includegraphics[width=0.55\columnwidth]{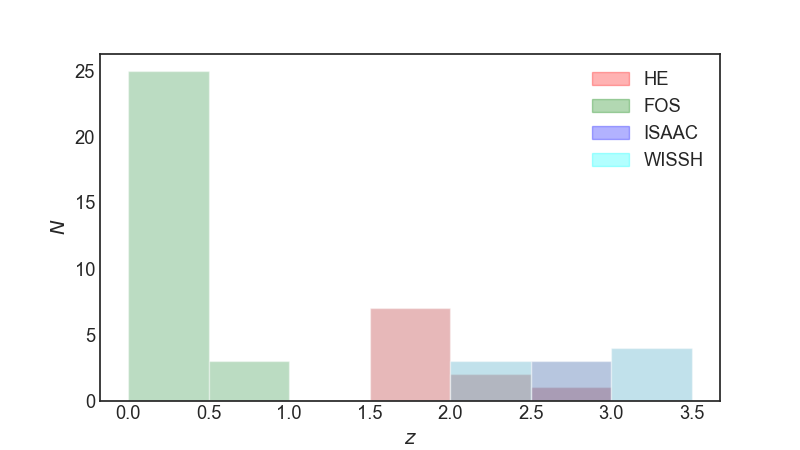}  \includegraphics[width=0.5\columnwidth]{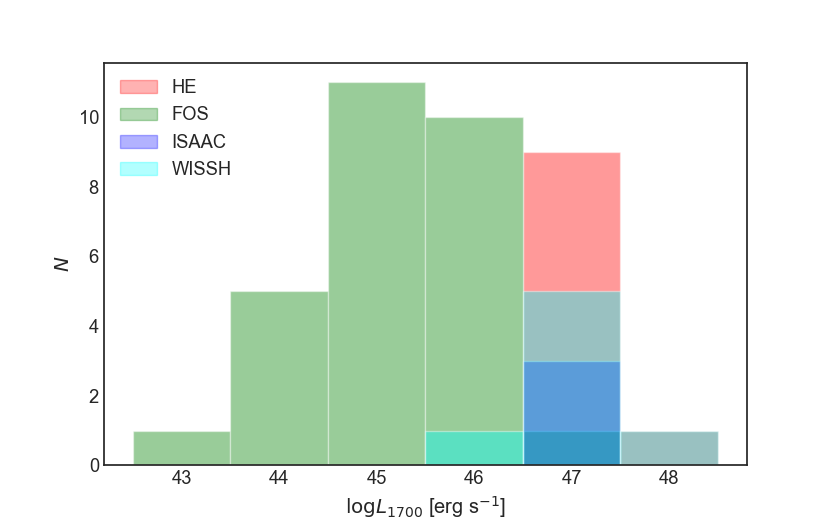}\\ 
\caption{Distribution of redshift (left) and of luminosity at 1700 \AA\ (right) for the four sub-samples considered in this paper.   \label{fig:distr}}
\end{figure} 

\newpage
 \begin{table}
 \begin{center}
\tablenum{1}\tiny\tabcolsep=2pt
\caption{Sample properties \label{tab:sample}}
\begin{tabular}{lllccclccl}\hline\hline
{IAU code} & {Common name} & {z} &  
{$f_{\lambda,1700}$\tablenotemark{\tiny a}} &  {S/N$_{1700}$} &   {Ref.\tablenotemark{\tiny b}} & {$f_{\lambda,5100}$\tablenotemark{\tiny a}} & {S/N$_{5100}$} & {Ref.\tablenotemark{\tiny b}} &  Notes\\
\hline
\tiny
\\
\hline
\multicolumn{8}{c}{FOS$^\star$\ sample}\\
\hline
J00063+2012	&	MRK 0335	    &	0.0252	&	60.8	&	15	&	S07	& 5.92	 	          					           &	55	&	M03 	    \\
J00392-5117	&	[WPV85] 007	    &	0.0290	&	2.5		&	15	&	S07	& 2.02 \tablenotemark{\tiny c}                           &	45	&	6dF		    \\
J00535+1241	&	UGC 00545	    &	0.0605	&	28.2	&	40	&	S07	& 5.77	 	                                       &	70	&	M03  	    \\
J00573-2222	&	TON S180	    &	0.0620	&	31.8	&	45	&	S07	& 14.6   	                                       &	55	&	M03 	    \\
J01342-4258	&	HE 0132-4313	&	0.2370	&	15.2	&	50	&	S07	& 1.44 \tablenotemark{\tiny d}	                       &	15	&	6dF	 & RI 	\\
J02019-1132	&	3C 057		    &	0.6713	&	17.7	&	25	&	S07	& 1.90	 	                                 	   &	35	&	S15	 & CSS	\\
J06300+6905	&	HS 0624+6907	&	0.3702	&	51.7	&	30	&	S07	& 5.04 \tablenotemark{\tiny e}	                       &	40	&	M03		\\
J07086-4933	&	1H 0707-495  	&	0.0408	&	22.2	&	20	&	S07	& 2.14 \tablenotemark{\tiny f}	                       &	35	&	6dF		\\
J08535+4349	&	[HB89] 0850+440 &	0.5149	&	5.7		&	20	&	S07	& 0.50 	                            &	15	&	M03		\\
J09199+5106	&	NGC 2841 UB3	&	0.5563	&	11.0	&	30	&	S07	& 1.25 	                            &	40	&	SDSS	\\
J09568+4115	&	PG 0953+414	    &	0.2347	&	17.1	&	30	&	S07	& 2.15 	                            &	75	&	M03		\\
J10040+2855	&	PG 1001+291     &	0.3298	&	17.2	&	25	&	S07	& 1.92 	                            &	45	&	M03		\\
J10043+0513	&	PG 1001+054	    &	0.1611	&	4.9		&	10	&	S07	& 1.50 	                            &	30	&	M03  	\\
J11185+4025	&	PG 1115+407	    &	0.1536	&	11.7	&	20	&	S07	& 0.46 	                            &	30	&	M03		\\
J11191+2119	&	PG 1116+215	    &	0.1765	&	41.3	&	40	&	S07	& 2.62 	                            &	50	&	M03		\\
J12142+1403	&	PG 1211+143	    &	0.0811	&	31.0	&	20	&	S07	& 5.45 	                            &	40	&	M03		\\
J12217+7518	&	MRK 0205	    &	0.0711	&	23.6	&	35	&	S07	& 1.73 	                            &	55	&	M03		\\
J13012+5902	&	SBS 1259+593	&	0.4776	&	19.1	&	25	&	S07	& 0.59 	                            &	50	&	M03		\\
J13238+6541	&	PG 1322+659	    &	0.1674	&	9.5		&	40	&	S07	& 0.71 	                            &	35	&	M03		\\
J14052+2555	&	PG 1402+262	    &	0.1633	&	22.6	&	25	&	S07	& 1.54 	                            &	45	&	M03		\\
J14063+2223	&	PG 1404+226	    &	0.0973	&	5.8		&	15	&	S07	& 1.12 	                            &	60	&	M03		\\
J14170+4456	&	PG 1415+451	    &	0.1151	&	10.2	&	25	&	S07	& 0.86 	                            &	35	&	M03		\\
J14297+4747	&	[HB89] 1427+480 &	0.2199	&	7.6		&	30	&	S07	& 0.30 	                            &	55	&	M03		\\
J14421+3526	&	MRK 0478	    &	0.0771	&	28.2	&	25	&	S07	& 2.04 	                            &	55	&	M03		\\
J14467+4035	&	[HB89] 1444+407 &	0.2670	&	18.7	&	45	&	S07	& 1.02 	                            &	20	&	M03		\\
J15591+3501	&	UGC 10120	    &	0.0313	&	7.3		&	20	&	S07	& 2.29 	                            &	55	&	SDSS	\\
J21148+0607	&	[HB89] 2112+059 &	0.4608	&	14.9	&	25	&	S07	& 0.81 	                            &	50	&	M03		\\
J22426+2943	&	UGC 12163	    &	0.0245	&	10.9	&	40	&	S07	& 0.67 	                            &	25	&	M03		\\
\hline
\multicolumn{8}{c}{HE sample}\\
\hline
J00456--2243	&	HE0043-2300	&	1.5402	&	15.5	&	115	&	S17	&	3.2	&	70	&	S17	& RL \\
J01242--3744	&	HE0122-3759	&	2.2004	&	21.7	&	95	&	S17	&	2.2	&	30	&	S17	\\
J02509--3616	&	HE0248-3628	&	1.5355	&	24.2	&	200	&	S17	&	0.8	&	50	&	S17 & RI, inv. radio sp.	\\
J04012--3951	&	HE0359-3959	&	1.5209	&	12.1	&	105	&	S17	&	1.8	&	40	&	S17	\\
J05092--3232	&	HE0507-3236	&	1.5759	&	11.7	&	160	&	S17	&	2.1	&	25	&	S17	\\
J05141-3326	&	HE0512-3329	&	1.5862	&	7.7	&	40	&	S17	&	2.7	&	25	&	S17	\\
J11065--1821	&	HE1104-1805	&	2.3180	&	23.9	&	75	&	S17	&	3.0	&	15	&	S17	\\
J13506--2512	&	HE1347-2457	&	2.5986	&	48.0	&	75	&	S17	&	3.9	&	50	&	S17	\\
J21508--3158	&	HE2147-3212	&	1.5432	&	17.0	&	150	&	S17	&	1.7	&	20	&	S17	\\
J23555--3953	&	HE2352-4010	&	1.5799	&	35.5	&	85	&	S17	&	6.3	&	60	&	S17	\\
\hline
\multicolumn{8}{c}{ISAAC sample}\\
\hline
J00570+1437 	&	SDSSJ005700.18+143737.7 	    &	2.6635	&	14.0	&	55	&	       SDSS	&	2.78	&	40	&	D22	&  normalized at 5000 \AA\ \\
J13202+1420 	&	SDSSJ132012.33+142037.1       	&	2.5357	&	8.4	&	40	&	SDSS	&	1.32	&	25	&	        D22	&  normalized at 5000 \AA\ \\
J16149+1448 	&	SDSSJ161458.33+144836.9       	&	2.5703	&	15.3	&	50	&	SDSS	&	2.54	&	45	&	    D22	&  normalized at 5000 \AA\ \\
\hline
\multicolumn{8}{c}{WISSH sample}\\
\hline
0801+5210	&	SDSS J080117.79+521034.5	&	3.2565	&	29.4	&	30	&	SDSS	&	4.12	&	20	&	V18	\\
1157+2724	&	SDSS J115747.99+272459.6	&	2.2133	&	4.2		&	15	&	SDSS	 &	2.37	&	25	&	V18	 & HiBAL QSO\\
1201+0116	&	SDSS J120144.36+011611.6	&	3.2476	&	17.0	&	30	&	SDSS	&	3.31	&	20	&	V18	\\
1236+6554	&	SDSS J123641.45+655442.1	&	3.4170	&	22.9	&	45	&	SDSS	&	2.30	&	25	&	V18	\\
1421+4633	&	SDSS J142123.97+463318.0	&	3.4477	&	20.9	&	25	&	SDSS	&	3.28	&	15	&	V18	\\
1521+5202	&	SDSS J152156.48+520238.5	&	2.2189	&	59.8	&	80	&	SDSS	&	8.40	&	35	&	V18	\\
2123--0050	&	SDSS J212329.46-005052.9	&	2.2791	&	32.2	&	60	&	SDSS	&	5.75	&	45	&	V18	\\
\hline
\end{tabular}
\end{center}
\vspace{-0.4cm}
\tablenotetext{}{\scriptsize $^a$\ In units of $10^{-15}$ \ergss\ cm$^{-2}$ \AA$^{-1}$, in rest frame, not corrected for Galactic extinction.}
\tablenotetext{}{\scriptsize $^b$ Reference to the origin of the spectra: \citetalias{sulenticetal17}; \citetalias{sulenticetal07}; \citetalias{marzianietal03a}; D22: Deconto Machado et al., in preparation; SDSS: SDSS and BOSS, \citet{smeeetal13}; 6dF: \citet{jonesetal04}. }
\tablenotetext{}{\scriptsize $^c$ Uncalibrated 6dF spectrum; flux scale set by a quick-look magnitude as in \citetalias{sulenticetal07}. \citet{grupeetal98} luminosity yields $f_{\lambda,5100} \approx 3.1  \cdot 10^{-15}$ \ergss\ cm$^{-2}$ \AA$^{-1}$. Very low  $f_{\lambda,1700}/f_{\lambda,5100} \approx 1$ ratio. The $f_{\lambda,1700}$\ from the S07 HST/FOS spectrum corresponds to a deep minimum of the UV flux. Later observations show a more than 5-fold increase in the UV continuum  \citep{leighlyetal15}.  }
\tablenotetext{}{\scriptsize  $^d$ Uncalibrated 6dF spectrum; flux scale set by a quick-look magnitude as in \citetalias{sulenticetal07}. \citet{grupeetal10} $V$\ luminosity implies $f_{\lambda,5100} \approx  0.86 \cdot 10^{-15}$ \ergss\ cm$^{-2}$ \AA$^{-1}$.}
\tablenotetext{}{\scriptsize  $^e$ Uncalibrated   spectrum; flux scale set by a quick-look magnitude as in \citetalias{sulenticetal07}. \citet{decarlietal10} yield  $5.24 \cdot 10^{-15}$ \ergss\ cm$^{-2}$ \AA$^{-1}$, in close agreement.}
\tablenotetext{}{\scriptsize  $^f$ Uncalibrated 6dF  spectrum; flux scale set by a quick-look magnitude as in \citetalias{sulenticetal07}. \citet{giannuzzostirpe96} yield  $1.80 \cdot 10^{-15}$ \ergss\ cm$^{-2}$ \AA$^{-1}$, in close agreement.}
\tablecomments{\scriptsize RL: jetted,$ \log$ \rk$\gtrsim 100$\ following \citet{gancietal19}; RI: $10 \lesssim \log$ \rk$\lesssim 100$; CSS: compact steep spectrum radio-source.}
\end{table}


\section{Data analysis}
\label{dataan}

\subsection{The quasar main sequence as an interpretative aid}

In the following the framework of the quasar MS to make assumptions on line shapes, both in the optical and in the UV spectral ranges. There are several papers that provide a description of the main trends associated with the MS. \citet{fraix-burnetetal17b} reviews  the main multifrequency trends.   \citet{sulenticetal00a,sulenticetal11} review the  case for two different quasar Populations: Population A (at low $z$, FWHM \hb\  $\lesssim 4000$ \kms), and  Population  B (FWHM \hb\  $\gtrsim 4000$ \kms). The limit is luminosity-dependent \citepalias{sulenticetal17}, and reaches FWHM $\gtrsim$ 5500 \kms\ at high luminosity $\log L \sim 48$). In the optical plane of the MS defined by FWHM \hb\ vs \rfe\   Population A has been subdivided into 4 spectral types (STs) according to \feii\ prominence: A1, with \rfe$\lesssim 0.5$; A2, with $0.5 \lesssim$ \rfe$\lesssim 1.0$;   A3, with $1.0 \lesssim$ \rfe$\lesssim 1.5$;  A4, with $1.5 \lesssim$ \rfe$\lesssim 2.0$ \citep[][see also \citealt{shenho14} for an analogous approach]{sulenticetal02}.  The condition \rfe\ $\gtrsim 1.0$\ restricts the MS to the tip of high \rfe\ values, and encompasses  10\%\ of objects (referred to as extreme Population A).  At low-$z$\ they are mostly narrow-line Seyfert-1 (NLSy1s) driving the MS correlations \citep{borosongreen92,sulenticetal00a,duetal16a}. Sources with \rfe$\gtrsim 2$ do exist \citep{liparietal93,grahametal96} but they are exceedingly rare (less than 1\%) in optically-selected samples \citep[][hereafter \citetalias{marzianietal13a}]{marzianietal13a}. We therefore group all sources  with \rfe$\gtrsim 1.5$\ in  A4.  

\subsection{Multicomponent $\chi^{2}$-minimization}

Resolution and S/N of the available  spectra are adequate for a multicomponent nonlinear fitting analysis using the IRAF routine {\tt specfit} \citep{kriss94}, involving an accurate deconvolution of \hb, \oiiiopt, \feii, \heiiopt\  in the optical, and of \aliii, \ciii\, and \siiii\ in the UV. A  $\chi^2$\  minimization analysis is necessary in all cases, since the strongest lines are heavily blended together, and the blend involves also features extended over a broad wavelength range, due to \feii\ (mainly optical, and  UV to a lesser extent) and \feiii\ (UV only).

\subsection{H$\beta$\ line}
\label{hb}

The \hb\ Balmer emission line is a reliable estimator of the ``virial'' broadening in samples of moderate-to-high luminosity \citep{wangetal09,trakhtenbrotnetzer12,shenliu12}. Typically, the \hb\ line profiles are fairly symmetric, and are thought to be dominated by a virialized component \citep[][\citetalias{sulenticetal17}]{petersonwandel99,petersonetal04}. Several previous works noted that \hb\ shows a Lorentzian-like profile  in sources belonging to Population A \citep[e.g., ][this is also seen in \mgii, \citealt{marzianietal13,popovicetal19}]{veroncettyetal01,sulenticetal02,craccoetal16}.  However, the \hb\ profiles can be affected by slight asymmetries and small centroid shifts. In Population A they are mostly due to blueshifted excess, often modeled with a blueward asymmetric Gaussian component (BLUE),   strongly affecting the \civ\ line profiles, and related to outflows \citepalias[e.g.,][ and references therein, \citealt{negreteetal18}]{sulenticetal17}.  In \hb,  BLUE is  detected as a faint excess on the blue side of the symmetric profile  assumed as the virialized  component of \hb, almost only in extreme Population A (several examples are shown in the Figures of the atlas of Appendix \ref{app:fits}). Even when the BLUE component is detected, its influence on the \hb\ FWHM is modest, leading at most to an increase of the broadening  $\approx 10$\%\  over the FWHM of the symmetric broad profile \citep{negreteetal18}.

To extract a profile that excludes the blueshifted excess, we considered a model of the { broad} \hb\ line with the following components (based on the approach of  \citealt{negreteetal18}): 
\begin{itemize}
\item the \hbbc, modeled with an (almost) unshifted Lorentzian profile;
\item  a blueshifted excess (BLUE) modeled with  a blueshifted Gaussian with free skew parameter \citep{azzaliniregoli12}. The skewed Gaussian function has no more outliers than the normal distribution, and retains the shape of the normal distribution on the skewed side. It is consistent with the suppression of the receding side of an optically thin  flow obscured by  an optically thick structure (i.e., the accretion disk);
\item the \hbnc, modeled with a Gaussian,   unshifted with respect to rest frame;
\item  \feii\ emission, modeled with a scaled and broadened template \citep[e.g.,][]{borosongreen92}, as defined by \citet{negreteetal18};
\item \oiiiopt, modeled by a core-component (assumed Gaussian and symmetric)  and a semi-broad component (assumed Gaussian but with the possibility of being skewed). This approach has been followed in several previous work \citep[e.g.,][]{zhangetal11};
\item \heiiopt, broad and narrow component. \heiiopt\ is not always detected in the spectra, especially in the case of strong \feii\ emission, but the line was included in the fits. 
\end{itemize}

\subsection{The 1900 blend}

The range 1700 -- 2000  is dominated by the 1900\AA\ blend which includes \aliii, \siiii, \ciii, as well as \feii\ and  \feiii\ lines.   On the blue side of the blend Si{\sc ii}$\lambda$1816 and N{\sc iii}]$\lambda$1750 are also detected. The relative intensity of these lines (apart from N{\sc iii}]$\lambda$1750 that is not affecting the blend and for which further observations are needed) is known to be a function of the location along the quasar main sequence \citep[][]{bachevetal04}. The line profiles and relative intensities are systematically different not only between Population A and B, but also within Pop. A there is a systematic trend of increasing \aliii\ and decreasing \ciii\ prominence  with increasing \feii\ emission \citep{bachevetal04}. 

Our interpretation of the 1900 blend for Population A sources closely follows previous analyses \citep{baldwinetal96,willsetal99,baldwinetal04,richardsetal11,negreteetal12,martinez-aldamaetal18a}. The fits include the following components (described in detail by \citealt{martinez-aldamaetal18a}):
\begin{itemize}
\item  \aliii, \siiii, and \ciii, modeled with a  Lorentzian profile.  We assume that the shapes of the strongest lines are consistent with the ones considered for the \hb\  broad components   (Lorentzian for Pop. A), and that  FWHM \aliii\ = FWHM \siiii\ \citep{negreteetal13}.  The fitting routine   may introduce a systematic blueshift to minimize $\chi^{2}$, in the case the profile is significantly affected by an unresolved blue shifted component, as observed in the case of \mgii\ 
\citepalias[][]{marzianietal13a};
\item \feiii\ emission, very intense in extreme Population A spectra,   modeled with an empirical template \citep{vestergaardwilkes01}.  Recent photoionization calculations indicate a more  significant  contribution of \feiii\ emission in correspondence of \siiii\ \citep{templeetal20}. However, the new \feiii\ model spectrum is consistent with the empirical template of \citet{vestergaardwilkes01}. 


The \feiii\ template  is usually  included with the    peak shift of \ciii\ free to vary in the interval 1908 -- 1915 \AA\ (see \citealt{martinez-aldamaetal18a}). In the case the peak shift is around $1914 $ \AA, the \feiii\ component may be representing more the $\lambda$1914 line anomalously enhanced by \lya\ fluorescence than \ciii. Considering the severe blending of these two lines, and the   weakness of \ciii\ in Population A, the relative contribution of \ciii\ and \feiii\ $\lambda$1914\ cannot be  measured properly. However, if the peak wavelength of the blend around \ciii\ is close to 1914 \AA, the \feiii\ $\lambda$1914 line was included in the fit;

\item \siii, usually fainter than \aliii. This line is expected to be stronger in extreme Population A \citep{negreteetal12};  

\item  \feii\ emission, modeled with a scaled and broadened theoretical template    \citep{bruhweilerverner08,martinez-aldamaetal18a}. The \feiiuv\ emission is never very strong around 1900 \AA, and at any rate  gives rise to an almost flat pseudo-continuum that is not affecting the relative intensity ratios of the \aliii, \siiii, \ciii\ lines. A spiky feature around 1780 \AA\ is identified with UV \feii\ multiplet \#\ 191 (\feii$\lambda1785$). In several extreme cases, attempting to scale the \feii\ template to the \feii$\lambda1785$\ intensity required   large \feii\ emission \citep{martinez-aldamaetal18a}. In such cases the \feii$\lambda1785$\ feature  may have been  selectively enhanced by \lya\ fluorescence over the expectation of the \citet{bruhweilerverner08} template. Considering the difficult assessment of the \feiiuv\ emission, no  measurements are reported in the present paper.


\item  a blueshifted excess (BLUE) modeled with  a blueshifted skew Gaussian.  At high luminosity in the HE sample, there are 2 cases (\object{HE0359-3959} and \object{HE1347-2457}) where a strong blueshifted component is obviously affecting the profile of the 1900 blend. Other cases are also detected in the WISSH sample (see \S \ref{xA} for the interpretation of the 1900 blend profiles  involving a blue shifted excess).  For two objects, the BLUE emission is overwhelming and masking the emission of the individual \aliii, \siiii, \ciii\ broad components (Section \ref{xA}). Otherwise,  the appearance of the blend is not suggesting, even at the highest luminosity, the presence of an outflow component spectroscopically resolved (i.e., of significant blueshifted emission as detected in \civ). Small in \aliii\ blueshifts do occur, but with amplitude $\ll$ than their FWHM.  
\end{itemize}

\subsection{Full profile measurements}


We assume that the symmetric and unshifted \hbbc\  is the representative line components of the virialized part of the BLR. It is expedient to define a parameter $\xi$\ as follows: 

\begin{equation}
 \xi_\mathrm{line} = {{\rm FWHM}_\mathrm{vir}/{\rm FWHM}} \label{eq:xi},   
\end{equation}

where the FWHM$_\mathrm{vir}$\ is the FWHM of the ``virialized'' component, in the following assumed to be \hbbc, and the FWHM\ is the FWHM measured on the full profile (i.e., without correction for asymmetry and shifts)  of any   line. In the case of \hb,  FWHM \hb\ $\approx$ FWHM \hbbc, and $\xi_\mathrm{H\beta} \approx 1$ (Section \ref{imme}).  For the sake of this work,  \hb\ and \hbbc\ can be considered almost equivalent, so that we will rely on the \hbbc\ — \hb\ decomposition obtained with {\tt specfit} only in a few instances. The blue excess  is usually faint with respect to \hbbc\  and no empirical correction has been applied. 

 A goal of the present paper is to derive $\xi$\ for \aliii\ and \ciii.  Similarly as for \hb, the \aliii\ lines are fit by  symmetric functions. This approach has been  applied  in all cases and appears appropriate for the wide majority of spectra ($\approx 90$\%),  where there is no evidence of a strong BLUE in \aliii\ and the \aliii\ peak position is left free to vary to account for small shifts that might be due to a spectroscopically-unresolved outflowing component.  A few cases for which there is evidence of contamination by a strong blue shifted excess  are discussed in Sect. \ref{xA}. 
 



\subsection{Error estimates}

The data used in this paper come from an array of instruments yielding spectra with widely different  S/N. In addition, the comparison is between  two emission lines, one of which is relatively strong ($W$(\hb)$\sim 100$ \AA), and one faint ($W$(\aliii)$\lesssim 10$ \AA\ in most cases). To make things worse, at low $z$\ the \aliii\ line is also recorded on  lower S/N spectra. These and other systematic differences have to be quantitatively taken into account  in the error estimates. A quality parameter $\mathcal Q$\ has been defined for \aliii, \hb, and \ciii\ as the ratio between the line equivalent width and its FWHM multiplied by the S/N ratio measured on the continuum. The  $\mathcal Q$\ values can be computed using the parameters reported in Tables \ref{tab:sample} and \ref{longtable}. The systematic differences in the spectra covering \aliii\ and \hb\ are reflected in the distribution of  $\mathcal Q$: \hb\ and \aliii\ occupy two different domains (Figures in Appendix \ref{app:unc}).  The corresponding fractional  uncertainties in FWHM computed from dedicated Markov Chain Monte Carlo (MCMC) simulations or by defining a relation with the $\mathcal Q$\ parameter as detailed in Appendix \ref{app:unc}     are significantly different for the two  lines, being  just a few percent in the FWHM of the narrowest sources with strong and sharp \hb\ and at worst $\approx 10$\%, but in the range $\approx 10$\%  — 50 \%\ for \aliii.  


\section{Results}
\label{results}

\subsection{Immediate results}
\label{imme}
The {\tt specfit} analysis results are provided in form of an atlas (Appendix \ref{app:fits})\ for the FOS$^\star$, HE, ISAAC, and WISSH samples. The \aliii\ and \hb\ spectral range are shown after continuum subtraction, on a normalized flux scale (at 1700 and 5100 \AA).  The parameters measured with the {\tt specfit} analysis or on the full profiles for \hb, and \aliii\ are reported in Table \ref{longtable}. Table \ref{longtable} lists, in the following order: identification by IAU code name (Col. 1), rest-frame flux and equivalent width of the \hb\ line { (Cols. 2--3)}. {The following columns (Cols. 4--6) report the \hb\ profile parameters: FWHM \hb, FWHM \hbbc, and shift.
Here for shift $s$\ we intend  the  radial velocity of the line peak  with respect to the rest frame as defined from the redshift measured  in the \hb\ spectral range; parameter \rfe\ and spectral type (Cols. 7--8);   rest-frame flux, equivalent width, FWHM and shift of the \aliii\ line  (Cols. 9--12). The FWHM refers to the individual component of the doublet, whereas flux and equivalent width $W$\ are measured over the full doublet; flux of \siiii\ (Col. 13); \ciii\ flux and FWHM (Cols. 14--15).  The \feiii\ flux measurement (Col. 16) was obtained by integrating the template over the range 1800 $-$\ 2150 \AA. The upper limit of the wavelength range set at 2150 \AA\ allows the inclusion of a broad feature peaking at $\approx 2050 - 2080$ \AA\ and mostly ascribed to \feiii\ emission  \citep{martinez-aldamaetal18a}. Further information on the reported parameters are given in the Table footnotes.   Errors on line widths have been  computed from the numerical simulations described in Appendix \ref{app:unc} or from the data listed in Tables \ref{tab:sample} and \ref{longtable} that yield $\mathcal Q$. The same approach has been followed for errors on line intensities and line shifts. 

The values of the \hb\ FWHM for the WISSH quasars  are fully consistent
with those reported by \citet{vietrietal18}  for all but two targets,
namely \object{SDSS J152156.48+520238.5} and \object{SDSS J115747.99+272459.6},  for which a discrepancy   can
be explained in terms of a different fitting technique.    Intensity ratios computed  between lines in the UV and the optical should be viewed with extreme care. The observations are not synoptical and were not collected with the aim of photometric accuracy. 


\subsection{FWHM H$\beta$\ vs. FWHM \aliii}
\label{fwhm}

Fig. \ref{fig:hbal} shows the FWHM \aliii\ vs FWHM \hb\ full profile. The overall consistency in the FWHM of the two lines is rather obvious from the plot.  In the case of \aliii\ and \hb, the Pearson’s correlation coefficient is $\approx 0.785$\ ($P \sim 5 \cdot 10^{-8}$\ of a chance correlation). A best fit with the  ordinary least-squares (OLS) bisector yields 

\begin{equation}
\mathrm{FWHM\, {AlIII}}  \approx     (273 \pm  216) +     (0.933 \pm    0.059) \mathrm{FWHM \, H}\beta.
\end{equation}

The two lines are, on average, unbiased estimators of each other, with a 0 point offset that reflects the  tendency of the \aliii\ lines to be somewhat broader than \hb\ but is not statistically significant (the offset by 250 \kms\ is at less than 1 $\sigma $\ confidence level). An orthogonal LSQ fit yields  slope  and offset  consistent with the OLS. The normalized $\chi_\nu^2 \approx 1$ also indicates that the ratio between the FWHM of the two lines is 1 within the uncertainties.  The maximum FWHM $\approx 6000$ \kms\ is observed for the sources of the highest luminosity (Section \ref{lum}) and is below the luminosity-dependent FWHM limit of Population A.
 
 Fig. \ref{fig:hbal}  should be compared to Fig. 3 of \citet[][hereafter \citetalias{marzianietal19}]{marzianietal19}, where one can see that there is no obvious relation between the FWHM of \civ\ and the FWHM of \hb.  For the Pop. A sources   \civ\ is systematically  broader than \hb,\ apart from in  two  cases in the HE sample, and   FWHM(\civ) shows a broad range of values for similar FWHM \hb\ i.e., FWHM(\civ) is almost degenerate with respect to \hb.  The \civ\ line FWHM values are so much larger than the ones of \hb\  making it possible that the   \mbh\ derived from FWHM \civ\ might be higher by  even more than one order of magnitude than the one derived from the \hb\ FWHM, \textbf{as pointed out in several past works \citep{sulenticetal07,netzeretal07,marzianisulentic12,mejia-restrepoetal16}.} We remark again that the \aliii\ line may show a blueshifted  excess in 6 sources in our sample, with convincing evidence in only  two cases (Sect. \ref{xA}) but that the line profile is otherwise well represented by a symmetric Lorentzian. In the case of a blueshifted excess,  the good agreement between FWHM \aliii\ and FWHM \hb\ is in part due to the \siii\ emission that, if no blueshifted \aliii\ emission is allowed,  becomes  very strong in the fit of the 6 sources, and compensate for the blueshifted excess. \siii\ is expected to be enhanced in the physical conditions of extreme sources \citep[][Section \ref{xA}]{negreteetal12}. At the same time, including the \siii\ line in the fits allows for a standard procedure that does not require identification and  a screening for the sources with a strong blueshifted excess, which apparently follow the correlation between \aliii\ and \hb\ full profile in Fig. \ref{fig:hbalst}. 

\subsection{Dependence on spectral type and \rfe}
\label{sptype}

\subsubsection{FWHM}

In Fig. \ref{fig:hbal} the data points are color coded according to their original subsamples. Fig. \ref{fig:hbalst} shows the joint sample FWHM \aliii\ vs FWHM \hb\ full profile color-coded according to ST. There are  systematic differences between the various STs, in the sense that A1 sources have \aliii\ narrower than \hb\ (at the relatively high confidence level of 2$\sigma$), and \aliii\ and \hb\ FWHM are almost equal for ST A2. The  \aliii\ and \hb\ FWHM ratio is reversed, in the sense that \hb\ is narrower than \aliii, for STs A3 and A4 grouped together. The difference between STs is reinforced if only the BC of \hb\ is considered (Fig. \ref{fig:hbalstbc}), since the FWHM  \hbbc\ is slightly lower than the FWHM \hb\ of the full \hb\ broad profile, with $\xi_\mathrm{H\beta} \approx 0.97 \pm 0.05$ on average, but $\xi_\mathrm{H\beta} \approx 0.79$\ for the spectral types A3 and A4. 
If we define    $\xi_\mathrm{AlIII}$ = FWHM(\hbbc)/FWHM(\aliii), we have the following median values { ($\pm$ semi-interquartile range)}:

\begin{center}
\begin{tabular}{lc}\hline
ST & $ \mu_\frac{1}{2}(1/\xi_\mathrm{AlIII}) \pm$ SIQR \\
\hline
A3-A4 & 1.267 $\pm$ 0.223\\
A2 & 1.093 $\pm$ 0.176\\
A1 & 0.796 $\pm$ 0.158\\
\hline
\end{tabular}
\end{center}

A2 is the highest-prevalence ST in Population A, with $\xi \approx 1$. However, across Population A there is a significant trend that implies differences of $\pm 20 - 25$\%\ with respect to unity.     The A3-A4 result is after all not surprising, considering that quasars belonging to these spectral types with the strongest \feii\ emission are the sources most affected by the high-ionization outflows detected in the \civ\ line. The A1 result { i.e.,  \aliii\ lines narrower than \hb\ by $\approx 20$\%  }  comes more as a surprise, and it is intriguing that it is consistent that also for the B Population \aliii\ is narrower than \hb\ (del Olmo et al., in preparation; \citealt{marzianietal17}).  This result may hint at a small but systematic extra broadening not associated with virial motions in A2.



\begin{figure}[htp!]
\centering
\includegraphics[width=0.5\columnwidth]{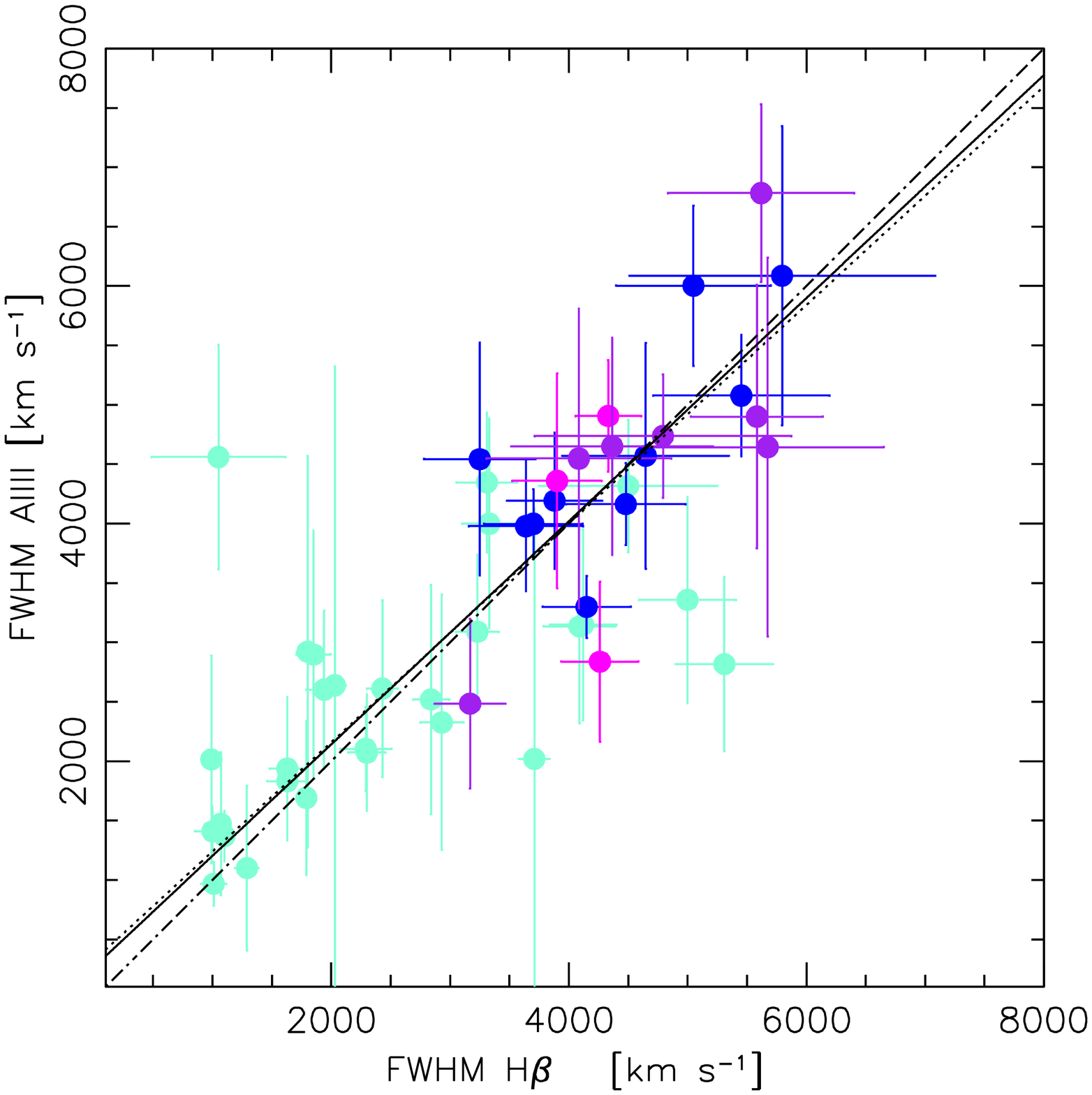}
\vspace{-1.5cm}
\includegraphics[width=0.5\columnwidth]{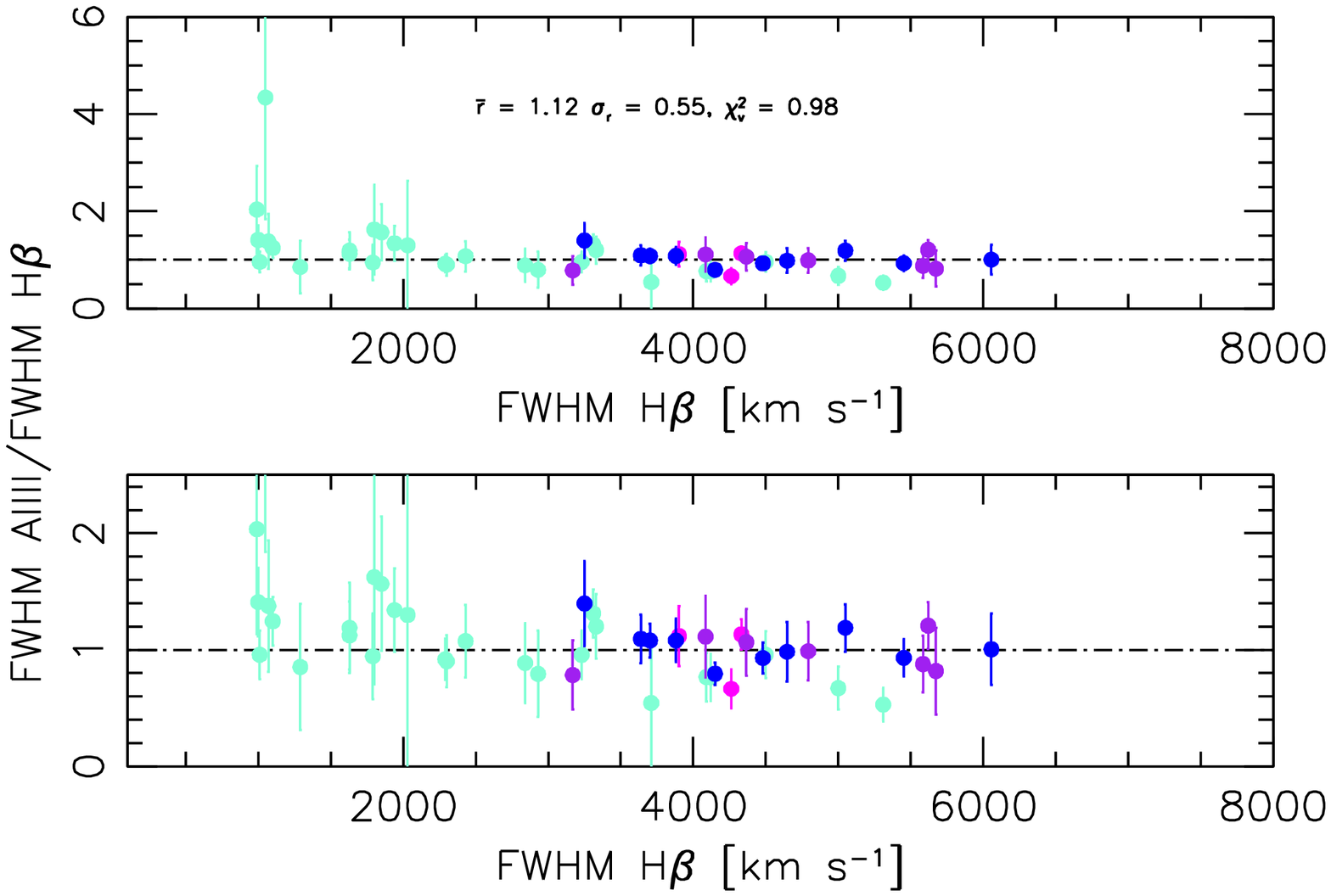}
\vspace{-.5cm}
\caption{Top panel: FWHM(\aliii) vs. FWHM(\hb) (full profiles) for the joint sample. Data points are color-coded according to sub-sample: FOS$^\star$ - aquamarine (\textcolor{aquamarine}{$\bullet$}), HE - blue    (\textcolor[rgb]{0,0,1}{$\bullet$}), ISAAC - magenta (\textcolor{magenta}{$\bullet$}), WISSH - purple (\textcolor{purple}{$\bullet$}). Dot-dashed line:  1:1 relation; filled line:   best fit obtained using the bisector technique; dotted line: best fit using the least orthogonal distance method \citep{pressetal92}. Middle panel:   ratio $r$ of FWHM \aliii\ over FWHM \hb. The average ratio  $\bar{r}$, the standard deviation  $\sigma_\mathrm{r}$  and the normalized $\chi^{2}_{\nu}$\  are reported.  Bottom: same as the middle panel,   with an  expanded scale along the $y$\ axis.   \label{fig:hbal} }
\end{figure}

\begin{figure}[htp!]
\centering
\includegraphics[width=0.5\columnwidth]{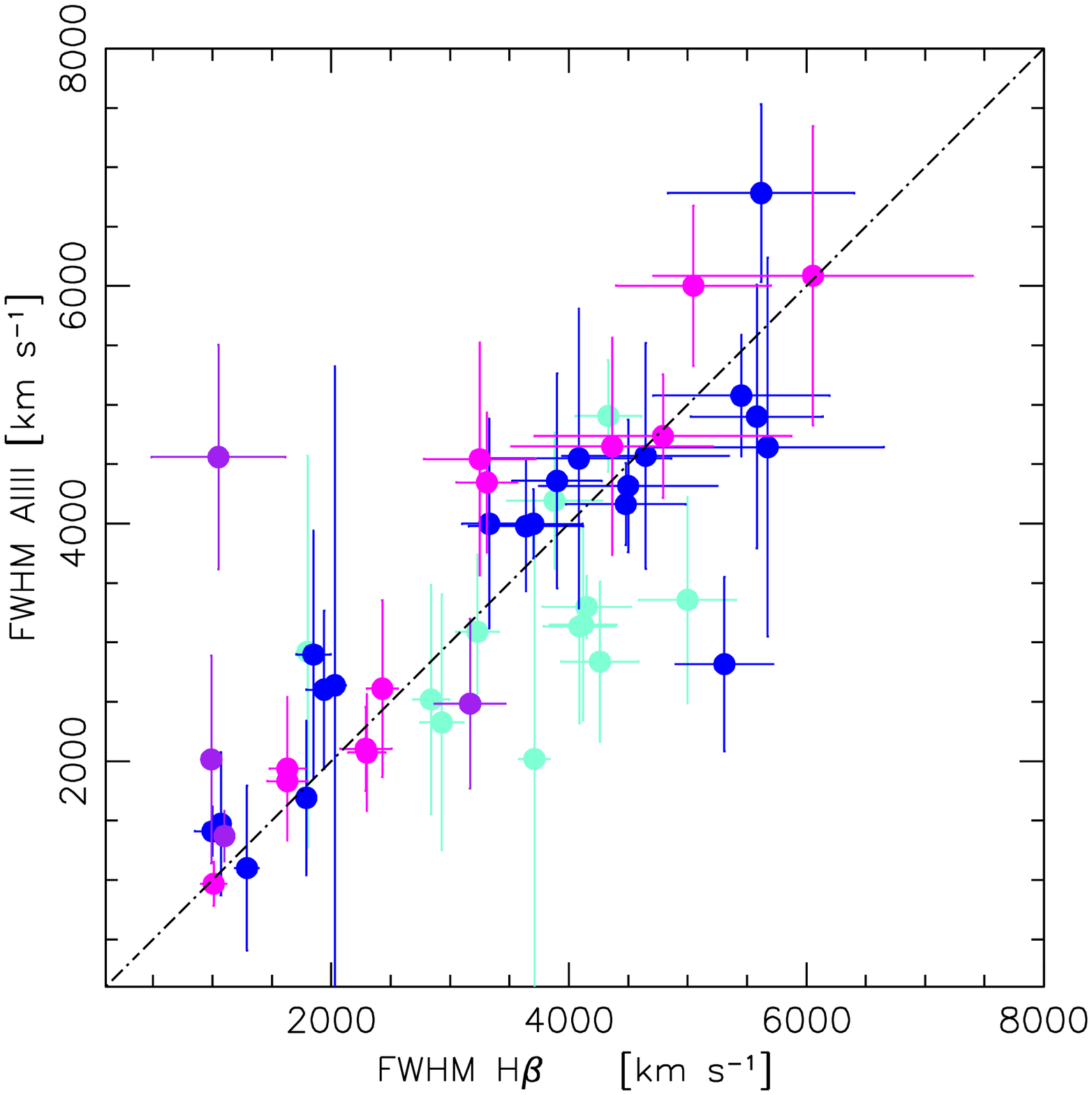}\\
\vspace{-.5cm}
\includegraphics[width=0.5\columnwidth]{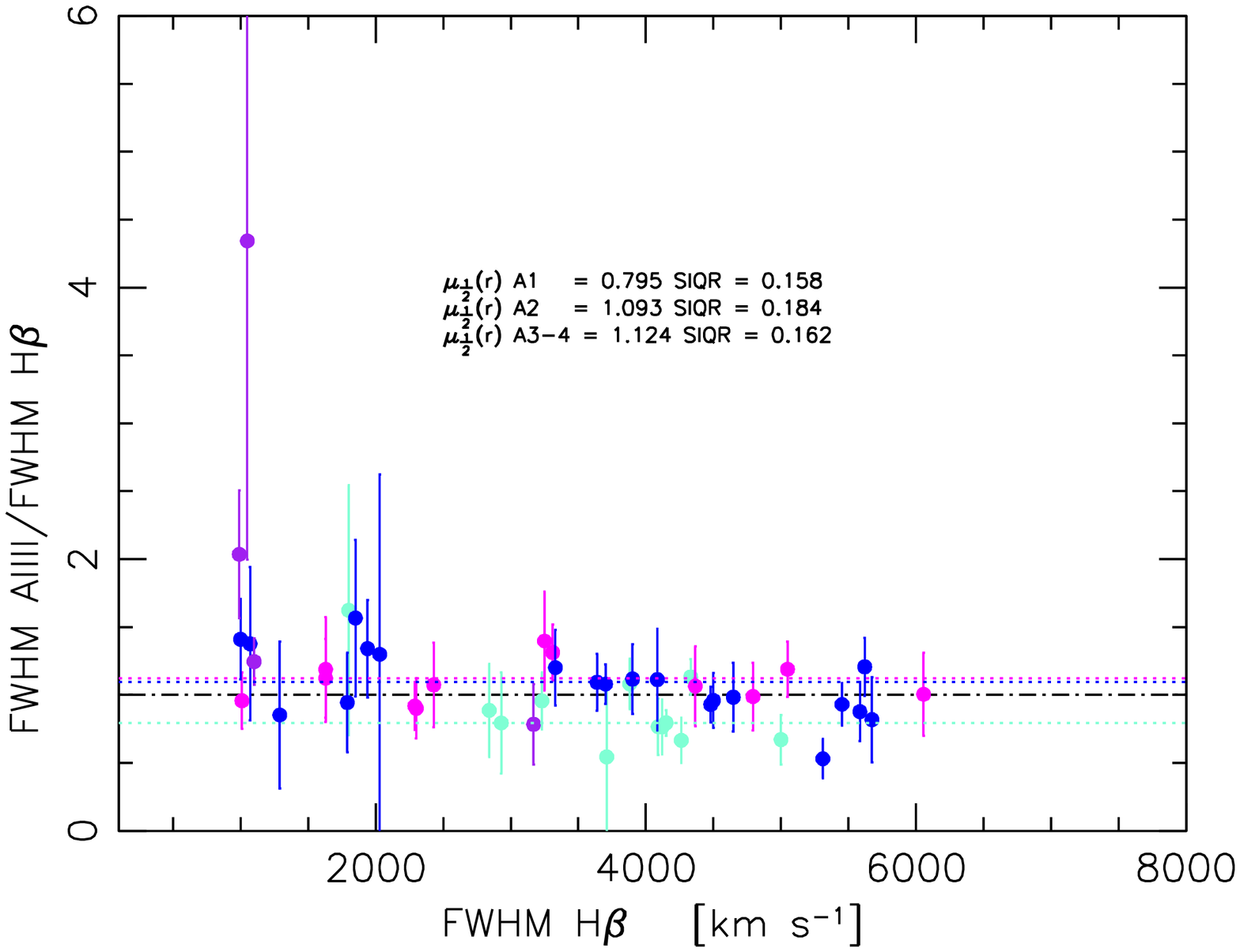}
\vspace{-1.5cm}
\caption{Top panel: FWHM(\aliii) vs. FWHM(\hb) (full profiles) for the joint  sample. Data points are color-coded according to spectral type: A1 - aquamarine (\textcolor{aquamarine}{$\bullet$}),  A2 - blue    (\textcolor[rgb]{0,0,1}{$\bullet$}), A3 - magenta (\textcolor{magenta}{$\bullet$}), A4  - purple (\textcolor{purple}{$\bullet$}).  Bottom: FWHM ratio between  \aliii\ and \hb.  The median ratio $\mu_\frac{1}{2}(r)$=FWHM \aliii/FWHM \hb,  and  SIQR   are reported  for spectral types A1 and A2, and the union of A3 and A4.  \label{fig:hbalst} }
\end{figure}

\begin{figure}[htp!]
\centering
\includegraphics[width=0.5\columnwidth]{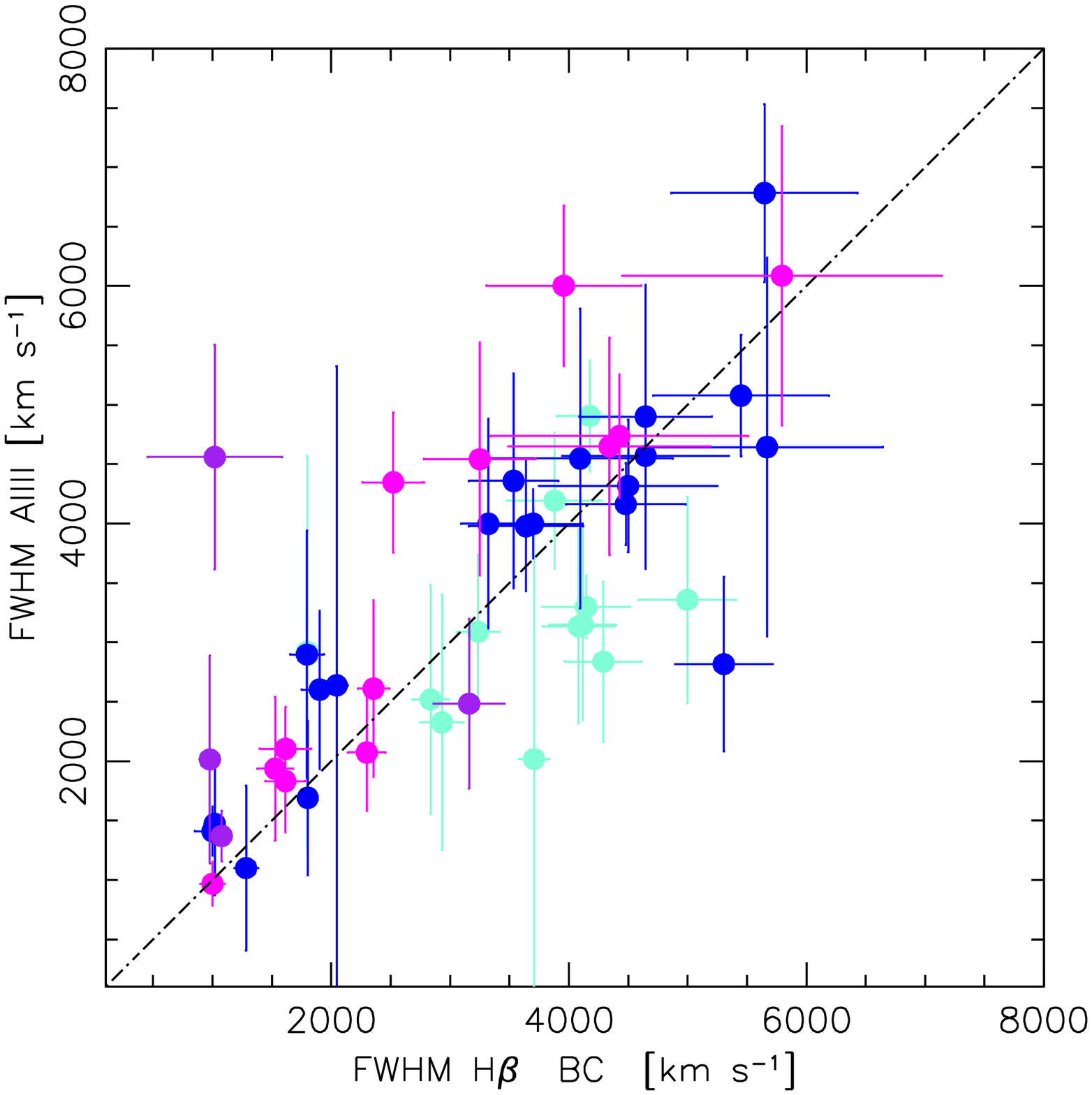}\\
\vspace{-.5cm}\hspace{0.1cm}
\includegraphics[width=0.5\columnwidth]{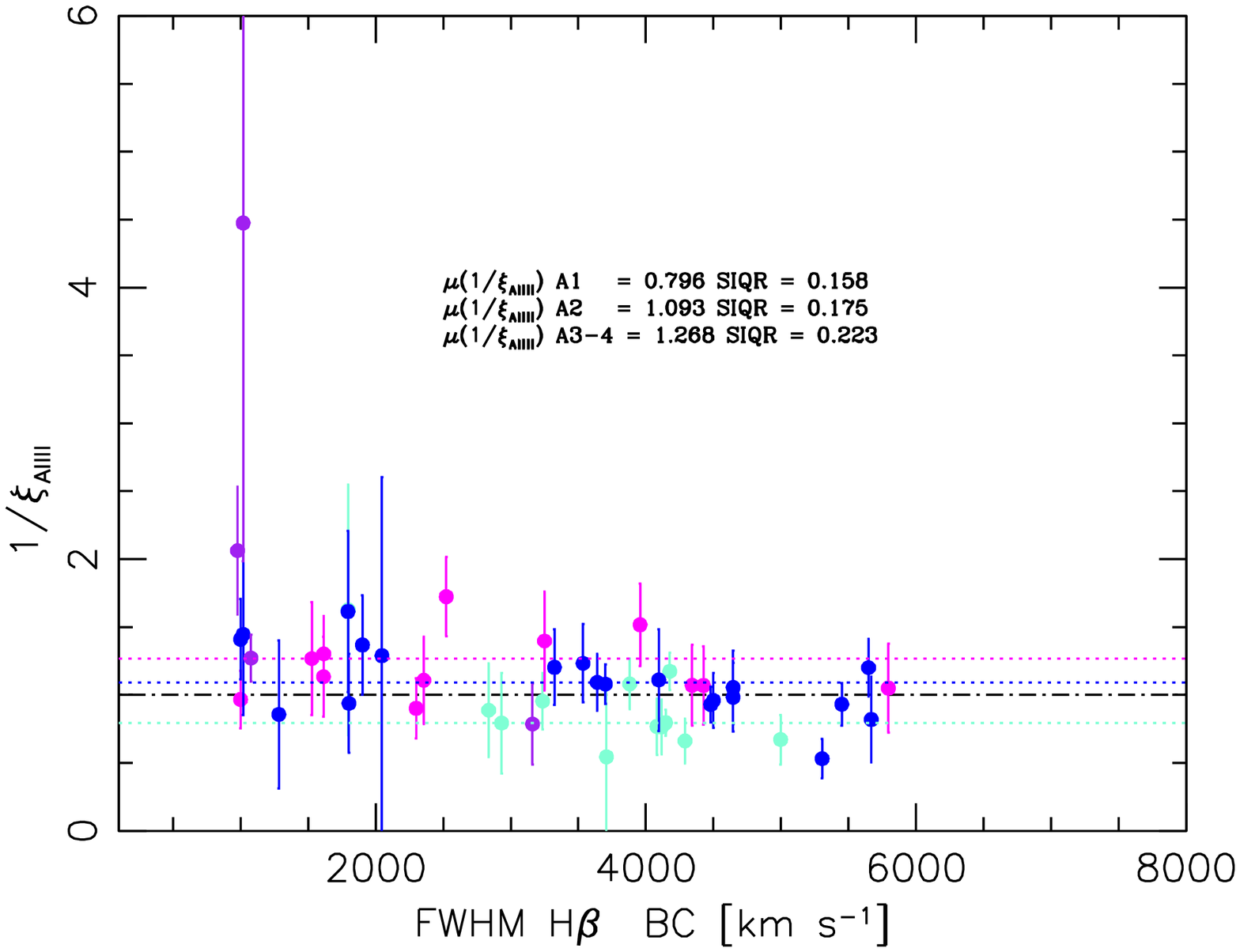}
\vspace{-0.5cm}
\caption{Top panel: FWHM(\aliii) vs. FWHM(\hbbc) for the joint sample. Data points are color-coded according to spectral type as in Fig. \ref{fig:hbalst}.  Bottom panel:  FWHM ratio between \aliii\ and \hbbc, $  1/\xi_\mathrm{AlIII}$.  The median ratio  $\mu_\frac{1}{2}(1/\xi_\mathrm{AlIII})$ and the SIQR     are reported  for spectral types A1 and A2, and for  the union of A3 and A4, as in Fig. \ref{fig:hbalst}.  \label{fig:hbalstbc} }
\end{figure}



\setcounter{table}{1}
\begin{sidewaystable*}[htp!]
\centering \tiny
\voffset=2cm
\hspace*{0cm}
\caption{Observed spectrophotometric quantities and main sequence classification\label{longtable}}	
{\fontsize{6.2}{6} \selectfont\tabcolsep=2pt
\begin{tabular}{lcccccccccccccccccc}	
		\hline\hline 
IAU code	& \multicolumn{5}{c}{\hb} &  \rfe\ & SpT &    \multicolumn{4}{c}{\aliii} & \siiii\ &  \multicolumn{2}{c}{\ciii} &  &   \feiii\  \\ \cline{2-6} \cline{9-12}\cline{14-15}
               &       F  & W      & FWHM & {FWHM BC}  &     Shift &              &      		&   F       &      W  & FWHM  & Shift &  F  & F   & FWHM         \\			
\multicolumn{1}{c}{(1)} &	(2)	&(3)	& (4)	&	(5)	& (6)	&	(7)	&	(8)	&	(9)	&	   (10) 	&	(11)  &    (12)	&    (13)	& (14) & (15)  & & (16)  	\\ \hline
\\
\hline
\multicolumn{15}{c}{FOS sample}\\
\hline
J00063+2012	&	544	$\pm$	14	&	99	&	1790	$\pm$	82	&     1802  &            	-130	$\pm$20	&	0.527	$\pm$	0.07	&	A2	&	92.5	$\pm$	36.2	&	1.7	&	1691	$\pm$	650	&	110	$\pm$	200	&	316.8	$\pm$	50.6  &	345.0	$\pm$	29.0  &	800 	$\pm$	77	&		&1579.8	$\pm$	328.9 \\
J00392-5117	&	103	$\pm$	6	&	50	&	1290	$\pm$	100	&     1283  &            	-10	$\pm$	20	&	0.842	$\pm$	0.09	&	A2	&	8.9		$\pm$	3.0		&	4.6	&	1100	$\pm$	693	&	-90	$\pm$	230	&	21.3	$\pm$	5.1   & 50.9	$\pm$	7.2    &  1634	$\pm$	268	 &		 &90.1	$\pm$	12.3  \\
J00535+1241	&	294	$\pm$	12	&	45	&	1100	$\pm$	64	&     1077  &            	-30	$\pm$	20	&	1.619	$\pm$	0.06	&	A4	&	184.4	$\pm$	17.2	&	7.2	&	1370	$\pm$	212	&	-300 $\pm$	160	&	321.5	$\pm$	32.5  &	277.7	$\pm$	102.3 &	1225	$\pm$	324	&		&857.8	$\pm$	148.8 \\
J00573-2222	&	741	$\pm$	23	&	47	&	1070	$\pm$	30	&     1019  &            	0	$\pm$	10	&	0.737	$\pm$	0.06	&	A2	&	50.2	$\pm$	9.0		&	1.6	&	1473	$\pm$	602	&	-270 $\pm$	240	&	125.8	$\pm$	21.3  &	289.7	$\pm$	46.3  &	1672	$\pm$	224	&		&184.9	$\pm$	27.3 \\
J01342-4258	&	35	$\pm$	5	&	25	&	1050	$\pm$	565	&     1019  &            	40	$\pm$	130	&	2.054	$\pm$	0.25	&	A4	&	71.7	$\pm$	16.6	&	5.2	&	4560	$\pm$	945	&	-520 $\pm$	570	&	48.6	$\pm$	22.8  &	11.7	$\pm$	3.0   &	5456	$\pm$	959	&		&250.6	$\pm$	35.9 \\
J02019-1132	&	122	$\pm$	14	&	66	&	4500	$\pm$	751	&     4500  &            	30	$\pm$	140	&	0.947	$\pm$	0.15	&	A2	&	41.5	$\pm$	5.5		&	2.4	&	4318	$\pm$	558	&	250	$\pm$	450	&	110.4	$\pm$	40.5  &	173.3	$\pm$	28.3  &	4318	$\pm$	965	&		&25.3	$\pm$	4.6 \\
J06300+6905	&	167	$\pm$	6	&	140	&	5000	$\pm$	410	&     4998  &            	90	$\pm$	80	&	0.281	$\pm$	0.11	&	A1	&	137.1	$\pm$	32.4	&	2.9	&	3358	$\pm$	869	&	-480 $\pm$	620	&	261.6	$\pm$	126.6 & 1119.4	$\pm$ 182.0  &  2962	$\pm$	504	&	   & 225.0	$\pm$	37.5 \\
J07086-4933	&	74	$\pm$	5	&	33	&	990		$\pm$	86	& 980   &    40	$\pm$	30	&	1.717	$\pm$	0.11	&	A4	&	48.9	$\pm$	17.3	&	2.2	&	2015	$\pm$	874	&	-140 $\pm$	620	&	120.0	$\pm$	43.6  &	39.4	$\pm$	10.4  &	2188	$\pm$	432	&		&429.1	$\pm$	50.6 \\
J08535+4349	&	73	$\pm$	6	&	125	&	4090	$\pm$	304	&     4080  &            	230	$\pm$	80	&	0.277	$\pm$	0.13	&	A1	&	18.3	$\pm$	3.6		&	3.5	&	3135	$\pm$	816	&	500	$\pm$	260	&	50.4	$\pm$	18.5  &	59.2	$\pm$	13.7  &	1889	$\pm$	342	&		&63.4	$\pm$	12.7 \\
J09199+5106	&	162	$\pm$	9	&	106	&	5310	$\pm$	411	&     5310  &            	270	$\pm$	80	&	0.878	$\pm$	0.08	&	A2	&	23.6	$\pm$	6.2		&	2.2	&	2817	$\pm$	733	&	200	$\pm$	630	&	55.3	$\pm$	12.9  &	103.3	$\pm$	26.7  &	3626	$\pm$	507	&		&135.4	$\pm$	39.1 \\
J09568+4115	&	393	$\pm$	6	&	154	&	3710	$\pm$	129	&     3710  &             	180	$\pm$	30	&	0.185	$\pm$	0.08	&	A1	&	30.6	$\pm$	10.7	&	1.9	&	2019	$\pm$	2111&	160	$\pm$	430	&	50.1	$\pm$	19.8  &	188.8	$\pm$	31.2  &	1922	$\pm$	354	&		&157.2	$\pm$	23.8 \\
J10040+2855	&	135	$\pm$	7	&	63	&	1940	$\pm$	151	&     1900  &             	-40	$\pm$	40	&	0.953	$\pm$	0.09	&	A2	&	84.3	$\pm$	26.9	&	5.7	&	2602	$\pm$	666	&	-70	$\pm$	430	&	141.7	$\pm$	32.2  &	140.7	$\pm$	37.2  &	1943	$\pm$	361	&		&101.0	$\pm$	16.6 \\
J10043+0513	&	136	$\pm$	7	&	90	&	1850	$\pm$	143	&     1800  &             	-110$\pm$	40	&	0.694	$\pm$	0.08	&	A2	&	25.6	$\pm$	9.6		&	6.2	&	2898	$\pm$	1041&	-60	$\pm$	480	&	72.7	$\pm$	19.3  &	73.7	$\pm$	21.6  &	1828	$\pm$	344	&		&117.0	$\pm$	20.5 \\
J11185+4025	&	39	$\pm$	1	&	83	&	2030	$\pm$	95	&     2050  &             	50	$\pm$	20	&	0.588	$\pm$	0.07	&	A2	&	36.7	$\pm$	12.0	&	3.5	&	2638	$\pm$	2688&	230	$\pm$	390	&	55.7	$\pm$	21.8  &	95.5	$\pm$	22.4  &	1963	$\pm$	464	&		&190.9	$\pm$	24.6 \\
J11191+2119	&	406	$\pm$	17	&	147	&	3230	$\pm$	179	&     3230  &             	40	$\pm$	20	&	0.439	$\pm$	0.07	&	A1	&	180.9	$\pm$	50.0	&	5.3	&	3089	$\pm$	651	&	120	$\pm$	390	&	413.7	$\pm$	73.8  &	510.3	$\pm$	92.2  &	2450	$\pm$	437	&		&883.6	$\pm$	138.3 \\
J12142+1403	&	644	$\pm$	25	&	111	&	1800	$\pm$	92	&     1800  &             	80	$\pm$	20	&	0.424	$\pm$	0.06	&	A1	&	80.9	$\pm$	34.7	&	2.9	&	2922	$\pm$	1649&	-160 $\pm$	530	&	117.4	$\pm$	39.1  &	349.3	$\pm$	72.9  &	1609	$\pm$	291	&		&420.3	$\pm$	77.8 \\
J12217+7518	&	224	$\pm$	11	&	135	&	4120	$\pm$	280	&     4120  &             	300	$\pm$	30	&	0.255	$\pm$	0.07	&	A1	&	69.1	$\pm$	22.0	&	3.2	&	3149	$\pm$	804	&	-310 $\pm$	430	&	217.9	$\pm$	45.6  &	226.4	$\pm$	58.5  &	2246	$\pm$	416	&		&178.7	$\pm$	29.4 \\
J13012+5902	&	37	$\pm$	2	&	57	&	3310	$\pm$	257	&     2520  &             	130	$\pm$	60	&	1.365	$\pm$	0.09	&	A3	&	136.0	$\pm$	25.0	&	8.8	&	4347	$\pm$	590	&	-160 $\pm$	630	&	141.8	$\pm$	42.1  &	74.0	$\pm$	30.9  &	5215	$\pm$	519	&		&487.2	$\pm$	95.7 \\
J13238+6541	&	69	$\pm$	4	&	94	&	2930	$\pm$	182	&     2930  &             	-80	$\pm$	80	&	0.497	$\pm$	0.07	&	A1	&	12.7	$\pm$	6.9		&	1.4	&	2326	$\pm$	1075&	-320 $\pm$	770	&	57.3	$\pm$	19.0  &	160.5	$\pm$	30.0  &	2076	$\pm$	399	&		&42.2	$\pm$	6.9 \\
J14052+2555	&	73	$\pm$	4	&	91	&	2300	$\pm$	158	&     2300  &             	-90	$\pm$	30	&	1.15	$\pm$	0.08	&	A3	&	93.3	$\pm$	28.3	&	4.8	&	2073	$\pm$	491	&	480	$\pm$	410	&	184.2	$\pm$	38.2  &	42.7	$\pm$	14.7  &	801	    $\pm$	155	&		&559.7	$\pm$	90.4 \\
J14063+2223	&	70	$\pm$	6	&	59	&	1630	$\pm$	151	&     1530  &             	-10	$\pm$	30	&	1.1		$\pm$	0.11	&	A3	&	33.1	$\pm$	5.1		&	6.8	&	1937	$\pm$	604	&	-180 $\pm$	410	&	62.2	$\pm$	12.1  &	31.2	$\pm$	14.6  &	1703	$\pm$	496	&		&121.8	$\pm$	26.8 \\
J14170+4456	&	51	$\pm$	6	&	63	&	2290	$\pm$	217	&     1620  &             	0	$\pm$	70	&	1.182	$\pm$	0.13	&	A3	&	57.6	$\pm$	4.1		&	6.4	&	2104	$\pm$	353	&	-80	$\pm$	250	&	119.4	$\pm$	20.4  &	121.7	$\pm$	22.5  &	2010	$\pm$	234	&		&161.2	$\pm$	22.0 \\
J14297+4747	&	48	$\pm$	2	&	151	&	2840	$\pm$	153	&     2840  &             	300	$\pm$	30	&	0.392	$\pm$	0.05	&	A1	&	14.7	$\pm$	5.7		&	2.1	&	2519	$\pm$	965	&	280	$\pm$	490	&	29.5	$\pm$	8.3   & 92.4	$\pm$	19.5   &  1795	$\pm$	325	&	 	&79.2	$\pm$	14.0 \\
J14421+3526	&	114	$\pm$	7	&	55	&	1630	$\pm$	169	&     1610  &             	-140$\pm$	20	&	1.357	$\pm$	0.09	&	A3	&	105.8	$\pm$	32.1	&	4.0	&	1832	$\pm$	433	&	-260 $\pm$	410	&	229.5	$\pm$	46.9  &	448.2	$\pm$	101.3 &	2007	$\pm$	366	&		&559.7	$\pm$	90.4 \\
J14467+4035	&	77	$\pm$	3	&	74	&	2430	$\pm$	131	&     2360  &             	30	$\pm$	30	&	1.327	$\pm$	0.06	&	A3	&	62.4	$\pm$	6.6		&	3.6	&	2611	$\pm$	745	&	120	$\pm$	360	&	142.6	$\pm$	33.3  &	106.5	$\pm$	27.7  &	2059	$\pm$	435	&		&373.6	$\pm$	51.4 \\
J15591+3501	&	121	$\pm$	7	&	44	&	1010	$\pm$	105	&     1000  &            	100	$\pm$	20	&	1.424	$\pm$	0.10	&	A3	&	21.1	$\pm$	2.7		&	3.3	&	967		$\pm$	185	&	180	$\pm$	340	&	64.3	$\pm$	8.5   & 152.4	$\pm$	20.2   &  1534	$\pm$	239	&	     &150.4	$\pm$	17.9 \\
J21148+0607	&	86	$\pm$	4	&	119	&	3330	$\pm$	228	&     3320  &            	0	$\pm$	30	&	0.781	$\pm$	0.08	&	A2	&	119.0	$\pm$	34.2	&	9.3	&	3999	$\pm$	883	&	10	$\pm$	400	&	295.2	$\pm$	55.0  &	297.1	$\pm$	54.4  &  2995	$\pm$	535	&		&131.9	$\pm$	20.9 \\
J22426+2943	&	29	$\pm$	2	&	47	&	1000	$\pm$	149	&     1000  &            	-20	$\pm$	40	&	0.821	$\pm$	0.15	&	A2	&	18.2	$\pm$	2.5	    &	1.7	&	1410	$\pm$	207	&	-290 $\pm$	370	&	52.3	$\pm$	4.8   & 108.1	$\pm$	28.9   &  1507	$\pm$	344	&		 &75.2	$\pm$	10.3 \\
\hline                                                                                                                                                        
\multicolumn{15}{c}{HE sample}\\          
\hline                                                 
{ J00456--2243}	&	229	$\pm$	9	&	68	&	4150	$\pm$	370	&	4146   &          190	$\pm$	60	&	0.37	$\pm$	0.10	&	A1	&	54.0	$\pm$	4.7		&	4.1	&	3299	$\pm$	261	&	10	$\pm$	190	&	93.0	$\pm$	16.0	&	230.0	$\pm$	63.7	&	3295	$\pm$	474	&		&	82.0	$\pm$	15.1	\\
{ J01242--3744}	&	111	$\pm$	10	&	48	&	3250	$\pm$	469	&	3250   &          0	$\pm$	100	&	1.155	$\pm$	0.14	&	A3	&	102.0	$\pm$	28.8	&	5.3	&	4543	$\pm$	981	&	10	$\pm$	390	&	94.0	$\pm$	20.5	&	 0.0	 	 			&	 	\ldots		 	&		&	146.0	$\pm$	23.0	\\
{ J02509--3616}	&	39	$\pm$	3	&	44	&	4480	$\pm$	504	&	4480   &          -20	$\pm$	120	&	0.531	$\pm$	0.13	&	A2	&	69.0	$\pm$	4.9		&	3.1	&	4164	$\pm$	345	&	20	$\pm$	130	&	122.0	$\pm$	13.8	&	18.0	$\pm$	2.3		&	3990	$\pm$	537	&		&				0.0	 	 	\\
{ J04012-3951}	&	88	$\pm$	9	&	50	&	5049	$\pm$	652	&	3957   &          -20	$\pm$	80	&	1.103	$\pm$	0.14	&	A3	&	70.0	$\pm$	8.6		&	6.2	&	6003	$\pm$	673	&	-150	$\pm$380&	39.0	$\pm$	9.9		&	17.0	$\pm$	8.8		&	2052	$\pm$	356	&		&	122.0	$\pm$	16.9	\\
{ J05092--3232}	&	149	$\pm$	8	&	67	&	3880	$\pm$	403	&	3880   &          -130$\pm$	80	&	0.311	$\pm$	0.10	&	A1	&	48.0	$\pm$	6.8		&	4.3	&	4193	$\pm$	574	&	-50	$\pm$	450	&	96.0	$\pm$	32.3	&	56.0	$\pm$	15.4	&	2500	$\pm$	467	&		&	34.0	$\pm$	6.2		\\
{ J05141-3326}	&	228	$\pm$	15	&	82	&	3702	$\pm$	415	&	3700   &          0	$\pm$	80	&	0.656	$\pm$	0.07	&	A2	&	57.0	$\pm$	5.6		&	7.5	&	3998	$\pm$	291	&	20	$\pm$	550	&	129.0	$\pm$	39.0	&	67.0	$\pm$	17.2	&	2255	$\pm$	435	&		&	81.0	$\pm$	15.4	\\
{ J11065--1821}	&	366	$\pm$	35	&	121	&	4647	$\pm$	702	&	4646   &          -50	$\pm$	110	&	0.557	$\pm$	0.11	&	A2	&	108.0	$\pm$	29.6	&	4.9	&	4570	$\pm$	951	&	0	$\pm$	390	&	205.0	$\pm$	37.1	&	113.0	$\pm$	30.8	&	3138	$\pm$	576	&		&	95.0	$\pm$	14.8	\\
{ J13506--2512}	&	162	$\pm$	29	&	37	&	6087	$\pm$	1346&	5795   &          0	$\pm$	180	&	1.266	$\pm$	0.23	&	A3	&	404.0	$\pm$	101.9	&	9.5	&	6087	$\pm$	1259&	-1480	$\pm$620&	269.0	$\pm$	52.5	&	17.0	$\pm$	9.6		&	2190	$\pm$	652	&		&	543.0	$\pm$	67.0	\\
{ J21508--3158}	&	120	$\pm$	9	&	66	&	5452	$\pm$	741	&	5450 &     0	$\pm$	130	&	0.816	$\pm$	0.10	&	A2	&	104.0	$\pm$	13.1	&	7.0	&	5078	$\pm$	513	&	-360	$\pm$190&	139.0	$\pm$	43.3	&	66.0	$\pm$	19.0	&	4095	$\pm$	696	&		&	75.0	$\pm$	10.2	\\
{ J23555--3953 }	&	318	$\pm$	21	&	46	&	3639	$\pm$	481	&	3640 &     0	$\pm$	100	&	0.549	$\pm$	0.08	&	A2	&	99.0	$\pm$	8.9		&	3.2	&	3979	$\pm$	547	&	0	$\pm$	270	&	150.0	$\pm$	27.0	&	0.0	 					&    	\ldots	  		&		&   370.0	$\pm$	70.2	\\
\hline                                                                           
\multicolumn{15}{c}{ISAAC sample}\\                                                                                                                                     	
\hline                                                                                                                                                           
J00570+1437	&	219	$\pm$	15	&	86.3	&	3901	$\pm$	377	&	3535 &                      -20	$\pm$	70	&	0.838	$\pm$	0.10	&	A2	&	120.0	$\pm$	32.7	&	10.2	&	4360	$\pm$	904	&	-510	$\pm$	380	&	99.0	$\pm$	21.2	&	56.0	$\pm$	16.7	&	3370	$\pm$	637	&		&	57.9	$\pm$	9.0	\\
J13202+1420	&	122	$\pm$	7	&	97.0	&	4262	$\pm$	325	&	4290 &                      -150 $\pm$	80	&	0.282	$\pm$	0.05	&	A1	&	45.0	$\pm$	5.3	&	6.5	&	2838	$\pm$	674	&	-710	$\pm$	280	&	78.0	$\pm$	23.9	&	96.0	$\pm$	25.5	&	3282	$\pm$	523	&		&	48.0	$\pm$	7.8	\\
J16149+1448	&	231	$\pm$	21	&	91.5	&	4332	$\pm$	277	&	4177 &                      -80	$\pm$	40	&	0.423	$\pm$	0.10	&	A1	&	51.0	$\pm$	4.5	&	4.1	&	4906	$\pm$	470	&	30	$\pm$	310	&	169.0	$\pm$	21.9	&	128.0	$\pm$	32.6	&	3499	$\pm$	712	&		&	0.3	$\pm$	0.04	\\
\hline
\multicolumn{15}{c}{WISSH sample}\\                                                                                                                             
\hline                                                                                                                                                          
J0801+5210	&	294	$\pm$	36	&	71	&	5620	$\pm$	784	&	5648 &    -128	$\pm$	140	&	0.55	$\pm$	0.15	&	A2	&	269.8	$\pm$	22.6	&	10.7	&	6783	$\pm$	747 &	-270	$\pm$	400	&	285.3	$\pm$	33.8	&	171.6	$\pm$	47.0	&	4576	$\pm$	934	&		&	49.3	$\pm$	9.7 \\
J1157+2724	&	87	$\pm$	7	&	37	&	3169	$\pm$	298	&	3161 &    -128	$\pm$	100	&	1.68	$\pm$	0.17	&	A4	&	26.4	$\pm$	3.1		&	6.7	    &	2484	$\pm$	715 &	-560	$\pm$	590	&	95.2	$\pm$	11.4	&	66.6	$\pm$	15.9	&	2913	$\pm$	512	&		&	1.6		$\pm$	0.2 \\
J1201+0116	&	211	$\pm$	23	&	59	&	4085	$\pm$	779	&	4095 &    -134	$\pm$	140	&	0.60	$\pm$	0.15	&	A2	&	126.1	$\pm$	16.6	&	9.0	    &	4548	$\pm$	814 &	-710	$\pm$	620	&	141.0	$\pm$	21.8	&	94.4	$\pm$	23.8	&	3100	$\pm$	770	&		&	2.4		$\pm$	0.4 \\
J1236+6554	&	141	$\pm$	15	&	60	&	5674	$\pm$	976	&	5669 &     225    $\pm$	130	&	0.52	$\pm$	0.12	&	A1	&	106.4	$\pm$	35.2	&	5.3	    &	4644	$\pm$	1262&	-520	$\pm$	440	&	163.5	$\pm$	39.5	&	3.5		$\pm$	2.7		&	4317	$\pm$	998	&		&	          0.0	 	\\
J1421+4633	&	214	$\pm$	27	&	66	&	5584	$\pm$	556	&	4646 &         -21	$\pm$	140	&	0.99	$\pm$	0.20	&	A2	&	174.5	$\pm$	63.2	&	10.0	&	4901	$\pm$	1594&	-1030	$\pm$	470	&	151.6	$\pm$	44.5	&	0.0	 	 		 		&	 	 	 		 	&		&	251.0	$\pm$	43.3\\
J1521+5202	&	188	$\pm$	23	&	21	&	4366	$\pm$	851	&	4343 &      81    $\pm$	150	&	1.15	$\pm$	0.18	&	A3	&	212.6	$\pm$	27.9	&	4.1	    &	4650	$\pm$	914 &	-510	$\pm$	760	&	271.8	$\pm$	26.0	&	1.8		$\pm$	0.5		&	1248	$\pm$	385	&		&	251.7	$\pm$	47.8\\
J2123$-$0050	&	296	$\pm$	24	&	51	&	4793	$\pm$	1082&	4426 &     -61	$\pm$	80	&	1.12	$\pm$	0.11	&	A3	&	179.3	$\pm$	20.6	&	6.6	    &	4737	$\pm$	518 &	-610	$\pm$	340	&	218.6	$\pm$	35.3	&	226.1	$\pm$	53.0	&	4519	$\pm$	703	&		&	23.7	$\pm$	3.8 \\
\hline
 \end{tabular}}
\hspace{3cm}\tablecomments{\scriptsize (1) JCODE identification  of the object; (2)   total flux of H$\beta$ (sum of all broad line components);  (3)  equivalent width of the full \hb\ profile, in \AA; (4) FWHM H$\beta$\ measured on the full profile, in \kms;  { (5) FWHM H$\beta$\ measured on the broad component (\hbbc)}, in \kms;  (6) peak shift of the \hb\ broad profile;  (7) \rfe\ parameter; (8) spectral type SpT following \citet{sulenticetal02}; (9) total flux of the \aliii\ doublet;  (10) \aliii\ equivalent width in \AA; (11) FWHM of the individual line components of the \aliii\ doublet; (12) shift of \aliii, in units of \kms; (13) flux of \siiii;  (14) flux of \ciii;  (15) \ciii\ FWHM; (16)    flux of \feiii\ computed over the range between 1800 -- 2150 \AA. All measurements are in the quasar rest frame and line fluxes are all in units of 10$^{-15}$ erg s$^{-1} $\ cm$^{-2}$. } 
\end{sidewaystable*}
\subsubsection{Shifts}

 In addition to shifts with respect to the rest frame, we consider also the shift $s^\star$\ defined as  the radial velocity difference between the peak position of the Lorentzian function describing the individual components of the \aliii\ doublet with respect to the quasars rest frame  and the peak position of \hb\ and  (reported in Table \ref{longtable}) i.e., $s^\star(\mathrm{AlIII}) = s(\mathrm{AlIII}) - s(\mathrm H\beta)  $. We adopt this definition  because   the spectral resolution and the intrinsic line width make it   difficult  to resolve the outflow in the  \aliii\ profile. In the  \civ\ and \hb\ profiles, and even in the ones of \mgii, it is possible to isolate a blue-shifted excess that contributes most of the \civ\ flux in several cases, superimposed to a symmetric profile originating in the low ionization part of the BLR. In the case of \aliii, this approach is more difficult, in part because of the low S/N (a blueshifted contribution as faint as the one of \hb\ would be undetectable for \aliii, especially in the FOS spectra), in part because of the intrinsic rarity of sources with a strong blue-shifted excess.  In addition,  the \hbnc\ is often overwhelmed by the \hbbc, making it difficult to accurately measure its width and shift.  The shifts of \hb\ broad profile with respect to the rest frame set by the \hbnc\ are usually modest, and the median is consistent with 0, $\mu(s(\mathrm H\beta))\approx 0 \pm 65$ \kms. The shift estimators $s^\star$\ and $s$\ therefore yield results that are in close agreement.

Fig. \ref{fig:trenda}  shows   that the $s^\star$\ amplitude  is relatively modest for spectral type A3. ST A4 shows a larger shift, associated with an increase in the FWHM \aliii\ with respect to \hb, although with a large scatter. This is due to  an outstanding case,  \object{HE0132-4313}, a NLSy1 with FWHM \aliii\ / FWHM \hb \ $\approx 5$, a behaviour frequent for the \civ\ in strong \feii\ emitters but apparently rarer for \aliii. 

Taken together, the FWHM and shift data suggest that major discrepancies are more likely for the relatively rare, stronger \feii\ emitters, ST A3 and A4. The left panel of Fig. \ref{fig:trenda} shows that only A4 has a median shift that is significantly different from 0.  A4 sources  are relatively rare sources (4 in our sample, and $\approx$ 3 \% in \citetalias{marzianietal13a}). In addition to the shift amplitude in \kms, the line shift normalized by line width may be a better description of the ``dynamical relevance'' of the outflowing component \citep{marzianietal13}. The    parameters  
$\delta(\frac{i}{4}) =   {c(\frac{i}{4})}/{\mathrm{FW}\frac{i}{4}\mathrm{M}}, i=1,2,3, $
yield the centroids $c(\frac{i}{4})$\ at fractional peak intensity $\frac{i}{4}$\  normalized by the full width at the same fractional peak intensity FW$\frac{i}{4}$M.  In the case of the \aliii\ shifts as defined above, $\delta$ can be approximated as $\delta(\mathrm{AlIII}) \approx
{s(\mathrm{AlIII})}/{\mathrm{FWHM(AlIII)}}. $ The values of the $|\delta$ \aliii $|$\  are $\lesssim 0.05$\ (right panel of Fig. \ref{fig:trenda}) save in the case of spectral  type A4 where $\delta \approx -0.1$.

\begin{figure}[htp!]
\centering
\includegraphics[width=0.43\columnwidth]{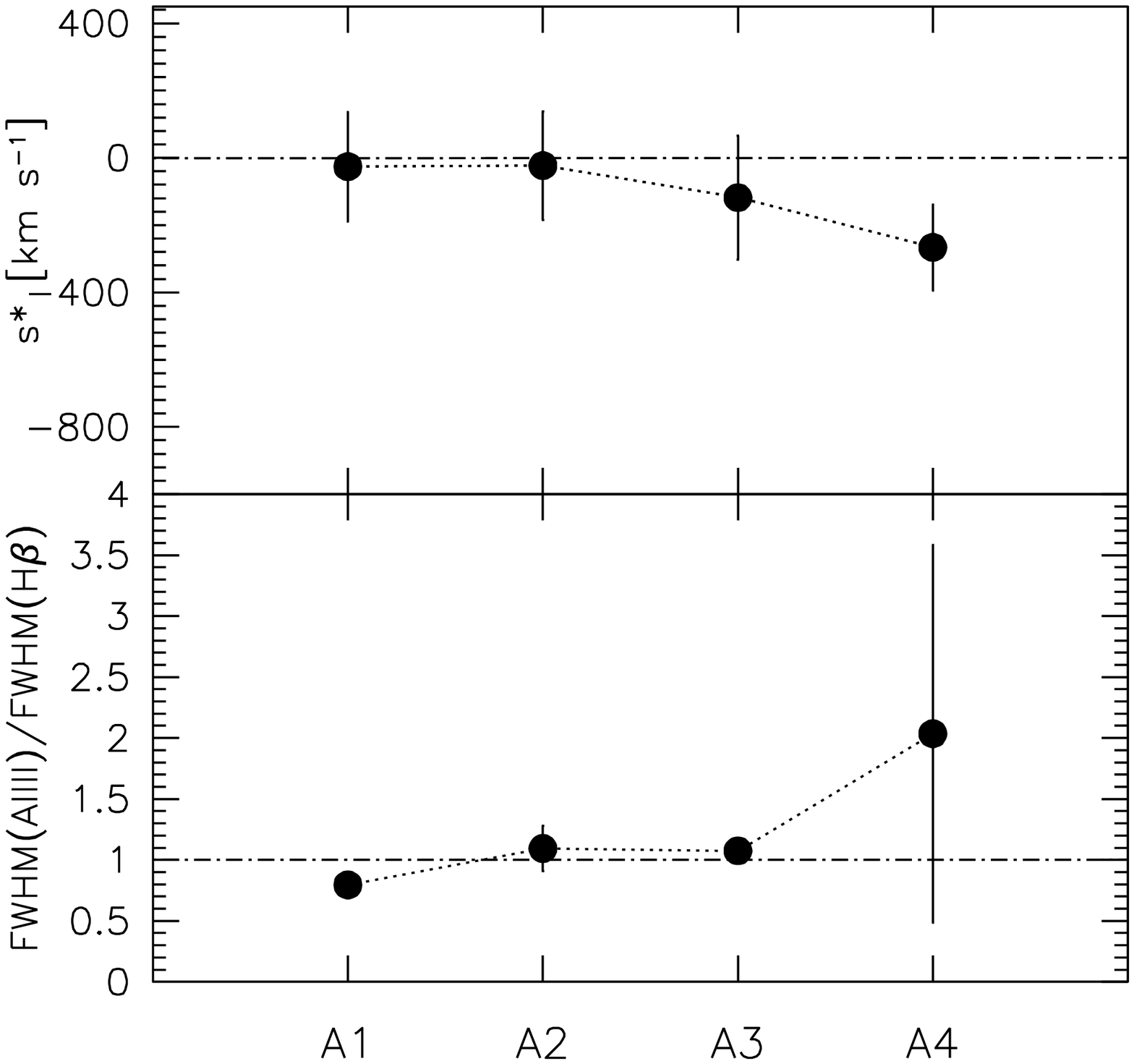}
\includegraphics[width=0.43\columnwidth]{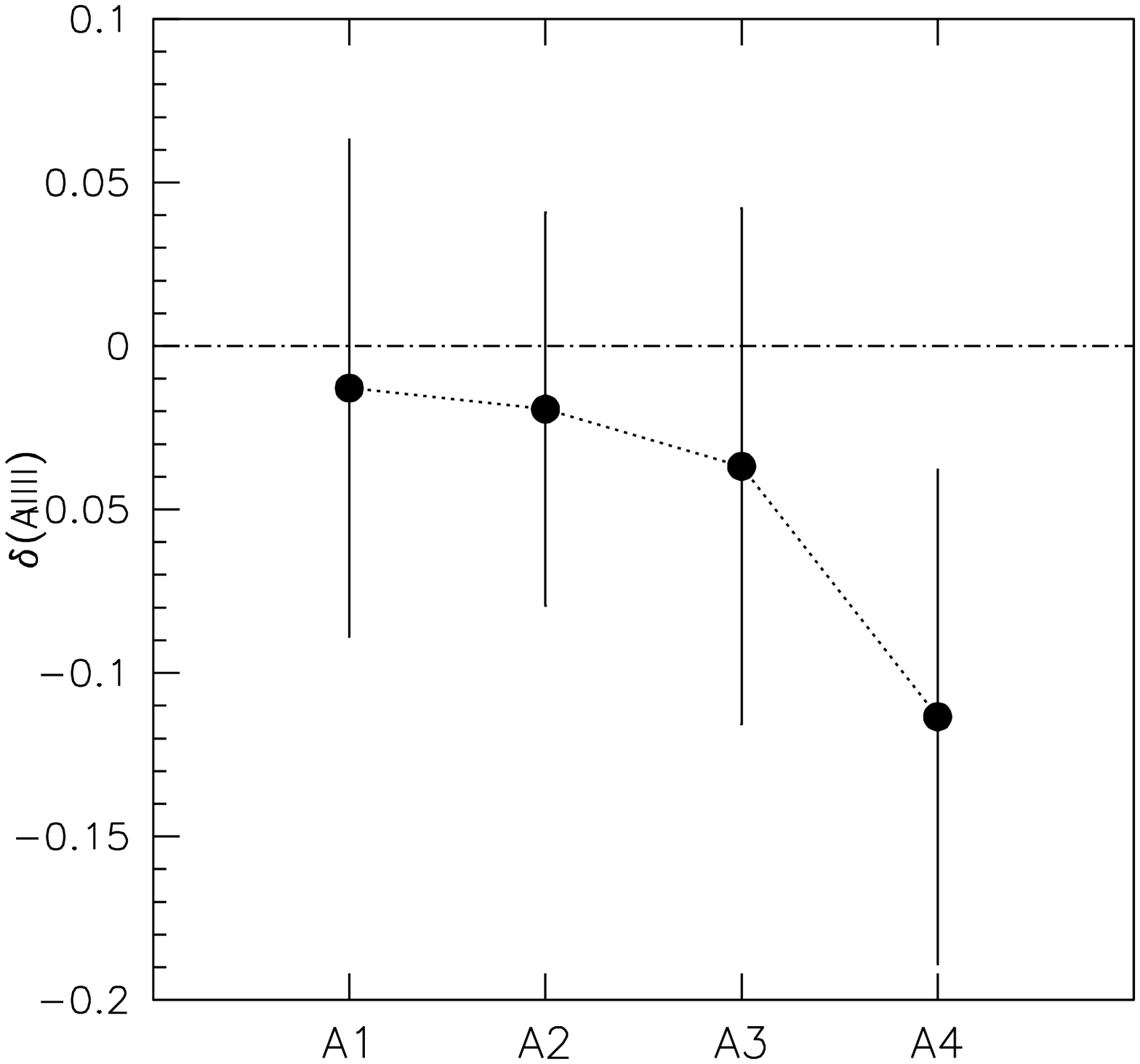}\\
\vspace{-0.5cm}
\caption{Left, upper panel: behavior  of shift $s^\ast$\ of \aliii\ with respect to rest frame as a function of the spectral type. Reported values are sample medians and  error bars are SIQR. Left bottom panel: same, for the ratio FWHM \aliii\ over FWHM \hb. Right: parameter $\delta$(\aliii) as a function of spectral type.    \label{fig:trenda} }
\end{figure}

\subsection{Dependence on luminosity}
\label{lum}
\subsubsection{FWHM}

An important clue to the interpretation of the \aliii\ broadening is provided by the trends with luminosity.  Fig. \ref{fig:hballum} shows FWHM \aliii\ vs. FWHM \hb\ with data points color-coded according to luminosity. There is no significant deviation from equality for the FWHM of \hb\ and \aliii.  At higher luminosity, both the \aliii\ and the \hb\ line become broader, and the largest line widths are measured on the ESO, ISAAC and WISSH samples.  The ratio $r$ = FWHM \aliii\ / FWHM \hb\  $\approx 1/\xi$    also does not depend on luminosity: dividing the sample by about one half at $\log L_{1700}$ = 46 [erg s$^{-1}$] yields median $r$\ values for the subsamples above and below this limit  that are very close to 1 (lower panel of Fig. \ref{fig:hballum}).  

The statistical equality between FWHM \hb\ and FWHM \aliii\ is not breaking down at any luminosity, at least within the limit sets by our sample and data quality. The explanation resides in the fact that both line FWHM increase in the same way with luminosity, as shown by the left, top panel  of Fig. \ref{fig:shiftl}. 
The trends  for \hb, \aliii, and \ciii\ alike (\ciii\ is discussed in Section \ref{hbciii})  can be explained if the broadening of the line is basically set by the virial velocity at the luminosity-dependent radius of the emitting region, $r _\mathrm{BLR} \propto L^\frac{1}{2}$. Under the standard virial assumption, we expect that FWHM(H$\beta$) $\propto M_\mathrm{BH}^{\frac{1}{4}} \left( L/L_\mathrm{Edd}\right)^{-\frac{1}{4}} f_\mathrm{S}(\theta)^{-\frac{1}{2}}$, with $f_\mathrm{S}$\ assumed to be mainly dependent of the angle $\theta$\ between the accretion disk axis and the line of sight \citep{mejia-restrepoetal18a}. Equivalently, FWHM(H$\beta$) $\propto L^{\frac{1}{4}} \left(L/L_\mathrm{Edd}\right)^{-\frac{1}{2}} f_\mathrm{S}(\theta)^{-\frac{1}{2}}$. While  \lledd\ is changing in a narrow range (0.2 — 1) and $f_\mathrm{S}$\ is also changing by a factor a few,   \mbh\ is instead spanning around 4 orders of magnitude. Over such a broad ranges of masses or, alternatively,  luminosity, we might expect that the dominant effect in the FWHM increase is associated right with mass or luminosity.  {  In  Fig. \ref{fig:shiftl}  the trend expected for FWHM(H$\beta$) $\propto L_{1700}^\frac{1}{4}$ is overlaid to the data points, and is consistent with the data in the luminosity range $44.5 \lesssim \log L_{1700} \lesssim 47.5$\ [\ergss], where a fivefold increase of the FWHM of both \hb\ and \aliii\ is seen.}

\begin{figure}[htp!]
\centering
\includegraphics[width=0.5\columnwidth]{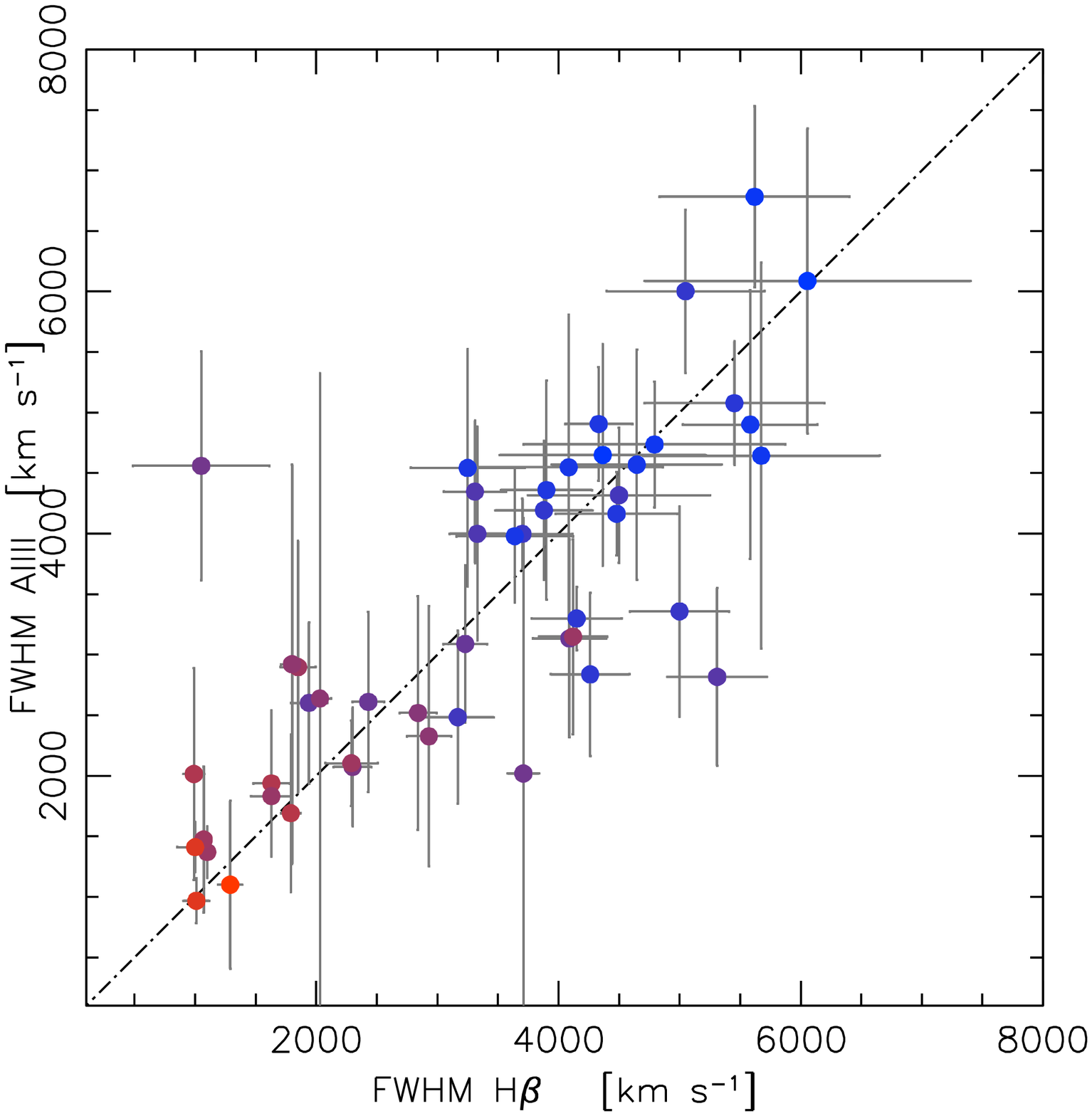}\\
\vspace{-.5cm}
\includegraphics[width=0.5\columnwidth]{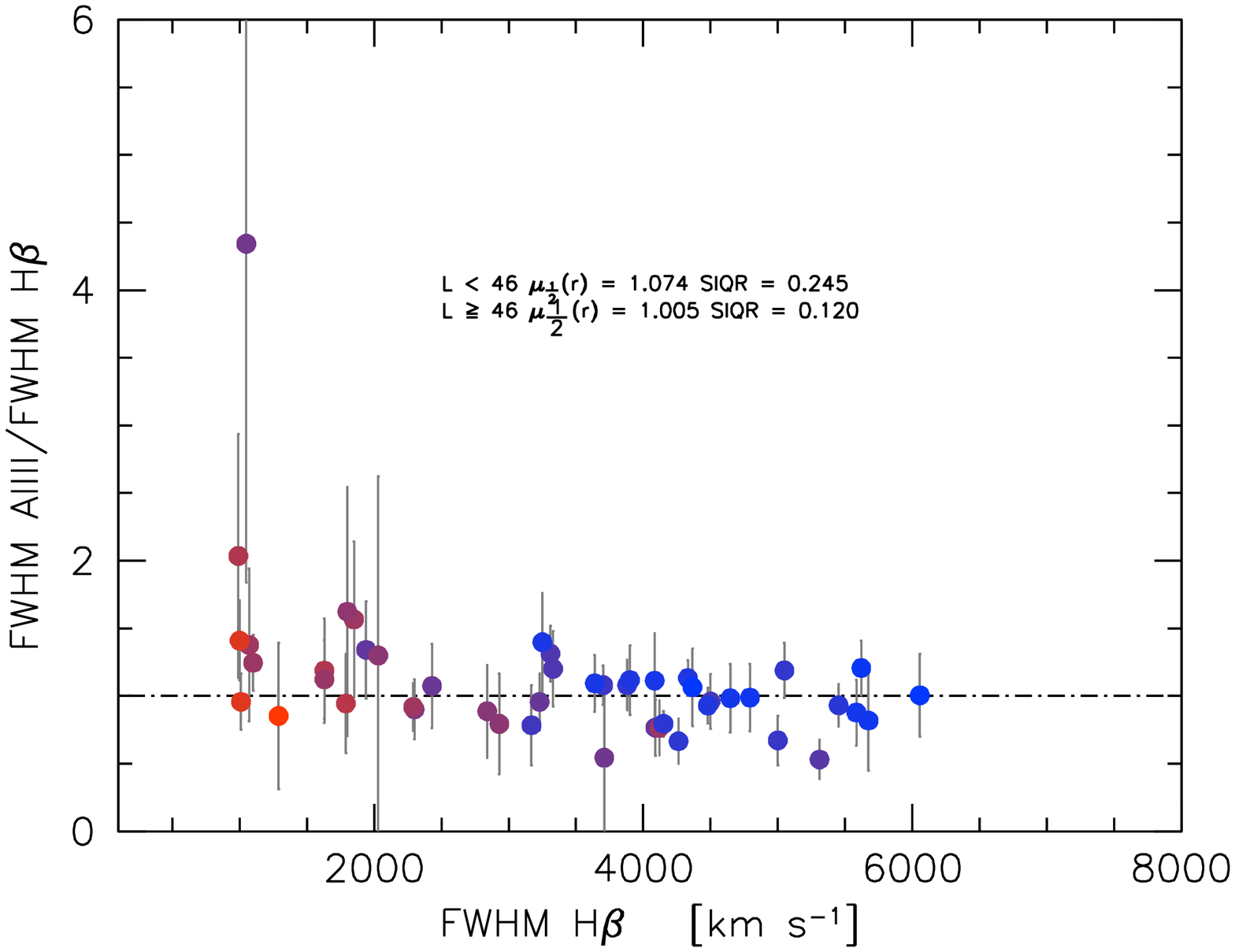}
\vspace{-1.5cm}
\caption{Top panel: FWHM(\aliii) vs. FWHM(\hb) for the joint sample. Data points are color-coded according to luminosity at 1700 \AA, from red (lowest; $\log L_{1700} \sim 44$) to blue (highest, $\log L_{1700} \sim 47$). Bottom panel:  FWHM ratio of \aliii\ and \hb.  The median ratios  $\mu_\frac{1}{2}{r}$ and the SIQR ${r}$  are reported   for more luminous  ($\log L \ge\ 46$ \ [erg/s]) and less luminous ($\log L < 46$) quasars.  \label{fig:hballum} }
\end{figure}

\begin{figure}[htp!]
\centering
\includegraphics[width=0.4\columnwidth]{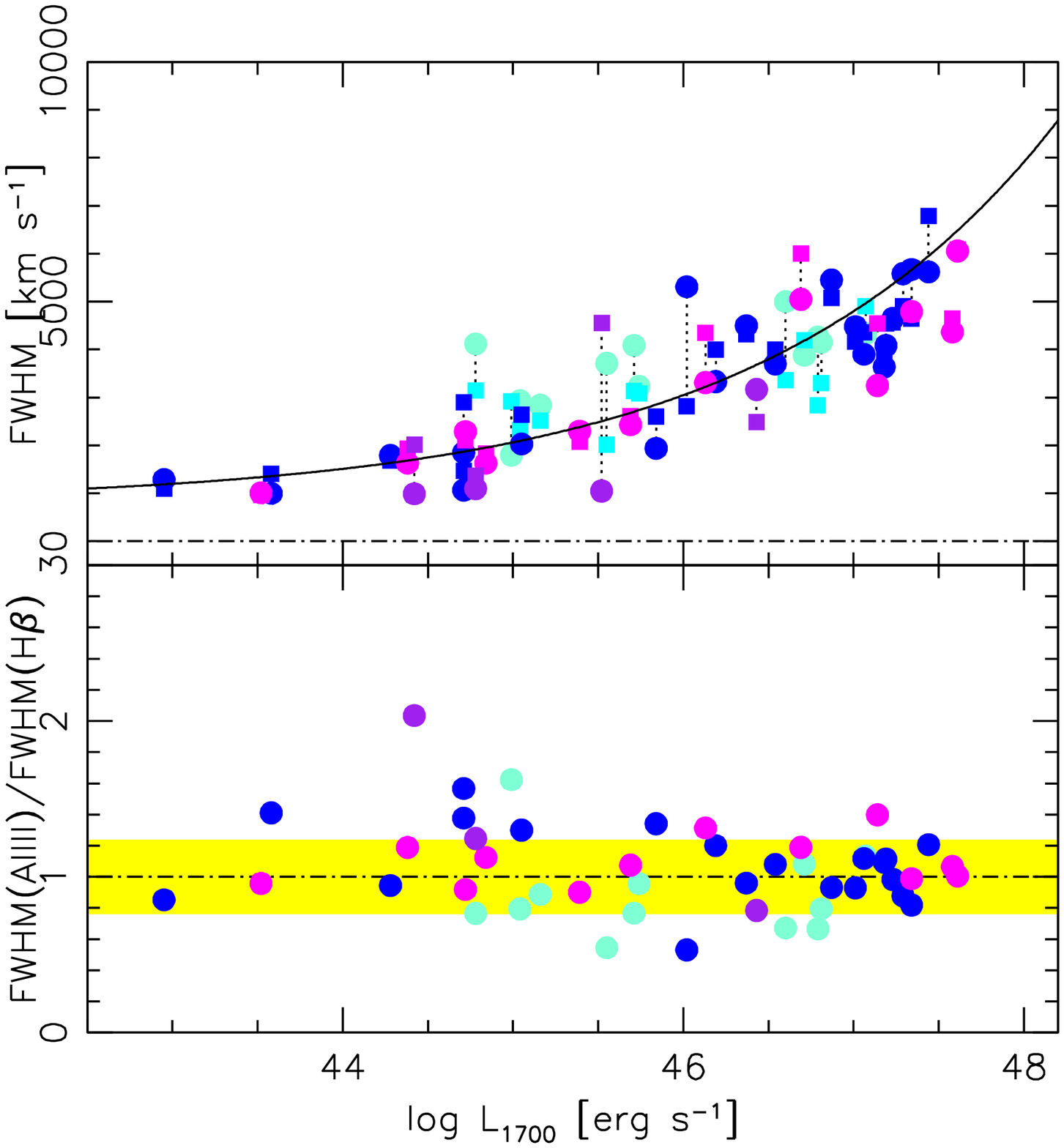}
\includegraphics[width=0.4\columnwidth]{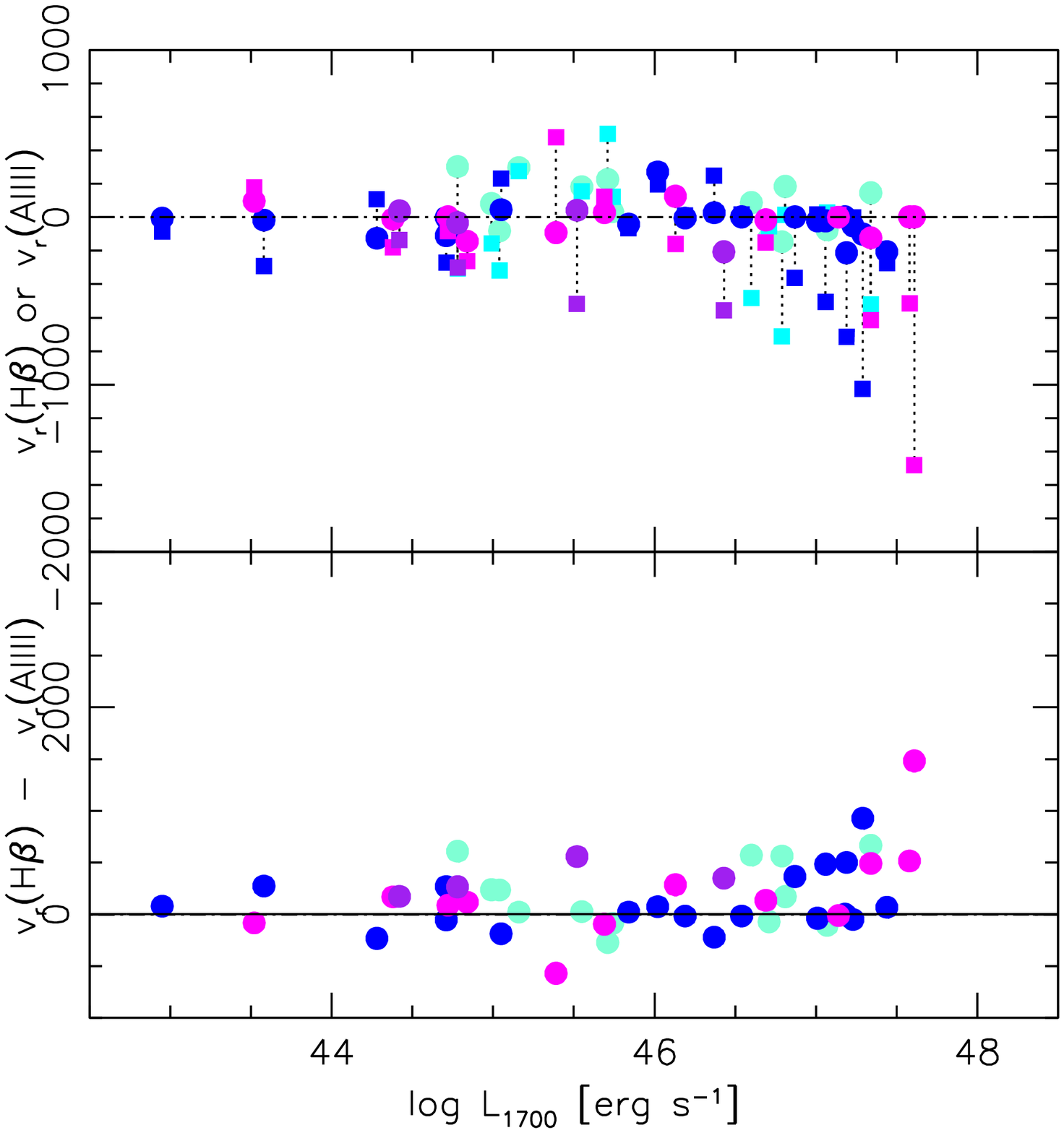}
\caption{\hb\ and \aliii\ profile parameter comparison as a function of luminosity.   Left panel: behavior of FWHM \aliii\ and \hb\   (upper half) and of the ratio FWHM(\aliii)/FWHM(\hb) as a function of $\log L$ at 1700 \AA. The filled line traces the expected increase of the FWHM in a virialized system with $L^\frac{1}{4}$.   The data point are identified on the basis  of spectral type as in Fig. \ref{fig:hbalst}. The yellow band marks the region where FWHM(\aliii)/FWHM(\hb)=1 within the errors: the median value of the uncertainty of the ratios, { $ \approx \pm $25\%}.  Right panels: shifts of \hb\   and \aliii\ (upper half), and their difference  $s^\star$ as a function of $\log L$(1700) (lower half).  The vertical dotted lines join \hb\ and \aliii\ parameters for the same object (e.g., they are not error bars).}
\label{fig:shiftl}
\end{figure} 

\subsubsection{Shift}

The  blueshifts  involve  radial velocities  that are relatively modest  (right panel of Fig. \ref{fig:shiftl}).  \aliii\ shifts exceed 1000 \kms\ only in two cases of extreme luminosity.  Even if we see some increase toward higher $s$\ values in the highest luminosity range, several high-luminosity sources remain unshifted within the uncertainties.   If we consider the dependence of shifts on luminosity,  at high $L$ there is an increase in the range of blueshifts involving values that are relatively large (several hundred \kms; Fig. \ref{fig:shiftl}). The parameter $\delta$\ as a function of $L$\ has a more erratic behaviour (Fig. \ref{fig:trendsl}), but only at the highest $L£$  $\delta \approx -0.1$, and  the effect of the line shift is at a 10\%\ level. Fig. \ref{fig:trendsl} is consistent with   \aliii\ blueshifts becoming more frequent and of increasing amplitude with luminosity.  

\begin{figure}[htp!]
\centering
\includegraphics[width=0.43\columnwidth]{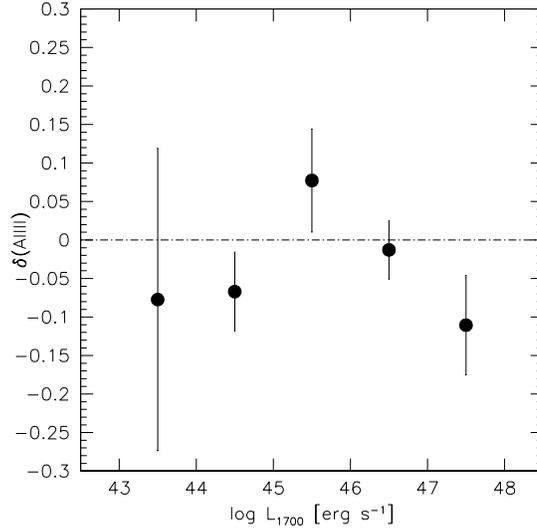}\\
\vspace{-0.5cm}
\caption{Parameter $\delta$(\aliii) as a function of luminosity. Median values are plotted for 1 dex luminosity intervals. Error bars are sample SIQR.    \label{fig:trendsl} }
\end{figure}

\subsection{FWHM \hb\ and FWHM \ciii}
\label{hbciii}

The \ciii\ line has been considered as a possible virial broadening estimator, and has been a target of reverberation mapping monitoring \citep{treveseetal07,treveseetal14,liraetal18,kaspietal21}. The problem in Population A is that \ciii\ shows a strong gradient in its intensity. In spectral type A1, \ciii\ is by far the strongest line in the 1900 \AA\ blend, but its prominence decreases with increasing \rfe, i.e., going toward spectral type A3 \citep{bachevetal04}. In spectral types A3 and A4  \ciii\ is  affected by  the severe blending with \siiii\ and \feiii\ emission (much more severe than for \aliii), and  may  become so weak to the point of being barely detectable or even undetectable\  \citep{vestergaardwilkes01,negreteetal14,templeetal20,bachevetal04,martinez-aldamaetal18a}. In addition, the \ciii\ line has a rather low critical density, and its intensity is exposed to the vagaries of density and ionization fluctuations much more than \aliii, whose emission is highly efficient in dense  gas over a broad range of ionization levels \citep[][and Sect. \ref{xA}]{marzianietal20}.  The measurements of the \ciii\  width   might be inaccurate in  extreme Population A if \feiii\ contamination is strong.  It is therefore legitimate to expect a greater dispersion in the width relation of \ciii\ with \hb. 

The FWHM of \ciii\ is shown against the FWHM of \hb\ full profile in Fig. \ref{fig:hbc3st}, upper panel. Error bars show uncertainties computed following the prescription of Appendix \ref{app:unc}.  Not surprisingly, the scatter in the FWHM ratio between \ciii\ and \hb\ is larger than in the case of \aliii.  The top and  middle  panels of Fig. \ref{fig:hbc3st} show that there is a significant deviation from unity, although for relatively  narrow  profiles around 2000 \kms\ the FWHM \ciii\  is close to the 1:1 line.  The $\chi_\nu^2$ is much higher than 1. The bottom panel shows the ratio between FWHM \ciii\ and FWHM \hb\  color-coded according to \ciii\ strength. The limit was set at normalized intensity (roughly equal to  equivalent width) $I = $ 10. The trend for sources above this limit implies  FWHM \ciii  = $\xi_\mathrm{CIII]}$ FWHM $\approx$  0.77 FWHM \hb. 

We didn’t detect any strong difference in the trend with luminosity of  \hb\ and \ciii\ FWHM, as it has been the case for \hb\ and \aliii. The two lines follow a similar trend with luminosity at 1700 \AA\ (Fig. \ref{fig:shiftlc3}).   
No shift analysis was carried out for \ciii\ due to the severe blending with \siiii\ and \feiii. 


The narrower profile of \ciii\ indicates a higher distance from the central continuum source than the one obtained from \hb, if the velocity field is predominantly virial \citep{petersonwandel00}. This result is also consistent with the findings of  \citet{negreteetal13} who, using \ciii\ intensity ratios in a photoionization  estimate of the emitting region radius, obtained  much larger radii than the ones obtained from reverberation mappings of \hb. 

\begin{figure}[htp!]
\centering
\includegraphics[width=0.5\columnwidth]{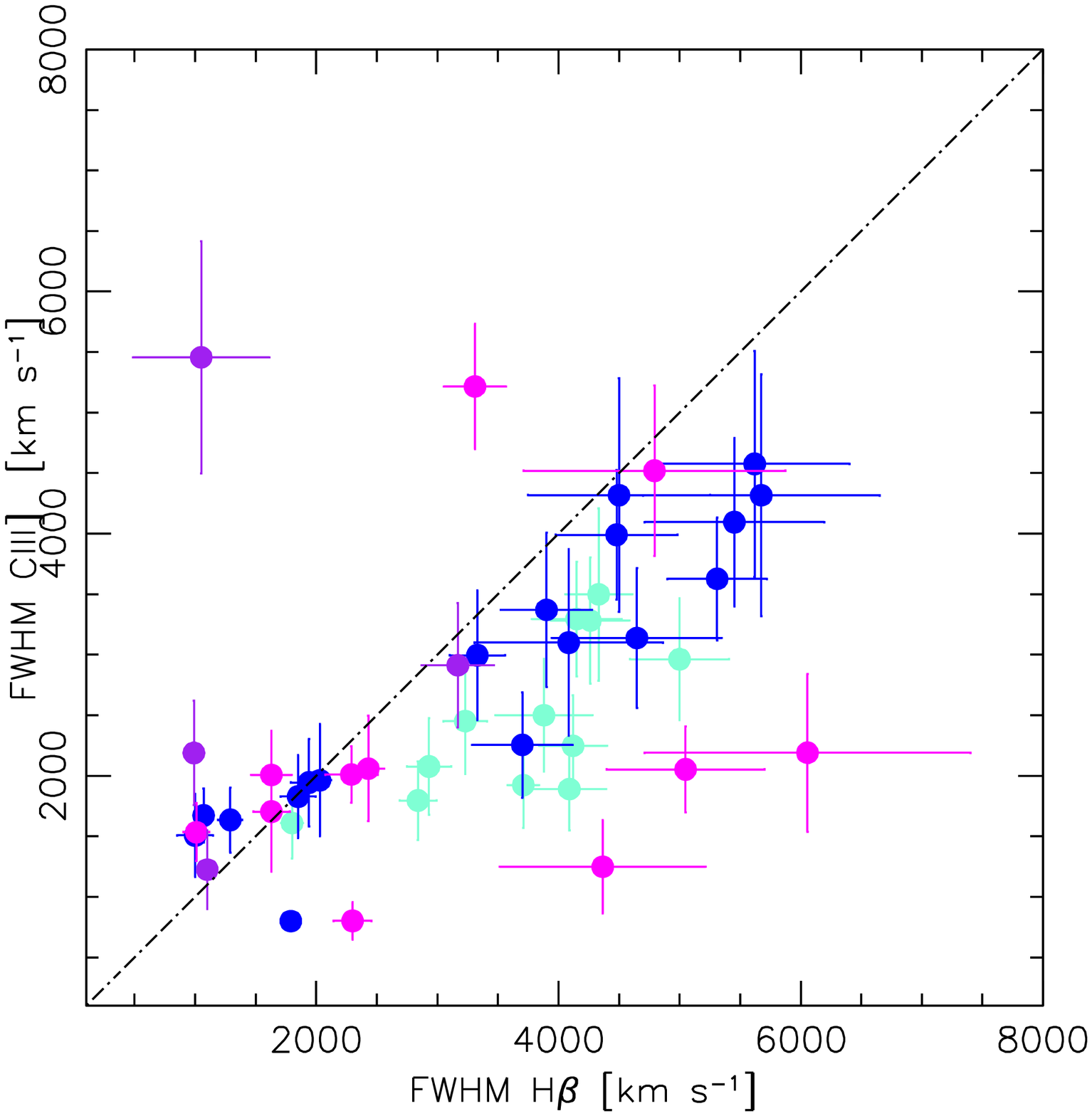}\\
\vspace{-.5cm}
\includegraphics[width=0.5\columnwidth]{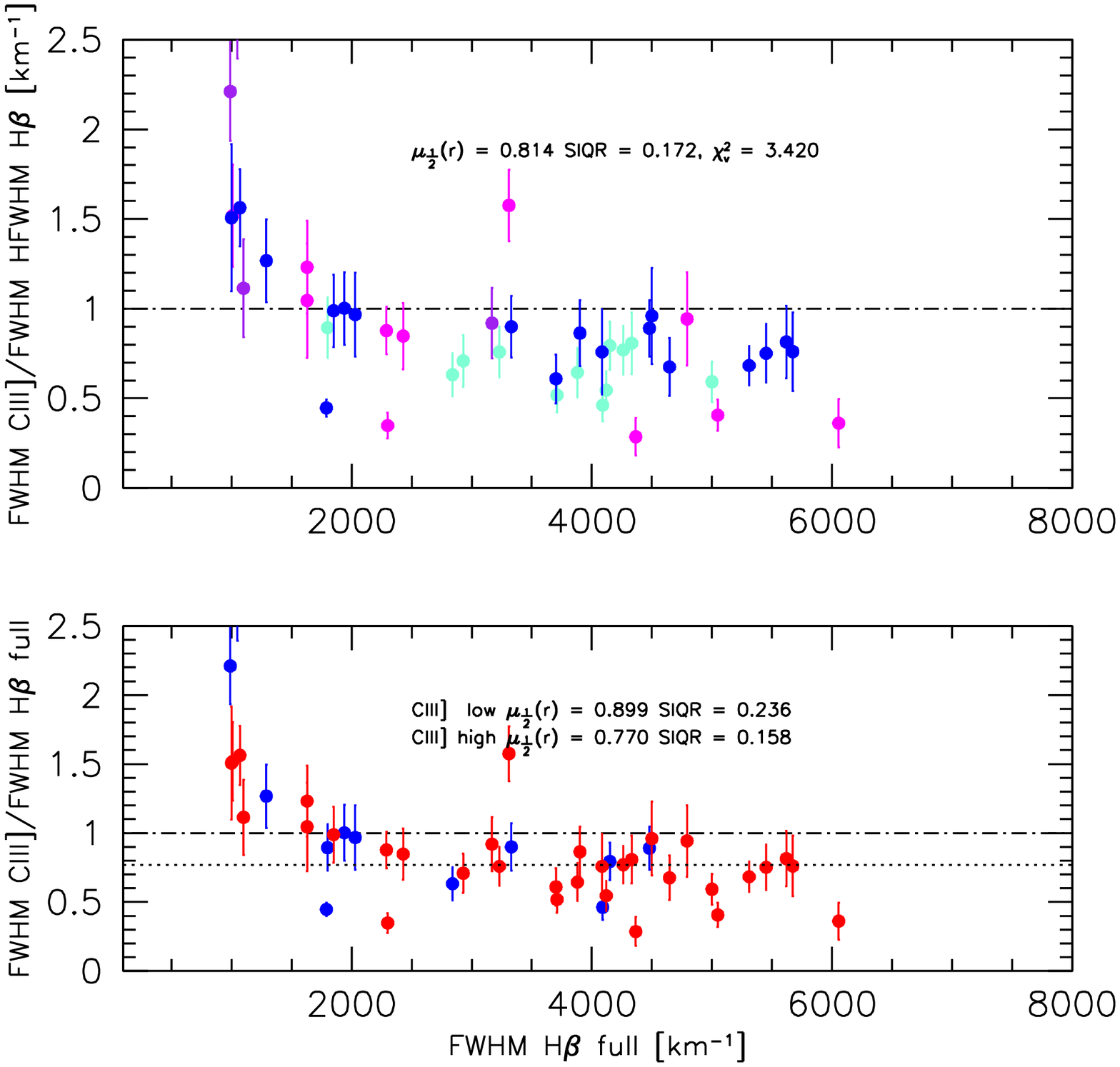}
\vspace{-0.5cm}
\caption{Top panel: FWHM(\ciii) vs. FWHM(\hb) for the joint sample. Data points are color-coded according to spectral types, as for Fig. \ref{fig:hbalst}. Middle panel:  FWHM ratio of \ciii\ and \hb,  The median ratio  $\mu_\frac{1}{2}({r})$, with $r =$ FWHM \ciii/FWHM \hb,   and the SIQR are reported. Bottom panel: same ratio with the data color-coded according to \ciii\ intensity: normalized intensity $> 10$\ (red)\ and $\le 10$\ (blue). Values of  $\mu_\frac{1}{2}({r})$\ and   SIQR  are reported for stronger  and weaker \ciii\  emitters.  \label{fig:hbc3st}}
\end{figure}

\begin{figure}[htp!]
\centering
\includegraphics[width=0.4\columnwidth]{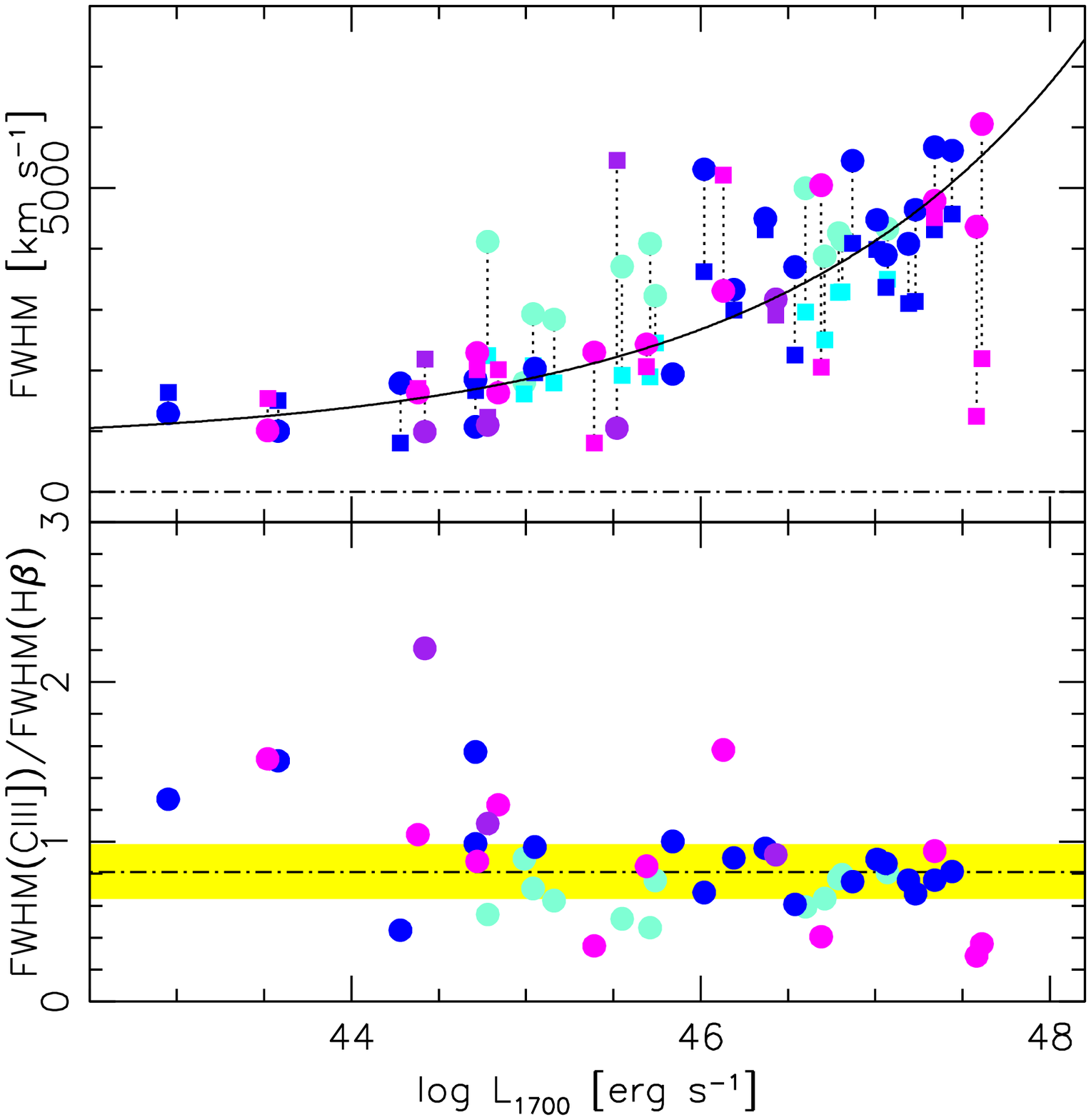}
\caption{\hb\ and \ciii\ profile parameter comparison as a function of luminosity.   Top panel: Behavior of FWHM \ciii\ and \hb\   (upper half) and of the ratio FWHM(\ciii)/FWHM(\hb) as a function of $\log L$ at 1700 \AA. 
The data point are color-coded on the basis  of spectral type.  As in Fig. \ref{fig:shiftl}, the filled line traces the expected increase of the FWHM in a virialized system with $L^\frac{1}{4}$.  The yellow band marks the region where FWHM(\civ)/FWHM(\hb) $\approx$ 0.8 within the errors. The width of the band has been  set by  the median value of the uncertainty of the ratios,  $\pm$ 0.17.   The vertical dotted lines join \hb\ and \aliii\ parameters for the same object.}
\label{fig:shiftlc3}
\end{figure}



\subsection{A \mbh\ scaling law based on \aliii}
\label{mass}

The goal is to obtain an \mbh\ estimator based on \aliii\ that is consistent with  the scaling law derived for \hb. In this context, the process is much simpler than in the case of \civ, where large blueshifts introduced a significant correction and a second-order dependence on luminosity of FWHM \civ\ could not be bypassed.   The \hb\ and \aliii\ widths of the two lines grow in a similar trend with $L$\ (Fig. \ref{fig:shiftl}). The \mbh\ scaling law can be derived in the form $\log M_\mathrm{BH} = \alpha \cdot \log L + 2 \cdot \log \mathrm{FWHM} + \gamma$ by minimizing the scatter and any systematic deviation of   \mbh\   estimated from \aliii\ with respect to the \hb-derived masses { from the \citet{vestergaardpeterson06} scaling law}:

\begin{equation}
    \log M_\mathrm{BH}\, (\mathrm{H\beta})  \approx  0.5 \log L_{5100,44} + 
  2 \log \left(\mathrm{FWHM}(\mathrm{H\beta}) \right) + 0.91,
\end{equation}

{ where   $L_{5100,44}$ is the rest frame luminosity  $\lambda L_\lambda$\ at 5100 \AA\ in units of $10^{44}$ \ergss, and the FWHM \hb\ is in \kms},   
considering that no correction is needed  to FWHM \aliii \  (i.e., $\xi_\mathrm{AlIII} \approx 1$).  { The \citet{vestergaardpeterson06} scaling law is a landmark that has been applied in hundreds of quasar studies. However, the \citet{vestergaardpeterson06} \hb\ scaling law  provides individual estimates with large error bars in relative estimates  ($\approx$0.5 dex at 1$\sigma$, see the discussion in their paper).  This is a limit to the precision of any scaling law based on the match with the one based on \hb. The large error bars of individual mass estimates can be mainly explained on the basis of   orientation effects \citepalias{marzianietal19} and of trends along the spectral types of the main sequence \citep{duwang19}.} 

If this condition is enforced, the relation between \mbh\ from \aliii\ and from \hb\  (Fig. \ref{fig:virialal3mass}) can be written as: 

\begin{equation}
\log M_\mathrm{BH}(\mathrm{AlIII})  \approx (1.000 \pm 0.043) \log M_\mathrm{BH}(\mathrm{H\beta}) - (0.001\pm 0.367). \label{eq:eqmass}
\end{equation}

The \aliii\ scaling law takes the form, with the FWHM in \kms:
\begin{eqnarray}\label{eq:masslaw}
\log M_\mathrm{BH}\, (\mathrm{AlIII}) & \approx & (0.579^{+0.031}_{-0.029}) \log L_{1700,44} + \\ \nonumber 
 && 2 \log \left(\mathrm{FWHM}(\mathrm{AlIII}) \right) + (0.490^{+0.110}_{-0.060}),
 \end{eqnarray}

with an rms scatter  $\sigma \approx 0.29$.   Figure \ref{fig:virialal3mass} suggests the presence of a well-behaved distribution with a few outlying points. The relation of Eq. \ref{eq:masslaw} considers the FWHM of 47 sources.  One data point has been excluded applying a $\sigma$\ clipping algorithm  (the one A4 outlier, \object{HE0132-4313}). This selective procedure is justified by the fact that only some of the most extreme sources of Population A (not all of them) show large blueshifts and only one (\object{HE0132-4313}) a  FWHM in excess to \hb\ by a large factor, deviating at more than 3 times the sample rms. Removal of \object{HE0132-4313} provides however only a minor, not significant change in the fitting parameters.   The scaling law parameter uncertainties have been estimated computing the coefficients $\alpha$ and $\gamma$ over a wide range of values and defining the interval where agreement  between \mbh\ from \hb\ and \aliii\ (Eq. \ref{eq:eqmass}) is satisfied within  1.00$\pm 1 \sigma$ for the best-fitting slope and  $0.00 \pm 1 \sigma$\ for the intercepts (respectively $\approx$0.043 and $\approx$0.367, as per Eq. \ref{eq:eqmass}). { Due to some scatter in FWHM relation at FWHM \hbbc\   $\sim$ 1000 \kms\ and the possibility of predominantly face-on orientation { \citep{mejia-restrepoetal18a}} the \mbh\ estimates at \mbh\ $\lesssim 10^7$ \msol\ should be viewed with care { due to the paucity of data points}. }    

It is important to remark that this result, unlike the one on \civ\ \mbh\ estimates, is obtained without any correction on the measured FWHM. The \aliii\ and \hb\ relation should be considered equivalent with respect to \mbh\ estimates in large samples of Population A sources. No scaling law $r_\mathrm{BLR}  -  L$\ has been ever derived from reverberation mapping on the \aliii\ lines. However, it is reassuring that the luminosity exponent ($\approx  0.579^{+0.031}_{-0.029}$)\ deviates by about $1 \sigma$\ from the one entering the scaling law  $r_\mathrm{BLR}  -  L$\ derived by \citet{bentzetal13} for \hb.

\begin{figure}[htp!]
\centering
\includegraphics[width=0.5\columnwidth]{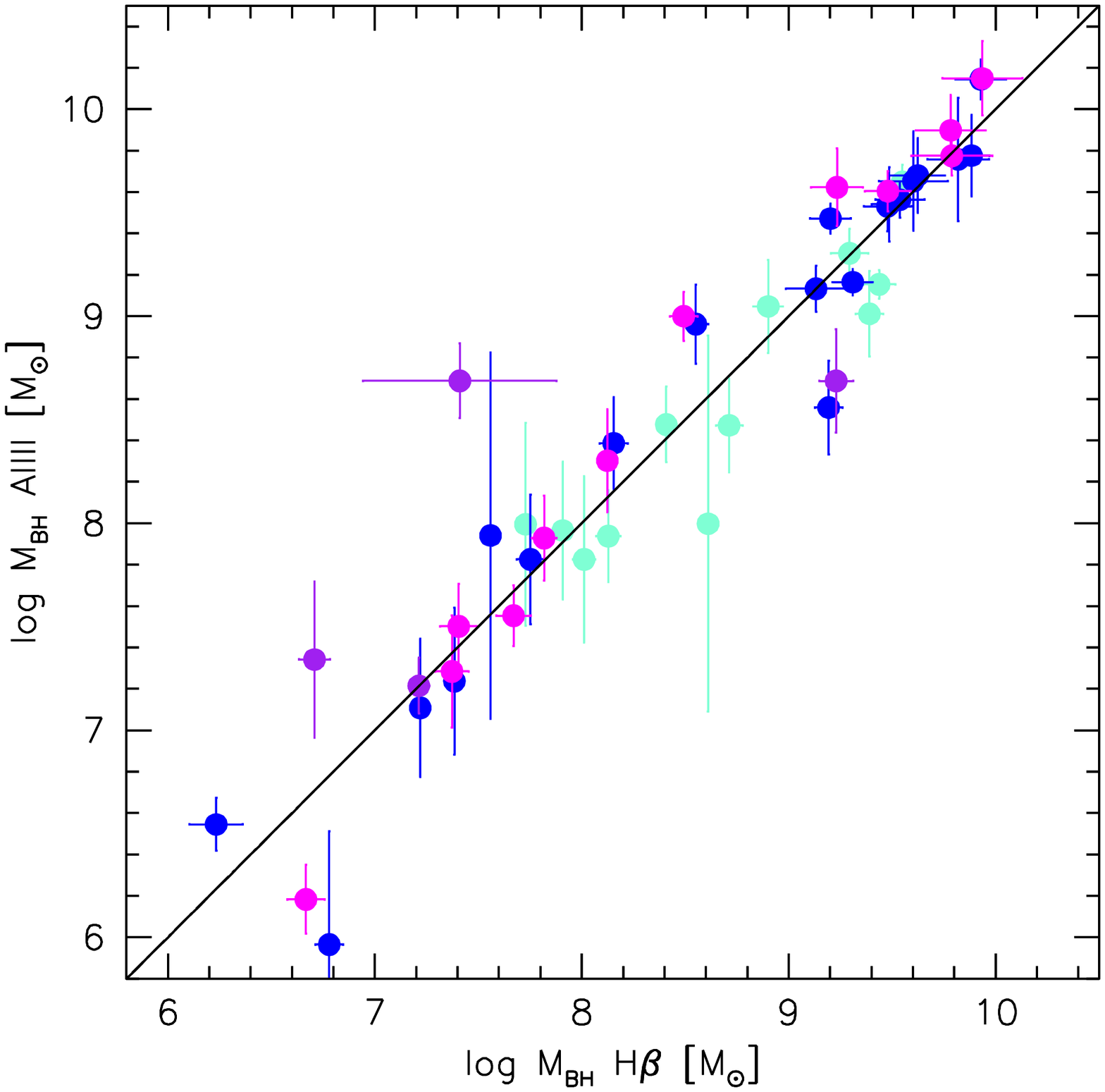}\\
\includegraphics[width=0.5\columnwidth]{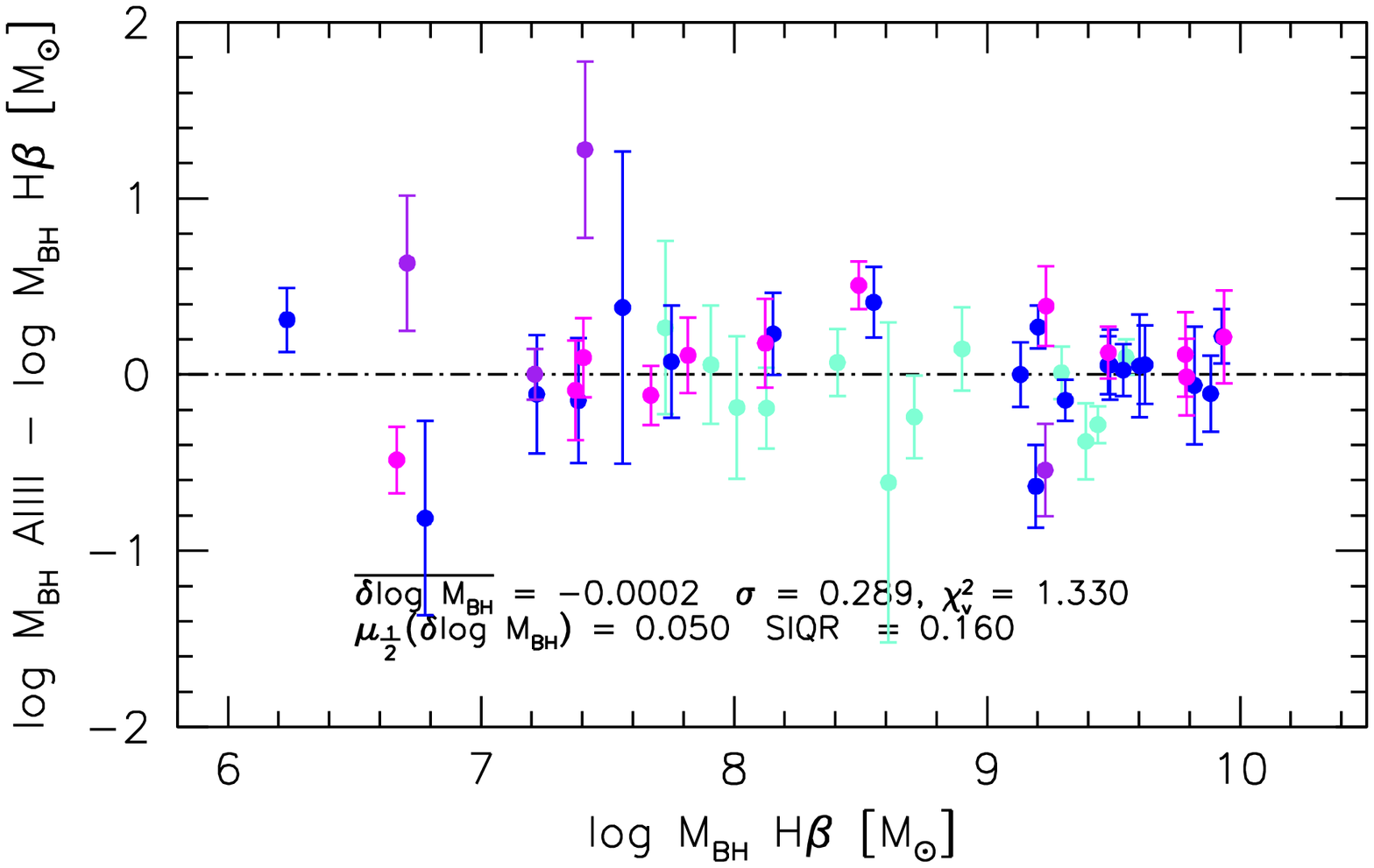}\\
\vspace{-3cm}
\caption{Decimal logarithm of black hole mass in units of solar masses  computed from the  relation of \citet{vestergaardpeterson06} based on FWHM \hb\ vs. the one computed from the \aliii\ FWHM following Eq. \ref{eq:masslaw}, with sub-samples identified by color according to their spectral types.  The filled line traces Eq. \ref{eq:eqmass}, while the dot-dashed line is the equality line.   The bottom panel shows the residuals as a function of the \mbh\ derived from  \hb.  Average values of the $\log $ \mbh\ differences, standard deviation and normalized $\chi_\nu^2$\ are reported  for the joint  sample minus one outlier with $\delta > 1$\  (top row). The bottom rows yields the median and the SIQR.   }
\label{fig:virialal3mass}
\end{figure}

\subsection{A \mbh\ scaling law based on \ciii}
\label{massciii}

An  approach analogous to the one adopted for \aliii\ was also applied to \ciii.  The goal is   to obtain an \mbh\ estimator based on \ciii\ that is consistent with  the scaling law derived for \hb. The process is again much simpler than in the case of \civ, where large blueshift introduced a significant correction and a second-order dependence on luminosity of FWHM \civ\ could not be bypassed.   Considering that only a very simple  correction is needed  to FWHM \ciii, $\xi_\mathrm{CIII]} \approx 1.25$, the \mbh\ scaling law can be derived in the form $\log M_\mathrm{BH} = \alpha \cdot \log L + 2 \cdot \log {\xi_\mathrm{CIII]}} \mathrm{FWHM} + \gamma$ by minimizing the scatter and any systematic deviation of   \mbh\   estimated from \ciii\ with respect to the \hb-derived masses.  Figure \ref{fig:virialc3mass}  suggests the presence of a well-behaved distribution with a few outlying points. The condition    

\begin{equation}
M_\mathrm{BH}(\mathrm{CIII]})  \approx (1.000 \pm 0.053) M_\mathrm{BH}(\mathrm{H\beta}) + (0.000\pm 0.454)  
\label{eq:eqmassc3}
\end{equation}

is satisfied if  the \ciii\ scaling law takes the form:
\begin{eqnarray}\label{eq:masslawc3}
\log M_\mathrm{BH}\, \mathrm{CIII]} & \approx & (0.6765 \pm 0.0450) \log L_{1700,44} + \\ \nonumber 
 && 2 \log \left({\xi_\mathrm{CIII]}}\mathrm{FWHM}(\mathrm{CIII]}) \right) + (0.332 \pm 0.120),
 \end{eqnarray}

with an rms scatter  $\sigma \approx 0.35$\ (excluding 1 outlier, for 44 objects). The scaling law parameter uncertainties have been estimated varying the coefficients $\alpha$ and $\gamma$\ as for the \aliii\ case. 


The  \mbh\ \ciii\ scaling law  is derived with a simple correction on the measured FWHM \ciii. The \ciii\ and \hb\ relation should be considered equivalent with respect to \mbh\ estimates in large samples of Population A sources. Care should however be taken to consider sources in which \ciii\ can be measured ($W \gtrsim 10$\AA) and to identify extreme objects, as discussed in Section \ref{xA}.   In addition, the heterogeneity of the sample and the possibility of different trends in different FWHM domains (FWHM (\ciii) $\approx$ FWHM(\hb) if FWHM $\lesssim$ 3000 \kms) makes the scaling law with \aliii\   more reliable.

\begin{figure}[htp!]
\centering
\includegraphics[width=0.7\columnwidth]{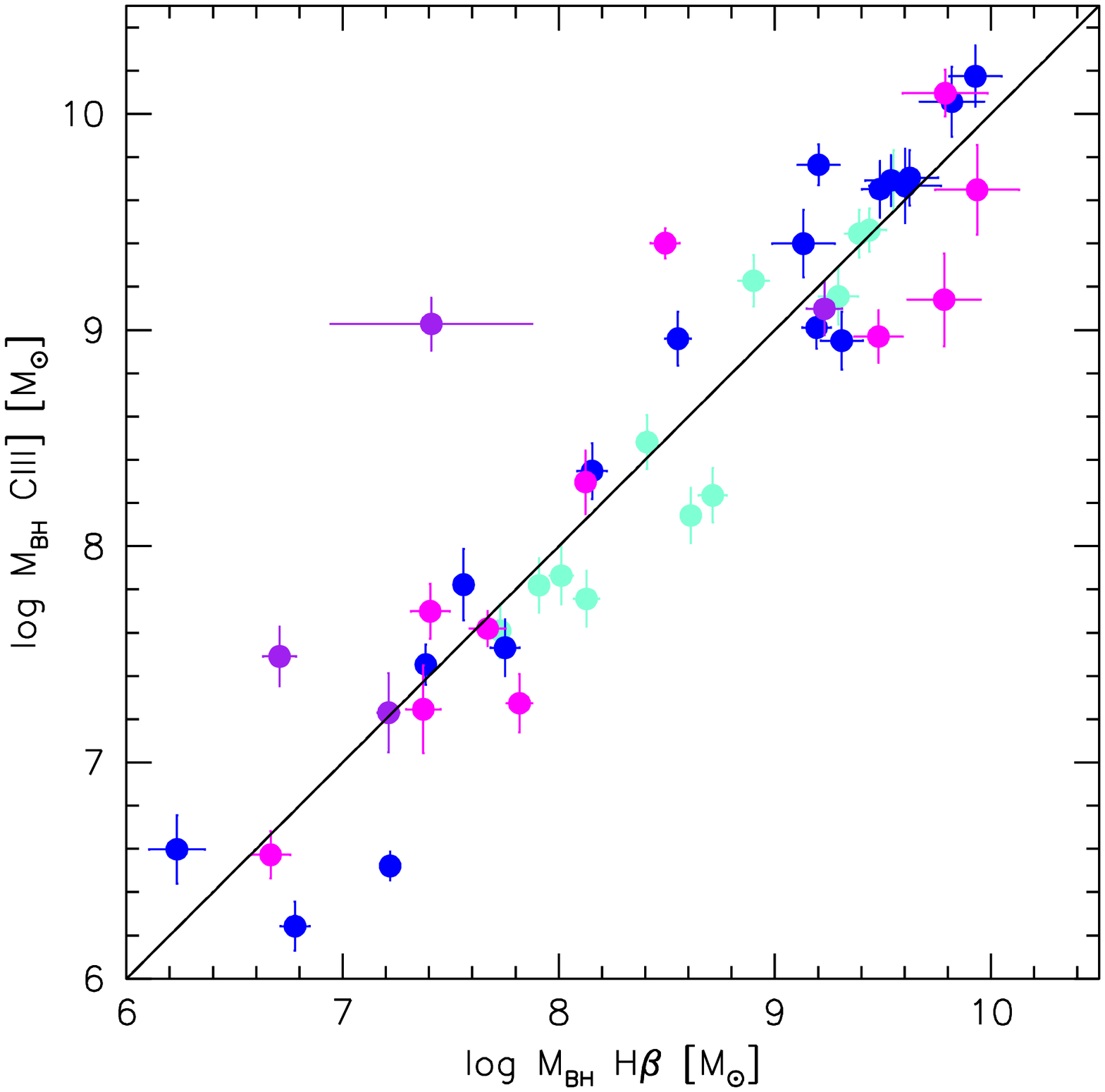}\\
\includegraphics[width=0.7\columnwidth]{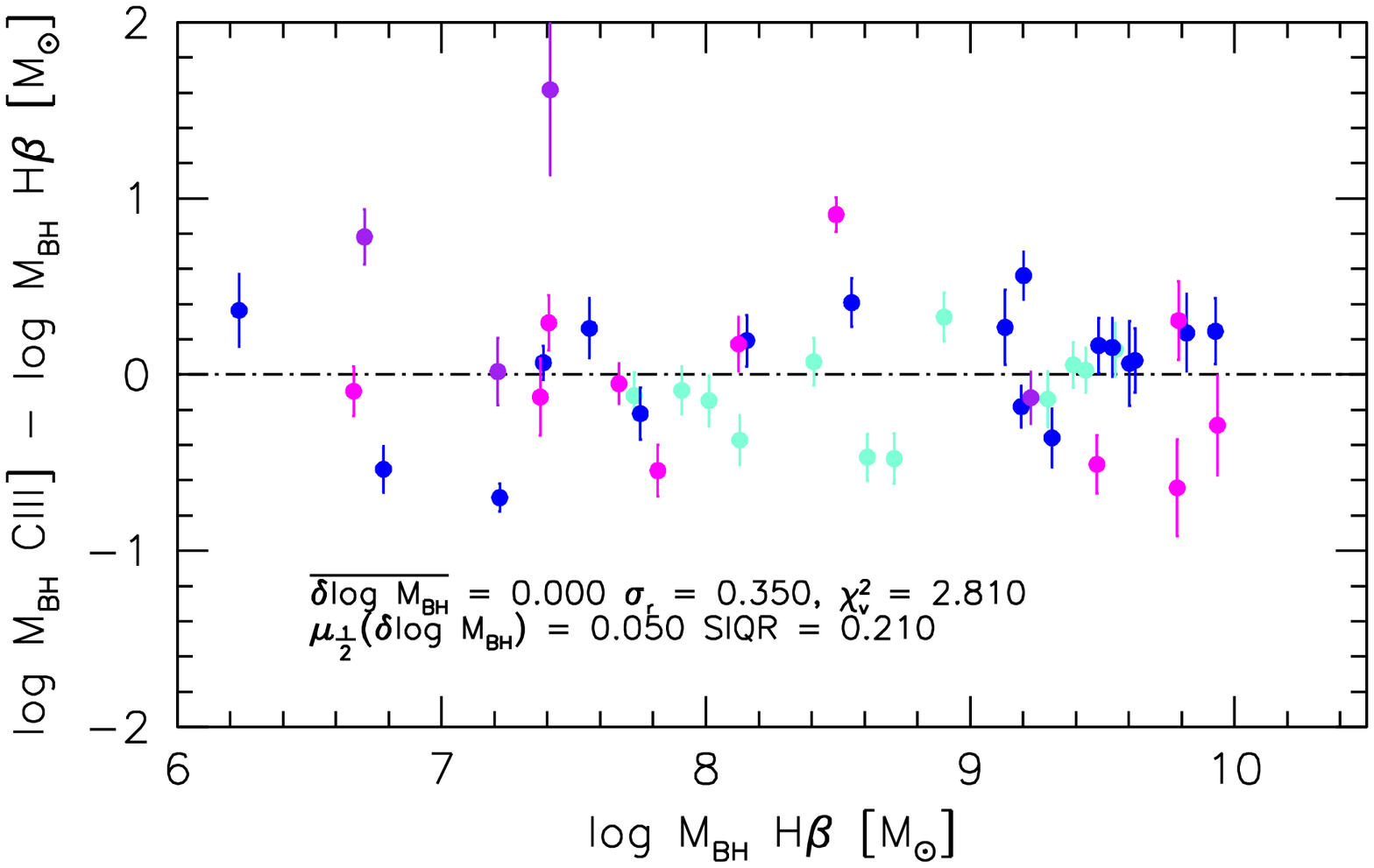}\\
\vspace{-4cm}
\caption{Decimal logarithm of black hole mass in units of solar masses  computed from the  relation of \citet{vestergaardpeterson06} based on the \hb\ spectral range vs. the one computed from the \ciii\ data following Eq. \ref{eq:masslawc3}, with spectral types color-coded as in Fig. \ref{fig:hbalst}.  As for Fig. \ref{fig:virialal3mass},  the filled line traces Eq. \ref{eq:eqmassc3}, while the dot-dashed line is the equality line.   The lower panel shows residuals as a function of \mbh\ derived from  \hb.  Average, standard deviation, and $\chi_\nu^2$\ of the  $\log$\ \mbh\ difference are reported on the top-row  of the inside caption; the bottom row reports median and SIQR. }
\label{fig:virialc3mass}
\end{figure}

\section{Discussion}
\label{disc}

\begin{figure}[htp!]
\centering
\includegraphics[width=0.5\columnwidth]{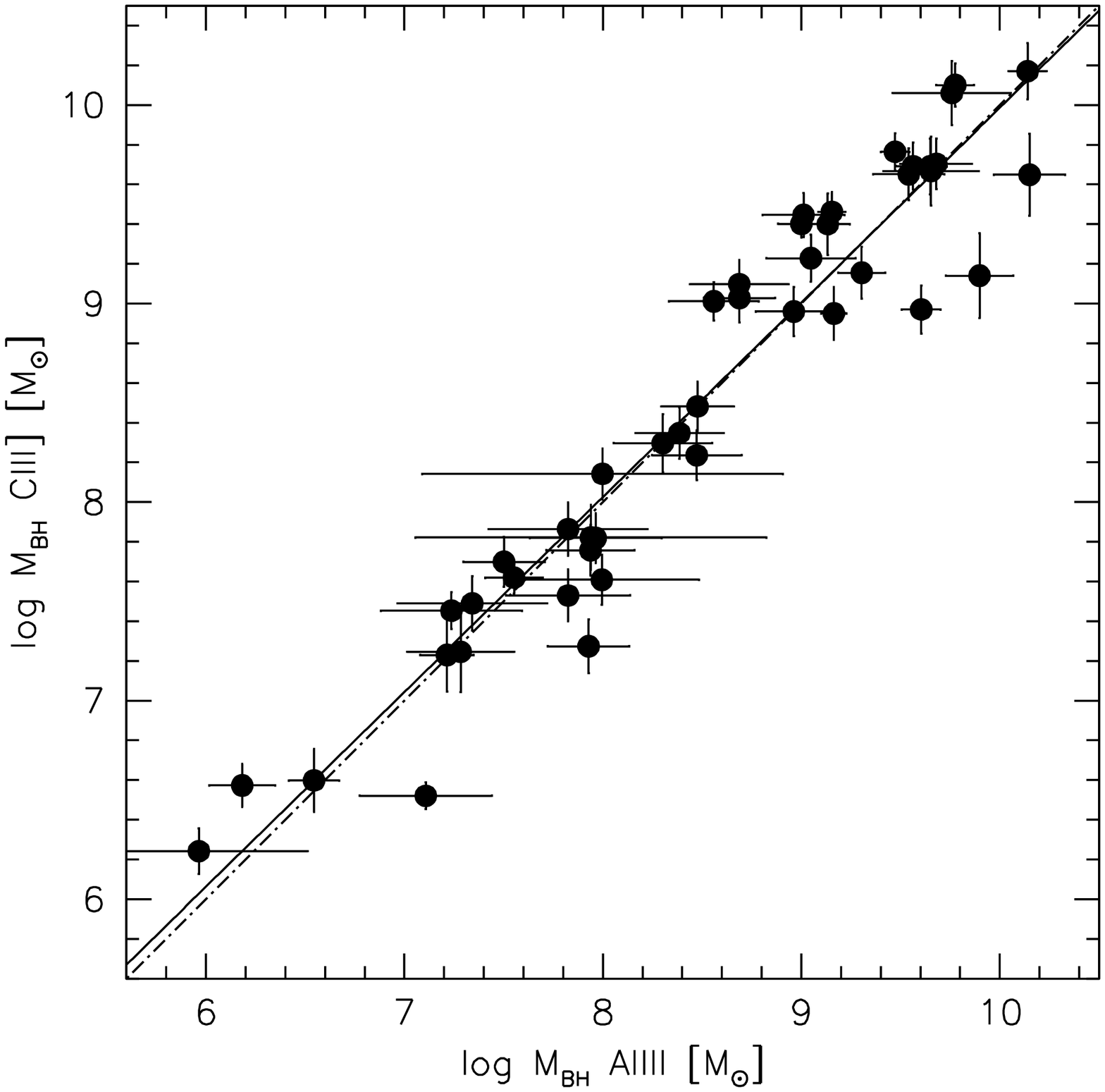}\\
\vspace{-0cm}
\caption{Decimal logarithm of black hole mass in units of solar masses  computed from the  scaling law based on FWHM \ciii\ (Eq. \ref{eq:masslawc3}) vs. the one computed from the \aliii\ FWHM following Eq. \ref{eq:masslaw}.  The filled line traces an unweighted least square fit; the dot-dashed line is the equality line.   }
\label{fig:mhbac}
\end{figure}

{ In the following we first compare the scaling laws obtained for \aliii\ and \ciii (Sect. \ref{consist}). 
The present work detects systematic shifts of \aliii\ toward the blue. Even if they are on average slightly above the uncertainties, it is interesting to analyze them in the context of systematic shifts affecting the prominent high and  low-ionization lines of \civ\ and \mgii\ (Section \ref{shifts}), and to consider more in detail sources in  the spectral type where \aliii\ are broader than the \hb\ ones  (Section \ref{xA}).}

\subsection{Consistency of scaling laws}
\label{consist}

The scaling laws derived for \aliii\ and \ciii\ are mutually consistent. Black hole mass estimates are shown in Fig. \ref{fig:mhbac}. An unweighted lsq fit yields slope $0.975 \pm  0.043$, consistent with unity, and intercept $0.246 \pm 0.371$, with an rms scatter $\approx$ 0.302. The median difference  of the mass values obtained with the two scaling laws is $\mu(\log M_\mathrm{BH} \mathrm{(CIII])} - \log M_\mathrm{BH} \mathrm{(AlIII)}) \approx 0.06 \pm 0.20$\ (SIQR). 



The  $r_\mathrm{BLR}  -  L$\ implicit in Eq. \ref{eq:eqmassc3} is consistent with the  $r_\mathrm{BLR}  -  L$\ relation derived for \civ\ in a previous study \citep{treveseetal14}, and only slightly higher than the one derived in the early study of  \citet[][$\alpha \approx 0.52 - 0.55$, c.f. \citealt{kaspietal21}]{kaspietal07}. More high S/N spectroscopic observations, including monitoring, should be carried out to explore  the full potential of the 1900 \AA\ blend lines and especially of \ciii\  as VBEs. 

The scaling law derived from the \aliii\ line is consistent with the \civ-based scaling law \citepalias{marzianietal19}: Fig. \ref{fig:malc} indicates only a slight bias   (less than the SIQR $\approx$ 0.2 dex, and the  rms $\approx $ 0.3 dex), as the \civ\ scaling law apparently overestimates the \mbh\ by  $\approx 20$\%\ and $\approx 40$\%\ with respect to the \mbh\ estimates based on  \aliii\ and \hb, respectively. Considering that the \civ-based scaling law requires a large correction ($\xi_\mathrm{CIV}$\ can be as low as $\approx$ 0.2) to the \civ\ FWHM dependent on both line shift and luminosity, the \aliii\ scaling law should be preferred in case observations of both \civ\ and \aliii\ are available. 

 \begin{figure}[htp!]
\centering
\includegraphics[width=0.7\columnwidth]{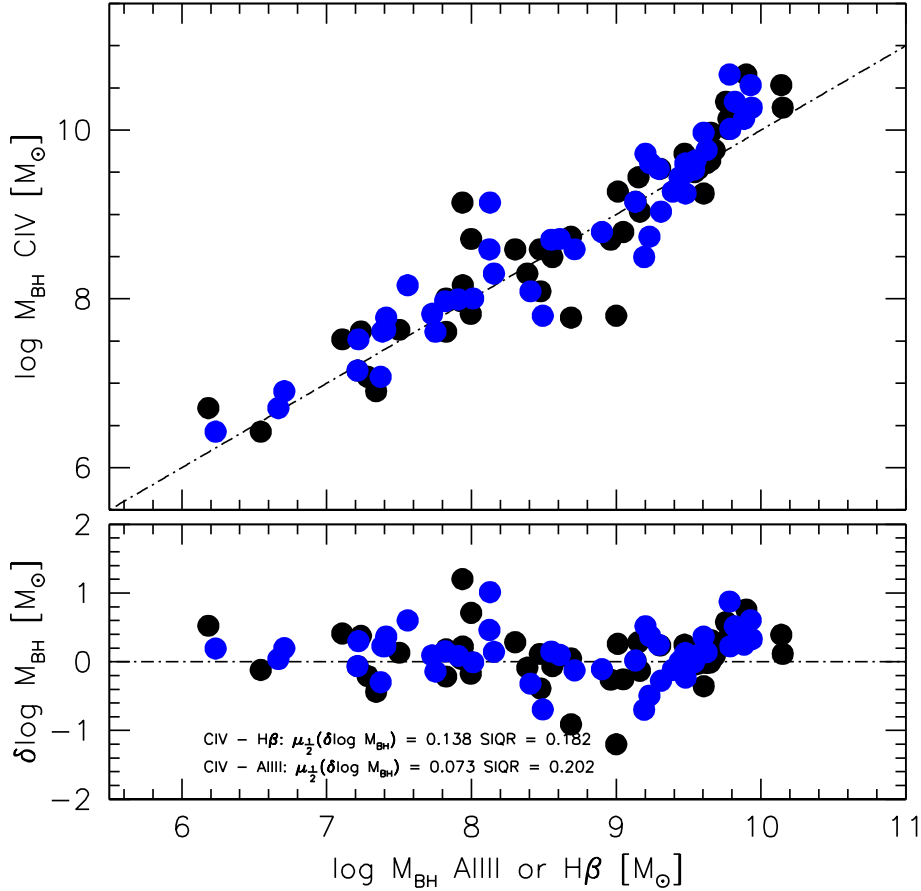}\\
\vspace{-0.2cm}
\caption{Decimal logarithm of black hole mass in units of solar masses computed from the \civ-based scaling law of \citetalias{marzianietal19} vs the ones   computed from the  relation of \citet{vestergaardpeterson06} based on FWHM \hb\ (blue) and  from the \aliii\ FWHM following Eq. \ref{eq:masslaw} (black). The dot-dashed is the equality line.   The bottom panel shows the residuals as a function of \mbh. Median  and SIQR  of the $\log $ \mbh\ differences $\delta \log M_\mathrm{BH} = \log M_\mathrm{BH}$ CIV $- \log M_\mathrm{BH}$ \aliii\ and $\delta \log M_\mathrm{BH} = \log M_\mathrm{BH}$ CIV $- \log M_\mathrm{BH}$ \hb\  are reported in the inside caption.   }
\label{fig:malc}
\end{figure}

\subsection{Shifts of \aliii\ vs shifts of \civ\ and \mgii}
\label{shifts}

Fig \ref{fig:shiftac} shows the \aliii\ peak shifts reported in Table \ref{longtable} vs. the centroid at half maximum of \civ\ from \citetalias{sulenticetal07}, \citetalias{sulenticetal17},  Deconto-Machado et al. (2022, in preparation), and  the ``peak" shift given by \citet{vietrietal18} for the four subsamples of the present work. The \aliii\ shift is usually very modest, and a factor of $\sim 10$ \ lower than the shift measured for \civ.  The \aliii\ and \civ\ shifts are however significantly correlated (Pearson's correlation coefficient $\approx 0.54$, with a significance $P \approx 1. - 2\cdot10^{-3}$): 

\begin{equation}
s(\mathrm{AlIII}) \approx (0.114 \pm 0.026) c({\frac{1}{2}})(\mathrm{CIV}) + (50.0 \pm 68.5) \mathrm{km\ s^{-1}}.   \end{equation}

\begin{figure}[htp!]
\centering
\includegraphics[width=0.7\columnwidth]{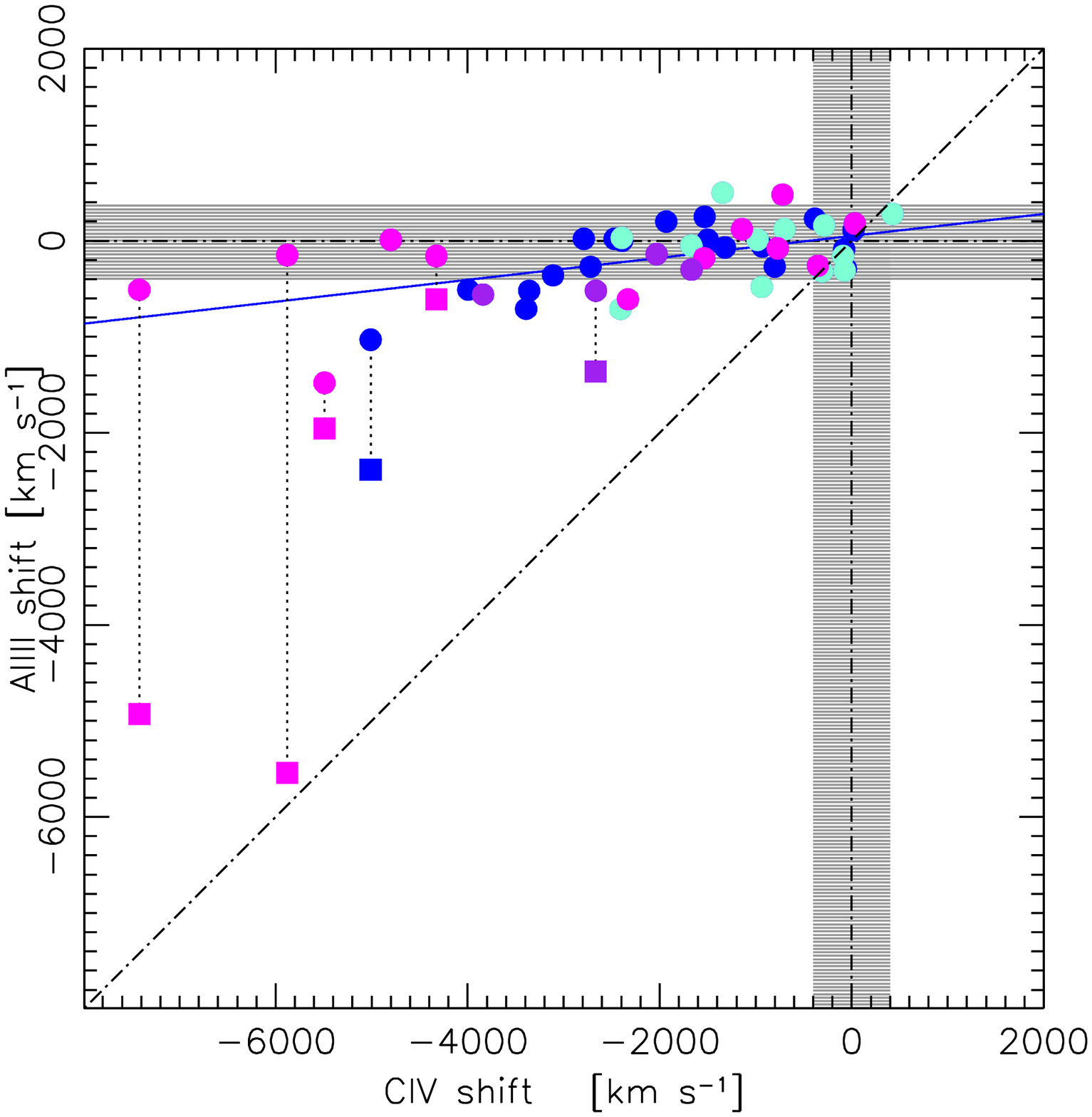}
\vspace{-0.5cm}
\caption{Shift $s$\ of \aliii\  vs shift of \civ\ with respect to rest frame. Data points are identified according to  the spectral type, with the same color code of Fig. \ref{fig:hbalst}. The filled and dot-dashed lines trace an unweighted least square fit and  the equality line, respectively. The grey bands identify uncertainties ranges in radial velocity $\pm 400$\kms\ where the lines are not significantly shifted. Squares refer to the c($\frac{1}{2}$)\ \aliii\ in the case the excess emission on the blue side of \aliii\ has been ascribed to \aliii. See Section \ref{xA}.\ \label{fig:shiftac} }
\end{figure}

The slope is shallow, but the correlation indicates that the shifts in \aliii\ and \civ\ are likely to be due to the same physical effect.  If we ascribe the small displacements observed in the peak shift $s$ of \aliii\ to the effect of  outflows, the outflow prominence is much lower than in the case of \civ, both in terms of radial velocity values and of $\delta$. 

As mentioned, \civ\ shifts and FWHM are correlated, implying that the broader the line, the higher the shift amplitude becomes \citep{coatmanetal16,sulenticetal17}.  
 The \aliii\ shows a   consistent behavior, but apparently masked by the much lower shift amplitudes; the presence of \aliii\  blueshifts  appears to be statistically  significant  at very high luminosity, and for spectral type A3 and A4.   When \civ\ shows large blueshifts i.e., for high \rfe\ or very high luminosity, the \aliii\ line becomes broader than \hb\ (even if the two lines remain in fair agreement). We see a relation between \aliii\ line shift and widths: in ST A3 and A4, where shifts are larger, the \aliii\ FWHM exceeds the one of \hb\ (Figure \ref{fig:trenda}). 
 

The \aliii\ and \civ\ results on line shifts are also consistent with  the ones obtained for \mgii. Small amplitude blueshifts of  a few hundreds \kms\ were measured on the full line profile of \mgii\  \citep{marzianietal13}. For \mgii\ the separation of a BLUE component and a symmetric Lorentzian  has been possible  on  median composite spectra because of the high S/N and of the peaky line core of the \mgii\ line.   The same operation is not feasible for individual \aliii\ profiles that are often significantly affected by noise, and in some cases even barely above noise.  The \aliii, \civ\ and \mgii\ are all resonance lines that may be subject to selective line-driven acceleration \citep{murraychiang97,proga07,risalitielvis10}. The different velocity amplitudes most likely reflect the difference in the line emitting region distance from the continuum source and in physical properties, such as ionization parameter, density and column density.

Rare sources with large shift amplitudes in \aliii\ 
are expected to be intrinsically infrequent even at the redshift where luminous quasars were fairly common ($z \approx 1.5 - 2$) and, even if over-represented because of a Malmquist-type bias, they  are outstanding and  pretty easily recognizable, especially in large samples of AGN (Sect. \ref{xA}).   The most striking case directly resembling \civ\ is the one of \object{HE0132-4313} \ which is an object of fairly low luminosity and an outlier in the plots FWHM \aliii\ vs FWHM \hb. Sources with large shift amplitude may be excluded or flagged if black hole mass estimates are being carried out.




\subsection{xA quasars}
\label{xA}

There are 16  sources   meeting the criterion \rfe $\gtrsim 1$ in the joint sample. The wide majority of these sources shows \aliii\ blueshift with respect to \hb.  The average shift is rather modest, $\approx -250$ although for 7 of them, $s \lesssim -250$ \kms. Six  objects show evidence of a strong excess on the blue side of the 1900 blend. Fits to the 1900 blend of these sources following the standard approach are shown in Appendix \ref{app:fits}. The fits have been repeated by allowing for an extra BLUE in the 1900 blend, represented as a skewed Gaussian (Fig. \ref{fig:xa}). The blueshifted excess  in some cases cannot be distinguished from   strong \siii\ emission. The \siii\ emission line can be of strength comparable to \aliii\ in the condition of low ionization and high density derived for the virialized component \citep{negreteetal12}. 

In two cases (e.g., \object{HE0359-3959},and  \object{SDSS J152156.48+520238.5}) the blueshifted excess is so overwhelming that \siii\ emission cannot account for the excess unless the \siii\ line itself is significantly blueshifted. These are perhaps the best cases supporting the evidence for a significant BLUE in \aliii. It is reasonable to assume  that the blueshifted excess is mainly due to \aliii, being \aliii\ a resonant line for which BALs are also observed. Broad absorption components are observed in the \aliii\ profile, even if  rarely, and with terminal radial velocity of the absorption through  much lower than the one of \civ\ \citep{gibsonetal09}. 


The Eddington ratio values derived from the continuum luminosity at 1700 \AA\ (after Galactic extinction correction) multiplied by a bolometric correction factor { $k_\mathrm{bol} \approx$} 3.5, and from the \mbh\ estimated from Eq. \ref{eq:masslaw}  range between \lledd $\approx 0.2$  and \lledd $\approx 1.5$. { If the luminosity-dependent $k_\mathrm{bol}$\ is applied to   the 1700 \AA\  continuum, $0.2 \lesssim $\lledd $\lesssim 1.1$. } { Both estimates confirm} that all quasars of the presented sample are within the range expected for Pop. A sources. The xA sources are at the high end of the \lledd\ distribution, with $\mu(\log$\lledd)$\approx -0.18$, and the \aliii\ BLUE sources are even more extreme with $\mu(\log$\lledd)$\approx -0.105$. Extreme radiation forces may  make it possible  to blow out rather dense/high column density gas from the virialized region associated with the emission of the low- and intermediate ionization lines \citep{netzermarziani10}. Sources showing a strong BLUE in \aliii\ could be the most extreme accretors,  perhaps in a particular “blow-out” phase of the quasar  evolution \citep[][]{donofriomarziani18}.

\begin{figure}
\includegraphics[width=0.225\columnwidth]{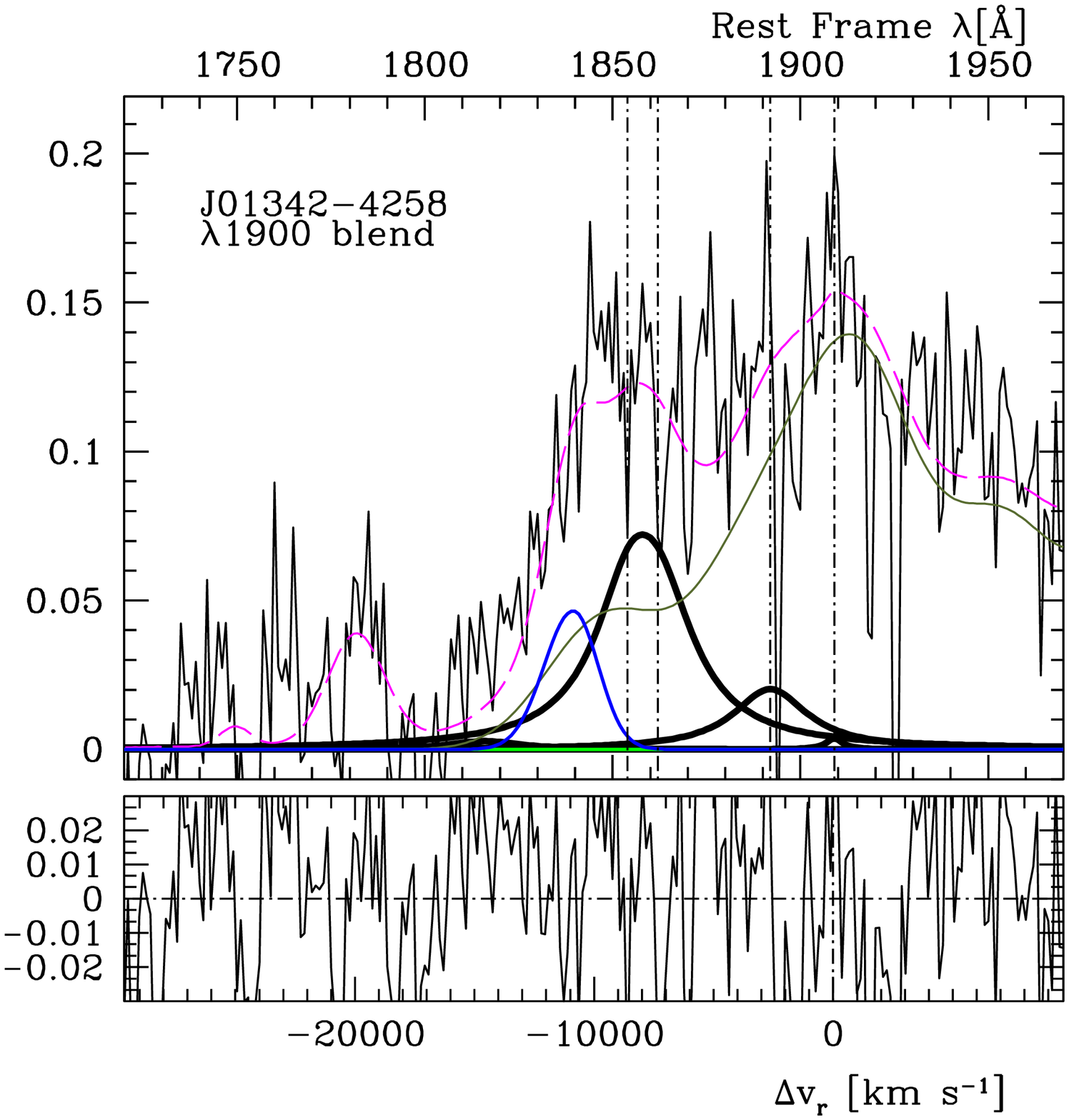}\includegraphics[width=0.225\columnwidth]{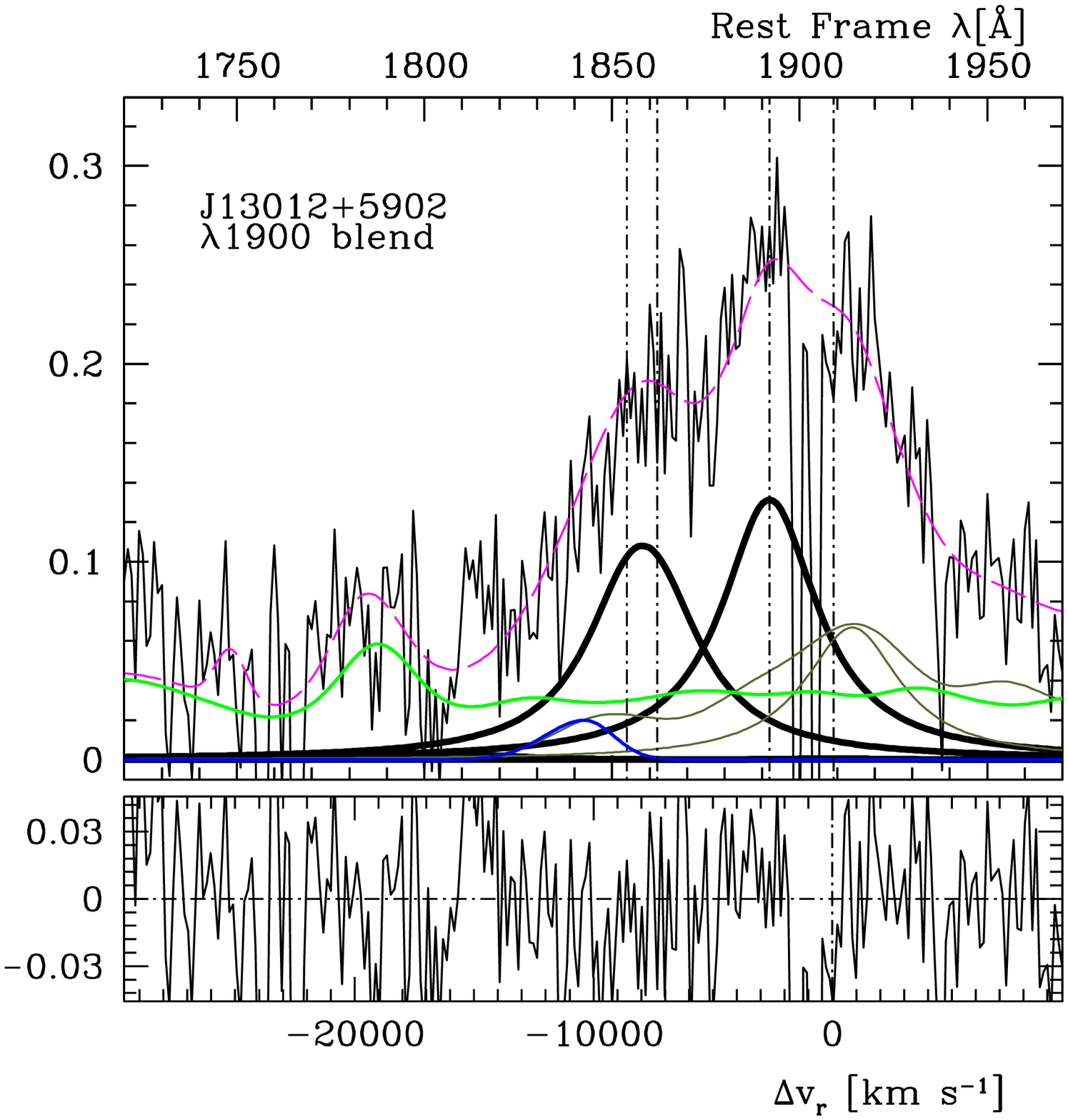} 
\includegraphics[width=0.225\columnwidth]{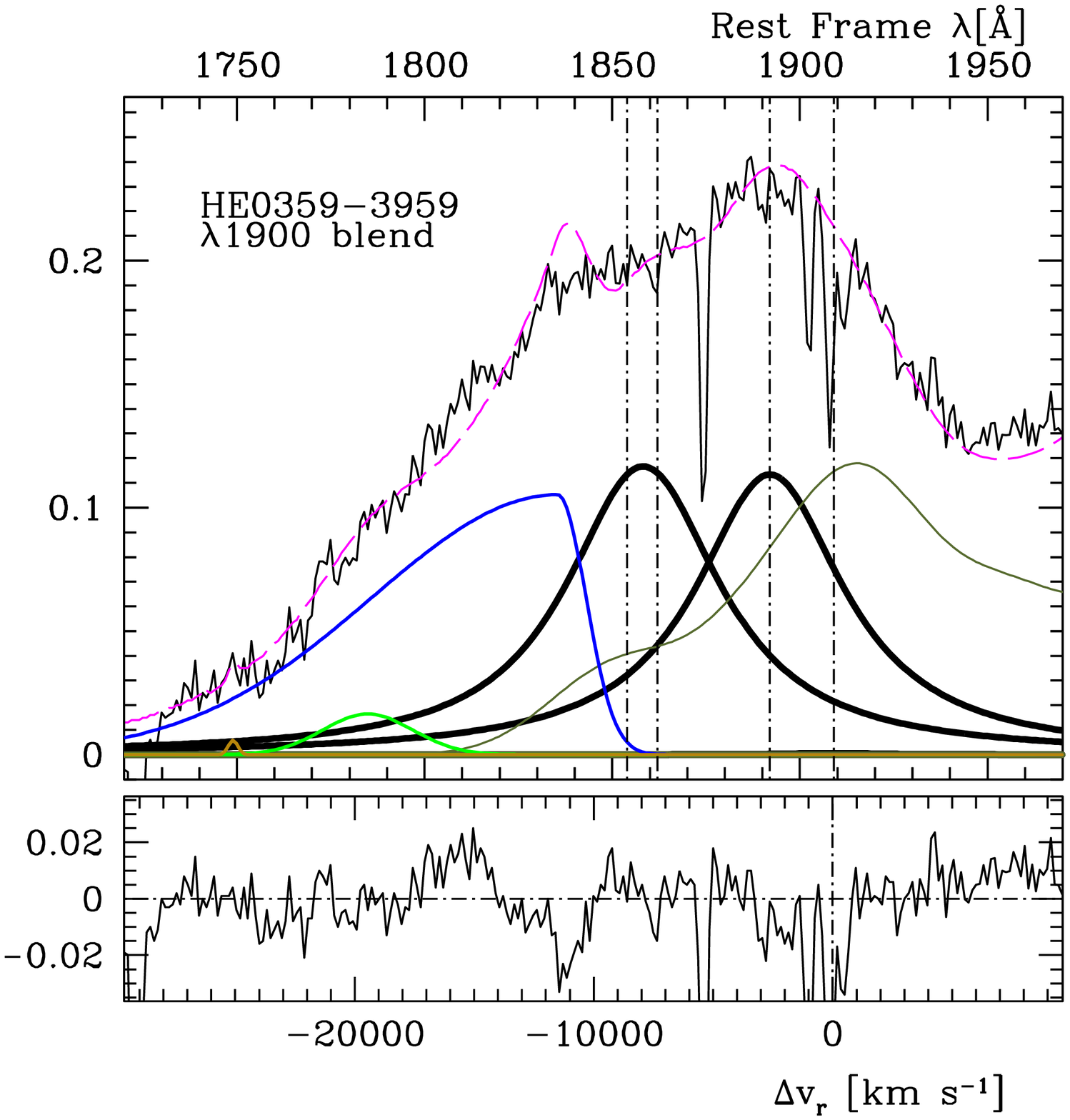}\includegraphics[width=0.225\columnwidth]{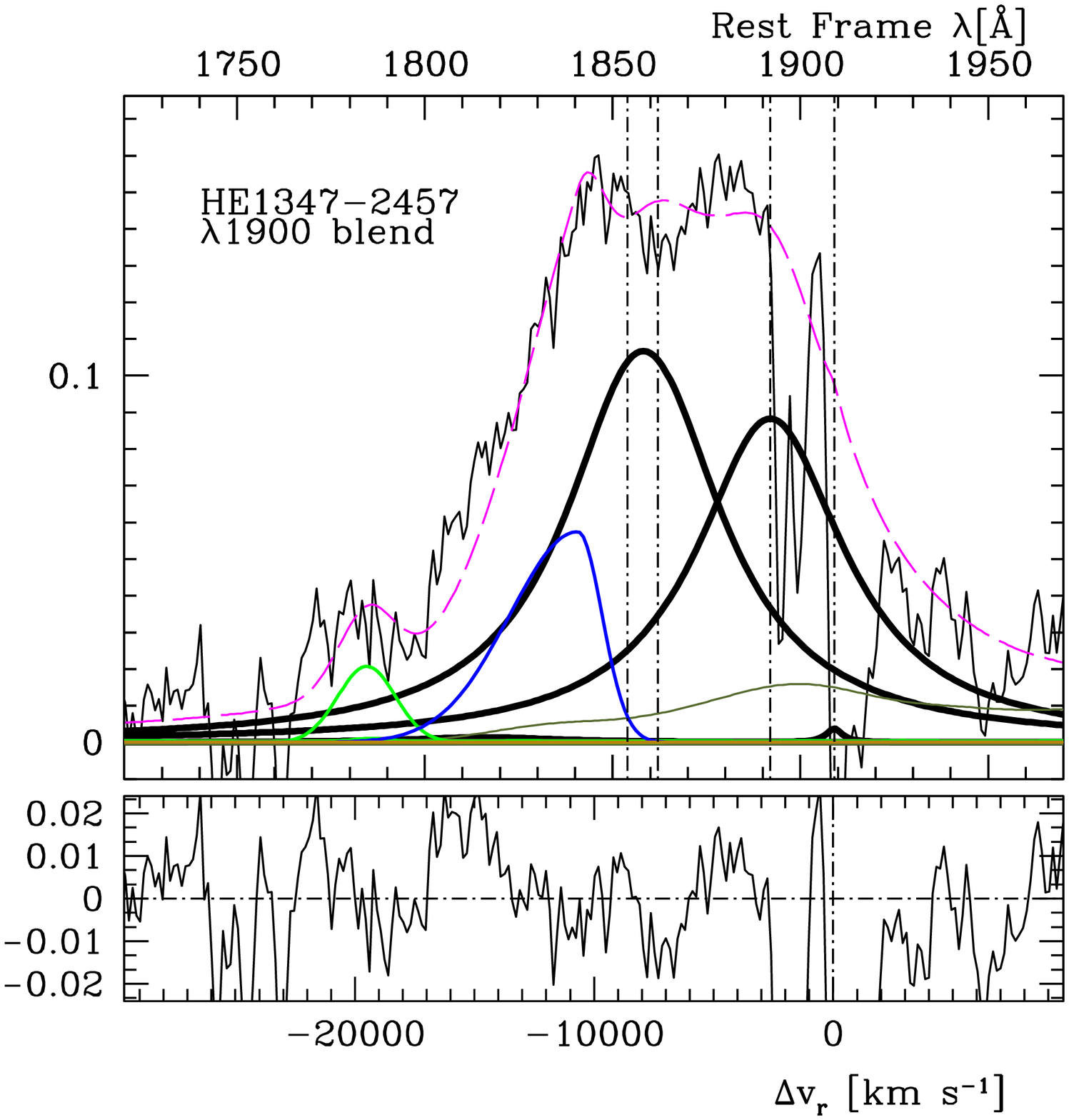}\\ 
\includegraphics[width=0.225\columnwidth]{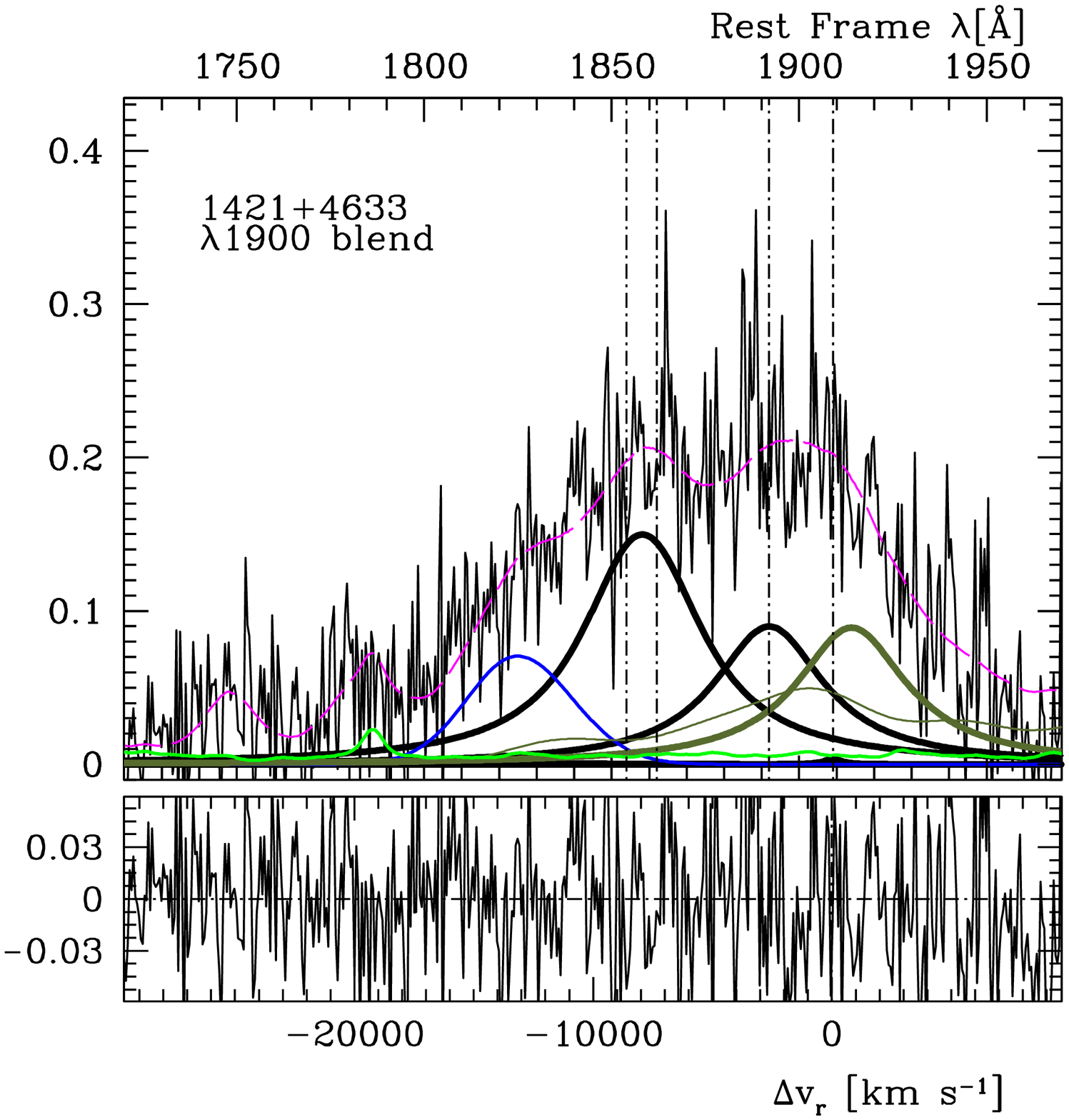}\includegraphics[width=0.225\columnwidth]{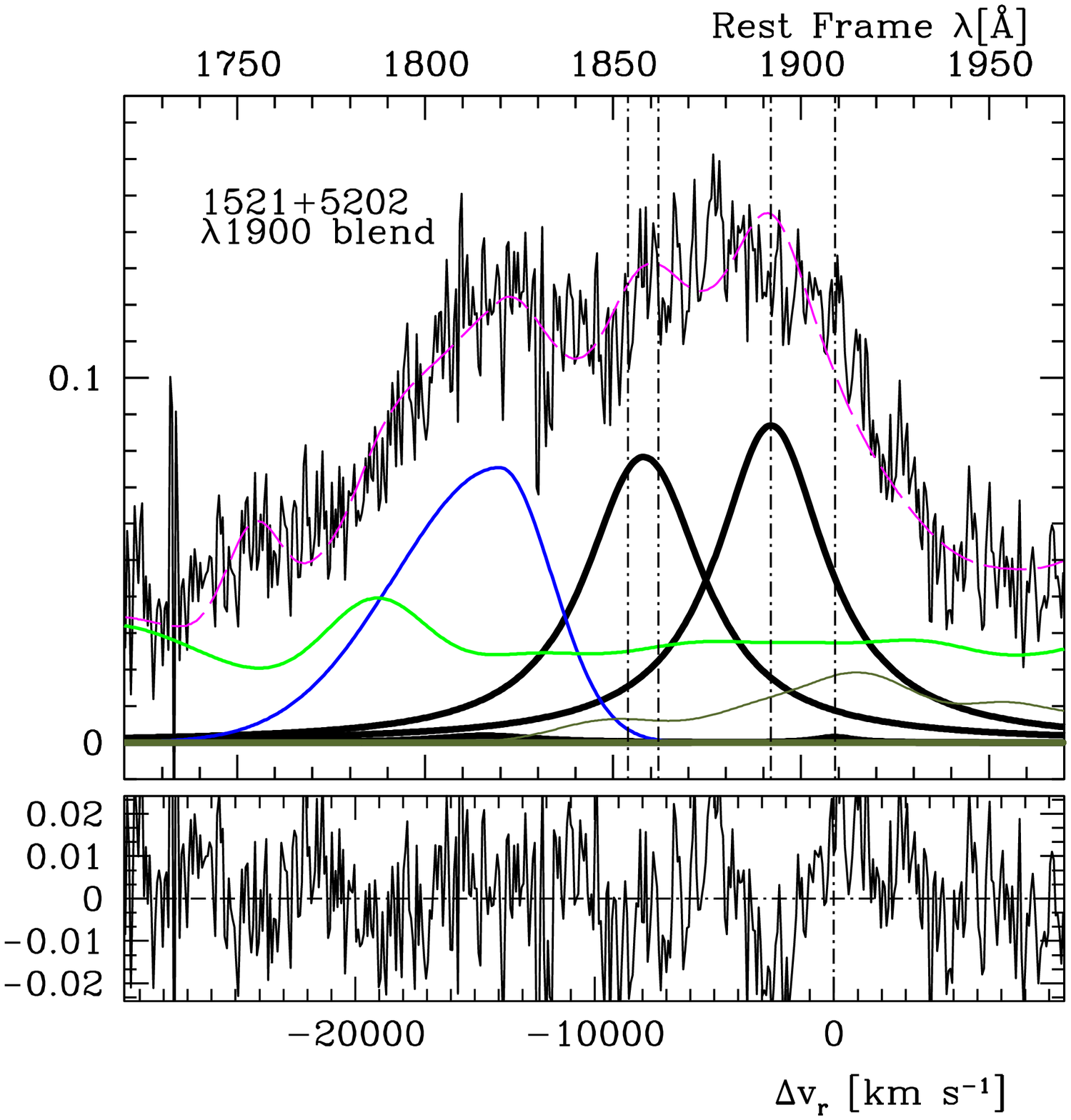} 
\caption{Analysis of the 1900 \AA\ blend  for the estreme Population A sources with excess emission on the blue side of \aliii.  Abscissa scales are rest-frame wavelength in \AA\ and radial velocity from rest frame wavelength of \ciii.  Ordinate scale is  normalized specific flux by the value at 1700 \AA.  The black lines identify \aliii, \siiii. The blueshifted excess BLUE is traced by a thick blue line. Green lines trace the adopted \feii\ (pale) and \feiii\ (dark) templates. Note that \ciii\ emission is almost absent, as all of the emission on the red side of \siiii\ can be ascribed to \feiii.   \label{fig:xa}}
\end{figure}


A related issue is whether sources with a strong blue-shifted component in \aliii\ can be empirically  distinguished from the rest of Pop. A quasars, without resorting to the knowledge of the rest frame. The FWHM of the whole blend (i.e., of the sum of all lines after continuum subtraction) is clearly affected by spectral type: going from A1 to A4 we see an overall decrease of prominence in \ciii, and an increase in \aliii\ with respect to the other line. The blue-shifted excess should further increase the FWHM of the blend. The parameter 

\begin{equation}
\mathcal{A} = \frac{\mathrm{FWHM_{1900}}}{\left(10^{\log (\lambda L_\lambda)_{1700} - 44}\right)^{0.25}}
\end{equation}

normalizes the FWHM of the whole blend $\mathrm{FWHM}_{1900}$\ (i.e., the sum of all line components) because of the increase of the line width with luminosity by a factor $L^\frac{1}{4}$. Fig. \ref{fig:a} shows the distribution of the $\mathcal{A}$ for the sources of ST A1 and A2, A3 and A4, and the 6 quasars for which the 1900  blend was fit with the addition of a blue-shifted excess. The distributions of A1+A2 and A3+A4 are significantly different at a 3$\sigma$\ confidence level according to a K-S test. However, there is considerable overlap around $\mathcal{A} \approx 3000$ \kms, making it difficult to unambiguously distinguishing between xAs, xAs with blue-shifted excess and other sources. 


\begin{figure}[htp!]
\centering
\includegraphics[width=0.85\columnwidth]{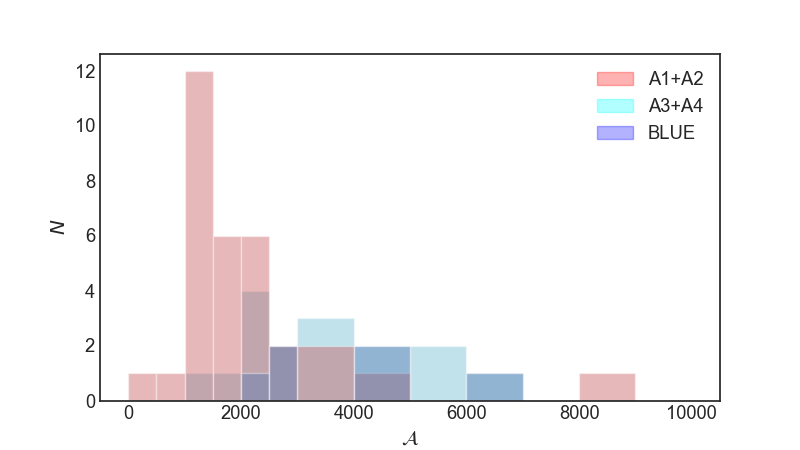}\\
\vspace{-0.25cm}
\caption{The distribution of the parameter $\mathcal{A}$, as defined in Sect. \ref{xA}, for three groups of quasars: spectral types A1+A2,  A3+A4, and the 6 quasars with a blue shifted excess fit to the 1900 \AA\ blend (Fig. \ref{fig:xa}). }
\label{fig:a}
\end{figure}

\subsubsection{Consistency of UV and optical classification for extreme Population A}

Extreme quasars can be identified by employing selection criteria in the optical and UV  \citepalias{marzianisulentic14}. The consistency of the selection criteria is however little tested, since observations covering both the 1900 \AA\ range and the \hb\ one are still rare. In the joint sample    most sources with \rfe $\gtrsim 1$\ also satisfy  the UV intensity-ratio conditions \aliii/\siiii $> 0.5$, and \siiii$>$\ciii. Figure \ref{fig:xabox} shows the location of the data points identified according to spectral type (defined by ranges of the optical parameter \rfe; A3 and A4 satisfy the condition \rfe $\gtrsim$ 1 by definition)  in  the plane defined by the UV ratios \ciii/\siiii\ vs. \aliii/\siiii.  There are several borderline cases,  but only one in which the criteria are  not satisfied: \object{J15591+3501} with intensity ratios \ciii/\siiii $\approx 2.38 \pm 0.45$, and \aliii/\siiii $\approx 0.33 \pm 0.06$. For \object{J14421+3526}, the feature at $\approx 1910$\ \AA\ is most likely a blend of \feiii\ and \ciii. In this case, only an upper limit can be assigned to the \ciii\ intensity, and the UV selection criteria may  not have been violated. The reason of the discordance for  \object{J15591+3501} is not clear.   The    majority of objects ($\approx 80$\%) in Figure \ref{fig:xabox}    supports the equivalence between the two xA selection criteria suggested by \citetalias{marzianisulentic14}.  Apart from borderline cases, five  A2 sources (4 if we exclude \object{J1421+4633} with \rfe $\approx$ 0.99) out of 19 enter    the domain of the xA (the grey shaded area of Figure \ref{fig:xabox}). These sources appear to be genuine xA in terms of the UV intensity ratios, but have lower than expected \rfe\ ($0.5 \lesssim $ \rfe $\lesssim$ 1). It is intriguing that the  four sources all belong to the high-$z$\ samples. The possibility of systematic differences as a function of redshift in the relative abundance of iron with respect to carbon and $\alpha$\ elements   should be further investigated \citep[e.g.,][and references therein]{martinez-aldamaetal21}.

\begin{figure}
\includegraphics[width=0.85\columnwidth]{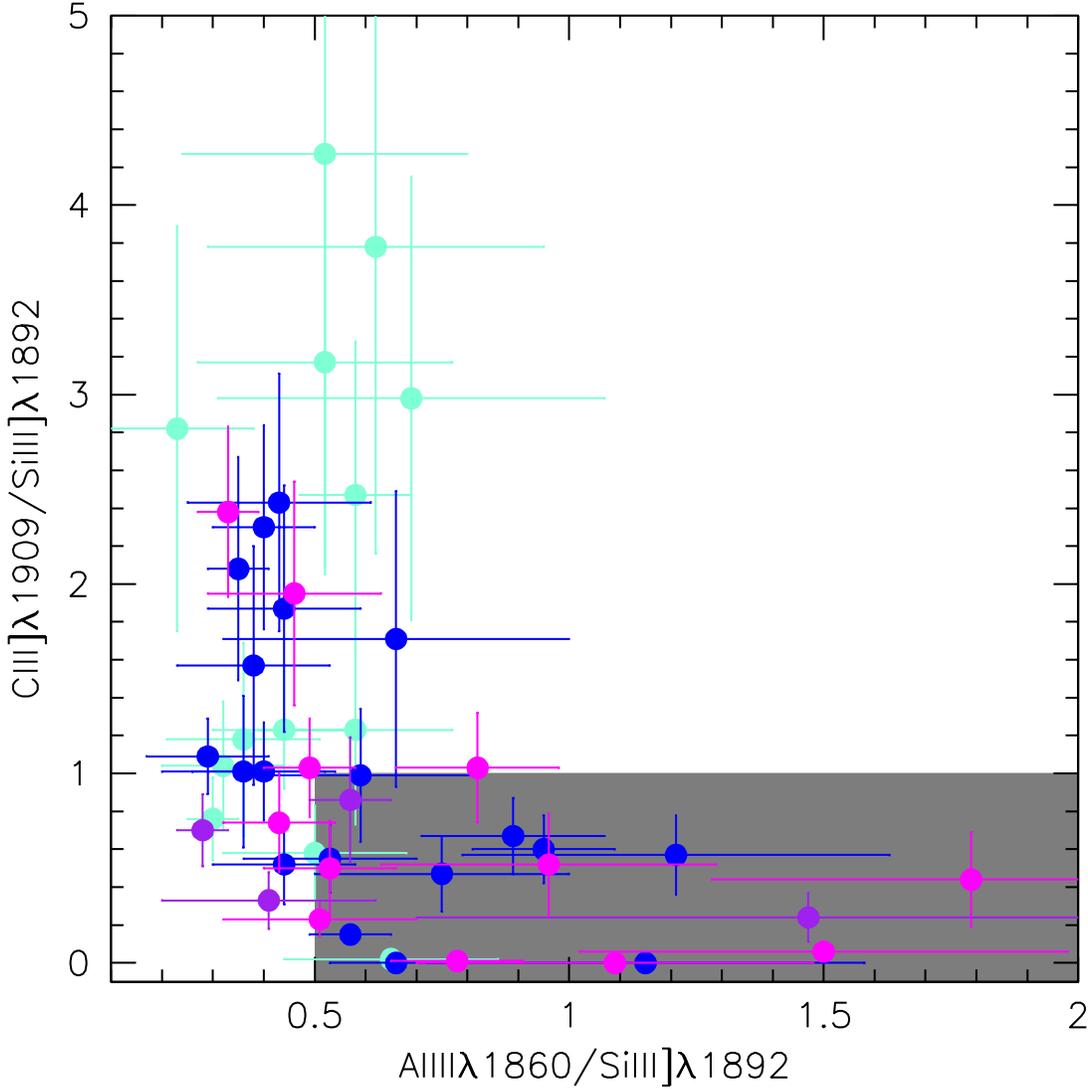} 
\caption{The distribution of xA   sources  in the plane defined by the intensity ratios  \ciii/\siiii\ vs \aliii/\siiii. The area associated with xA sources is the lower-left shaded box. Sources are color coded according to spectral type as in the previous Figures.   \label{fig:xabox}}
\end{figure}

 \section{Summary and conclusion}

The present investigation has shown a substantial equivalence of \hb\ and \aliii\ and \ciii\ as virial broadening estimators for Population A quasars, thereby providing a  tool suitable for \mbh\ estimates up to $z \lesssim 4$ from observations obtained with optical spectrometers. More in detail, the { salient } results of the present investigation  can be summarized as follows:


\begin{itemize}
\item the \aliii\ and \hb\ FWHM are highly correlated and, in the joint sample of 48 Population A sources, can be considered statistically equivalent  { over 4 orders of magnitude in luminosity (Sections \ref{imme} and \ref{fwhm})}. 

\item  The FWHM ratio between \aliii\ and \hb\ increases with increasing \rfe\ or, equivalently, from spectral type A1 to A4 { (Sect. \ref{sptype}).  Extreme Pop. A sources appear to be 20\%\ broader than the sample average, while spectral type A1 20 \%\ narrower than spectral type A2.} 

\item { Systematic blueshifts are revealed in \aliii; however, in most cases the amplitude of the blueshifts is modest or smaller than the uncertainties, reaching a sample median for spectral types A3 and A4 of just $\approx -160$ \kms\ (Sections \ref{sptype} and \ref{lum}). }  
\item { The line FWHM of \hb, \aliii\ and \ciii\ increases with luminosity as a function of $L^\frac{1}{4}$, as expected for a virial velocity field of the line emitting gas (Sect. \ref{hbciii}).  }


\item { The following scaling law between \mbh\ and luminosity and FWHM \aliii: $\log M_\mathrm{BH}\, (\mathrm{AlIII})   \approx   (0.579^{+0.031}_{-0.029}) \log L_{1700,44} +    2 \log \left(\mathrm{FWHM}(\mathrm{AlIII}) \right) + (0.490^{+0.110}_{-0.060}) $\ (Eq. \ref{eq:masslaw} in Sect. \ref{mass}) }  provides an estimate of \mbh\ with an rms scatter of $\approx$ 0.3 dex with respect to the \hb-derived masses,   {   for \mbh $\gtrsim 10^7$ \msol.}   
\item An { analogous} scaling law has been defined also for \ciii\ { (Eq. \ref{eq:masslawc3} in Sect. \ref{massciii}).} The measurement of the  \ciii\ FWHM is however more strongly affected by the severe blending and by the \ciii\ weakness  in sources with high \rfe. The \ciii\ scaling law requires a constant correction factor to the FWHM of \ciii, $\xi_\mathrm{CIII]} \approx  1.25$. { The scaling laws derived from \aliii\ and \ciii\ line width are mutually consistent (Sect. \ref{consist})}.


\item { Although \aliii\ shift amplitudes are $\approx \frac{1}{10}$ the shifts of \civ\ (Section \ref{shifts}), } it is   unclear  whether \aliii\ can be exploited as a virial luminosity estimator for extreme Population A sources { (Section \ref{xA}): the \aliii\ profile is strongly affected by a blueshifted excess in several extreme Pop. A sources (Sect. \ref{xA}).}  The majority of quasars show consistency between FWHM \aliii\ and \hb, and  a minority of sources that show FWHM \aliii $\gg$ FWHM \hb\ might be easy to recognize in large samples. The extent of systematic effects should however be analyzed  by a thorough study of a very large sample of xA sources with full coverage of the optical and UV rest-frame ranges from 1000 to 5500 \AA. 
 \end{itemize}
 
{ These results show that the \aliii\ line is a good UV substitute of \hb\ and can be used for black hole mass estimations  with the  advantage to be at the rest-frame of the source. }
  The results on FWHM \aliii\  should be compared with the ones obtained for \civ\ \citepalias{marzianietal19}, where the equivalence was obtained at the expense of corrections that were dependent on the accurate knowledge of the quasar rest frame, and therefore not fully achievable without additional measurements in spectral ranges distinct from the one of \civ: the \oii\ line is the narrow low ionization line closest in wavelength to \civ\ and offers a reliable rest frame estimator \citep{bonetal20}, but is very rarely covered  along with \civ.

\acknowledgments

 The authors are grateful for the contribution of Jack Sulentic to the early development of this paper.  
 { PM, AdO and ADM acknowledge financial support from from the Spanish grants MCI PID2019-106027GB-C41 and the State Agency for Research of the Spanish MCIU through the ``Center of Excellence Severo Ochoa" award for  the Instituto de Astrof\'{\i}sica de Andaluc\'{\i}a (SEV-2017-0709).} { ADM acknowledges the support of the INPhINIT fellowship from ``la Caixa" Foundation (ID 100010434).The fellowship code is LCF/BQ/DI19/11730018.}
 PM  was supported  by  the Hypatia of Alexandria visiting grant  during a two-month stay in  2020 at the Instituto de Astrof\'{\i}sica de Andaluc\'{\i}a  where part of the analysis was carried out. NB and EB acknowledge the support of Serbian Ministry of Education, Sci-ence and Technological Development, contract number 451-03-68/2022-14/200002. DD acknowledges support from grant PAPIIT UNAM IN-113719.

Funding for the Sloan Digital Sky Survey (SDSS) has been provided by the Alfred P. Sloan Foundation, the Participating Institutions, the National Aeronautics and Space Administration, the National Science Foundation, the U.S. Department of Energy, the Japanese Monbukagakusho, and the Max Planck Society. The SDSS Web site is http://www.sdss.org/.

The SDSS is managed by the Astrophysical Research Consortium (ARC) for the Participating Institutions. The Participating Institutions are The University of Chicago, Fermilab, the Institute for Advanced Study, the Japan Participation Group, The Johns Hopkins University, Los Alamos National Laboratory, the Max-Planck-Institute for Astronomy (MPIA), the Max-Planck-Institute for Astrophysics (MPA), New Mexico State University, University of Pittsburgh, Princeton University, the United States Naval Observatory, and the University of Washington.
\vfill

\vspace{5mm}





{NUMPY} (\citealt{numpy});  {R} (\citealt{r_stats});   {SuperMongo} (\citealt{supermongo}).

\clearpage
\appendix

\section{Estimation of  uncertainties}
\label{app:unc}

\subsection{Bayesian estimates}

Uncertainties were estimated following a Bayesian approach, considering the likelihood function 
\begin{equation}
\log {\mathcal L} \propto - \sum_{i} \frac{(f_\mathrm{i} - m_\mathrm{i}({\bf \Theta}))^2}{2\sigma_\mathrm{i}^2} \propto -\frac{1}{2} \log \chi^2,
\end{equation}
where $f_\mathrm{i}$ are the specific flux values as a function of wavelength (or of pixel number), $\sigma_\mathrm{i}$\ the uncertainty in $f_\mathrm{i}$\ (in practice from the S/N set constant over the spectrum), $m_\mathrm{i}({\bf \Theta})$\ the expectation value for the multicomponent model ${\bf \Theta}$ of the spectrum obtained via a {\tt specfit} analysis. The ${\bf \Theta}$ can be any set of free parameters employed in the fits: intensity, shift and width of each line, intensity, shift and width of each template.  Priors were specified for several parameters in terms of a range of permitted values. The posteriors of  the model parameters ${\bf \Theta}$\ (for instance, the distributions of FWHM \hb\ and \aliii\  given the data) were obtained by creating a random walk with a modified Metropolis-Hasting algorithm: a new candidate set of model parameters ${\bf \Theta}$\ was randomly generated, and screened by an accettance parameter $\alpha$. The set ${\bf \Theta}$\ included model parameters believed to significantly affect the line widths (in practice, most of the parameters included in the {\tt specfit} analysis). For example, the \oiiiopt\ lines were modeled with two components, a ``core" component represented by a symmetric Gaussian, and a semi-broad component modeled by a skew Gaussian. The template \feii\ emission was scaled, shifted and broadened as done in the {\tt specfit} procedure.    
The dispersion of  the   posterior distribution of each spectral  parameter was assumed to yield its uncertainty $\delta$\ at $1 \sigma$\ confidence level. 


\subsection{The quality parameter ${\mathcal Q}$}

 The next step was to connect the uncertainty in FWHM, shift and intensity to a quality parameter  ${\mathcal Q}$, which may turn useful in case very late samples of quasars are analysed.   The quality parameter 

\begin{equation}
{\mathcal Q} = \log_{10}\frac{W}{\mathrm{FWHM}} \cdot {\frac{\mathrm S}{\mathrm N}}
\end{equation}

defined as the product of the S/N times a line  equivalent width $W$ divided by its FWHM, increases with S/N and line prominence over the continuum and decreases with increasing line widths. The signal in each resolution element is proportional to the ratio $W$/FWHM, which is a measurement of the sharpness of the line, as  obviously ${\mathcal Q} \propto \log_{10}\frac{I_\mathrm{peak} \cdot \mathrm{FWHM}}{\mathrm{FWHM} I_{\mathrm{c}}} \cdot {\frac{I_\mathrm{c}}{\mathrm N}} \propto \frac{I_\mathrm{peak}}{\mathrm N}$.  The quality parameter ${\mathcal Q}$\
obviates to the inadequacy of the S/N measurement carried out on the continuum. By multiplying it by the ratio $W$/FWHM we compute a more apt average S/N for   a line depending on its strength and width. 
The parameter ${\mathcal Q}$\ is larger for sharp lines in spectra with high S/N in the continuum.  The large differences in S/N, line width and line strength between \aliii\ and \hb\ is reflected in the distribution of the ${\mathcal Q}$\ parameter, shown in Fig. \ref{fig:qhistos}. 

To be of any practical use, the  ${\mathcal Q}$\ parameter needs to be anchored to estimates of the uncertainties. The posterior distributions of the spectral parameters were computed for about 30 sources.    Fig. \ref{fig:predict} shows a well-defined trend between ${\mathcal Q}$ and the fractional uncertainty  $\delta$FWHM/FWHM for \hb,  \aliii, and \ciii\ derived from the MCMC simulations. Especially for large ${\mathcal Q}$\ values, the scatter is relatively modest, and the relation between the   parameter FWHM, flux and shift and  $\log {\mathcal Q}$\ can be written in a linear form, save for the fractional uncertainty of FWHM \aliii\ that is best fit by $\delta \mathrm{FWHM}/\mathrm{FWHM} \approx 1/(a + b\log {\mathcal Q}) $. Table \ref{tab:r} provides the coefficients $a$\ and $b$\ of the best fits along with ${\mathcal Q}$\ domain. The FWHM relations  were obtained by a non-linear fit algorithm implemented in {\tt R} \citep{r_stats}, and  are shown as the thick lines in Fig. \ref{fig:predict}. 

%
%
%


 
\begin{table*} 
\begin{center}
 \caption{Relation between fractional uncertainties and ${\mathcal Q}$ \label{tab:r}}
\tabcolsep=3pt
\begin{tabular}{lccc}
 \hline \hline
Parameter &   $a$  & $b $ &  $\log{\mathcal Q}$\ domain   \\
\hline
\\
\hline
 \multicolumn{4}{c}{\hb} \\ 
\hline
 FWHM\tablenotemark{$a$}  & 0.100 & --0.125 & $\approx -0.8 \ldots 0.5$ \\   
 F\tablenotemark{$b$} &     0.0666 & --0.07022 &     \\
 Shift\tablenotemark{$c$} & 66 &-0.112 &   \\
\hline
 \multicolumn{4}{c}{\aliii} \\ 
\hline
 FWHM\tablenotemark{$d$}  & 10.96 & 6.15 & $\approx -1.6 \ldots -0.6$ \\   
 F\tablenotemark{$b$} & 0.061 & $-0.2277$           \\
 Shift\tablenotemark{$c$} &  180 & $-219.7$  \\
\hline
 \multicolumn{4}{c}{\ciii} \\ 
\hline
 FWHM\tablenotemark{$a$}  & 0.160 & --0.026 & $\approx -2.5 \ldots 0.0$  \\   
 F\tablenotemark{$b$} &   $-0.026$ & $-0.292$ &        \\
\hline
\multicolumn{4}{c}{\feii}\\
\hline
F\tablenotemark{$b$} & -0.04755 & 0.07520 & $\approx -0.8 \ldots 0.5$ \\ 
\hline
\end{tabular}\\
\tablenotetext{a}{ $\delta \mathrm{FWHM}/\mathrm{FWHM} \sim a + b \log{\mathcal Q}$}  
\tablenotetext{b}{$\delta F/F \sim a + b \log {\mathcal Q}$}  
\tablenotetext{c}{$\delta s = a + b \log {\mathcal Q}$ [\kms]}
\tablenotetext{d}{ $\delta \mathrm{FWHM}/\mathrm{FWHM} \sim 1/(a + b 
\log{\mathcal Q})$}  
\end{center}
\end{table*}

\begin{figure}
\includegraphics[width=0.45\columnwidth]{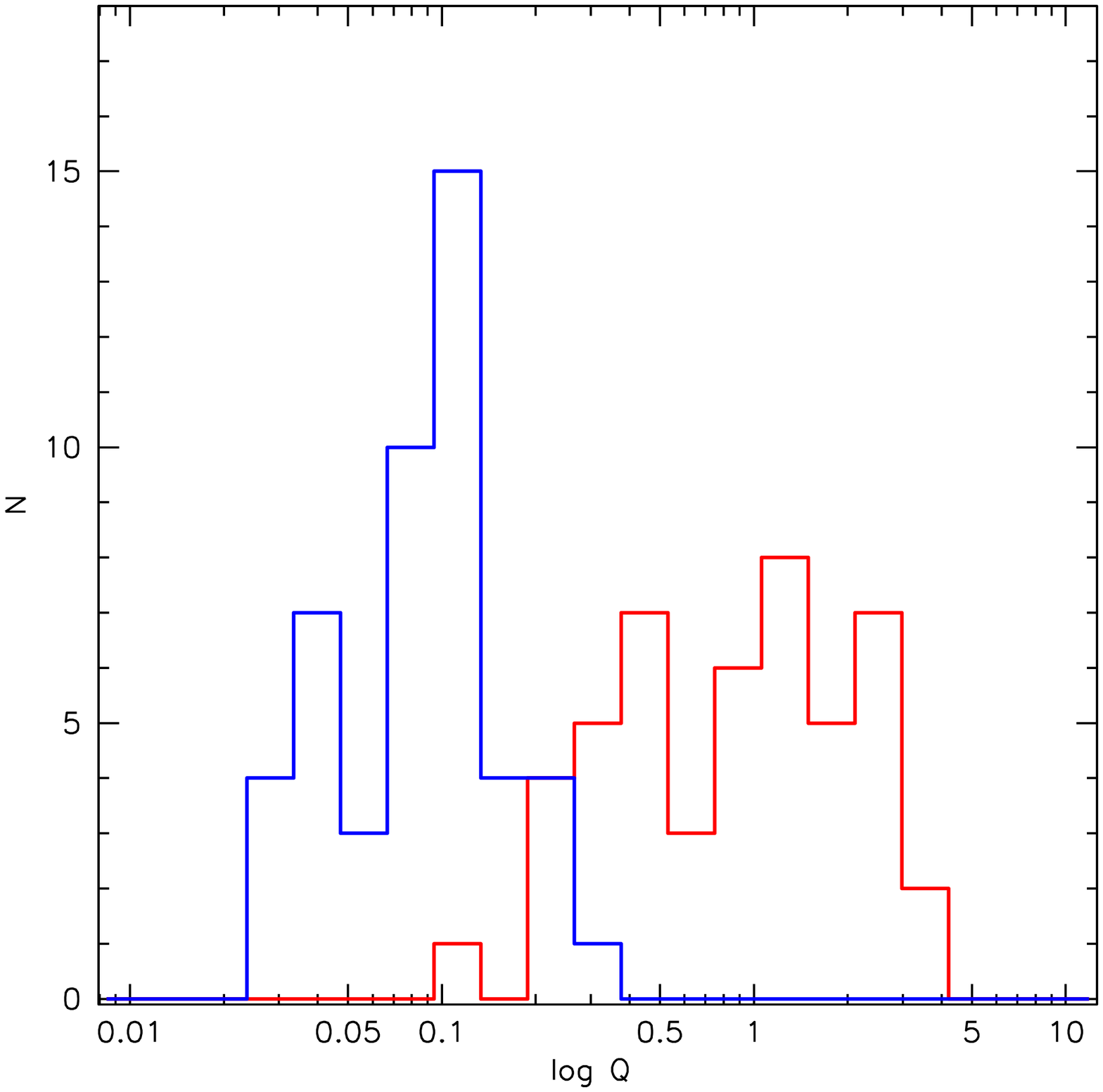} 
\includegraphics[width=0.45\columnwidth]{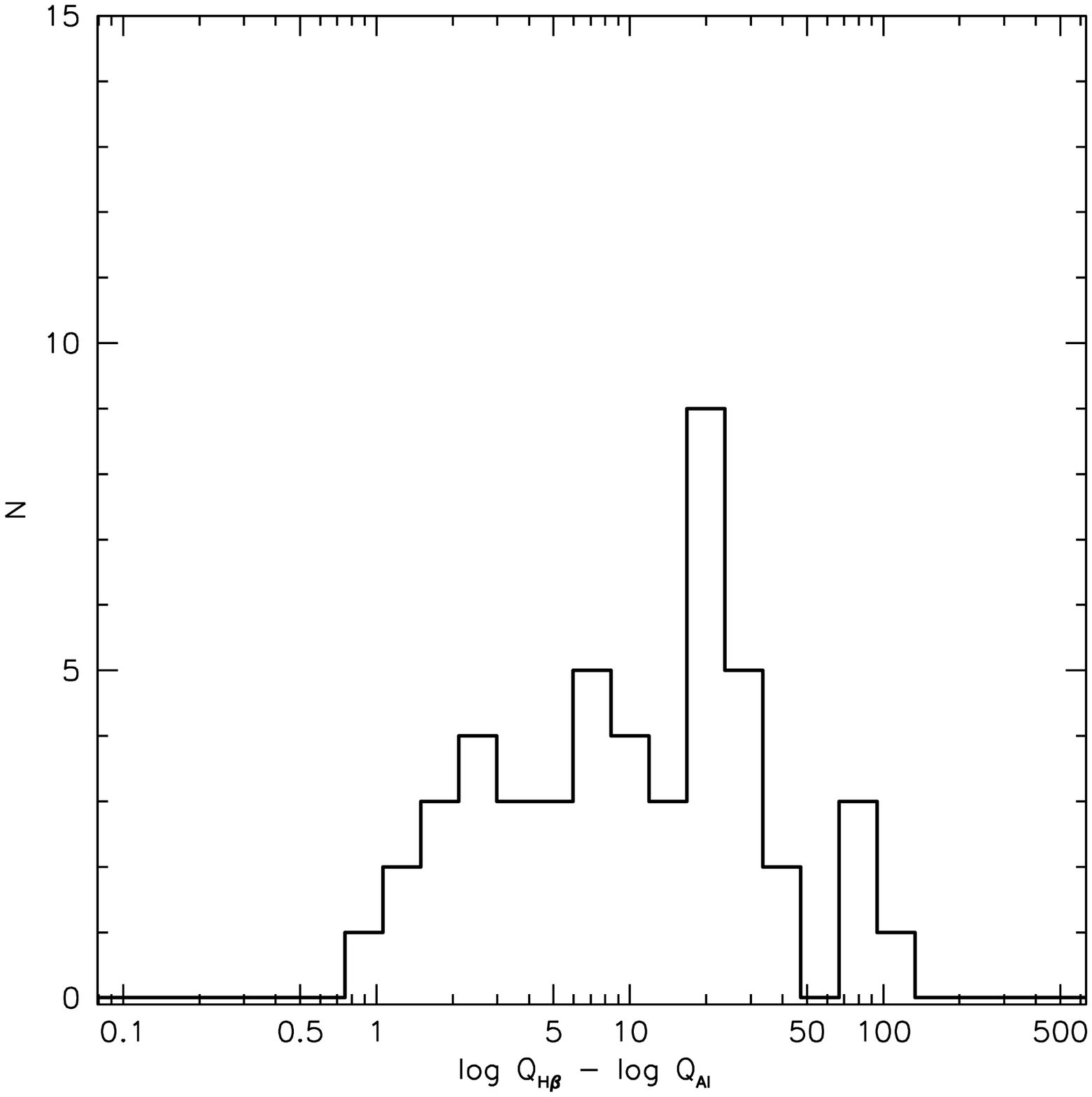}\\
\caption{Left: distribution of the $\log$\ of parameter ${\mathcal Q}$\ parameter for \aliii\ (blue) and \hb\ (red) for the joint sample considered in this paper. Right: distribution of  $\log {\mathcal Q}$\  differences between \hb\ and \aliii.\label{fig:qhistos}}
\end{figure} 
\vfill

\begin{figure}
\includegraphics[width=0.65\columnwidth]{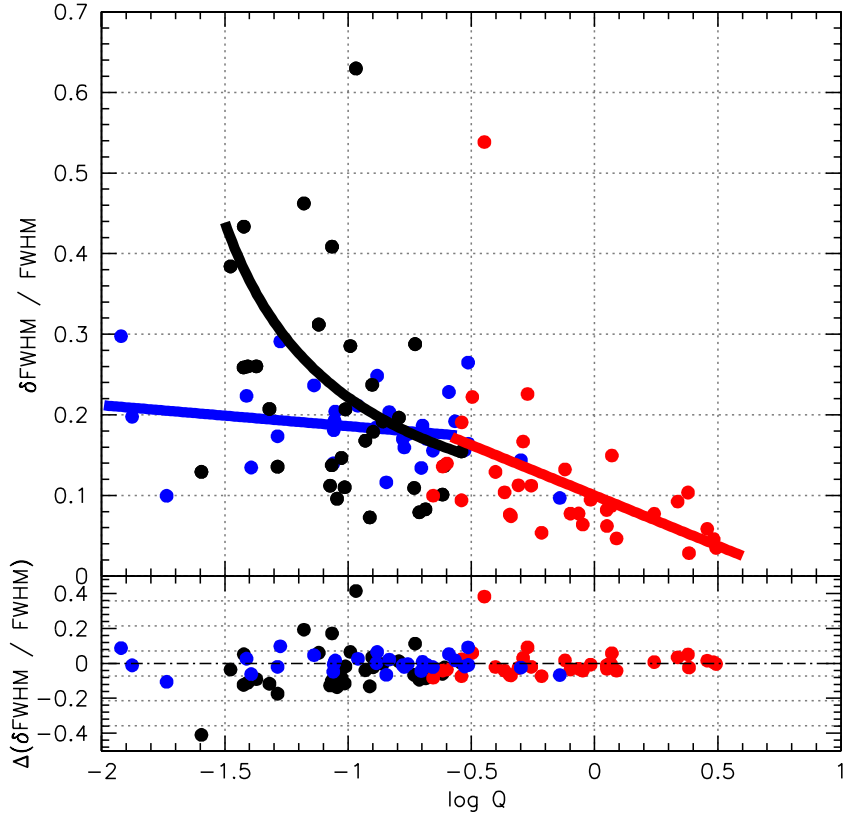} 
\caption{Relation between the fractional uncertainty ${\delta \mathrm{FWHM}}/{\mathrm{FWHM}}$ and the  logarithm of parameter ${\mathcal Q}$. Red spots are for \hb,  black for \aliii, blue for \ciii. The thick lines trace the relations reported in Table \ref{tab:r} { for the FWHM of \aliii\ (black) and \ciii\ (blue).}   \label{fig:predict}}
\end{figure} 
\vfill

\clearpage
\null\phantom{Pollo}

\section{\hb\ and 1900 blend paired comparison}
\label{app:fits}

The results of line profile fitting for the 1900 blend and \hb\ are shown in the following figures: Fig. \ref{fig:fosa} for the FOS sample, Fig. \ref{fig:hea} for the HE sample, Fig.  \ref{fig:isaaca} for the FOS sample, and Fig.  \ref{fig:wissha} for the WISSH sample. All spectra have been continuum subtracted and normalized by the 5100 \AA\ (\hb) and 1700 \AA\ (1900 blend) specific flux.

\begin{figure}
\includegraphics[width=0.225\columnwidth]{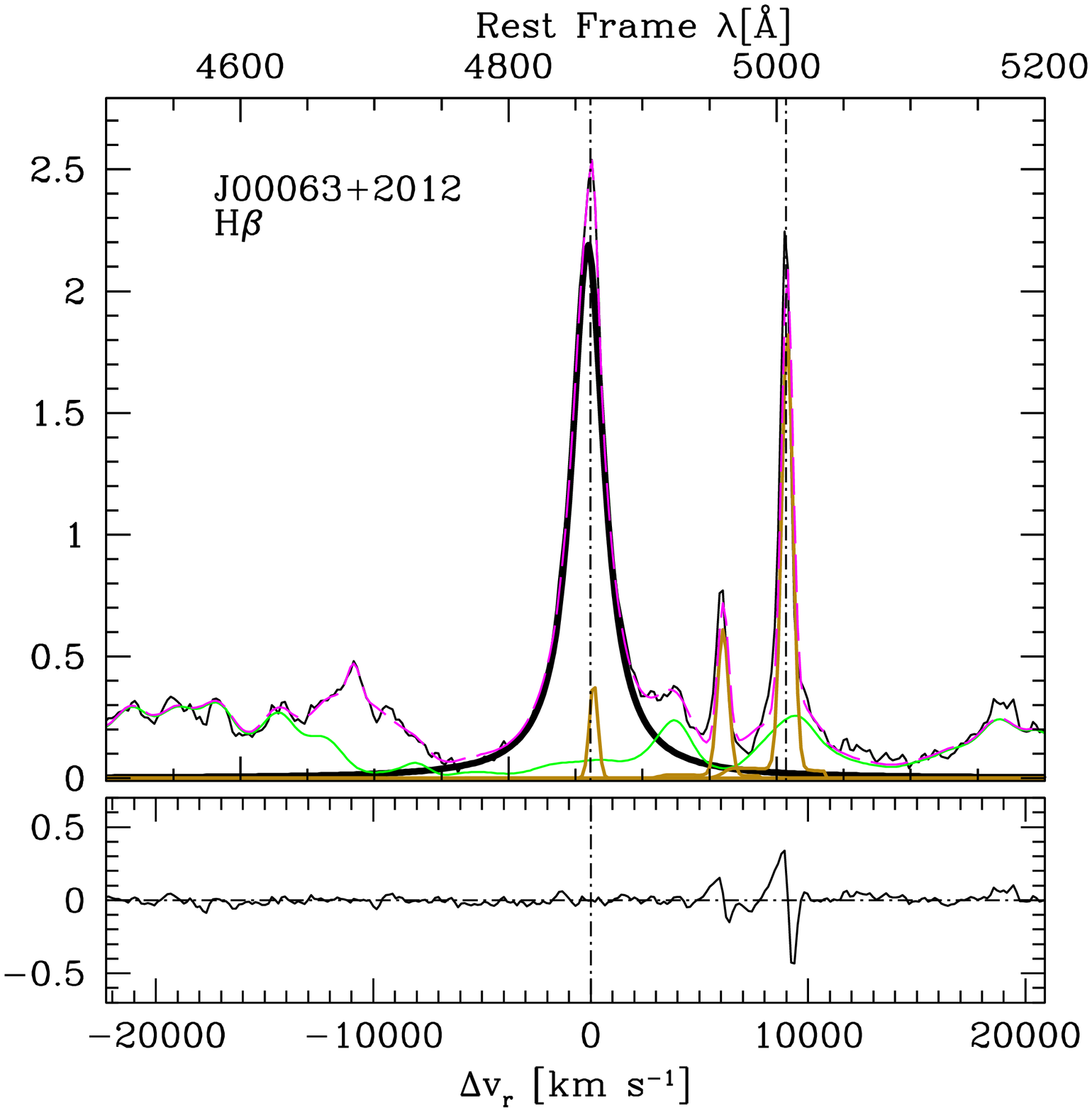}\includegraphics[width=0.225\columnwidth]{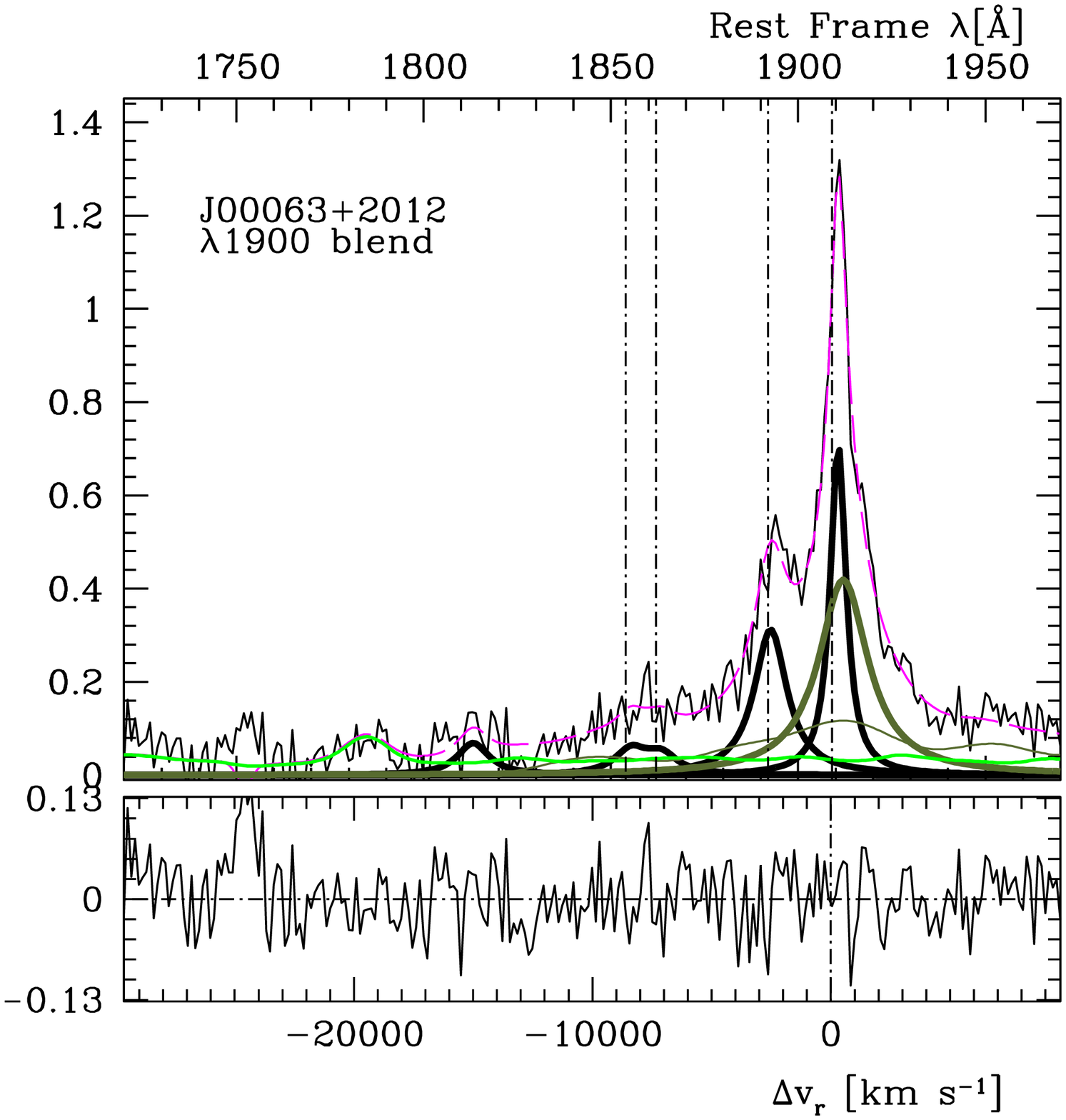} 
\includegraphics[width=0.225\columnwidth]{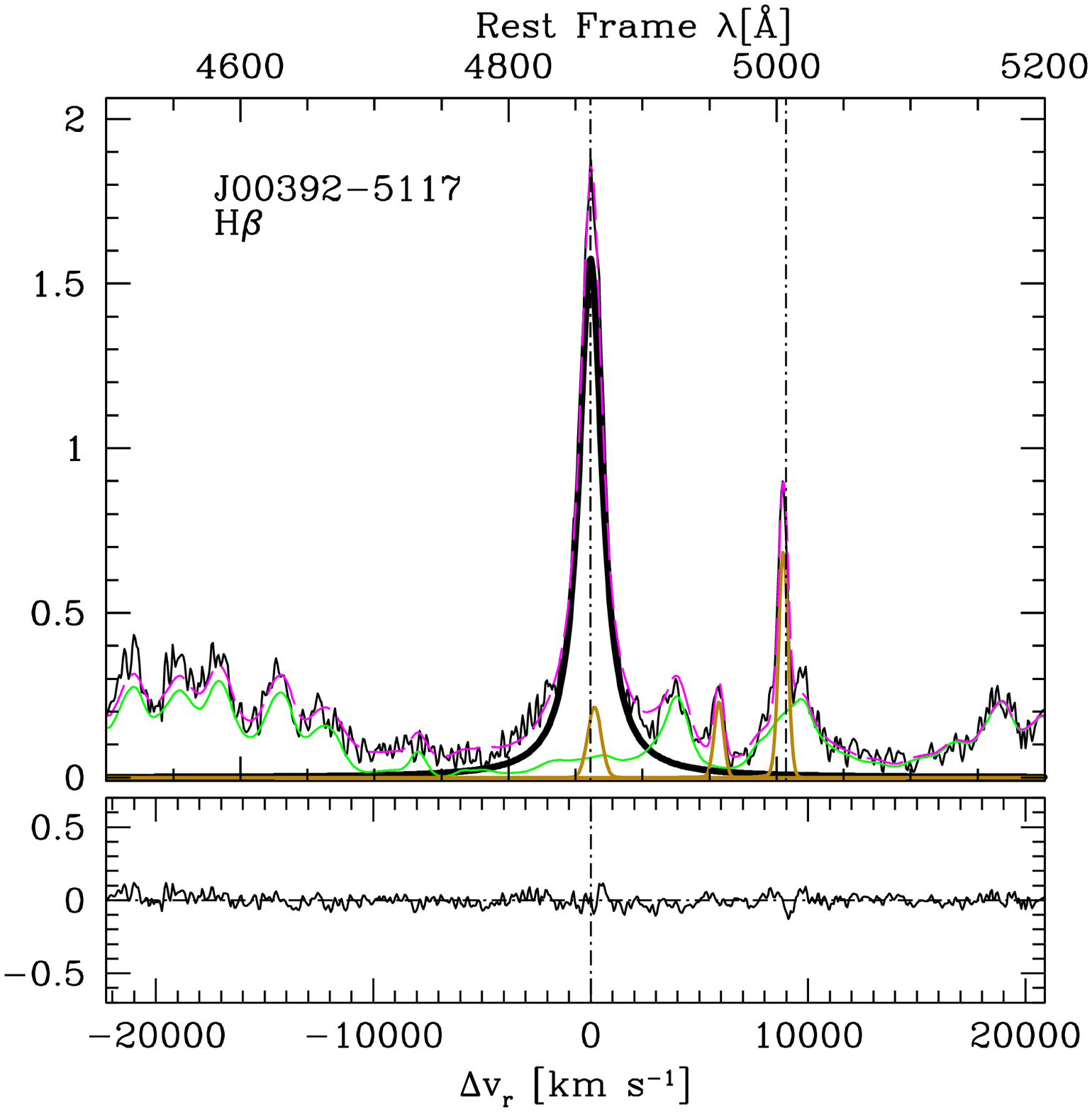}\includegraphics[width=0.225\columnwidth]{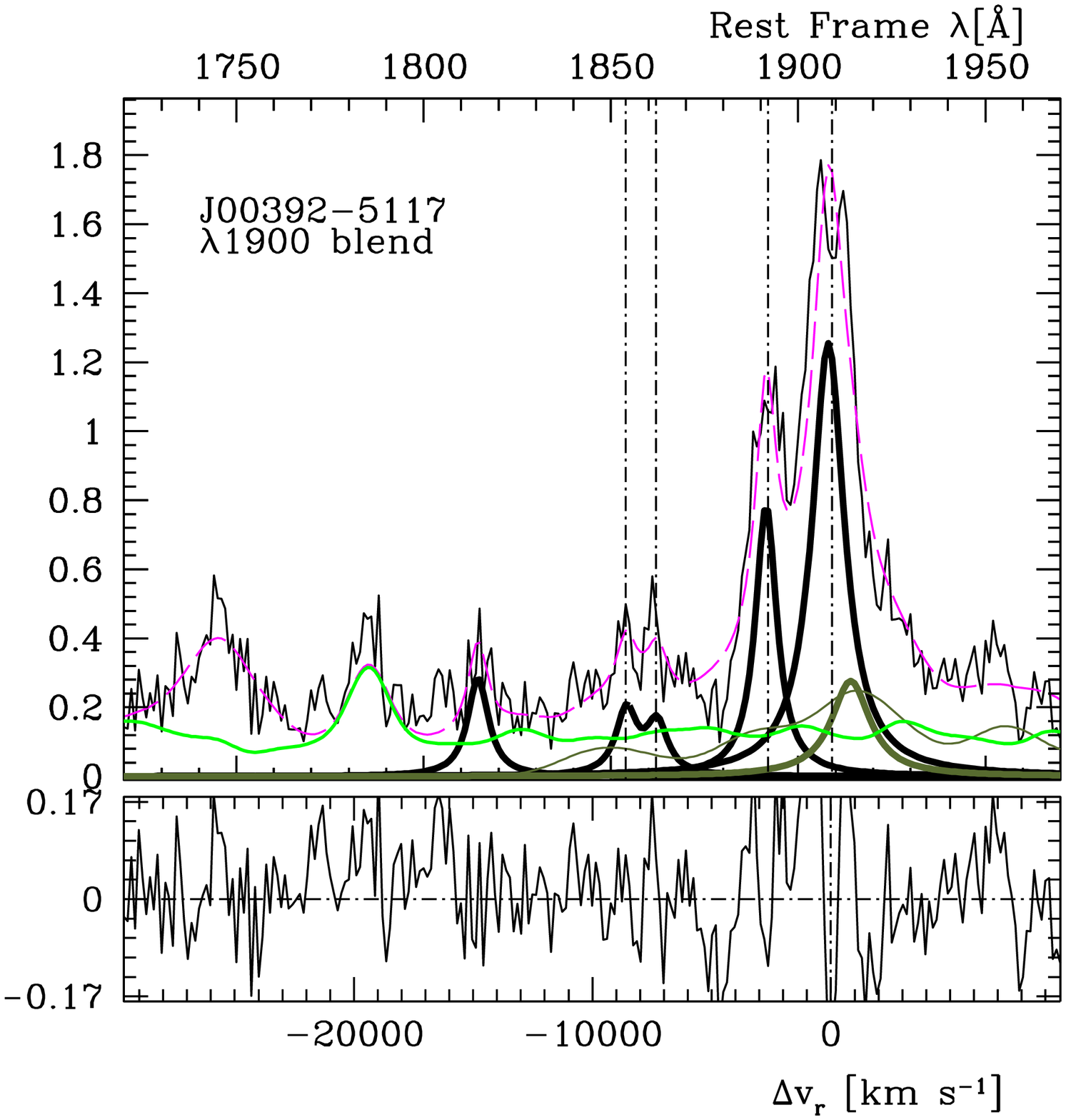}\\
\includegraphics[width=0.225\columnwidth]{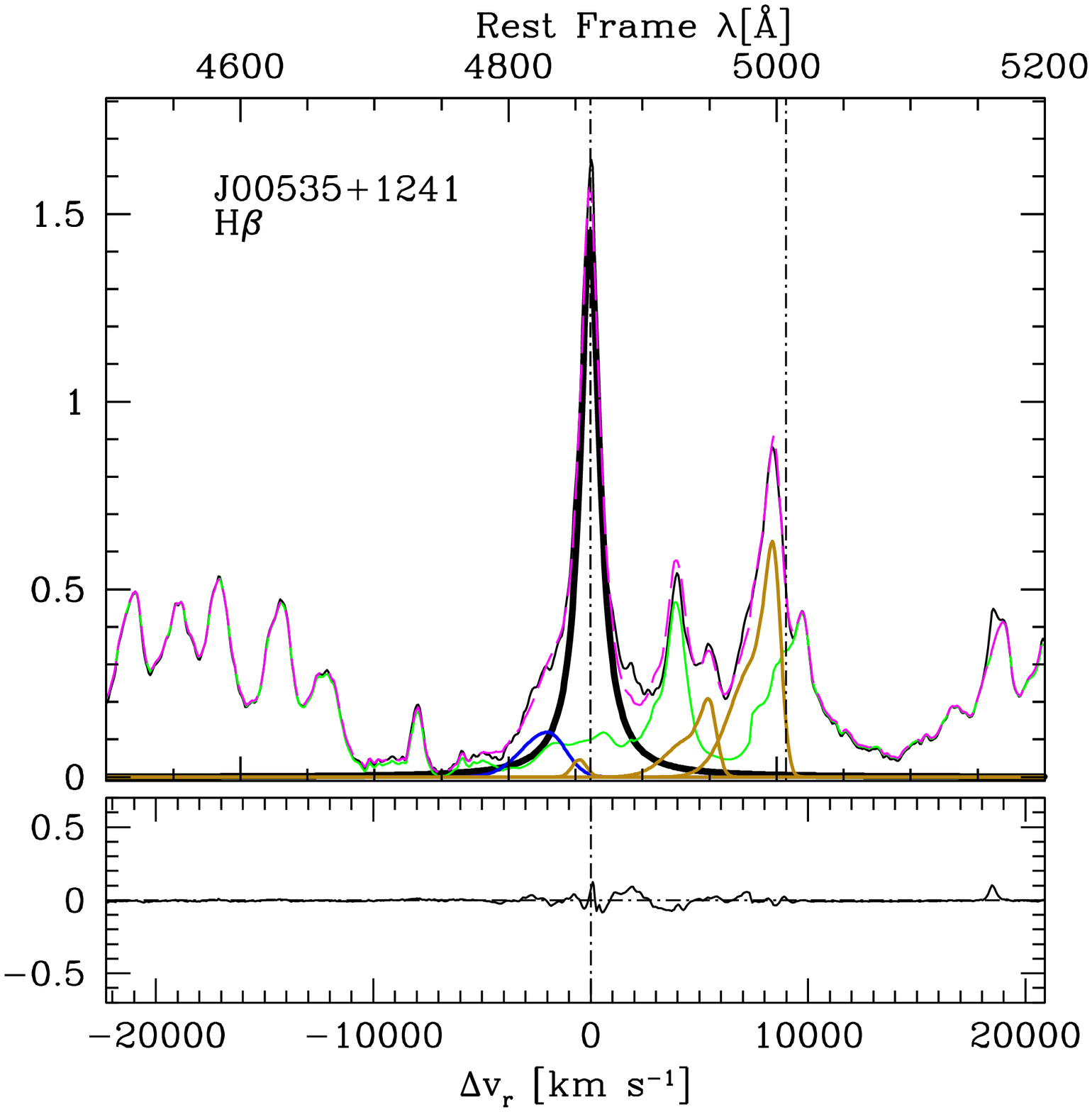}\includegraphics[width=0.225\columnwidth]{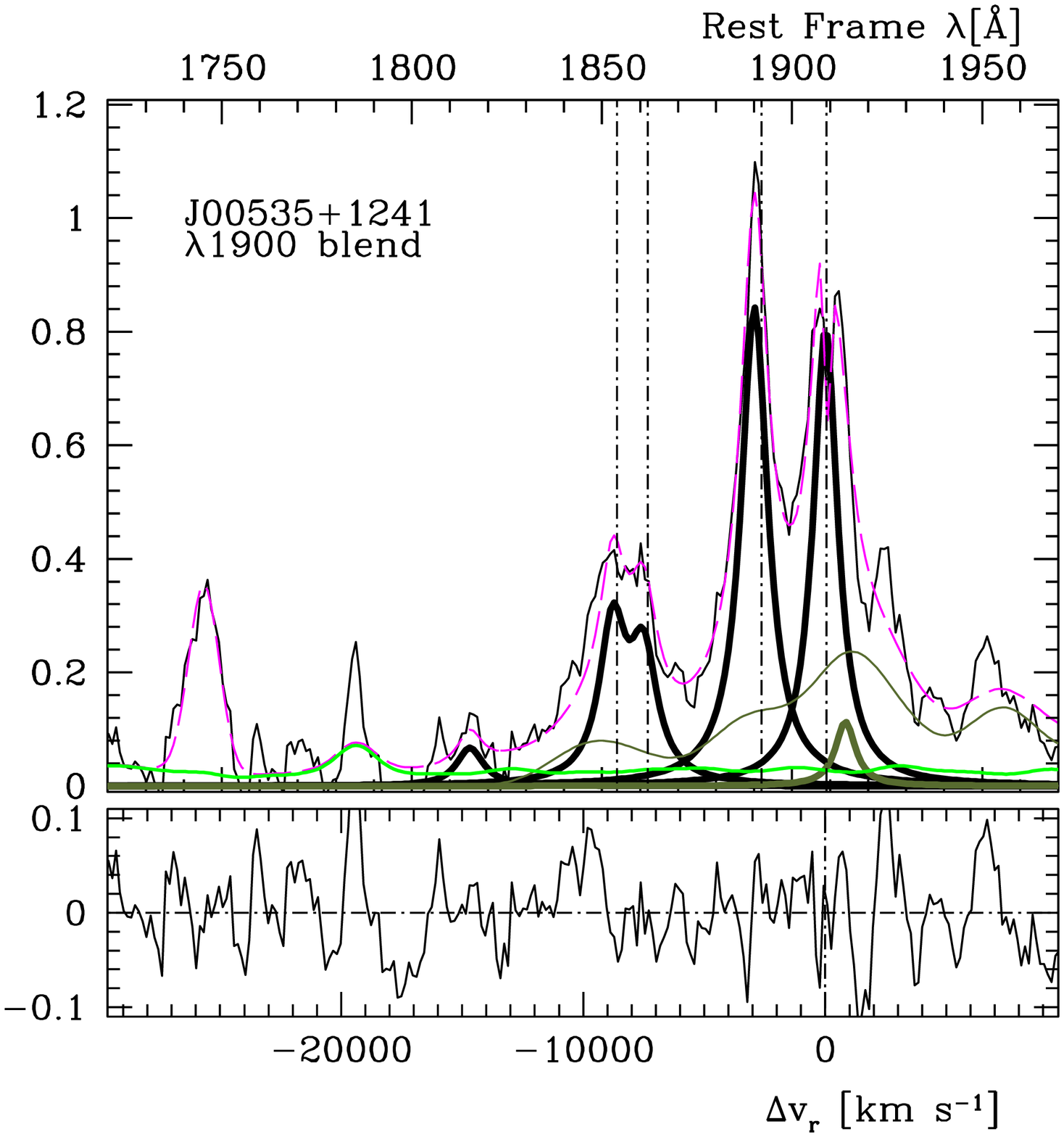} 
\includegraphics[width=0.225\columnwidth]{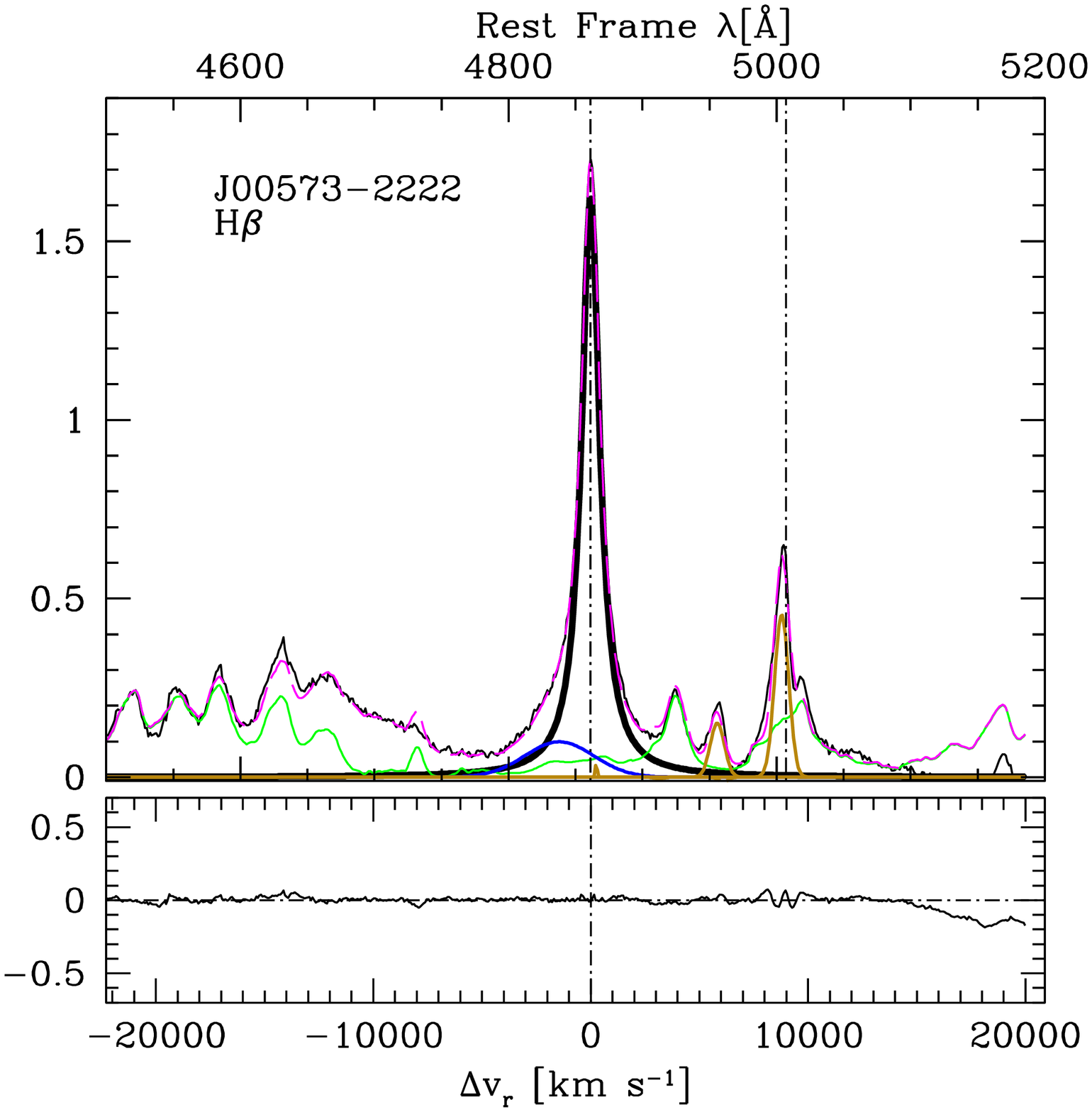}\includegraphics[width=0.225\columnwidth]{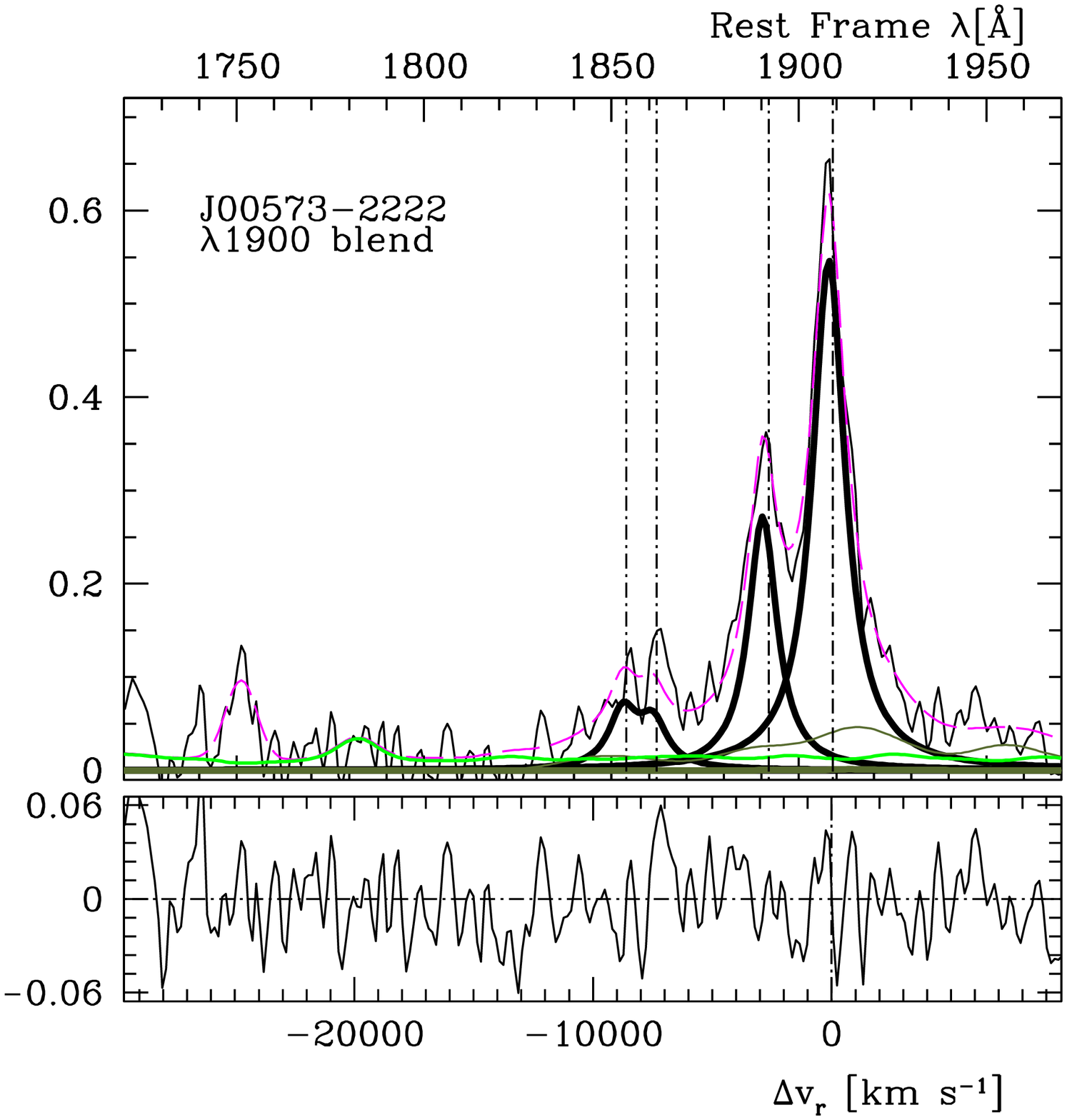}\\
\includegraphics[width=0.225\columnwidth]{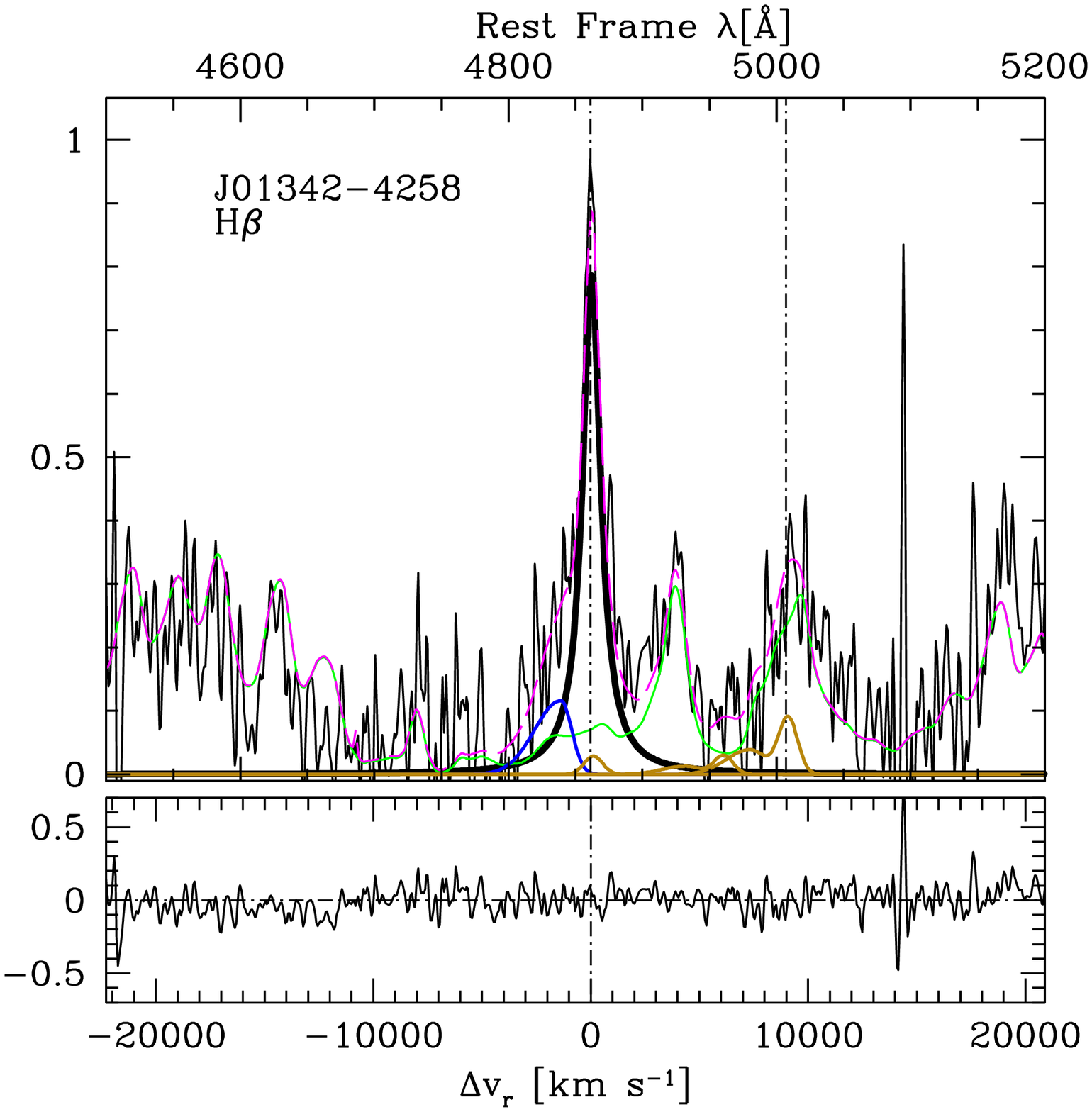}\includegraphics[width=0.225\columnwidth]{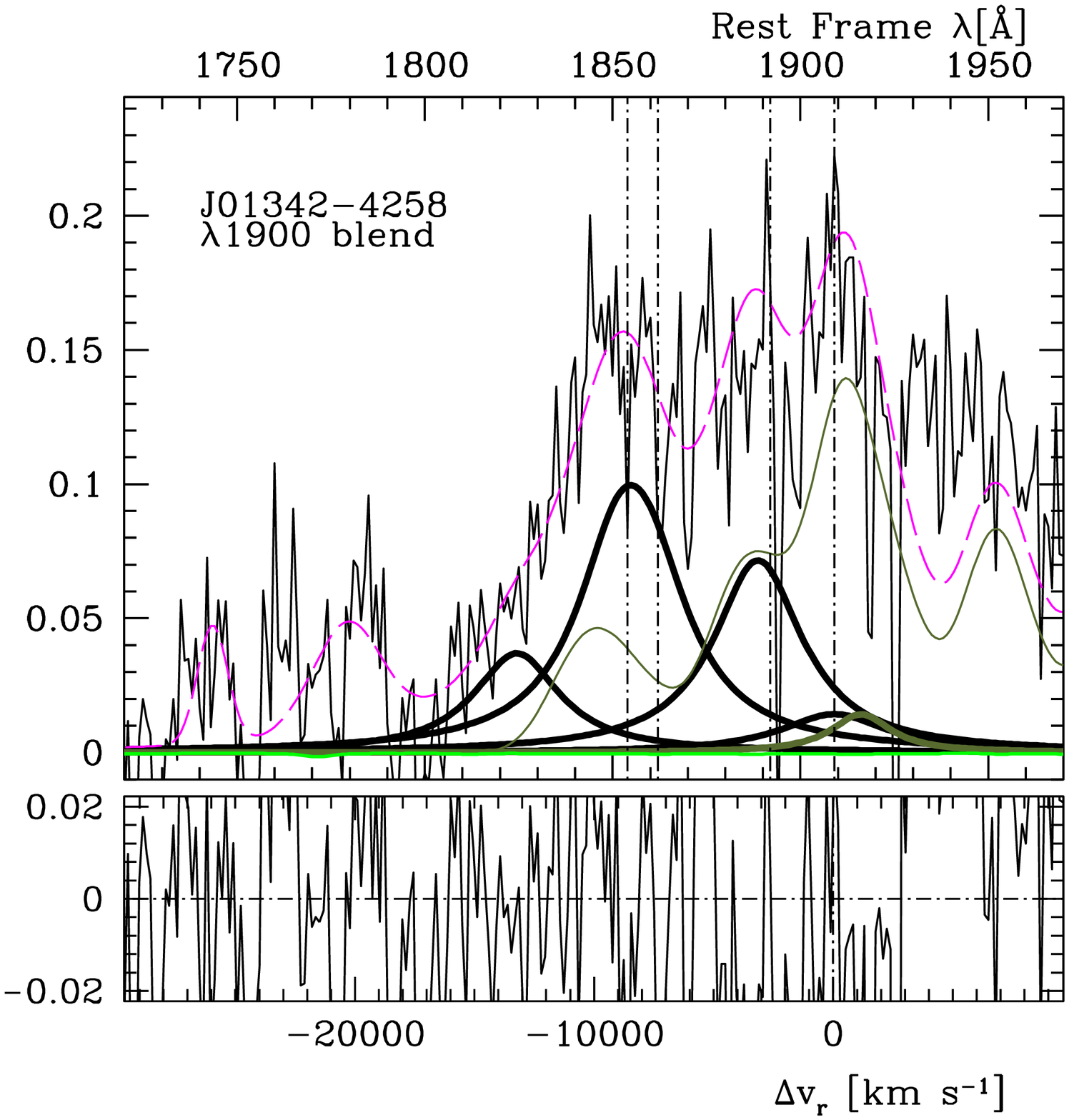} 
\includegraphics[width=0.225\columnwidth]{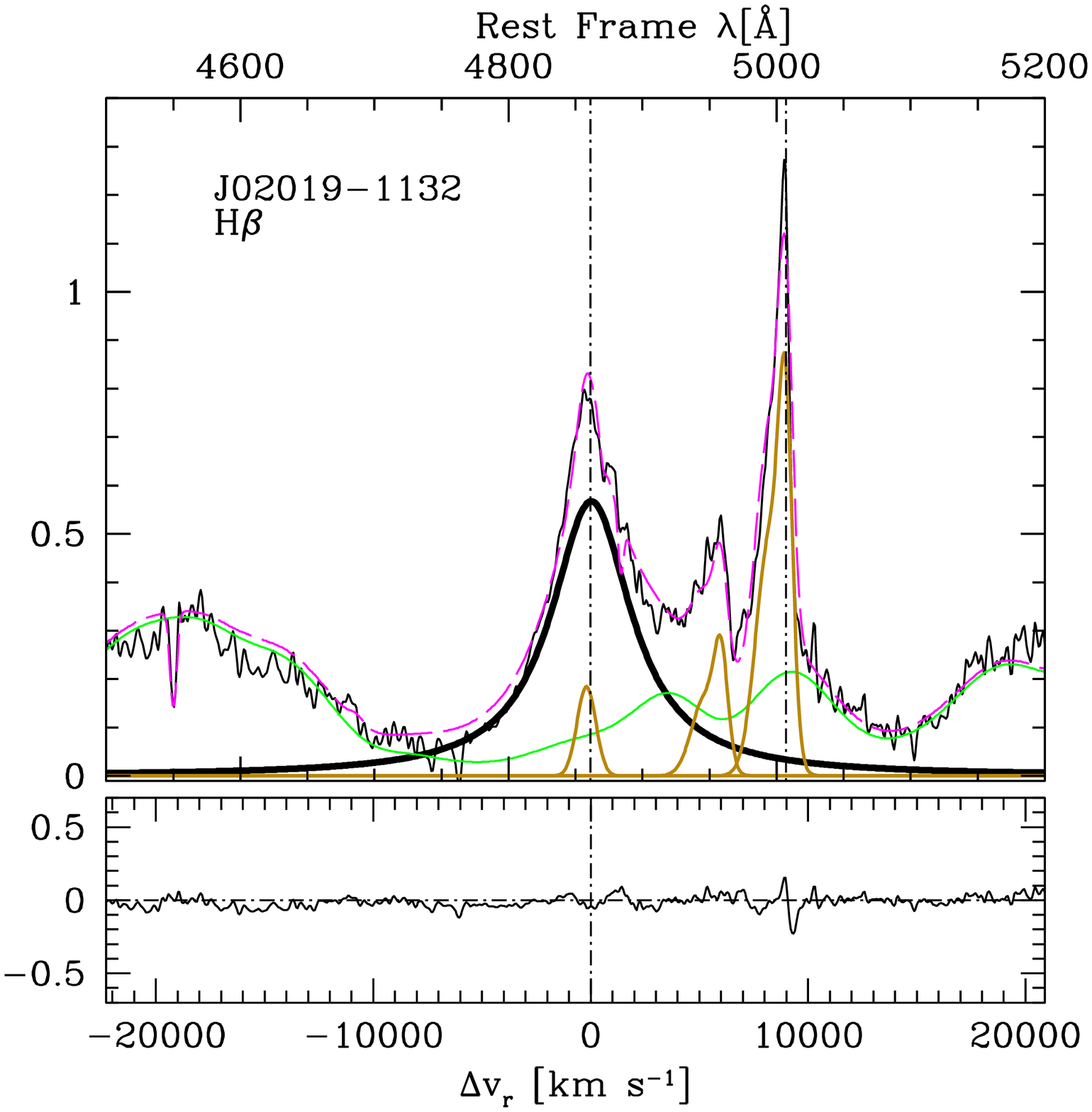}\includegraphics[width=0.225\columnwidth]{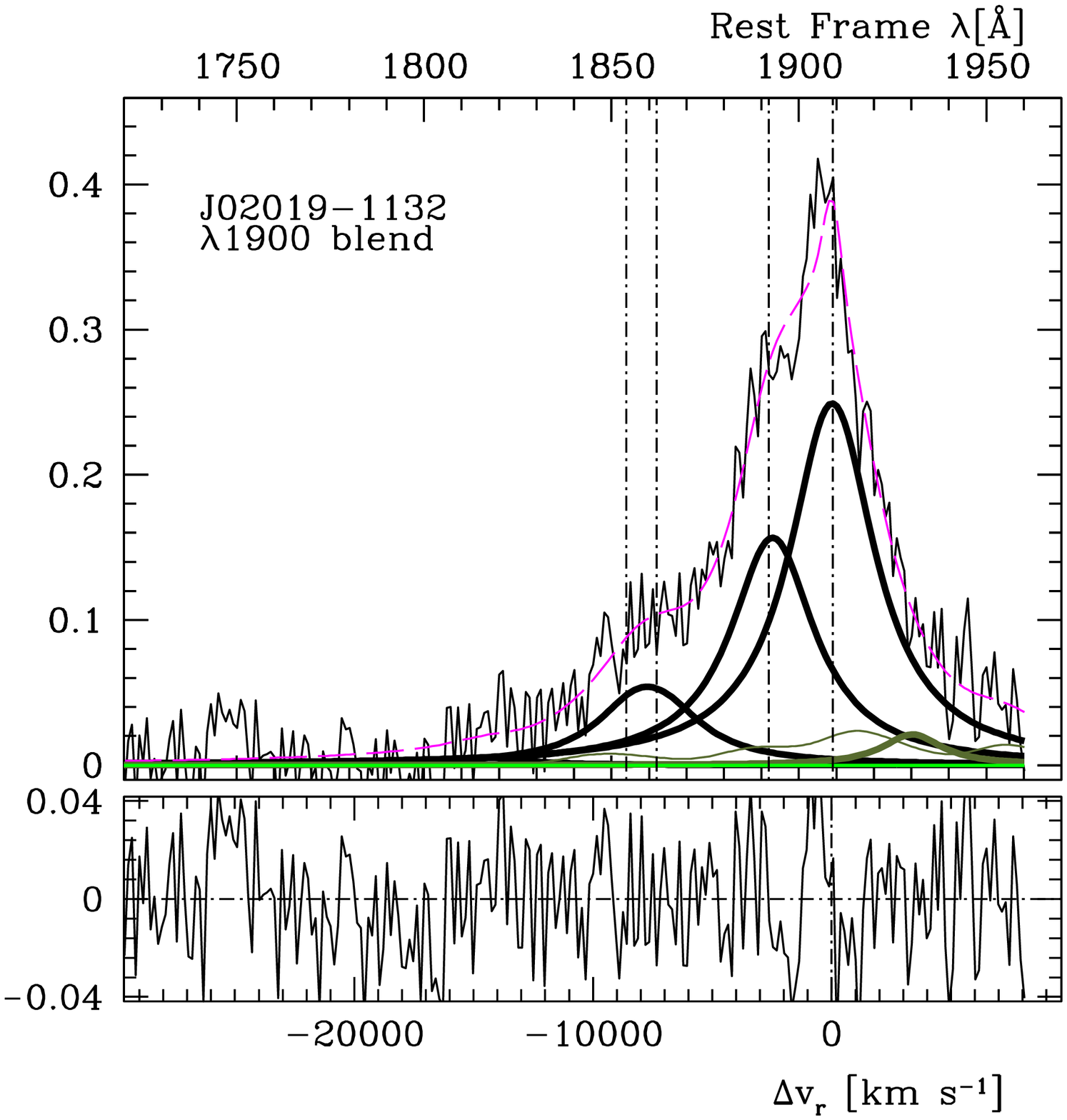}\\
\includegraphics[width=0.225\columnwidth]{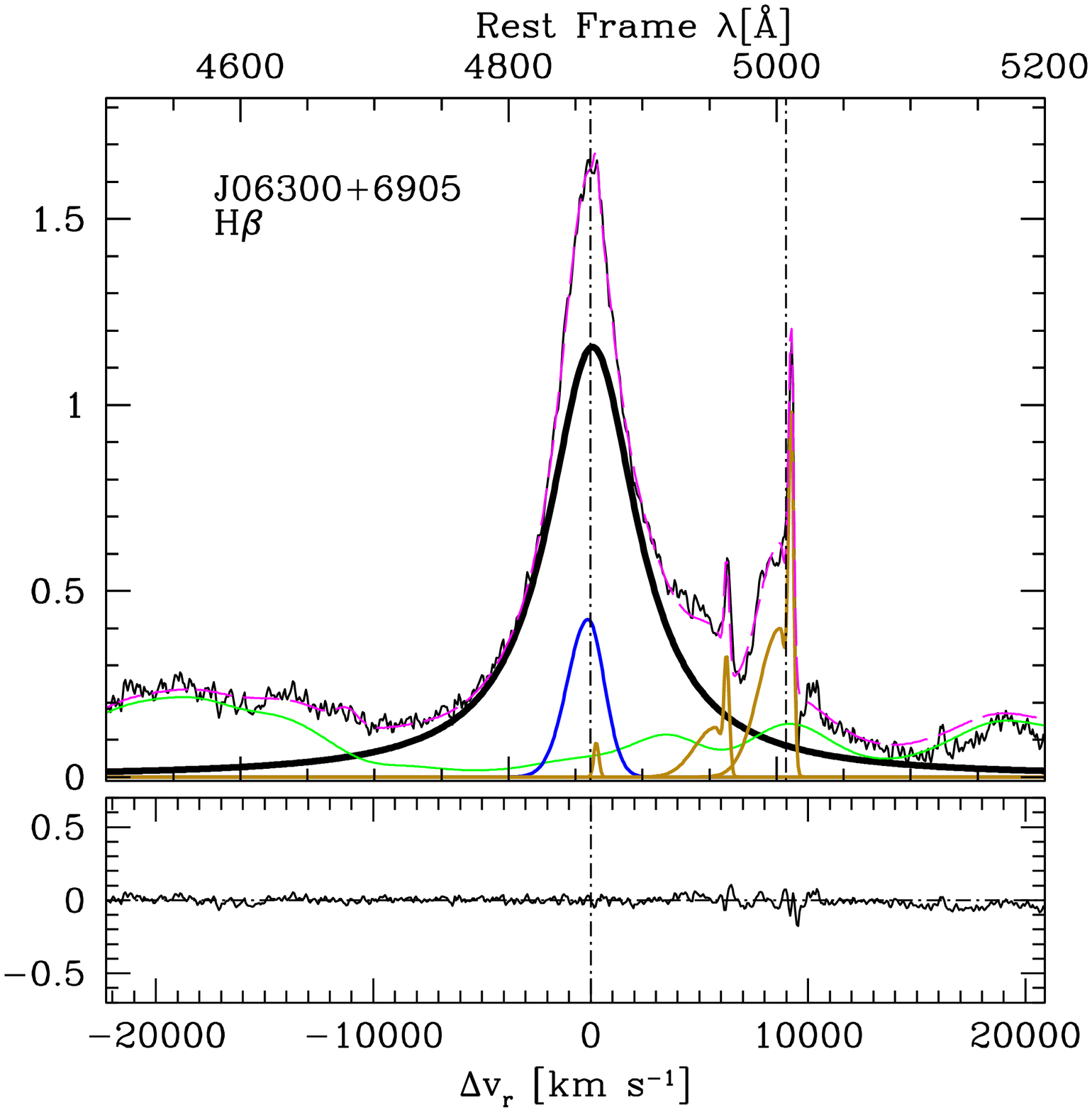}\includegraphics[width=0.225\columnwidth]{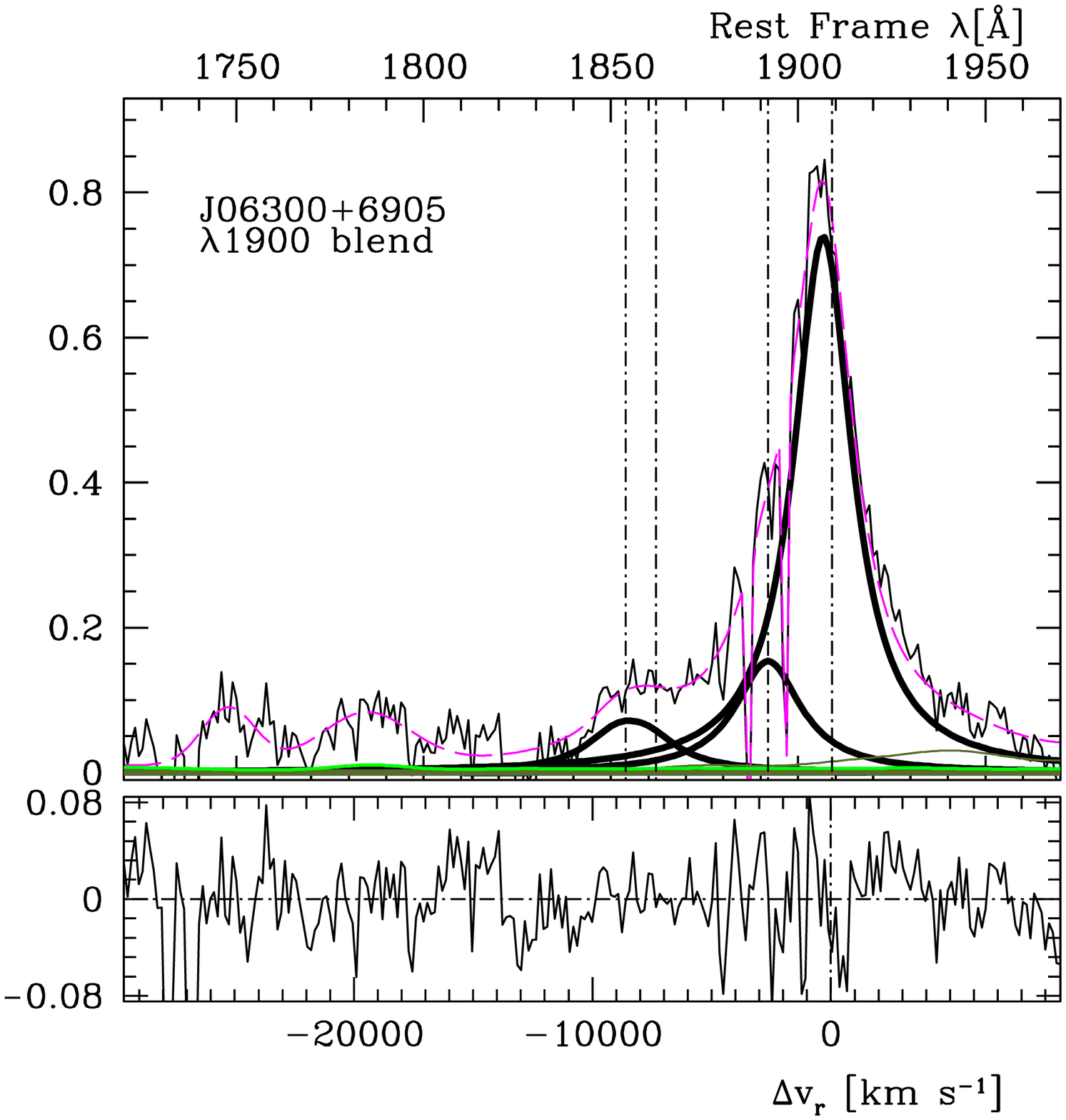}  
\includegraphics[width=0.225\columnwidth]{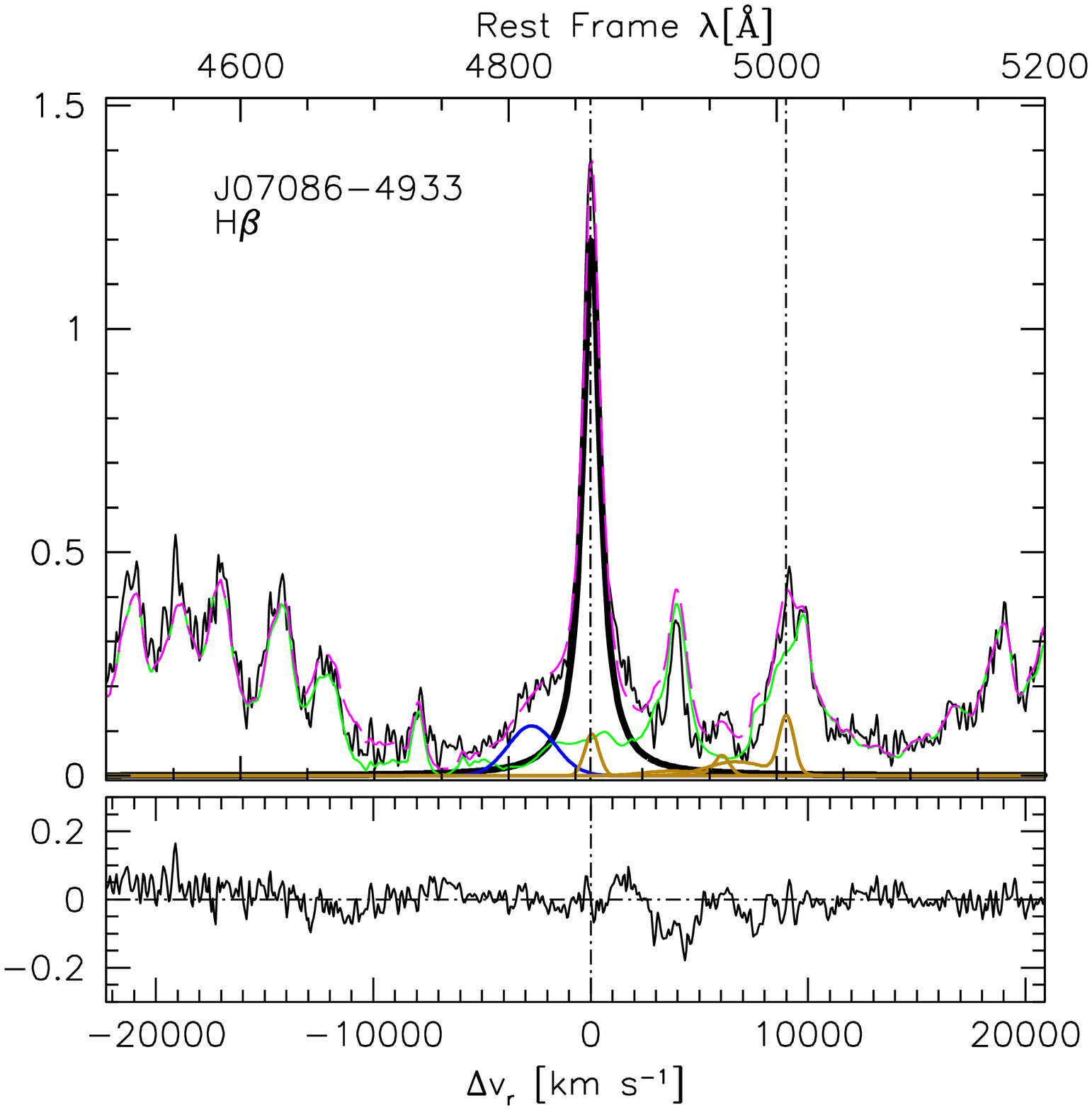}\includegraphics[width=0.225\columnwidth]{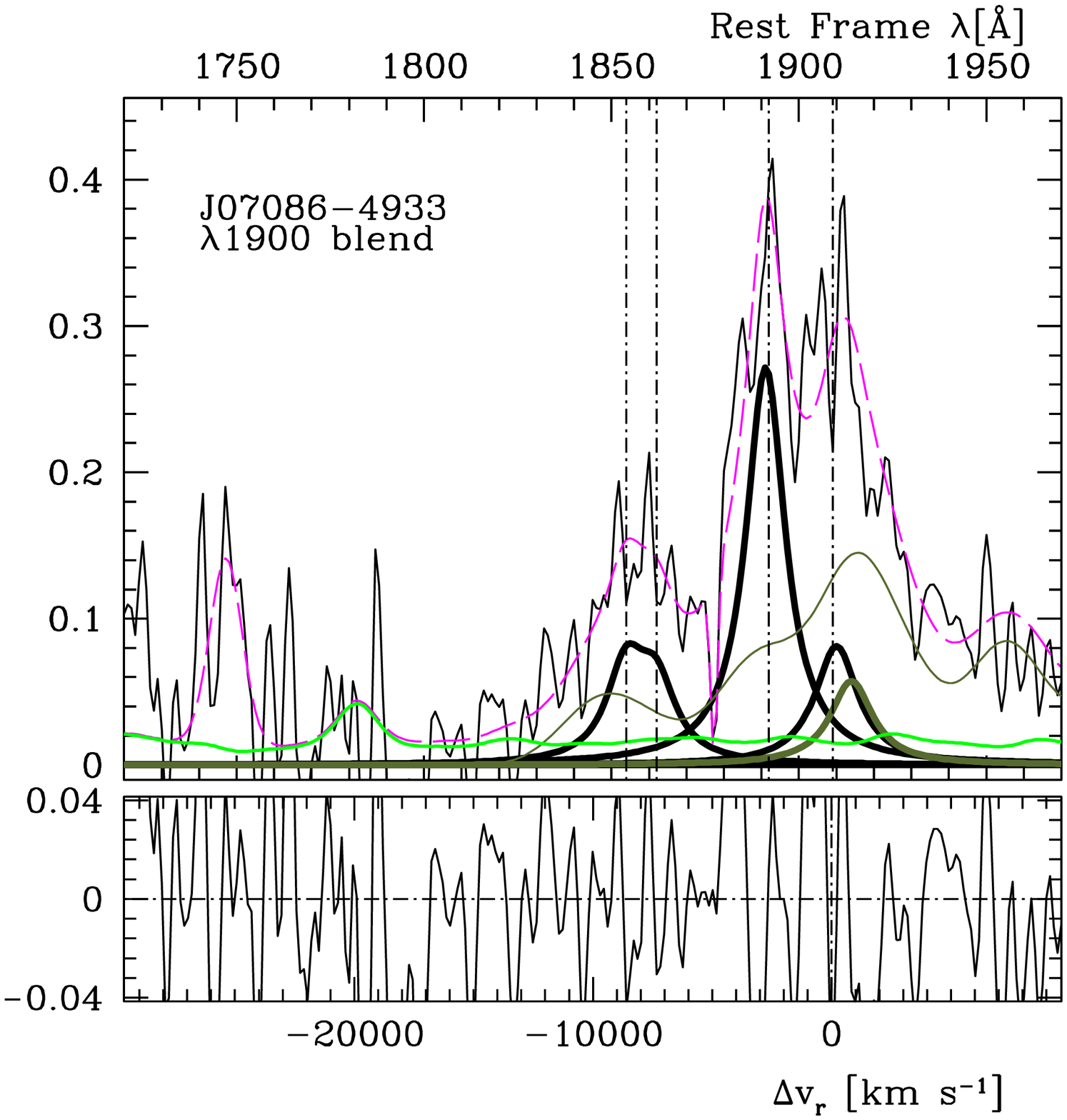}\\ 
\includegraphics[width=0.225\columnwidth]{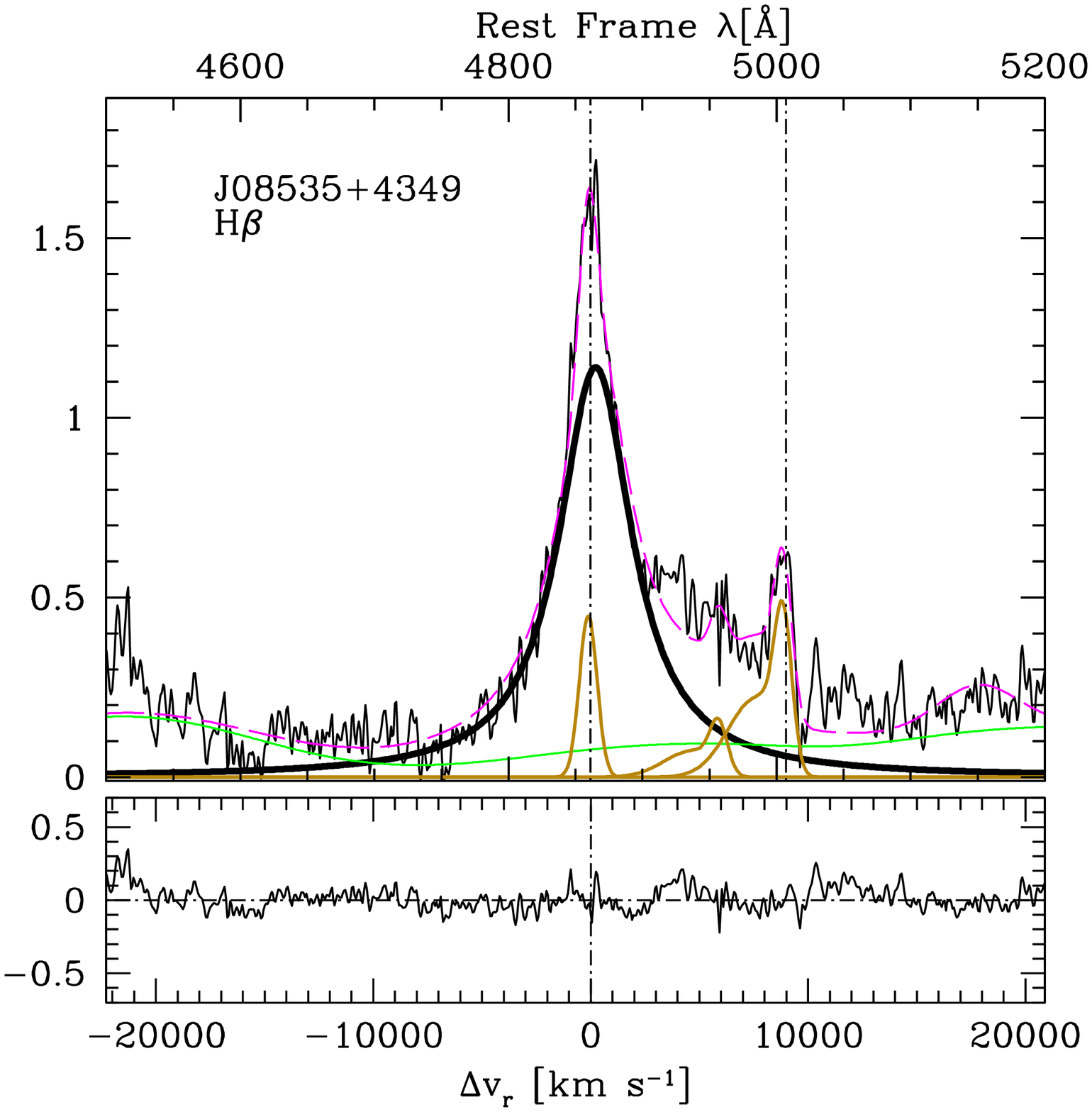}\includegraphics[width=0.225\columnwidth]{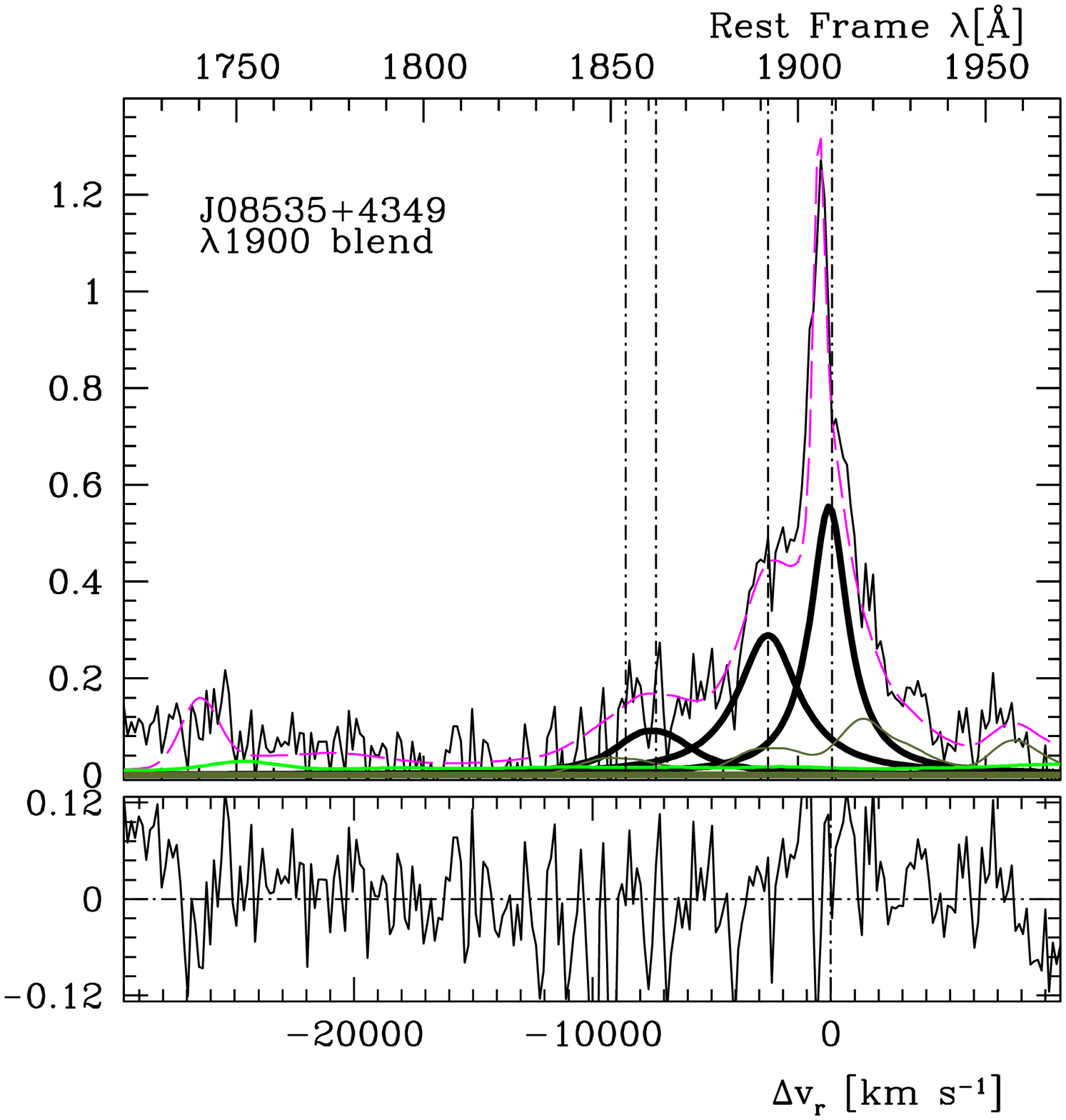}
\includegraphics[width=0.225\columnwidth]{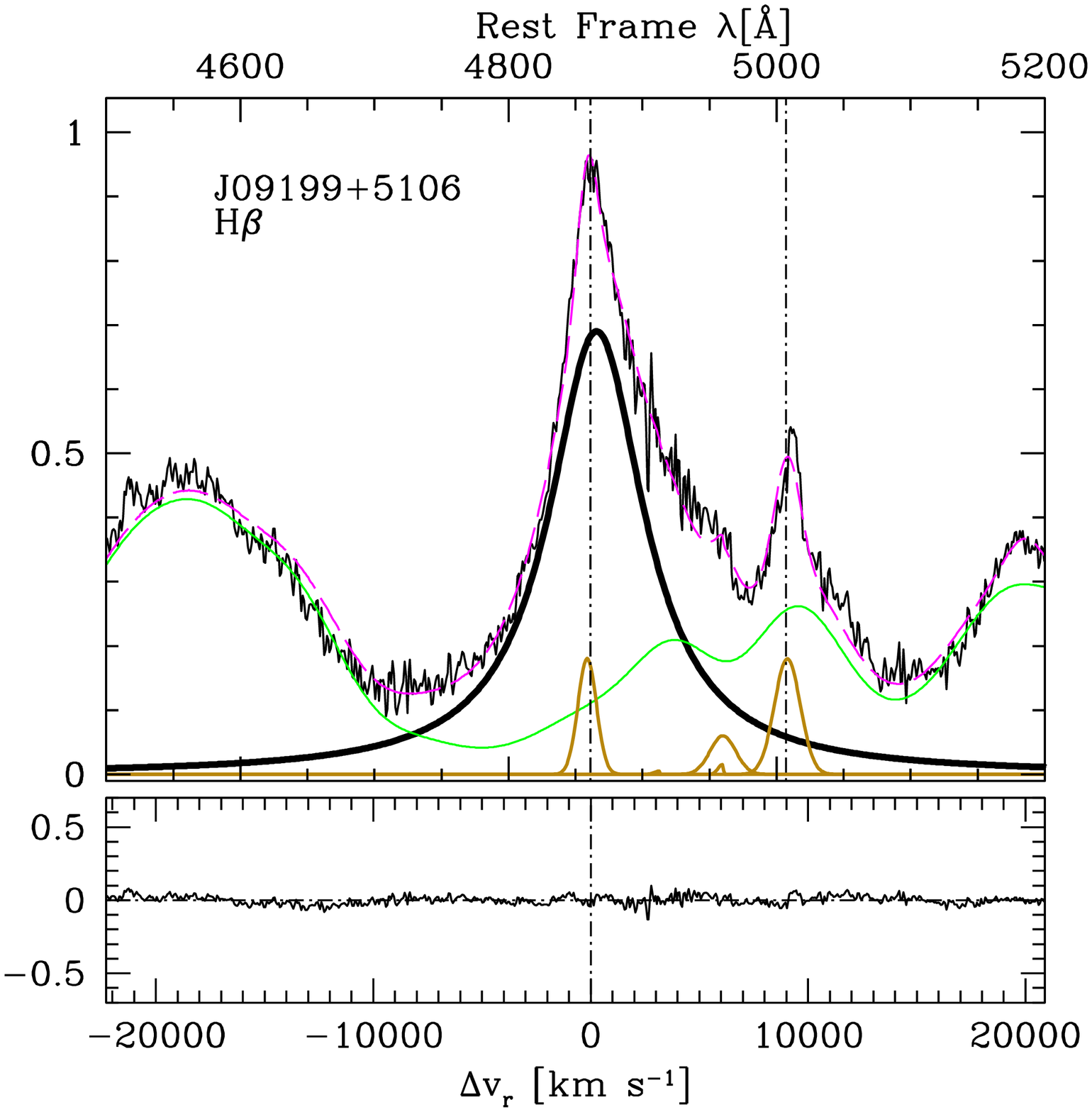}\includegraphics[width=0.225\columnwidth]{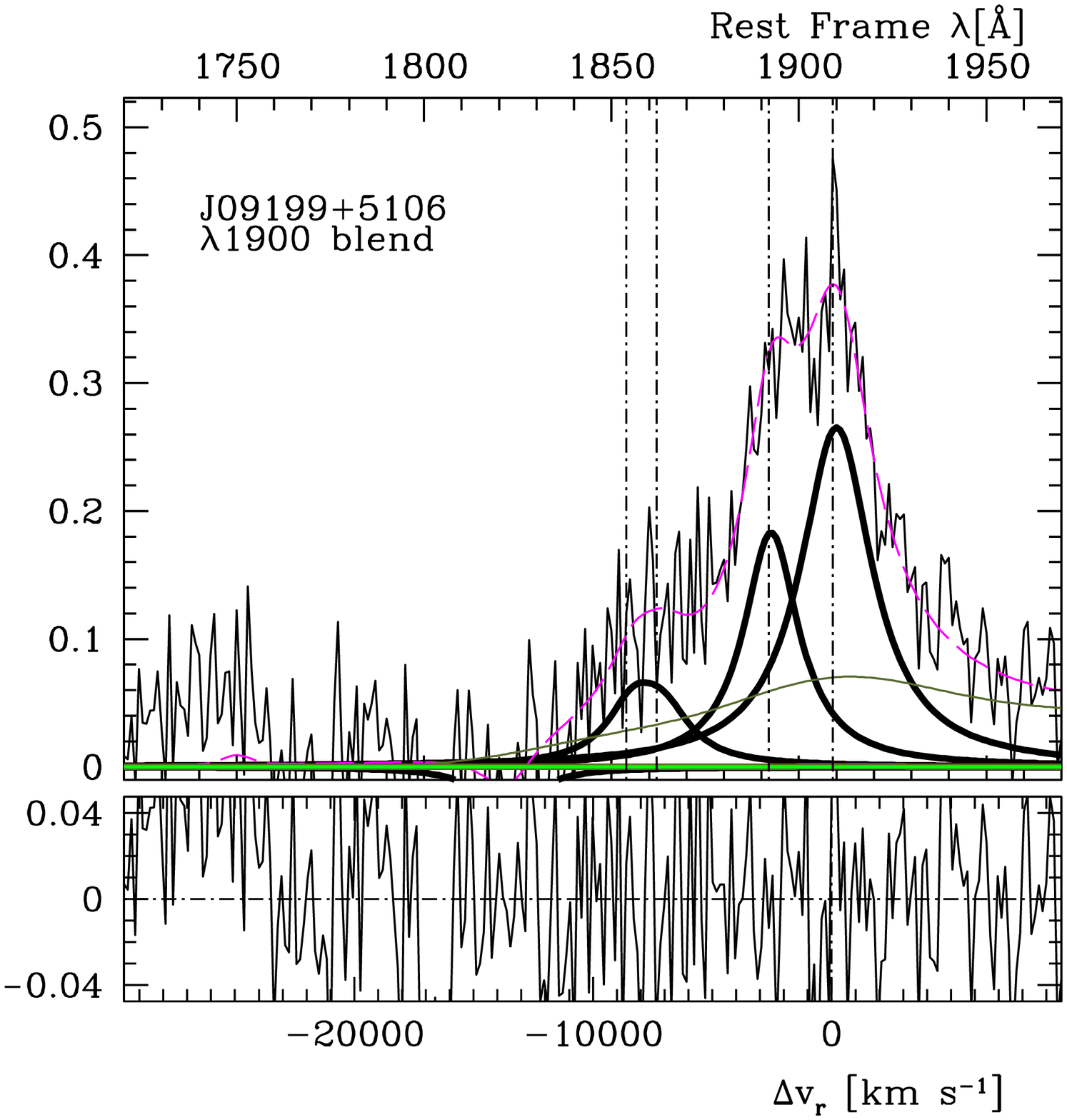}
\caption{Analysis of the \hb\ spectral range (left) and of  1900 \AA\ blend (right), for the sources of the $FOS^\ast$\ subsample.  Abscissa scales are rest-frame wavelength in \AA\ and radial velocity from rest-frame. Ordinate scale is  normalized specific flux by the value at 5100 \AA\ and at 1700 \AA. The dashed magenta lines trace the sum of all emission components of the model.  The black lines identify the \hb\ broad component \hbbc\ (left), and \aliii, \siiii, and  \ciii\ (right).  The blue line the blueshifted excess in the \hb\ profile. Green lines trace the adopted \feii\ (pale) and \feiii\ (dark) templates.  Golden  lines trace narrow emission lines or line components. \label{fig:fosa}}
\end{figure} 
\vfill
\addtocounter{figure}{-1}

\begin{figure}
\includegraphics[width=0.225\columnwidth]{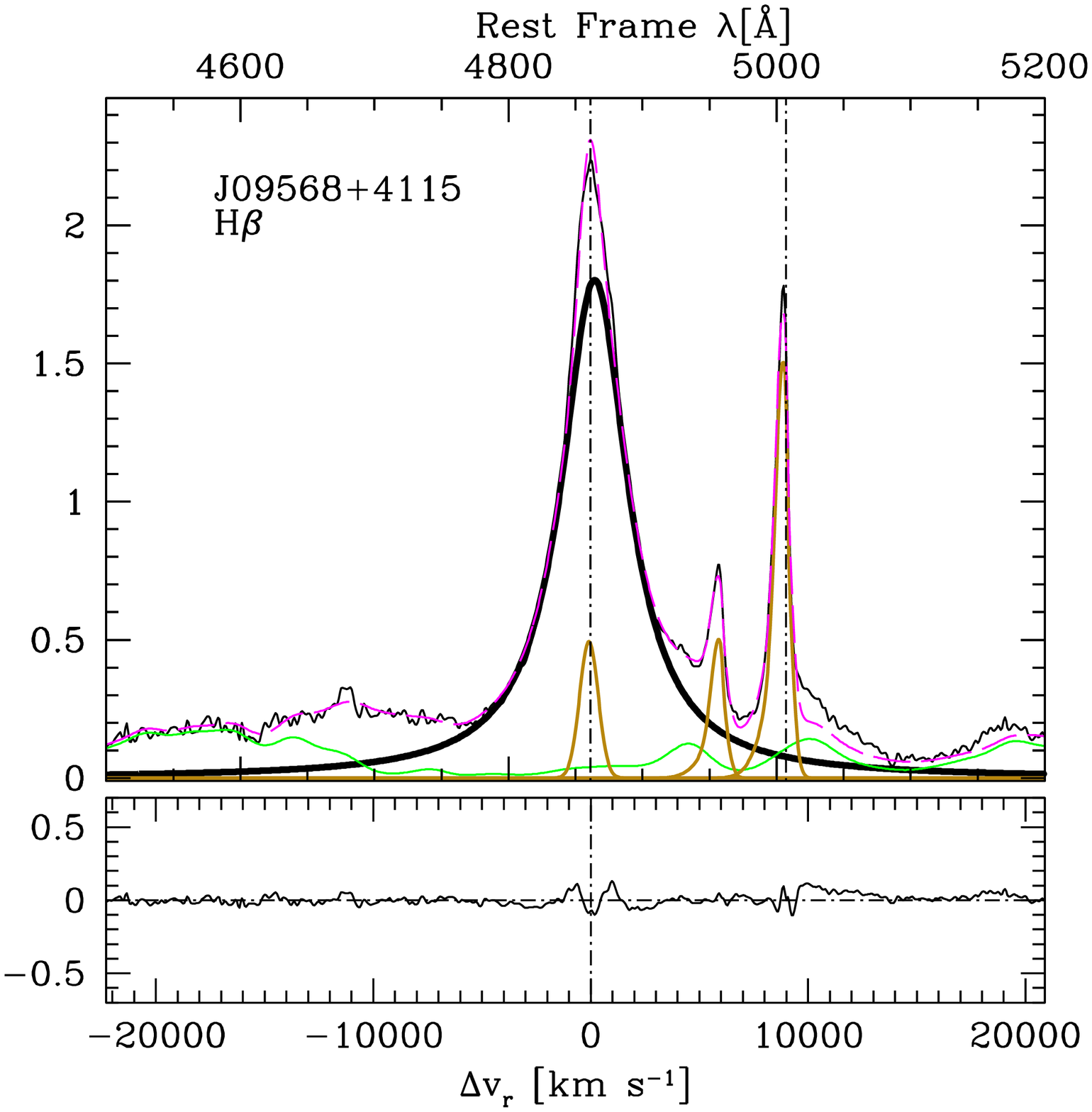}\includegraphics[width=0.225\columnwidth]{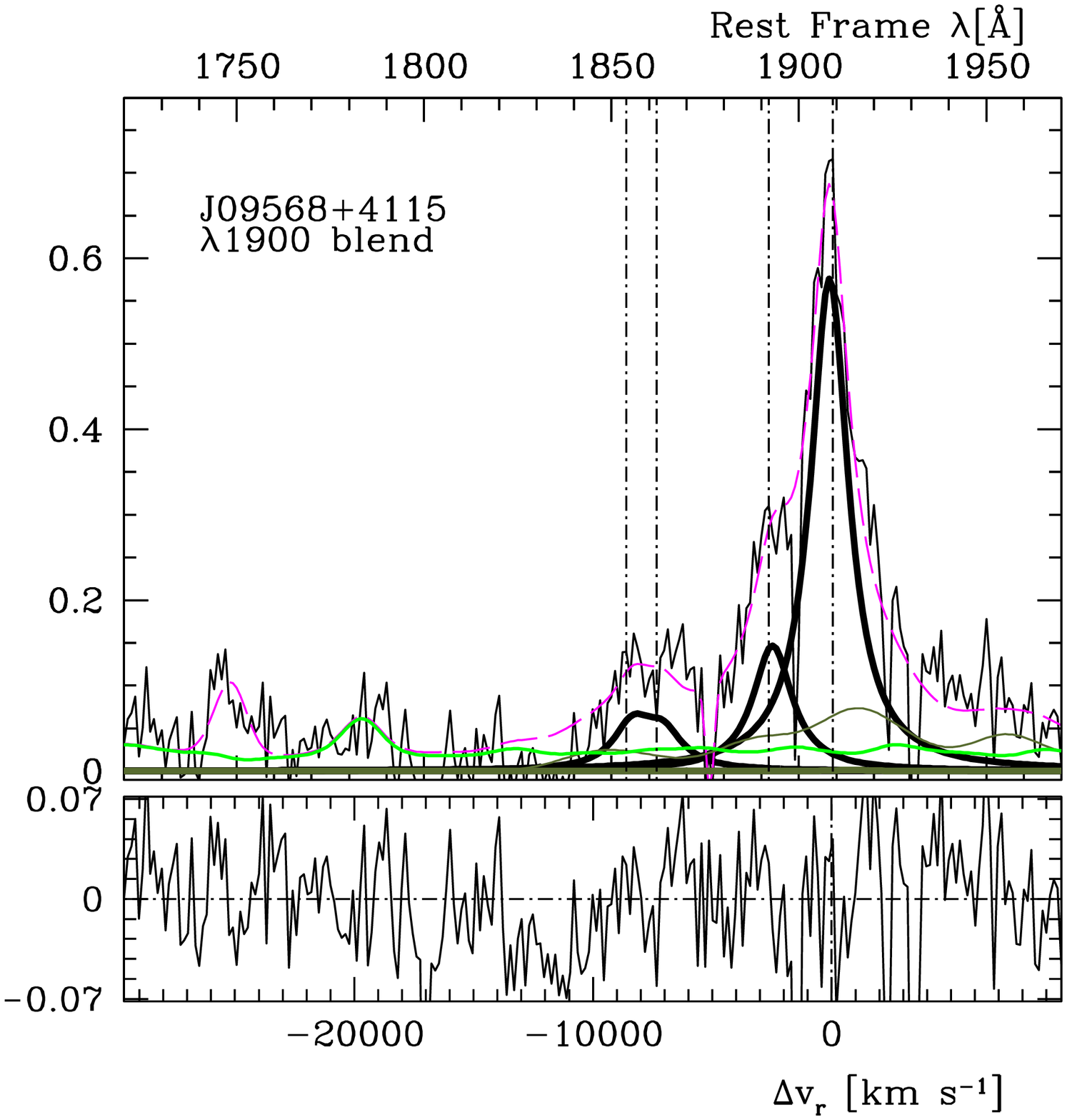}
\includegraphics[width=0.225\columnwidth]{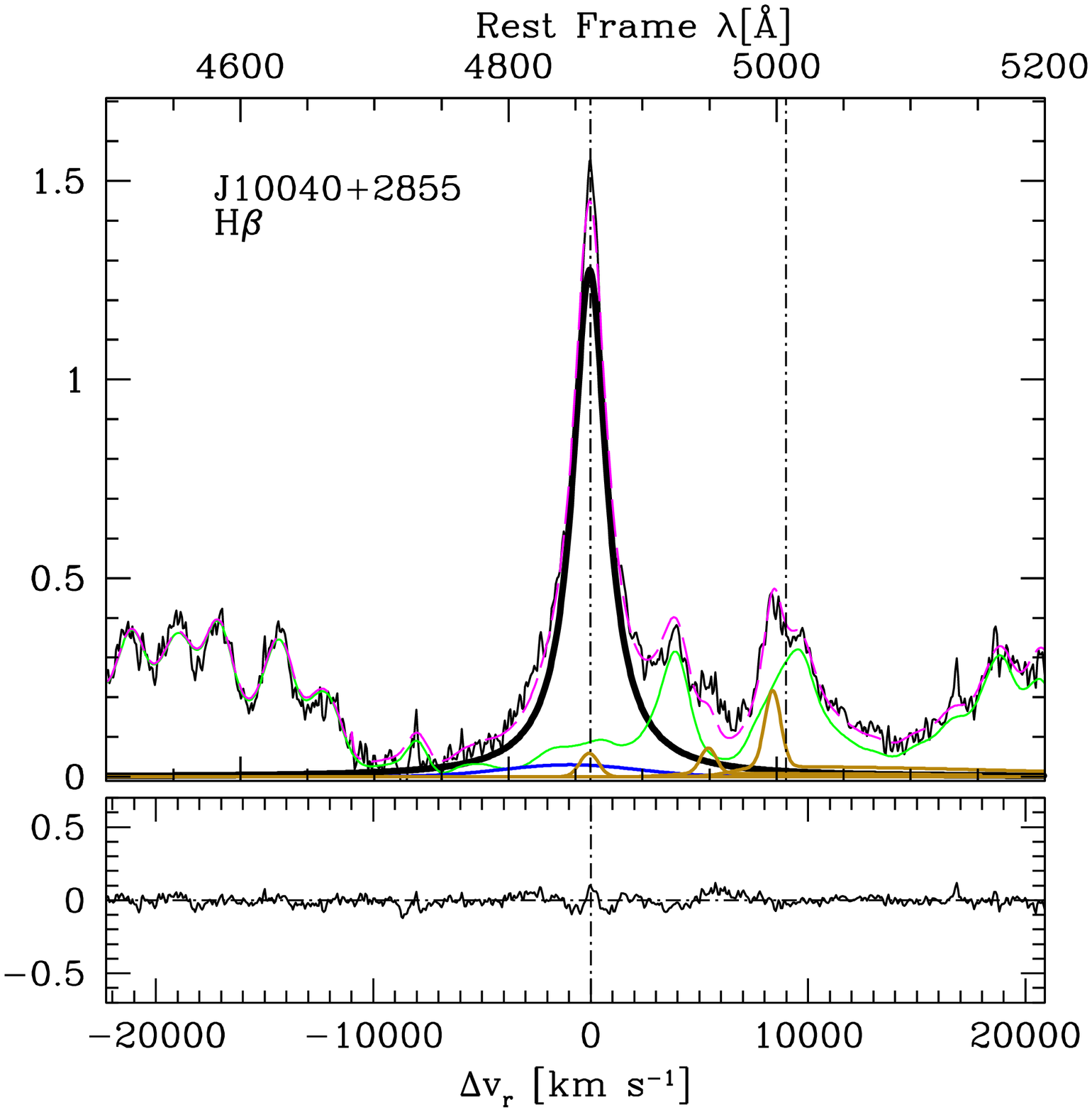}\includegraphics[width=0.225\columnwidth]{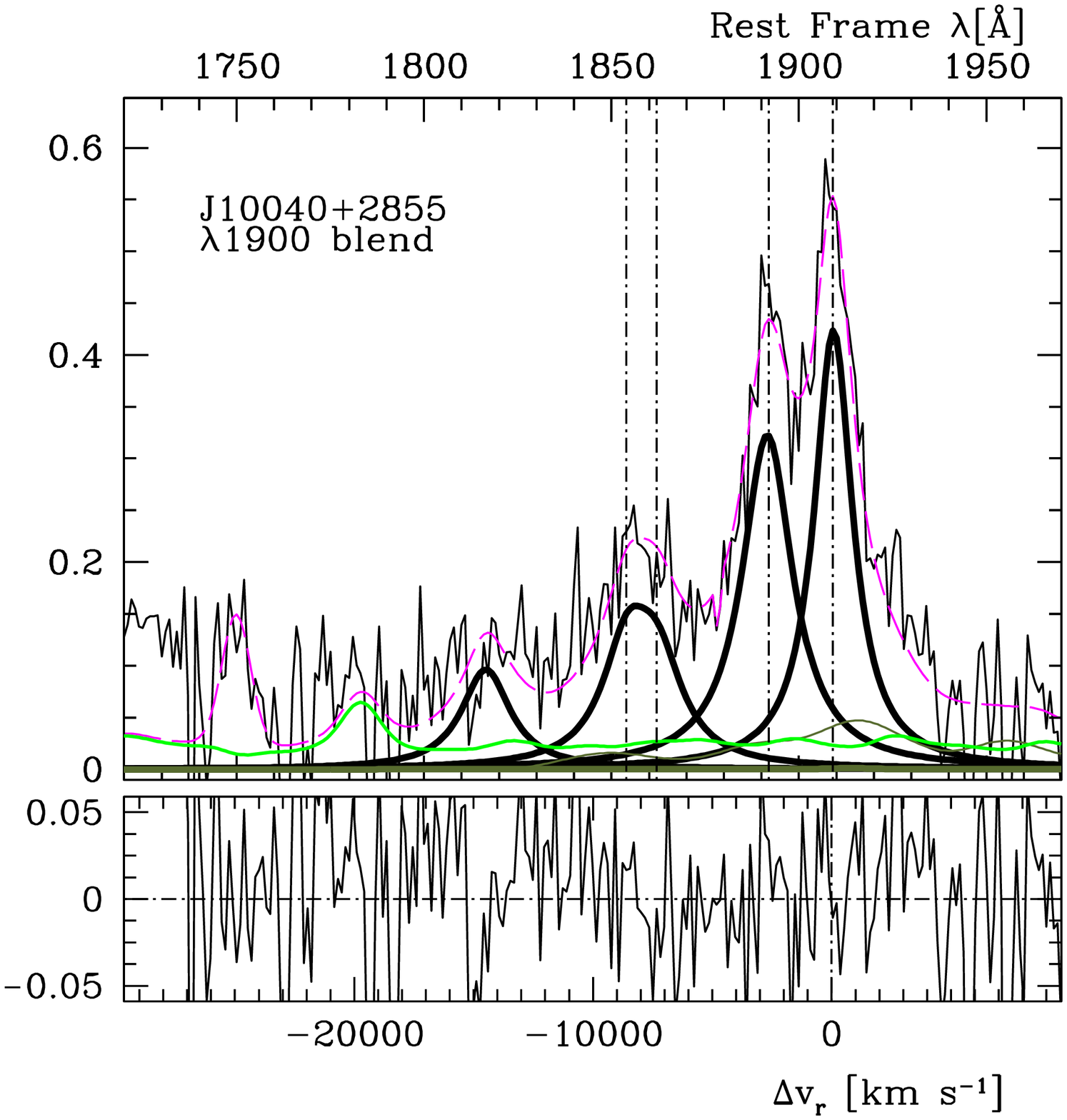}\\
\includegraphics[width=0.225\columnwidth]{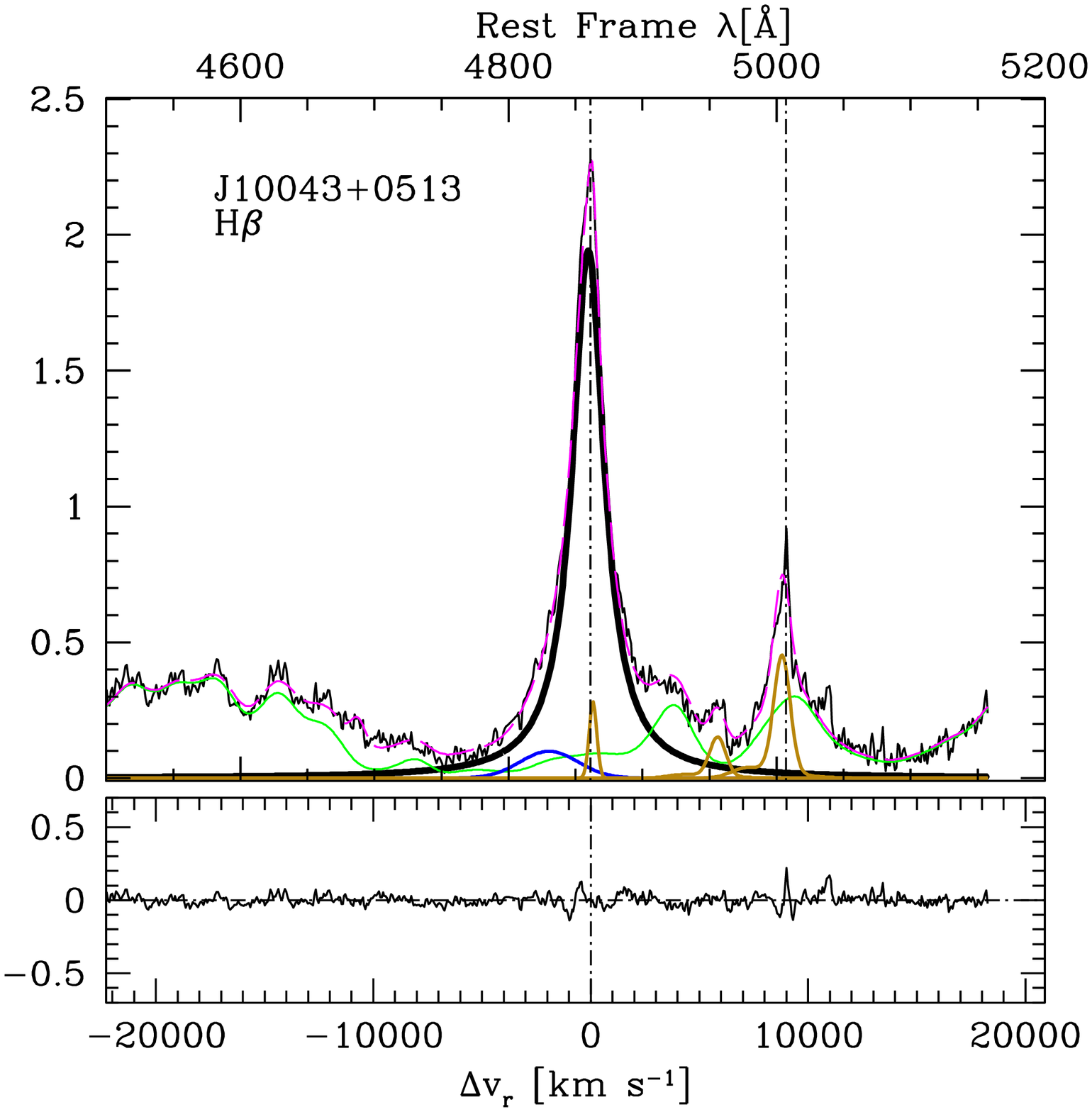}\includegraphics[width=0.225\columnwidth]{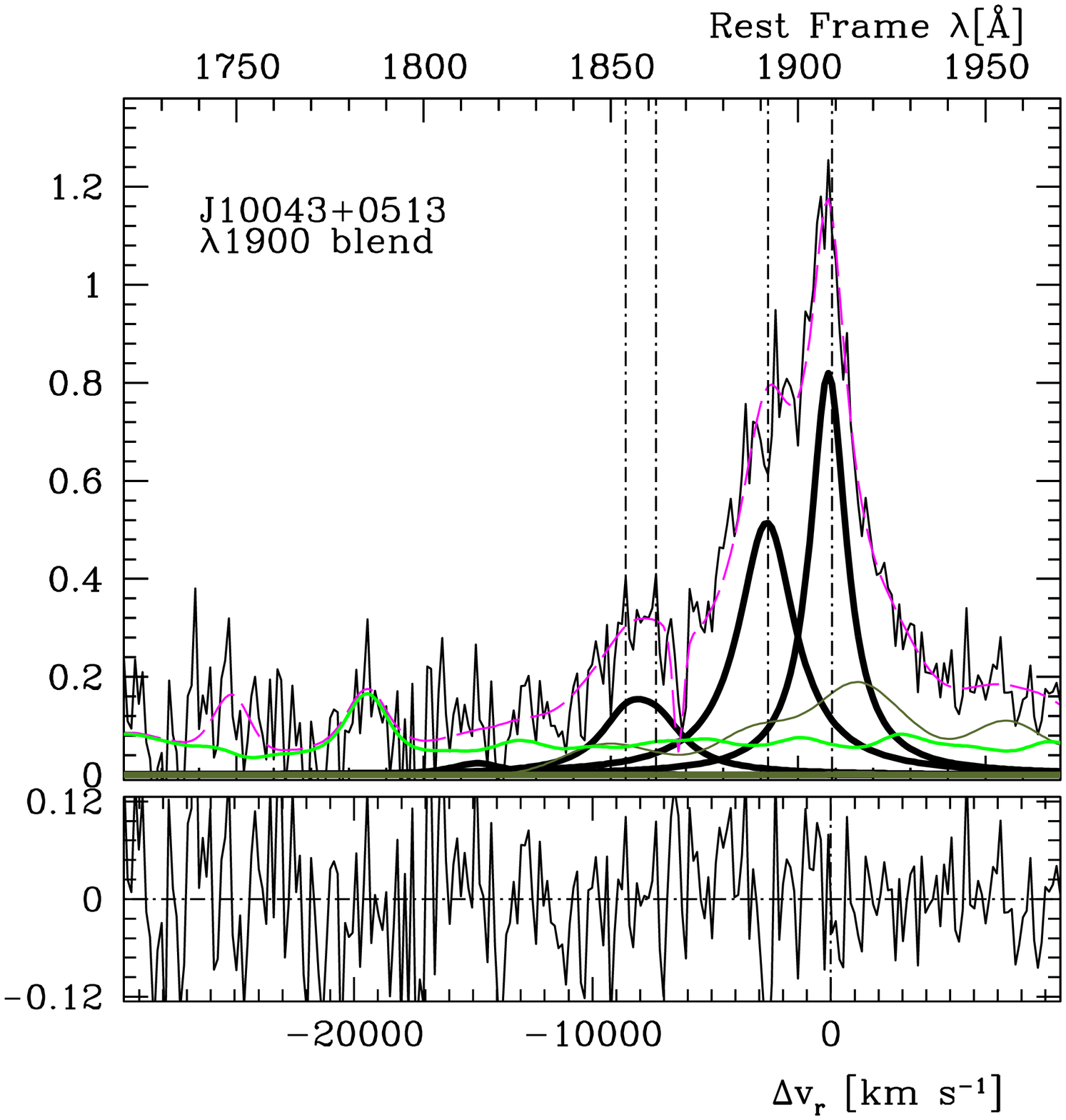}
\includegraphics[width=0.225\columnwidth]{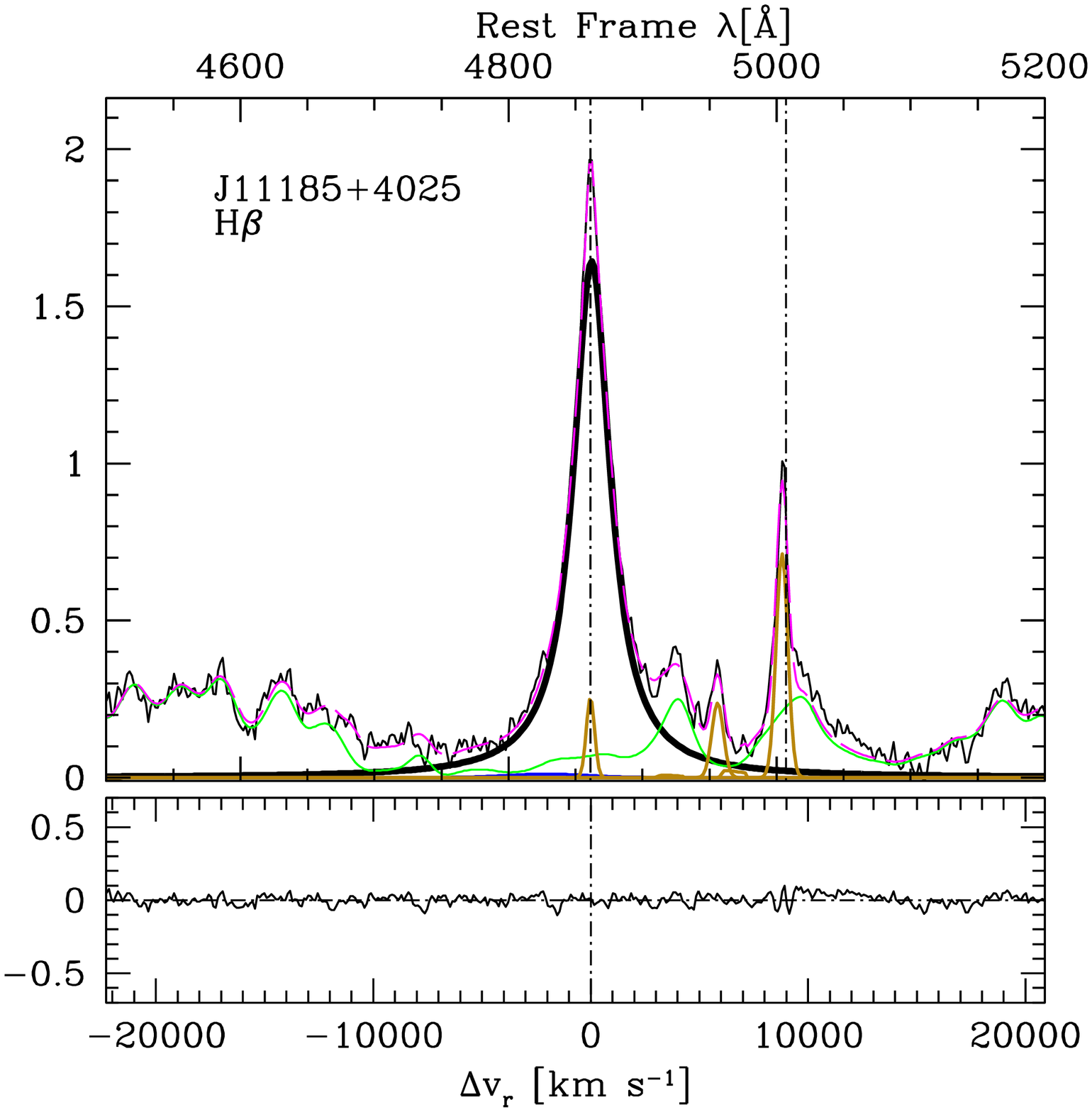}\includegraphics[width=0.225\columnwidth]{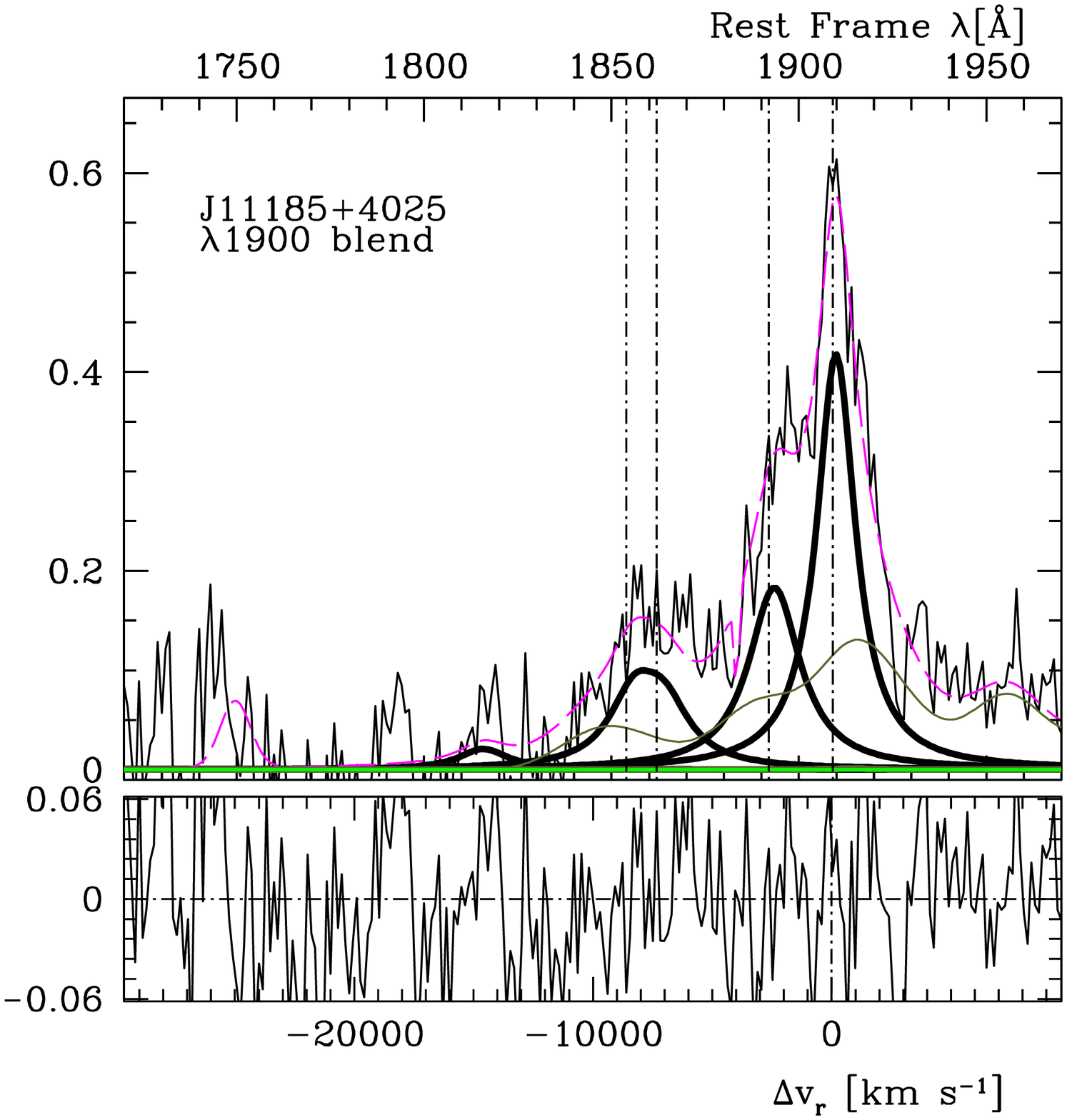}\\ 
\includegraphics[width=0.225\columnwidth]{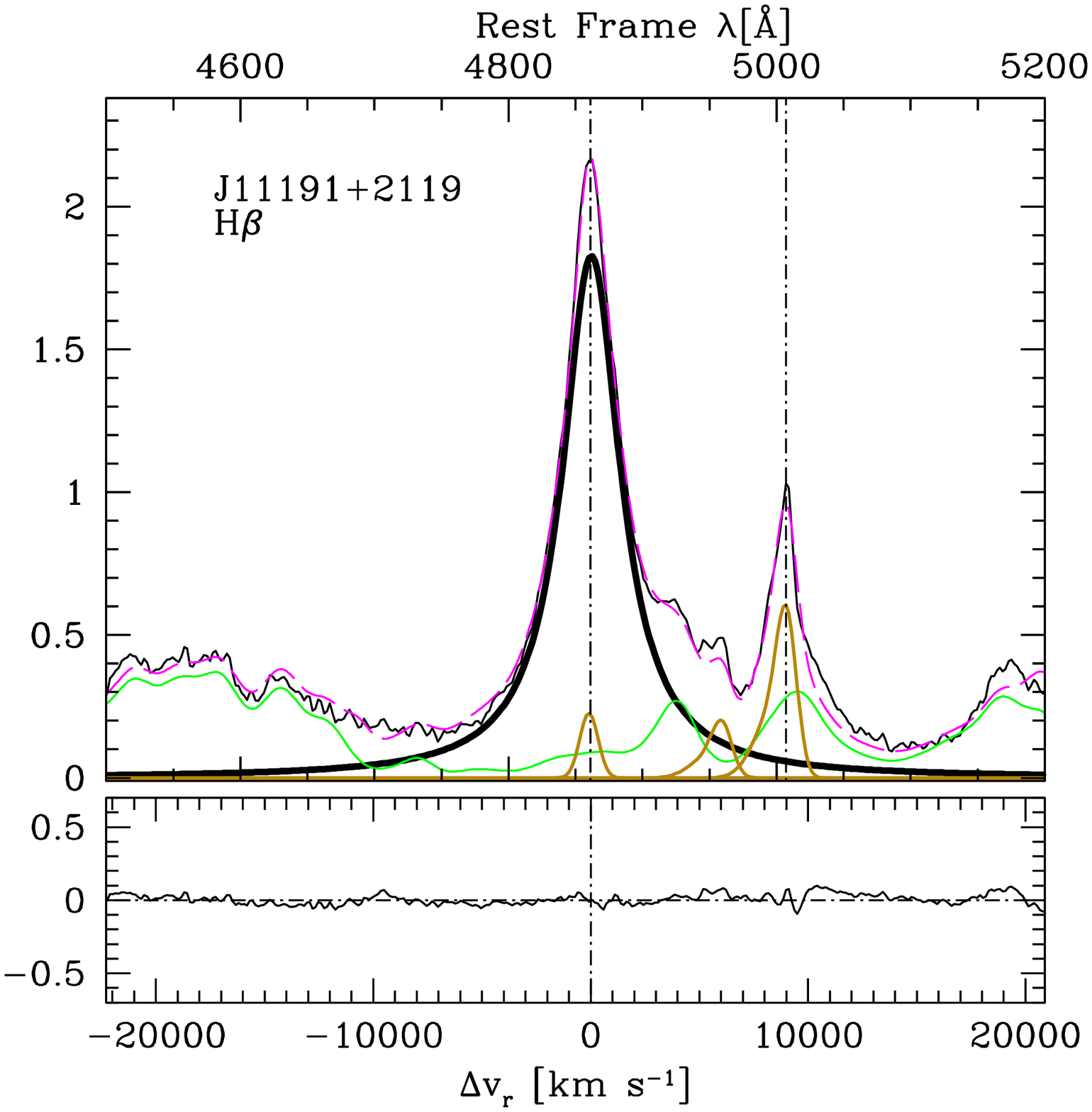}\includegraphics[width=0.225\columnwidth]{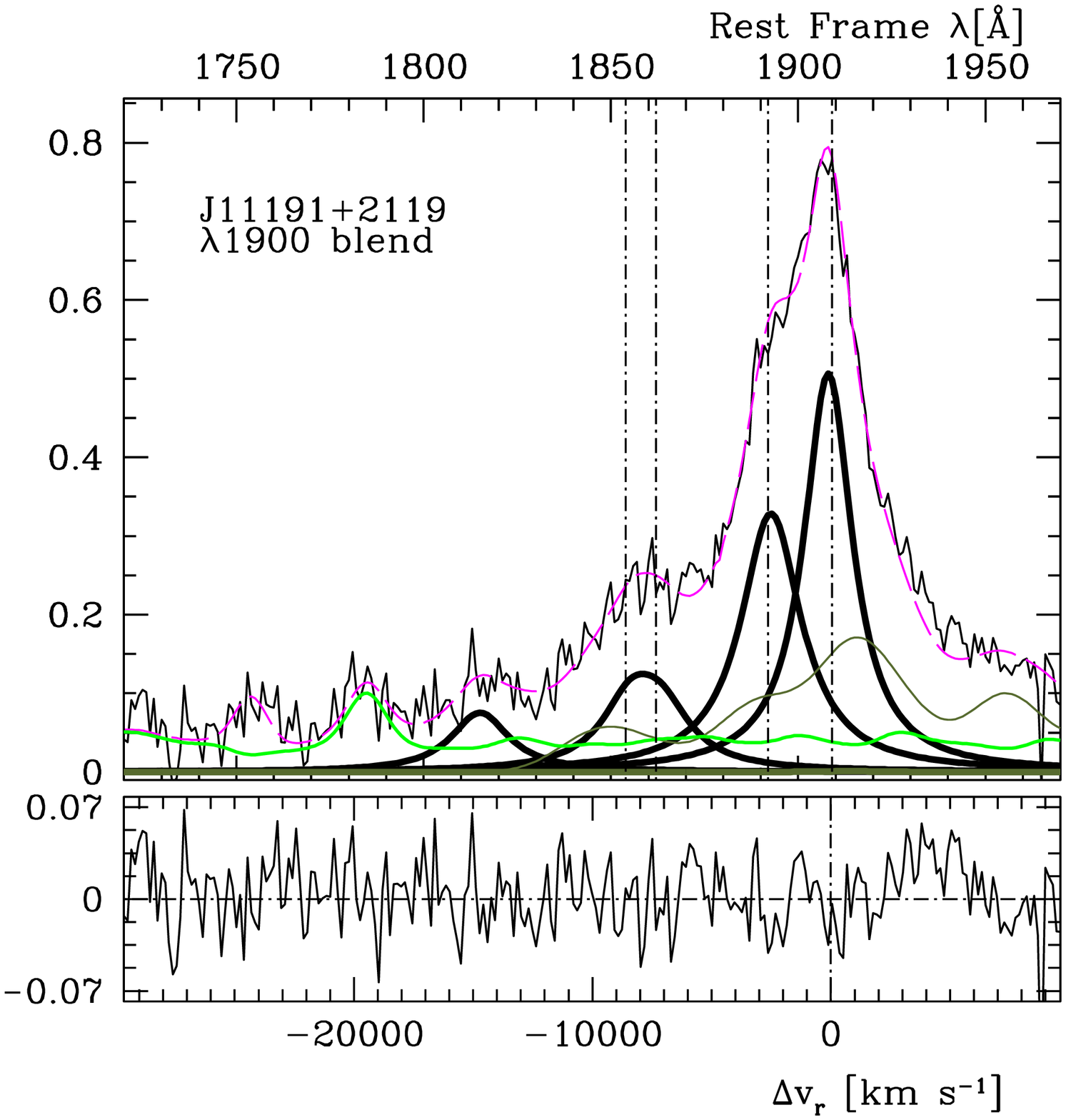}
\includegraphics[width=0.225\columnwidth]{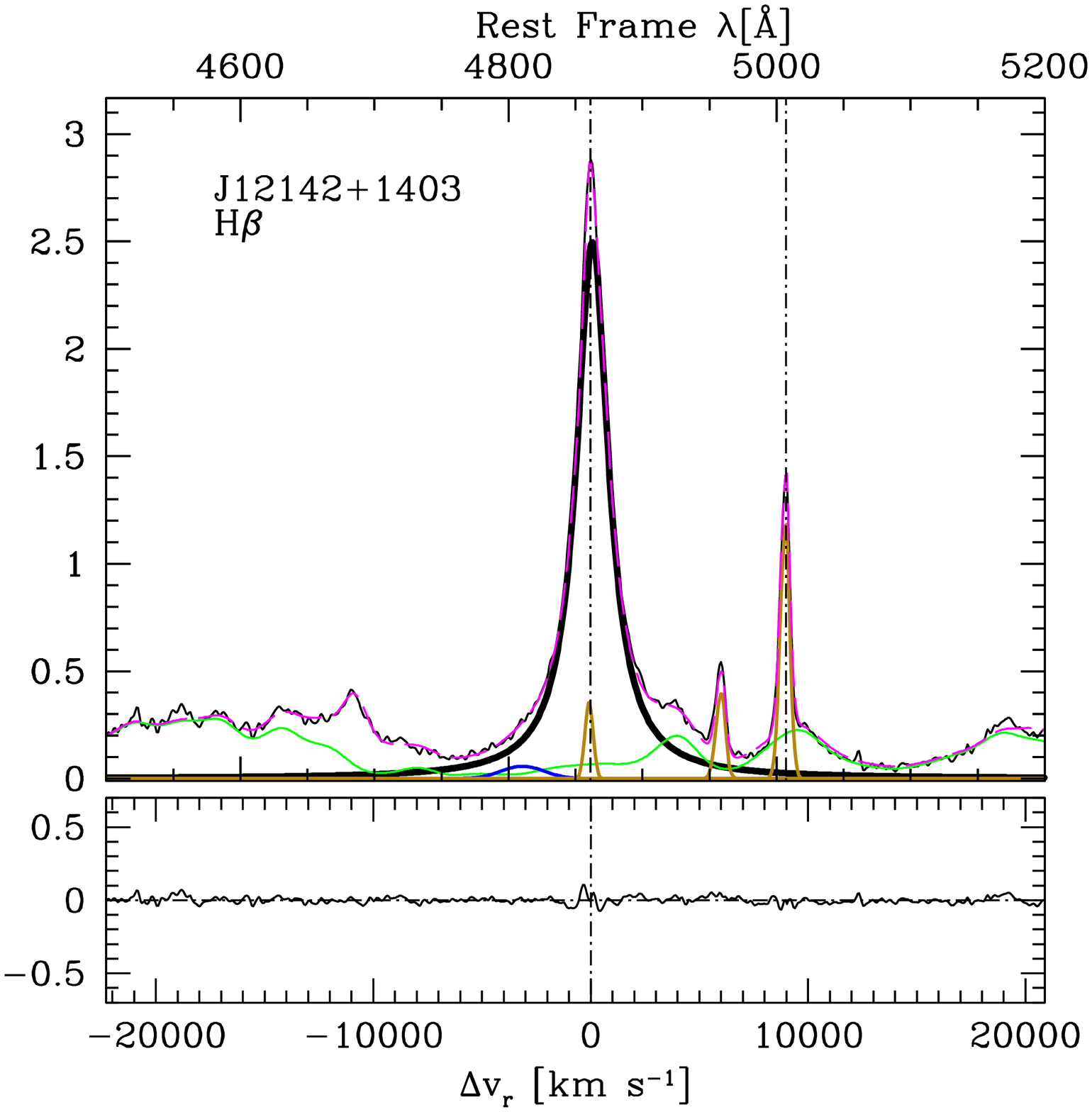}\includegraphics[width=0.225\columnwidth]{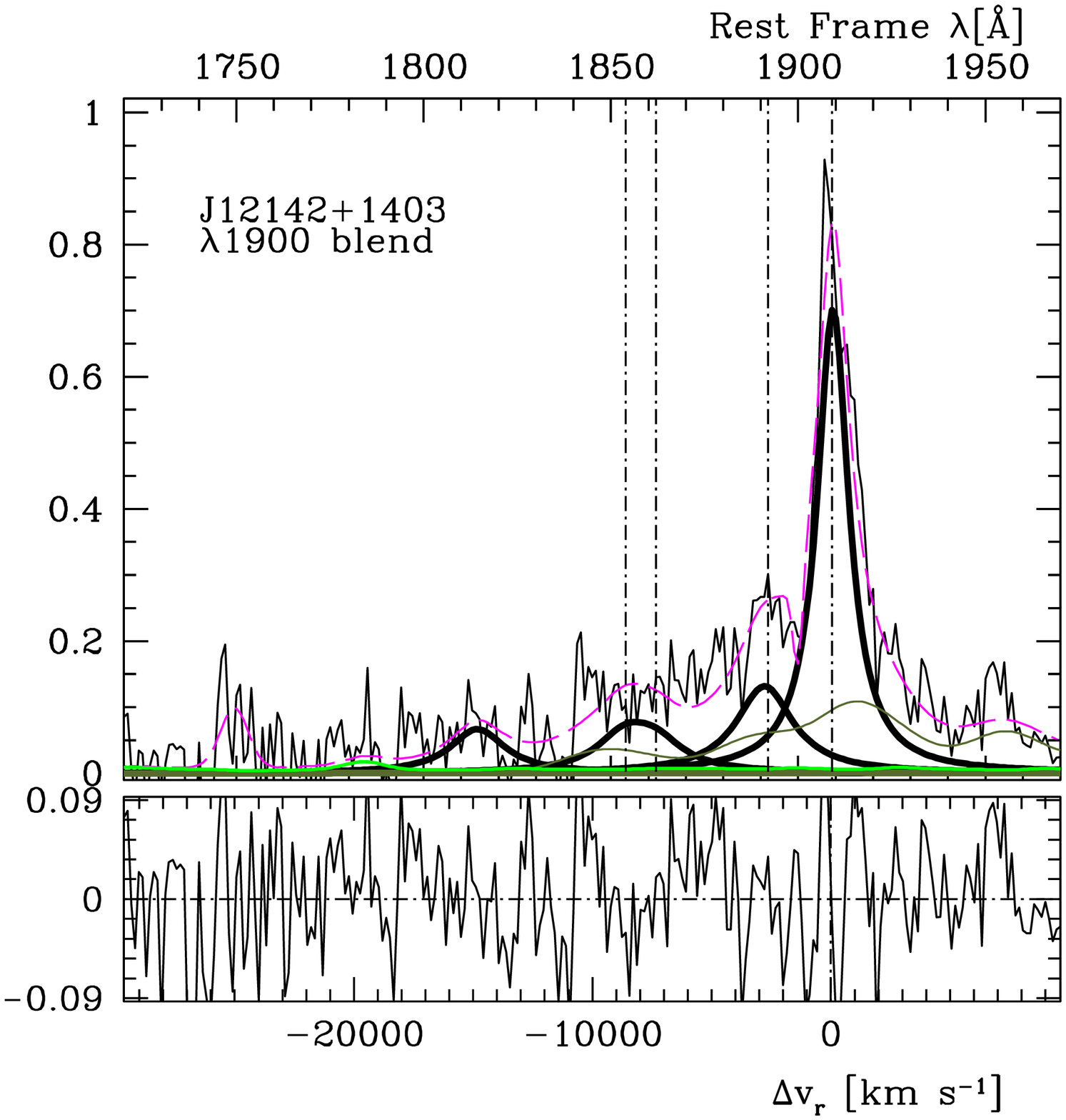}\\ 
\includegraphics[width=0.225\columnwidth]{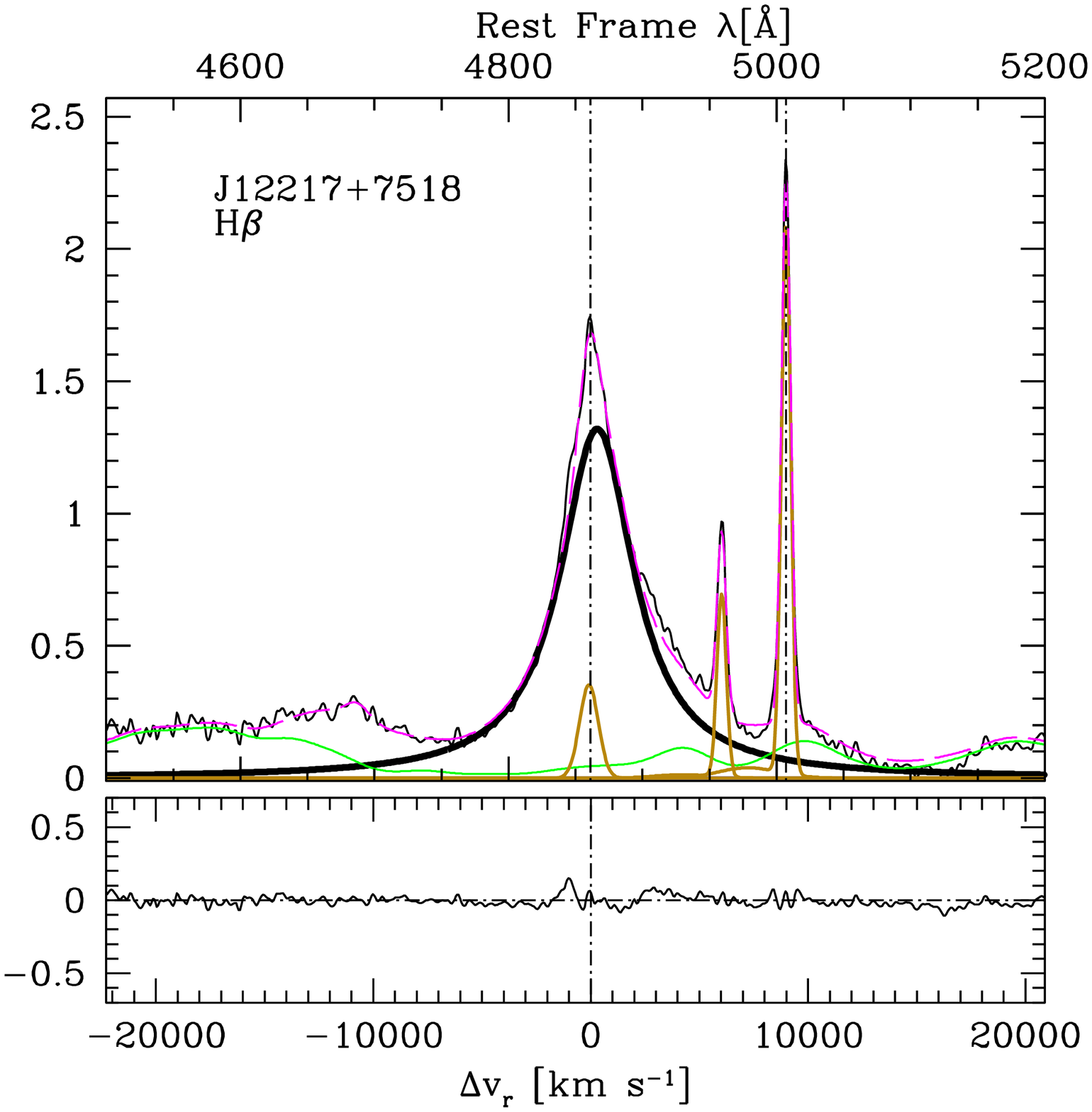}\includegraphics[width=0.225\columnwidth]{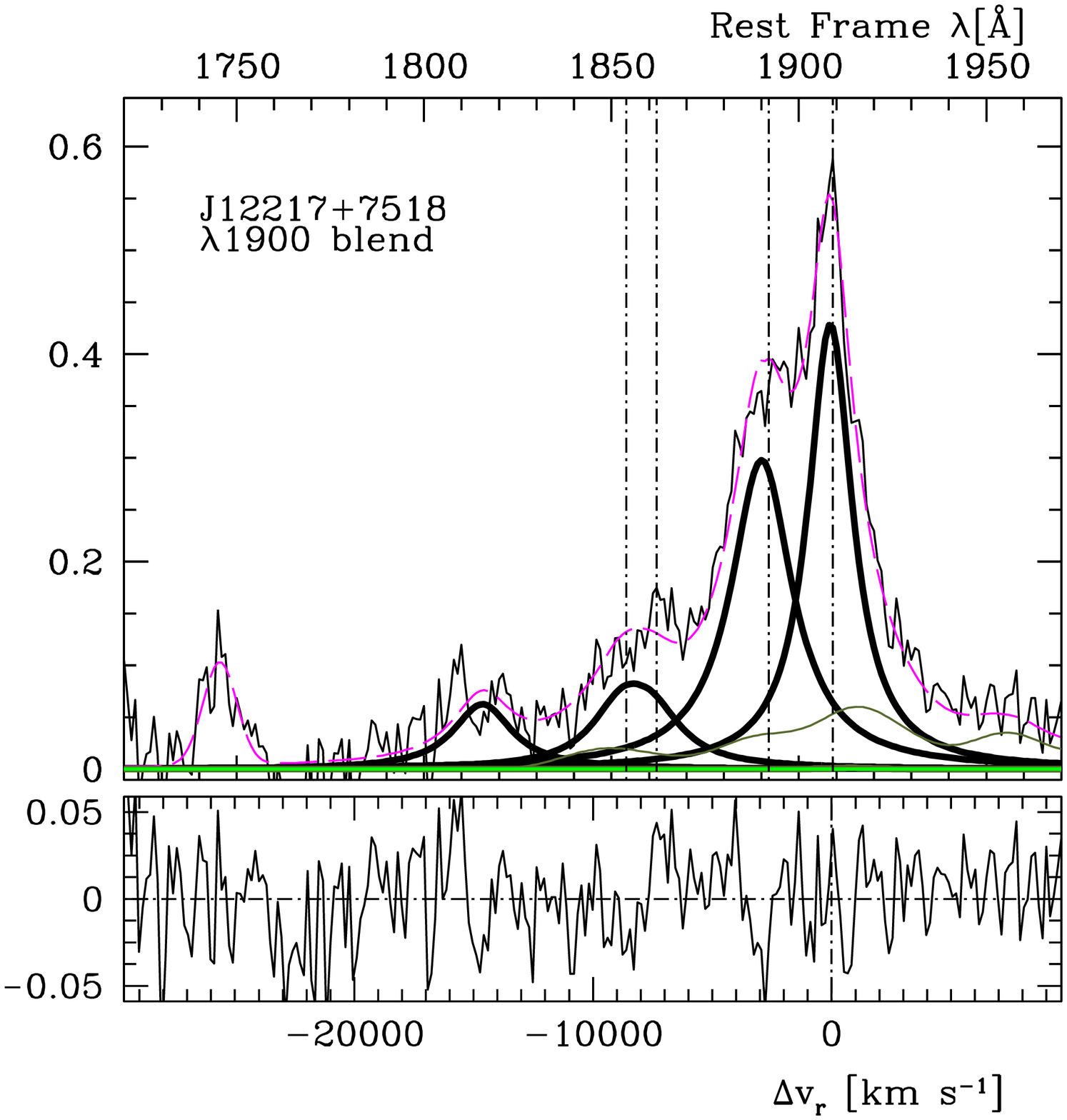}
\includegraphics[width=0.225\columnwidth]{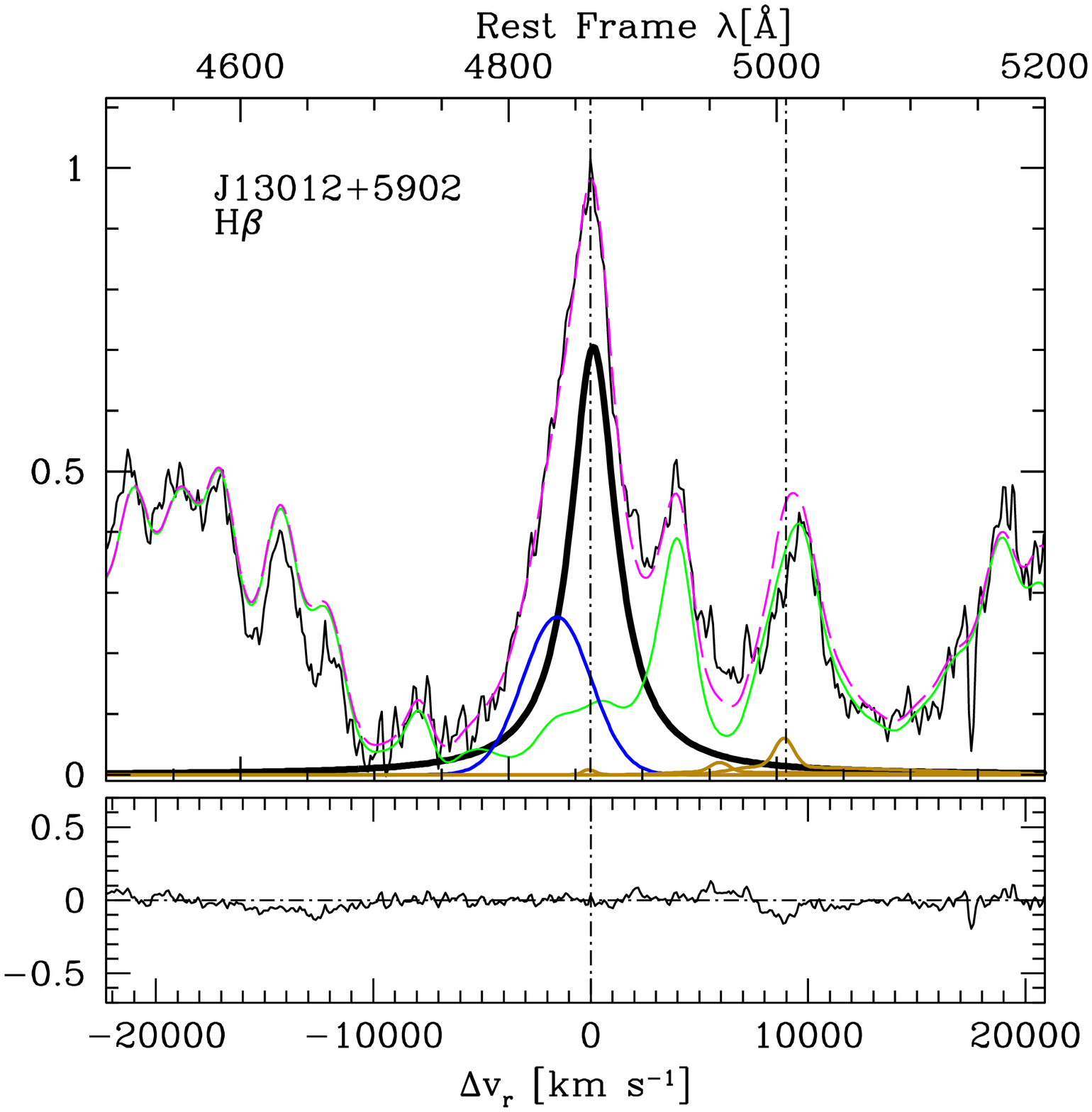}\includegraphics[width=0.225\columnwidth]{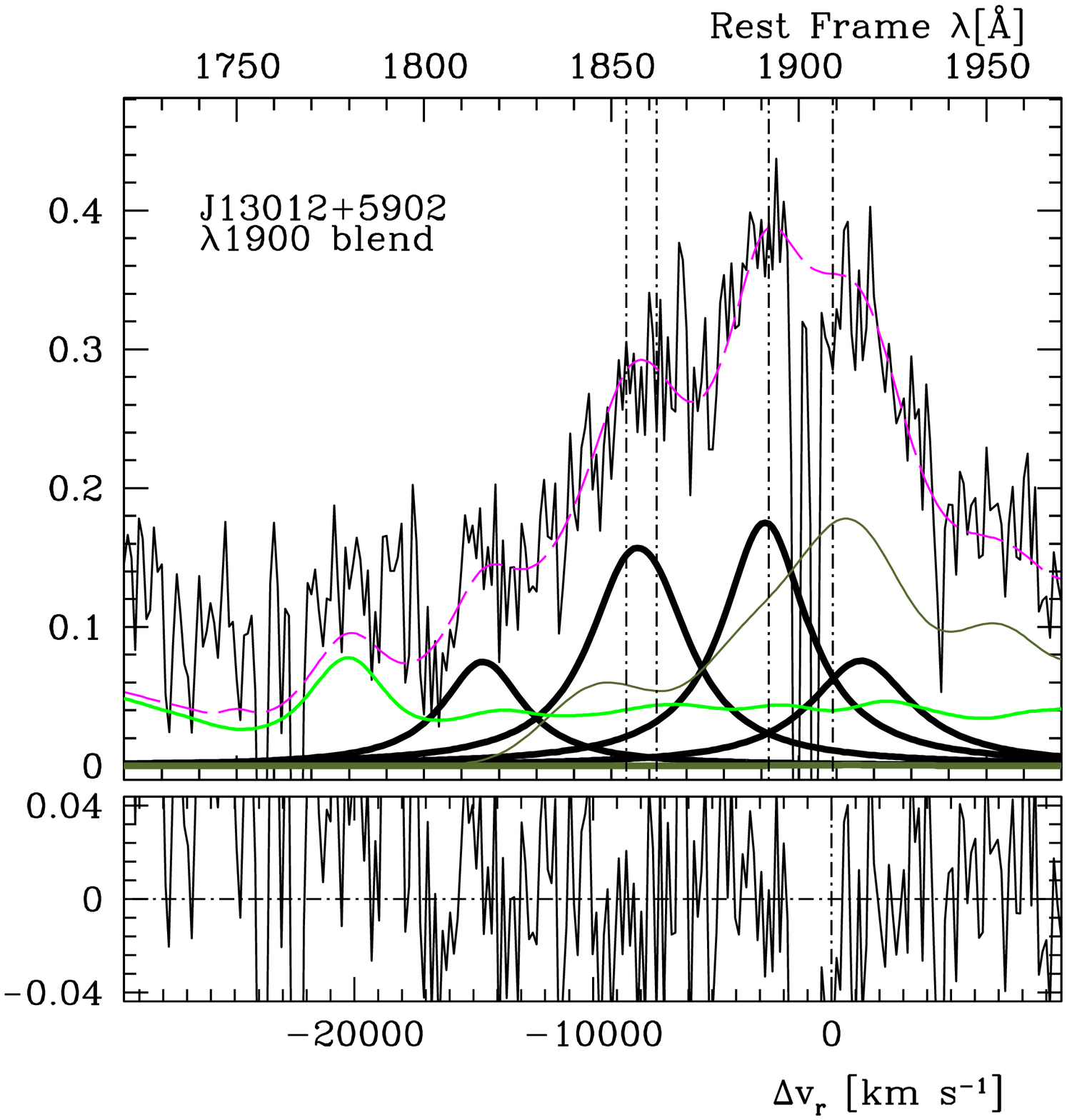}\\ 
\includegraphics[width=0.225\columnwidth]{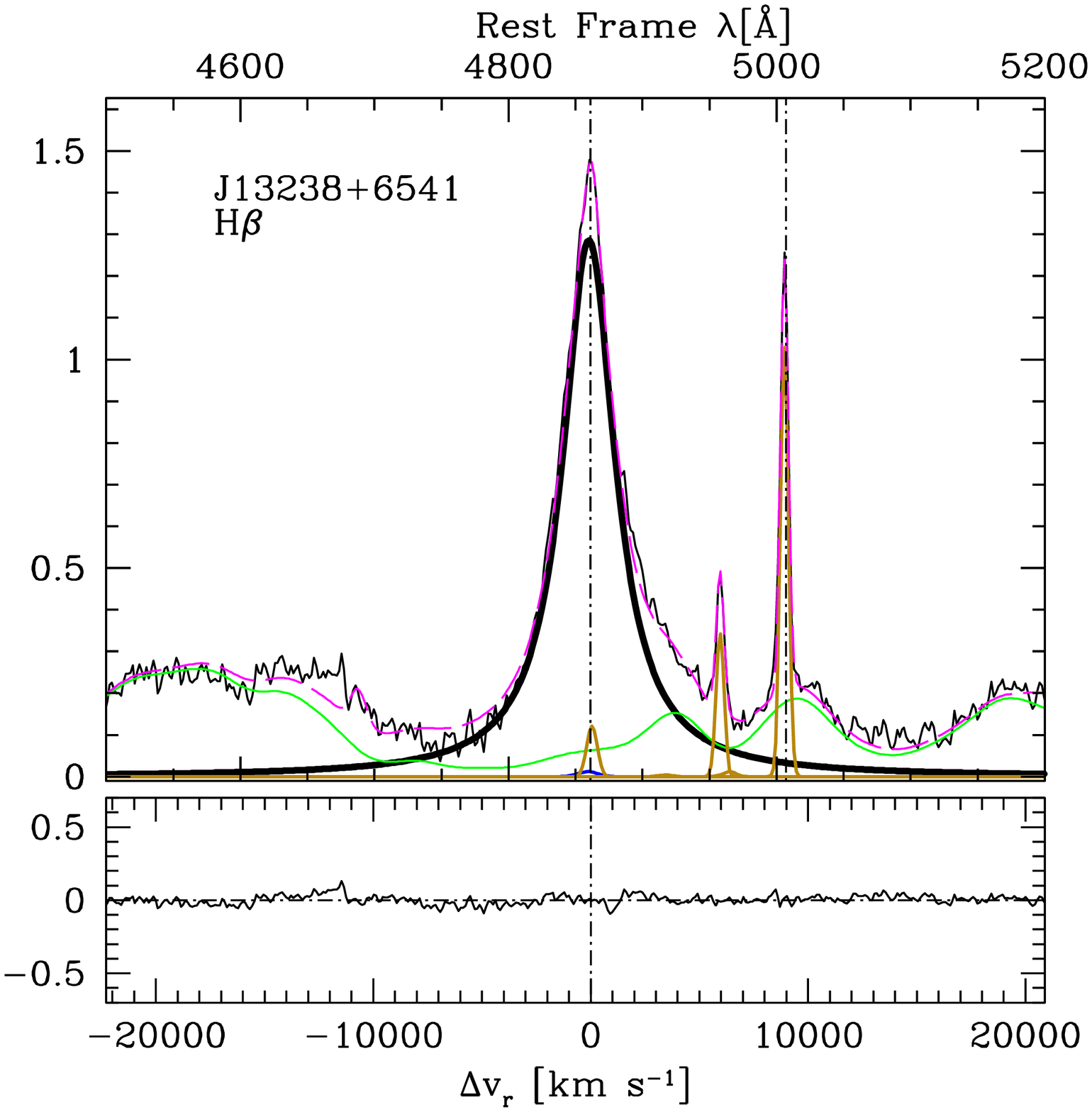}\includegraphics[width=0.225\columnwidth]{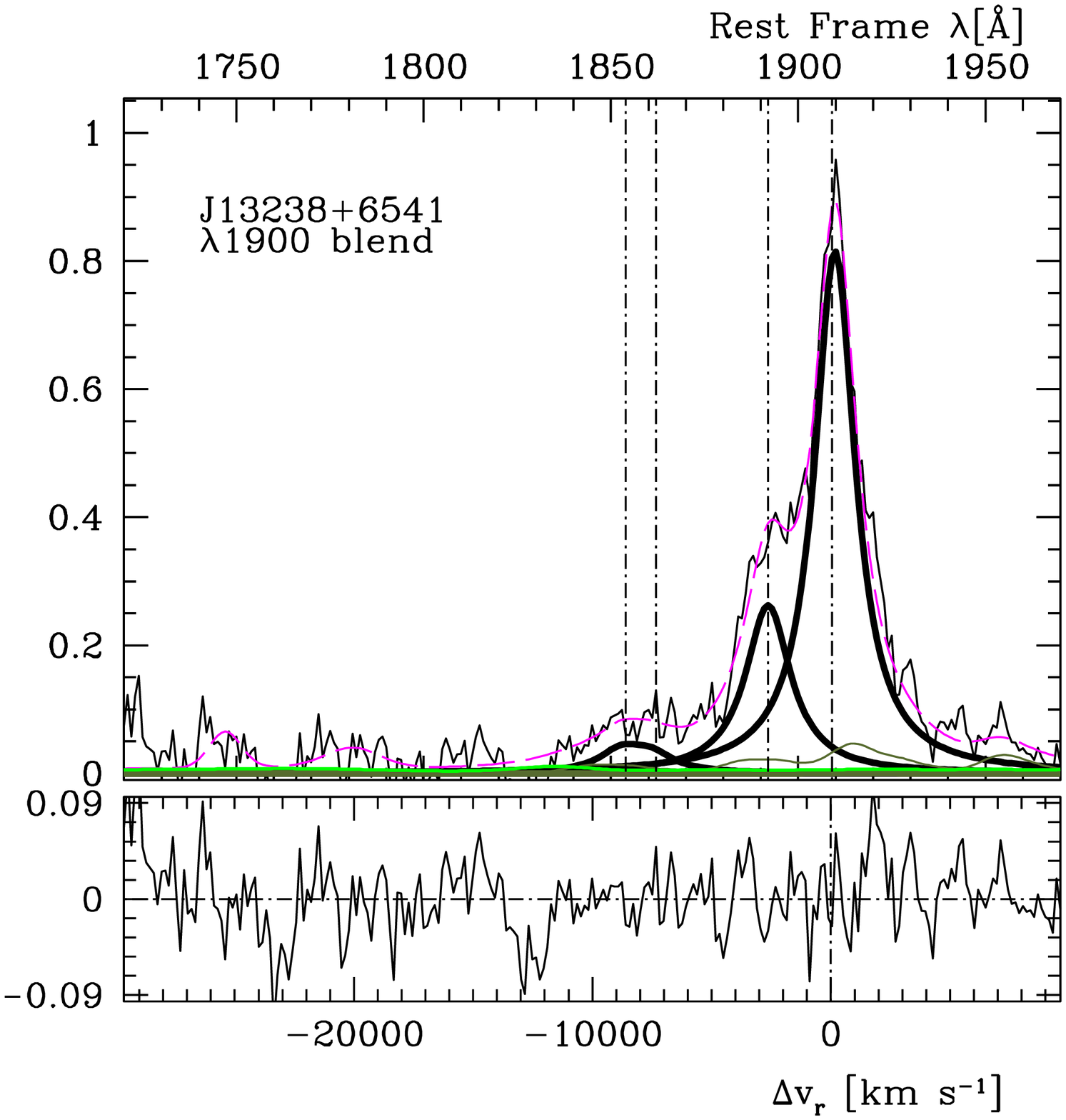}
\includegraphics[width=0.225\columnwidth]{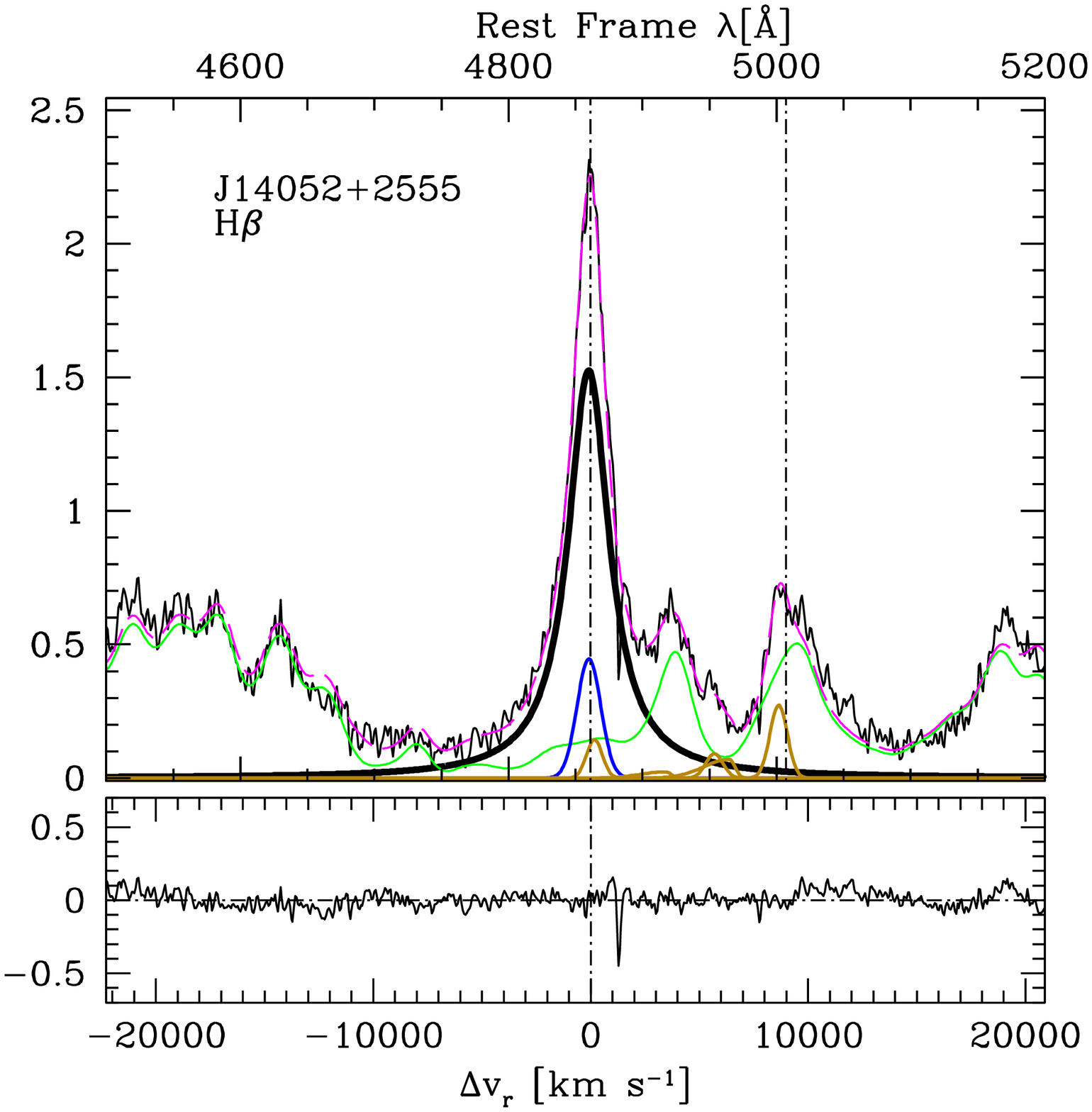}\includegraphics[width=0.225\columnwidth]{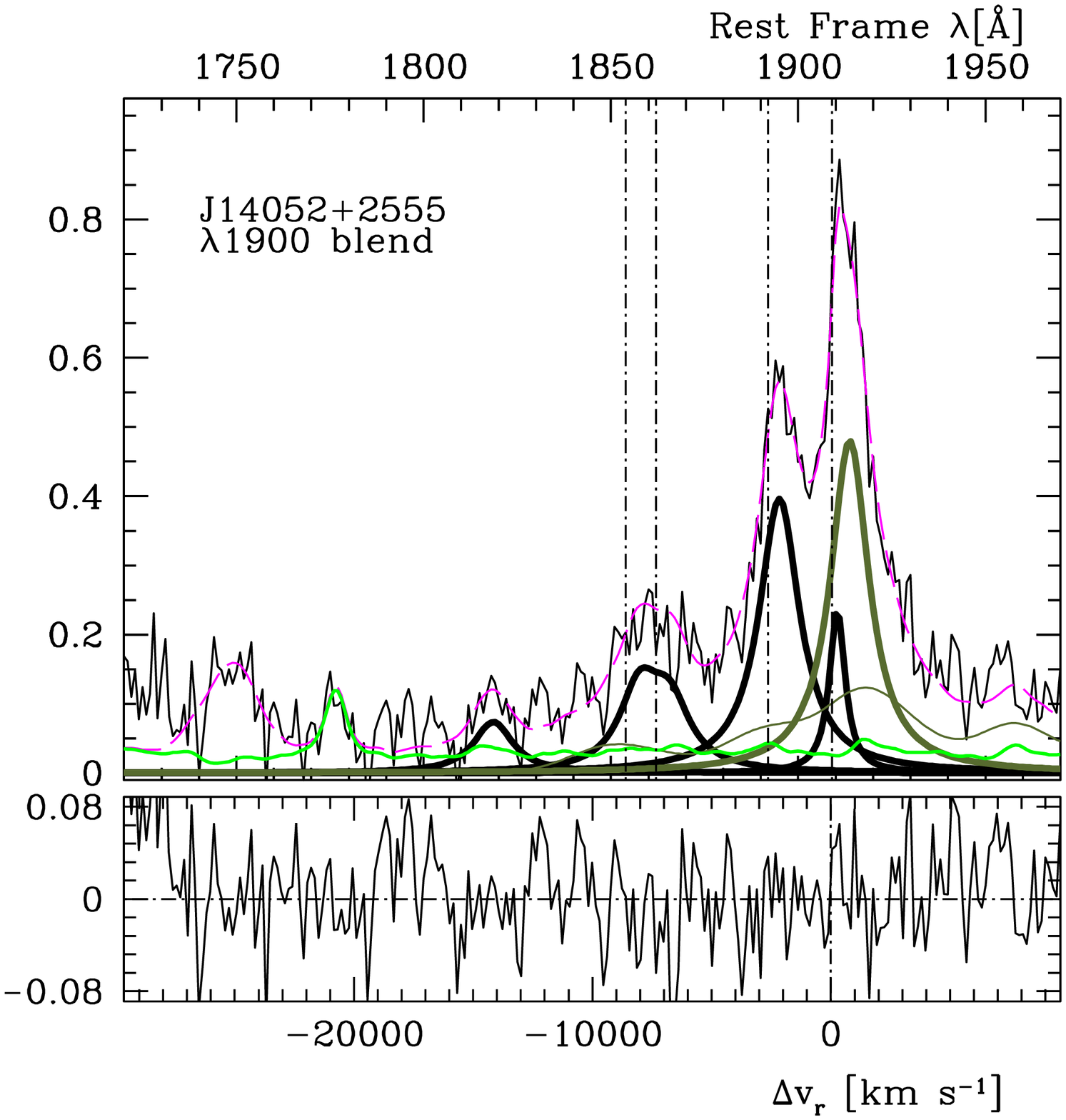} 
\caption{Analysis of \hb\ and of the 1900 \AA\ blend\ for Pop. A sources (cont.). \label{fig:fosa1}}
\end{figure} 
\addtocounter{figure}{-1}

\begin{figure}
\includegraphics[width=0.225\columnwidth]{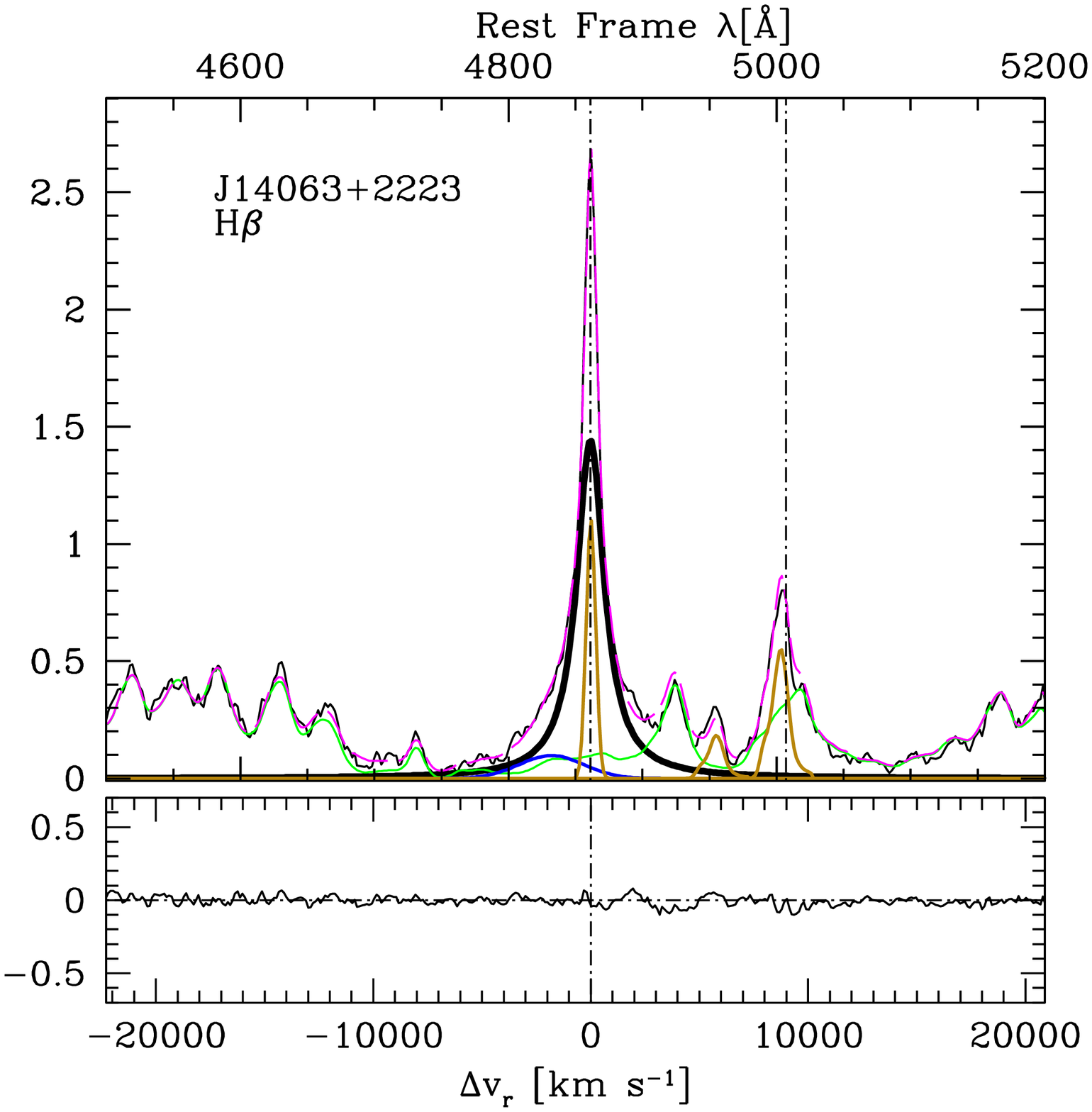}\includegraphics[width=0.225\columnwidth]{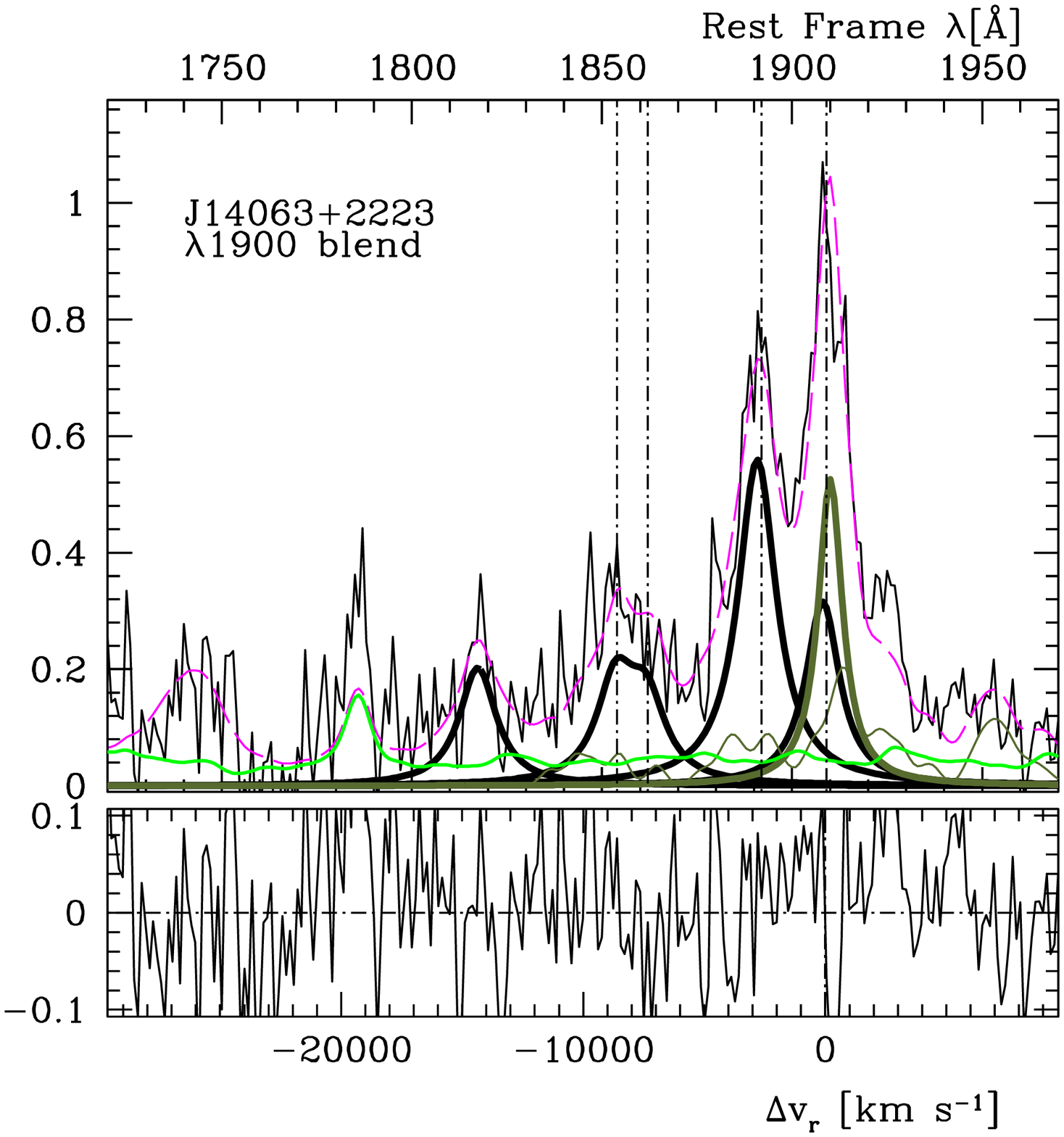}
\includegraphics[width=0.225\columnwidth]{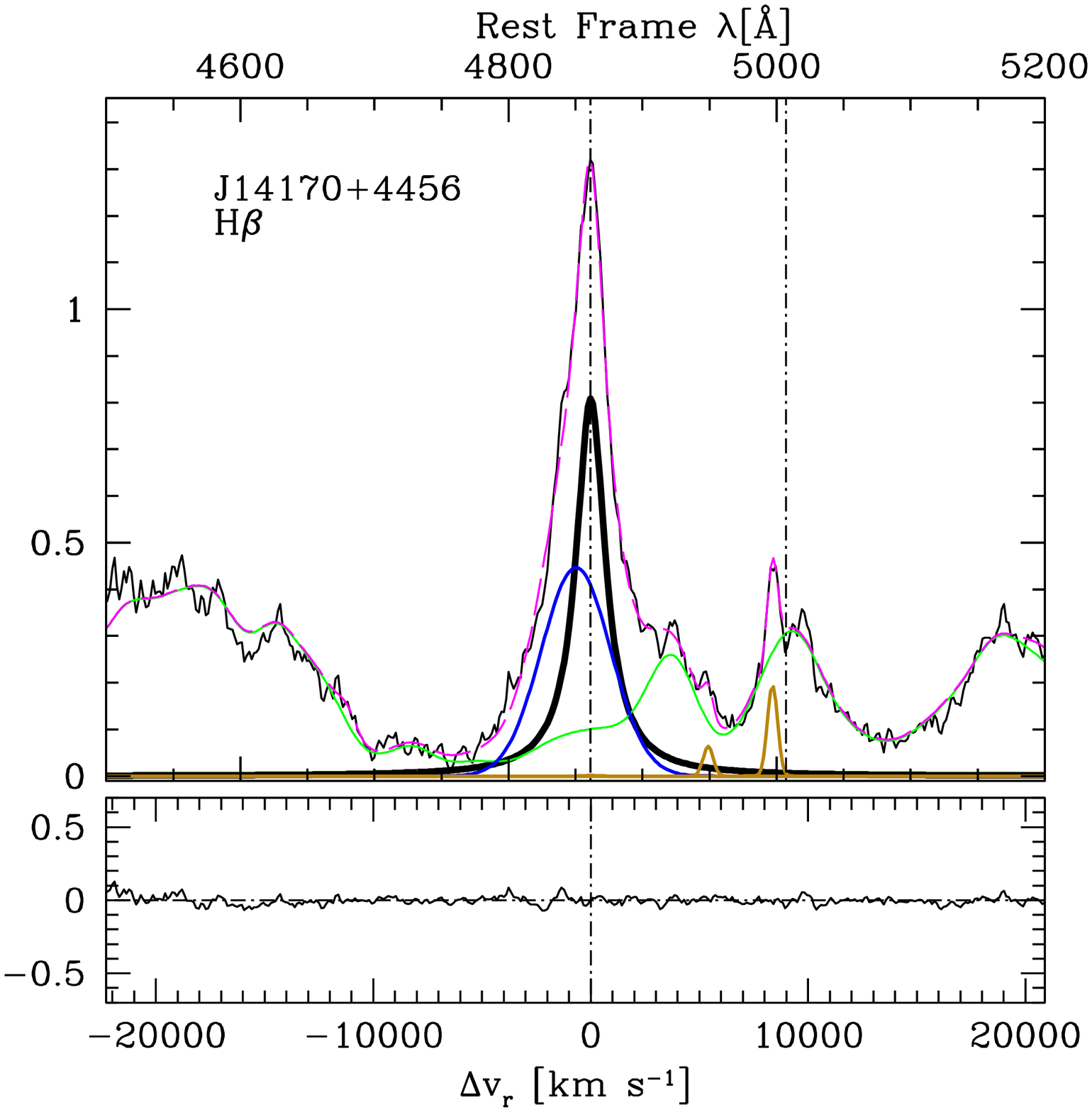}\includegraphics[width=0.225\columnwidth]{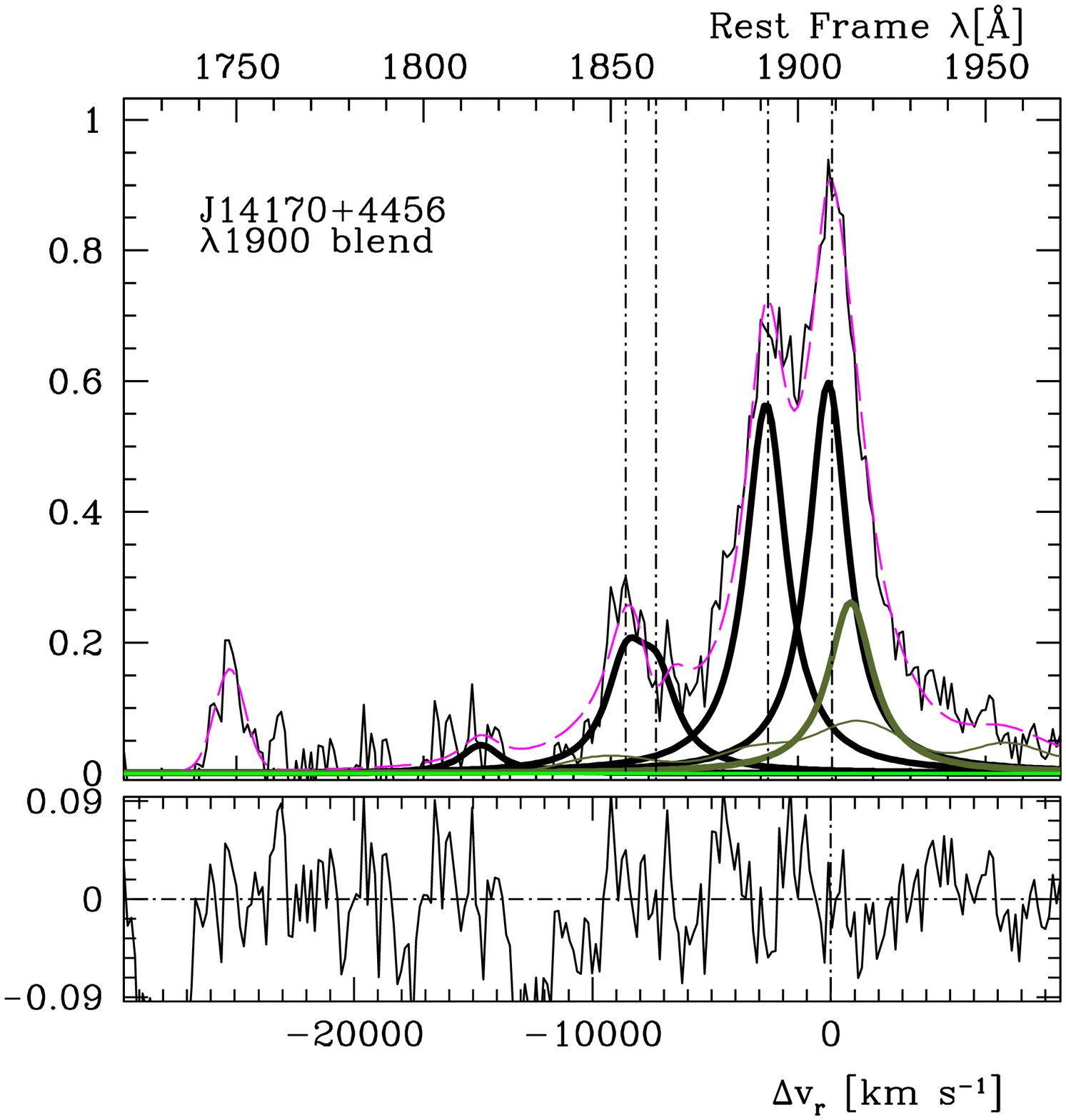}\\ 
\includegraphics[width=0.225\columnwidth]{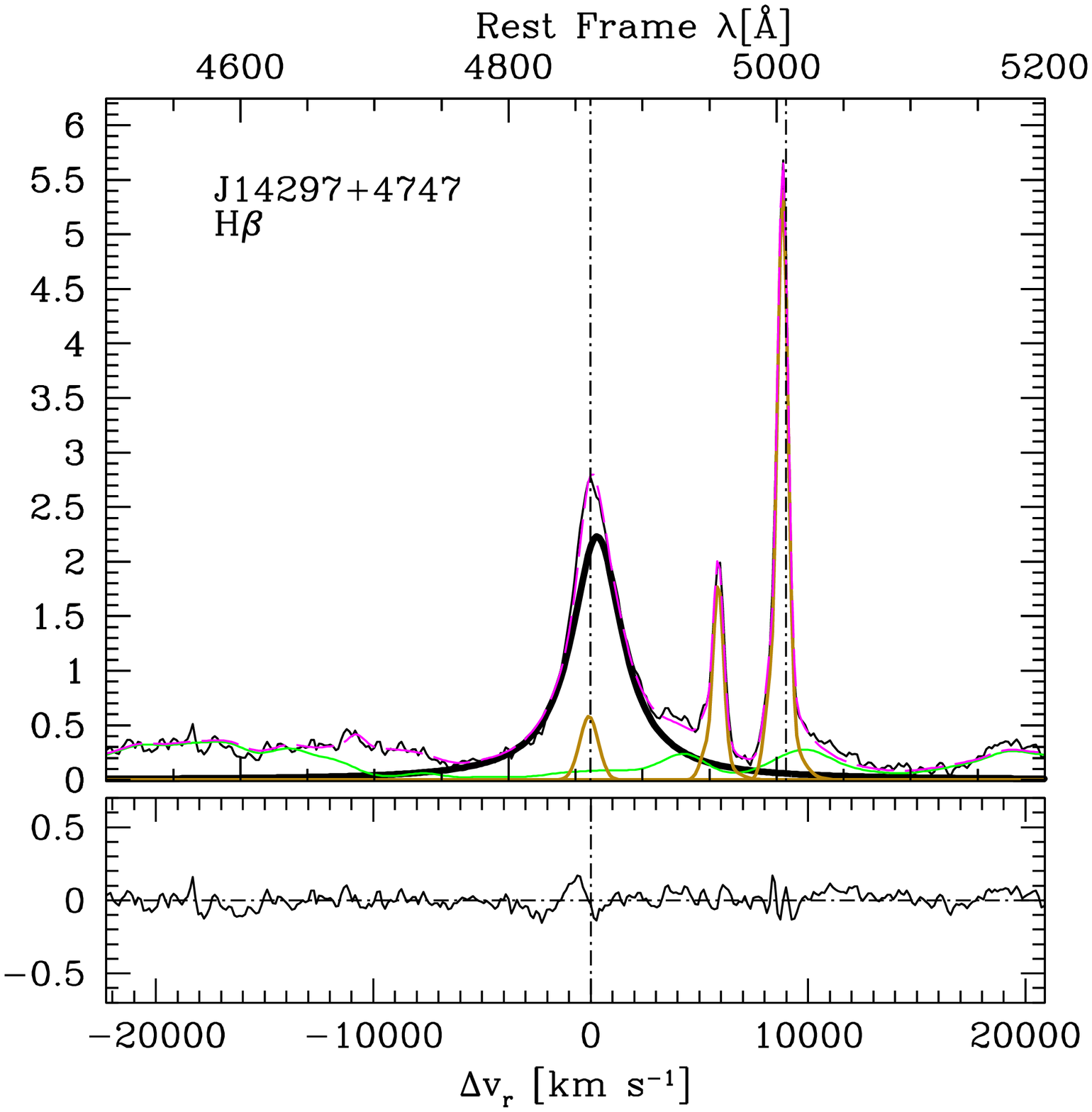}\includegraphics[width=0.225\columnwidth]{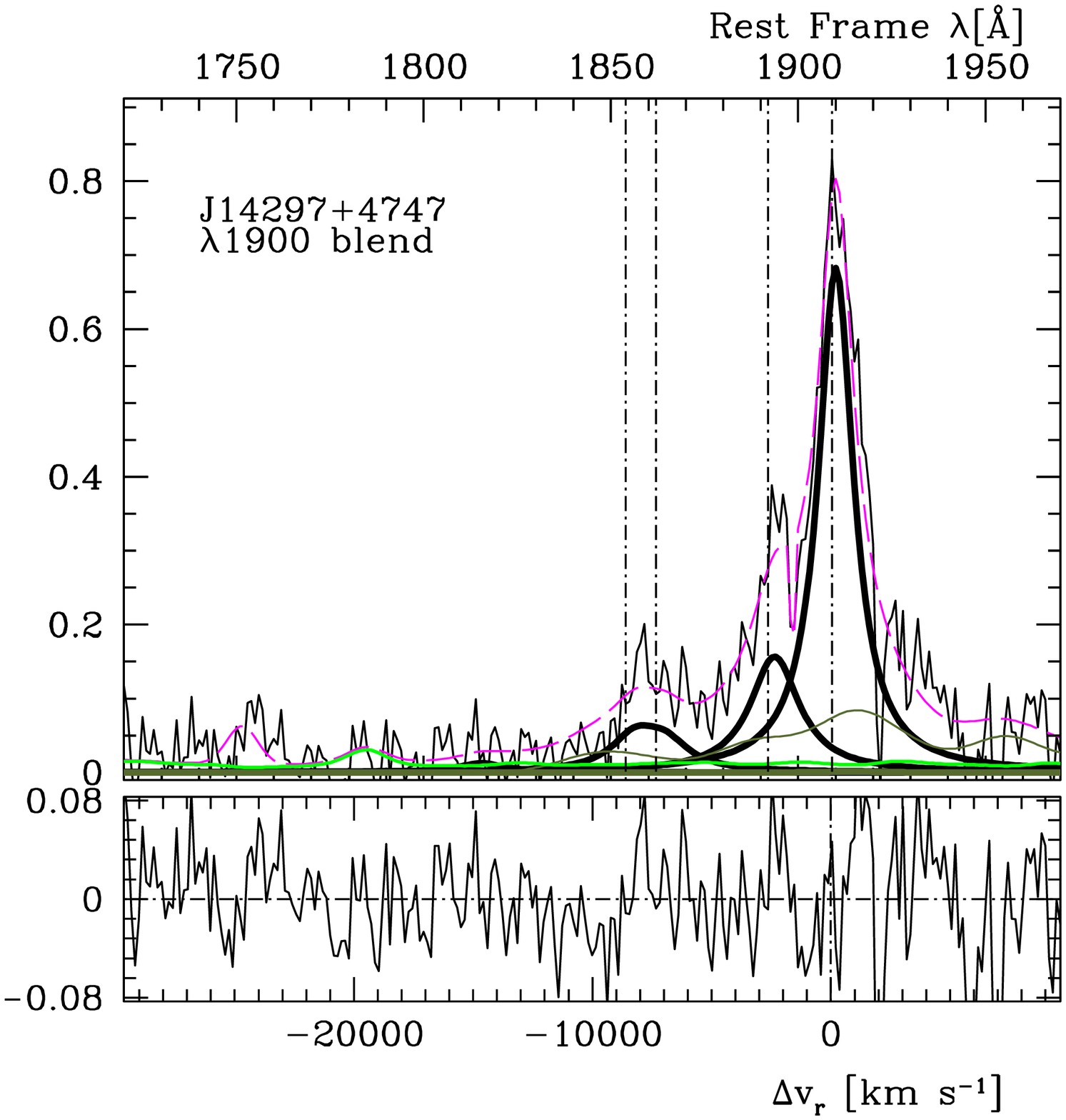}
\includegraphics[width=0.225\columnwidth]{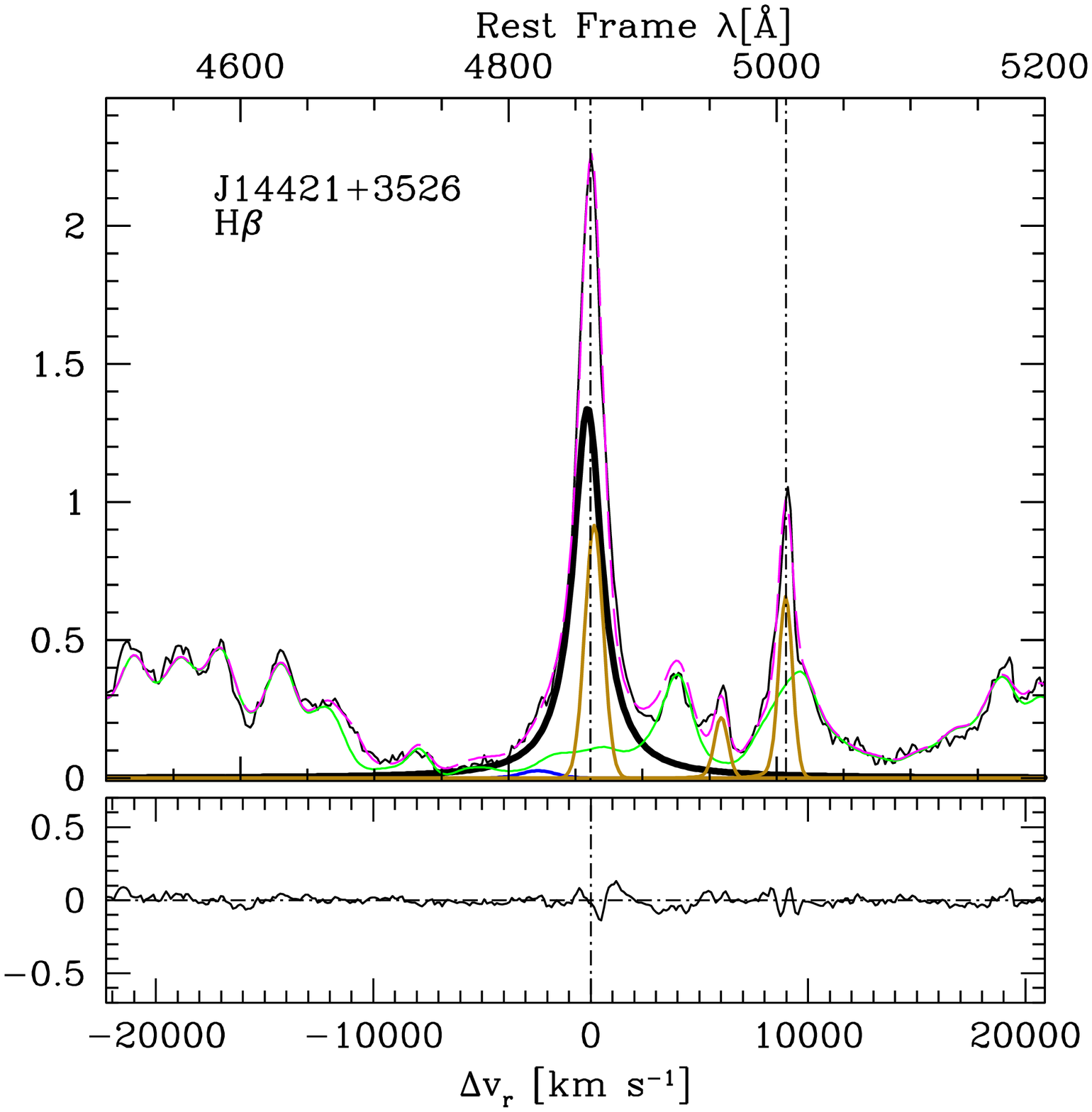}\includegraphics[width=0.225\columnwidth]{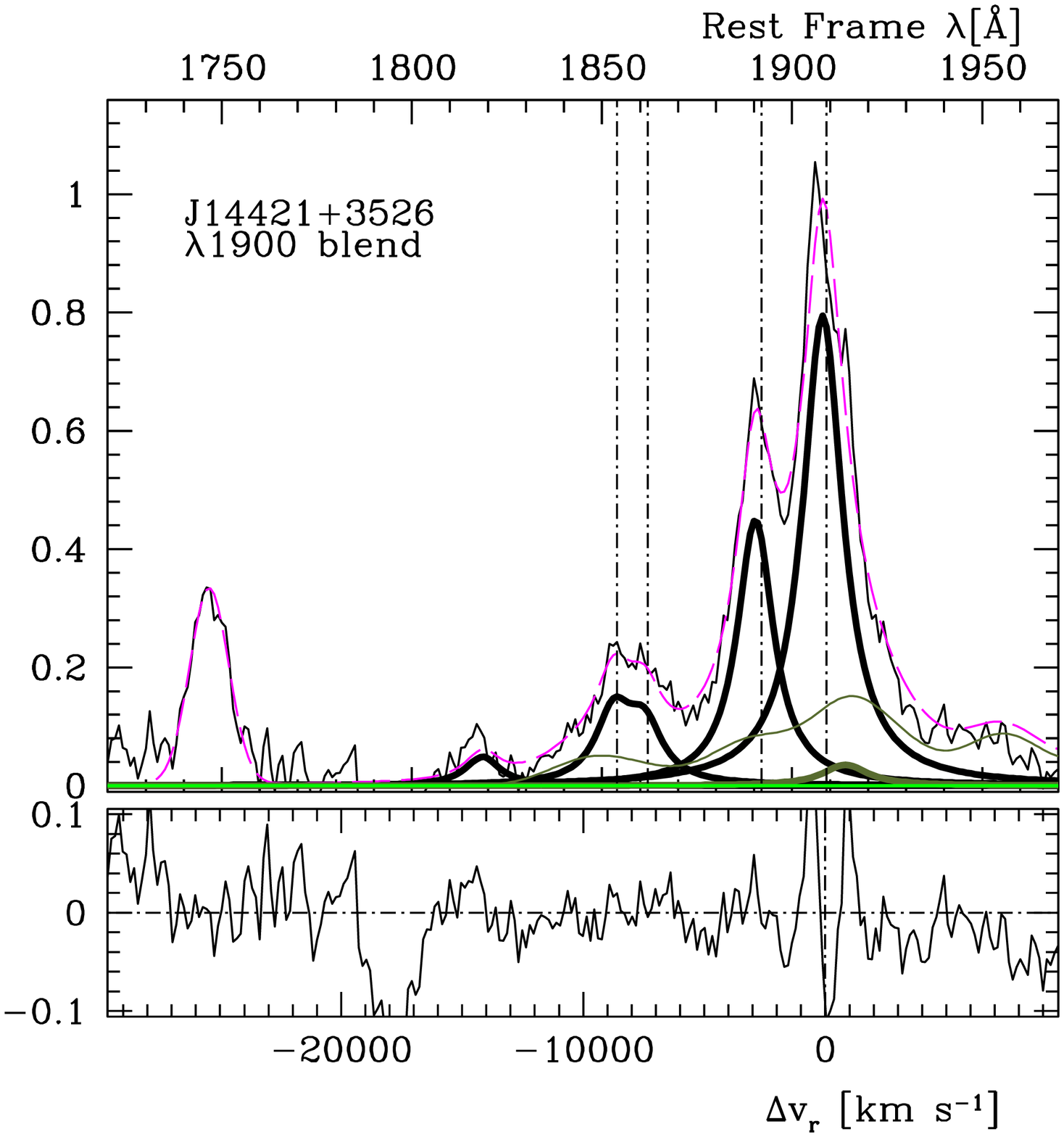}\\  
\includegraphics[width=0.225\columnwidth]{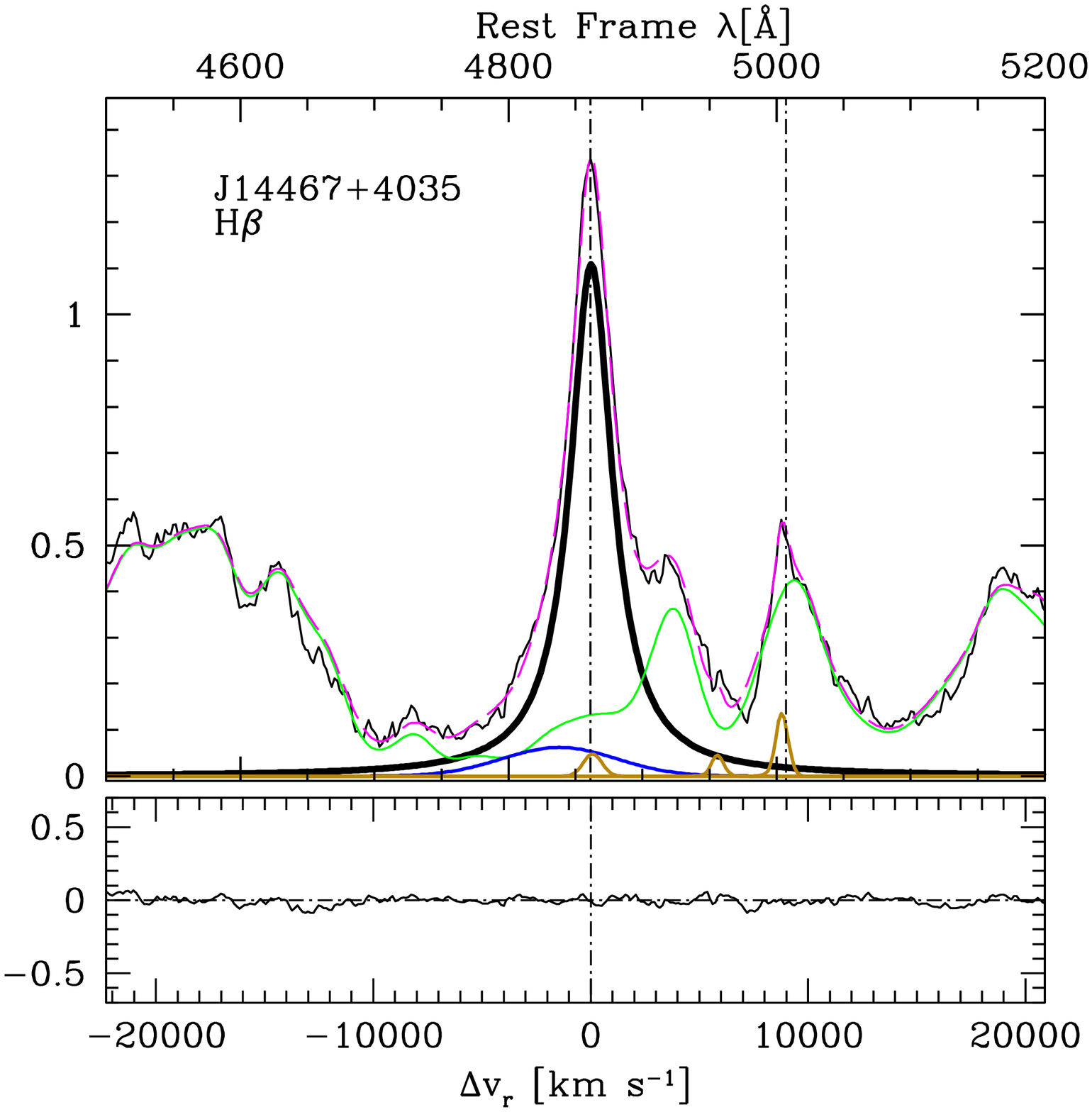}\includegraphics[width=0.225\columnwidth]{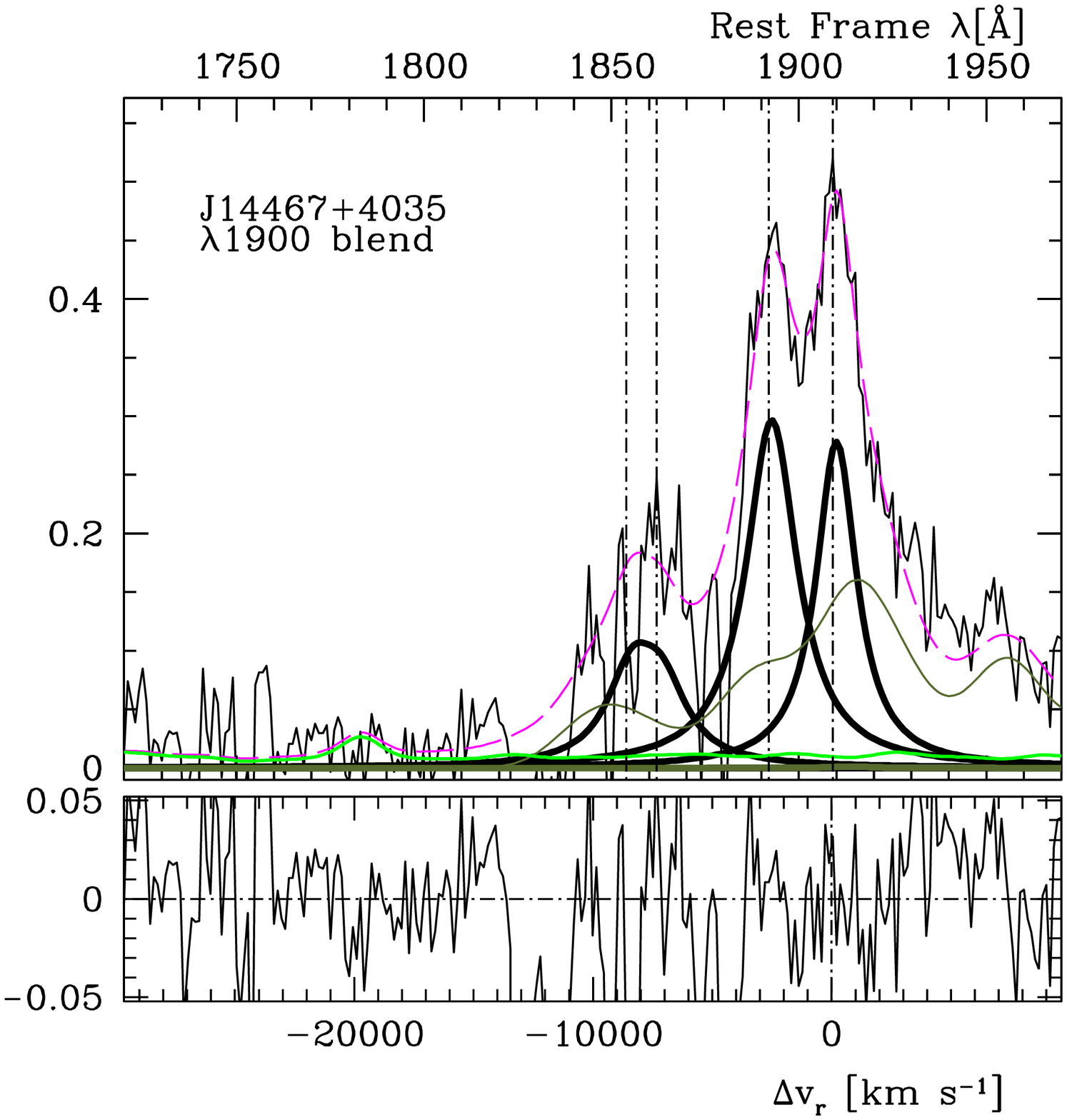}
\includegraphics[width=0.225\columnwidth]{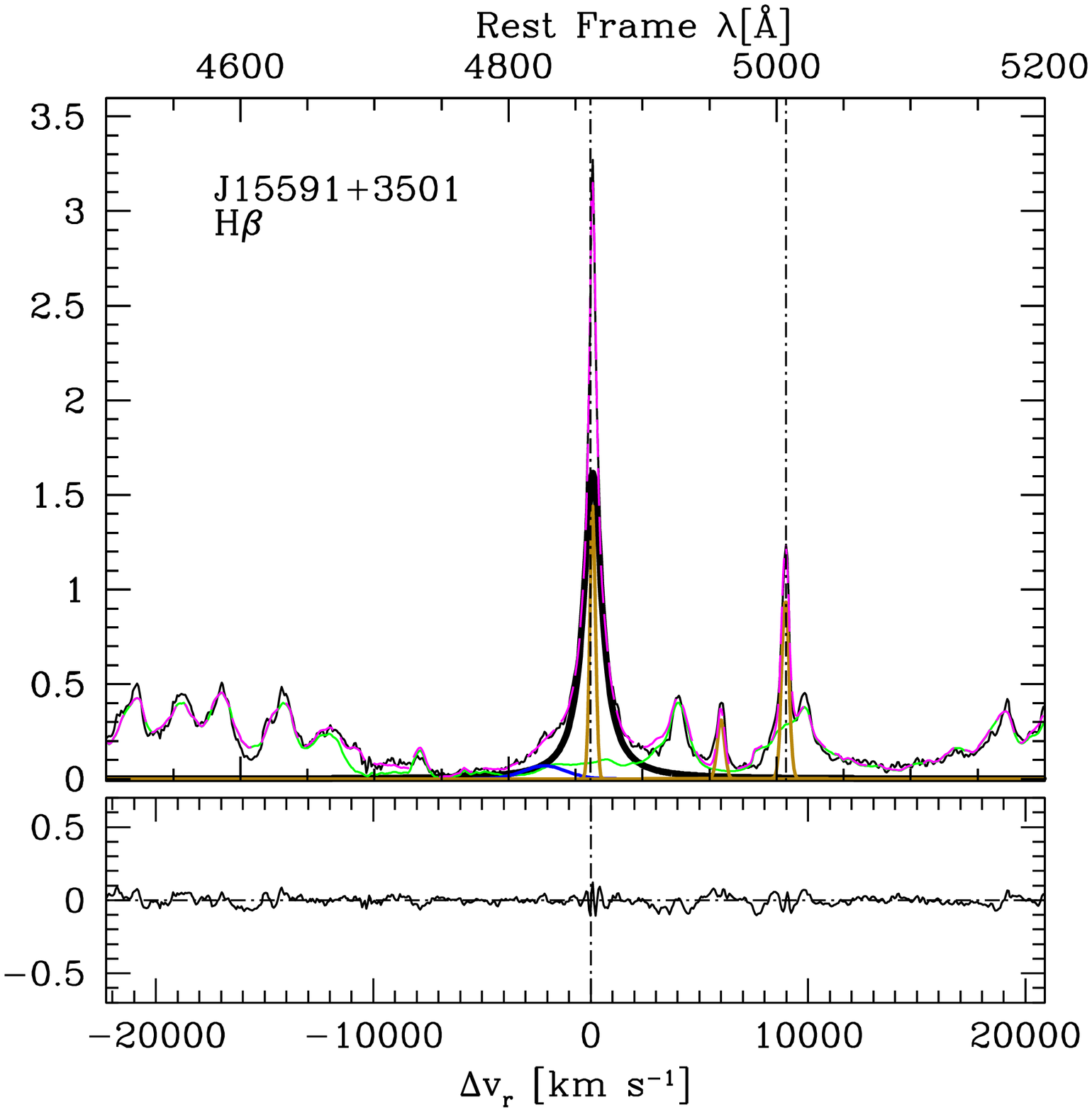}\includegraphics[width=0.225\columnwidth]{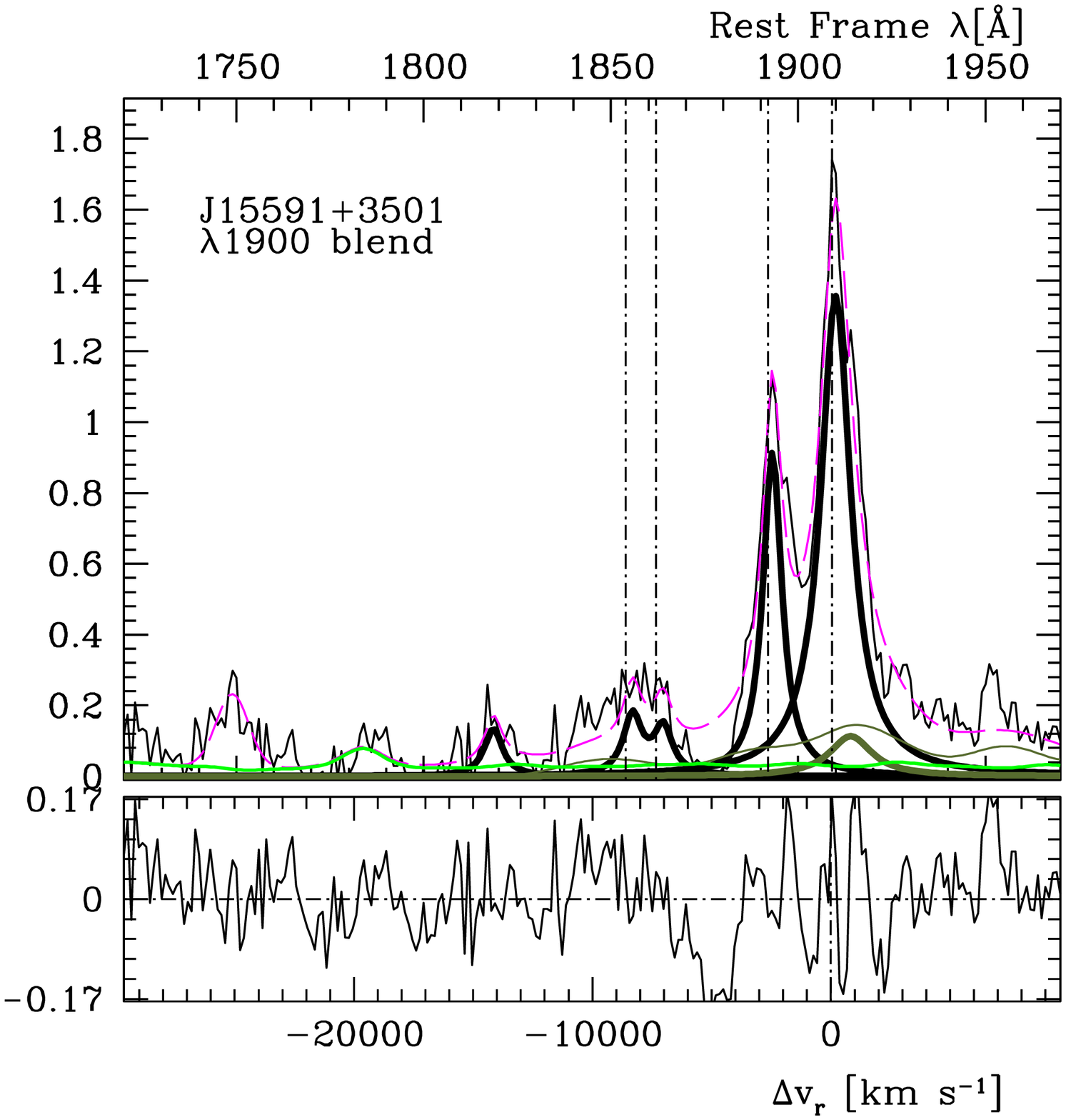}\\  
\includegraphics[width=0.225\columnwidth]{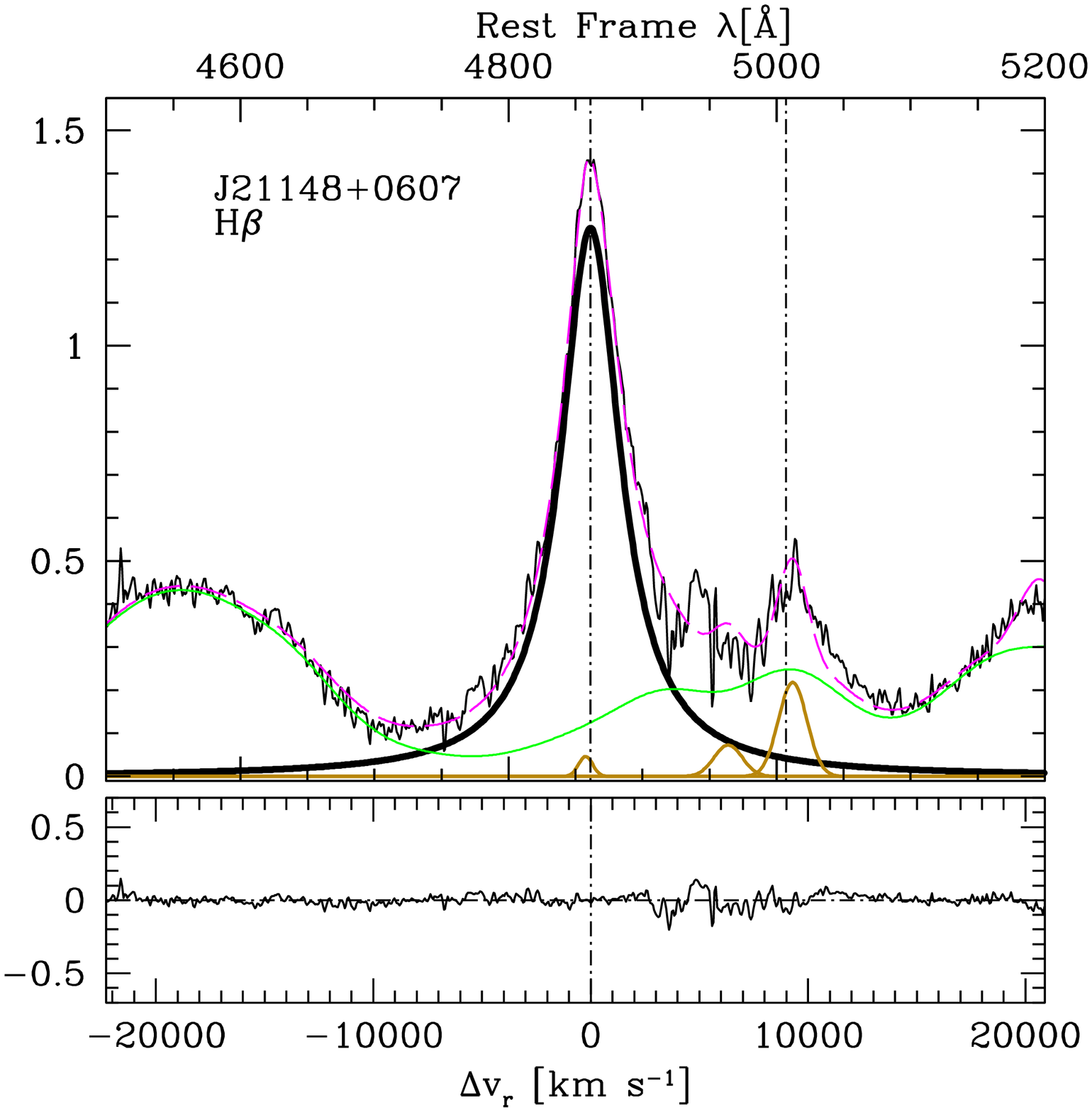}\includegraphics[width=0.225\columnwidth]{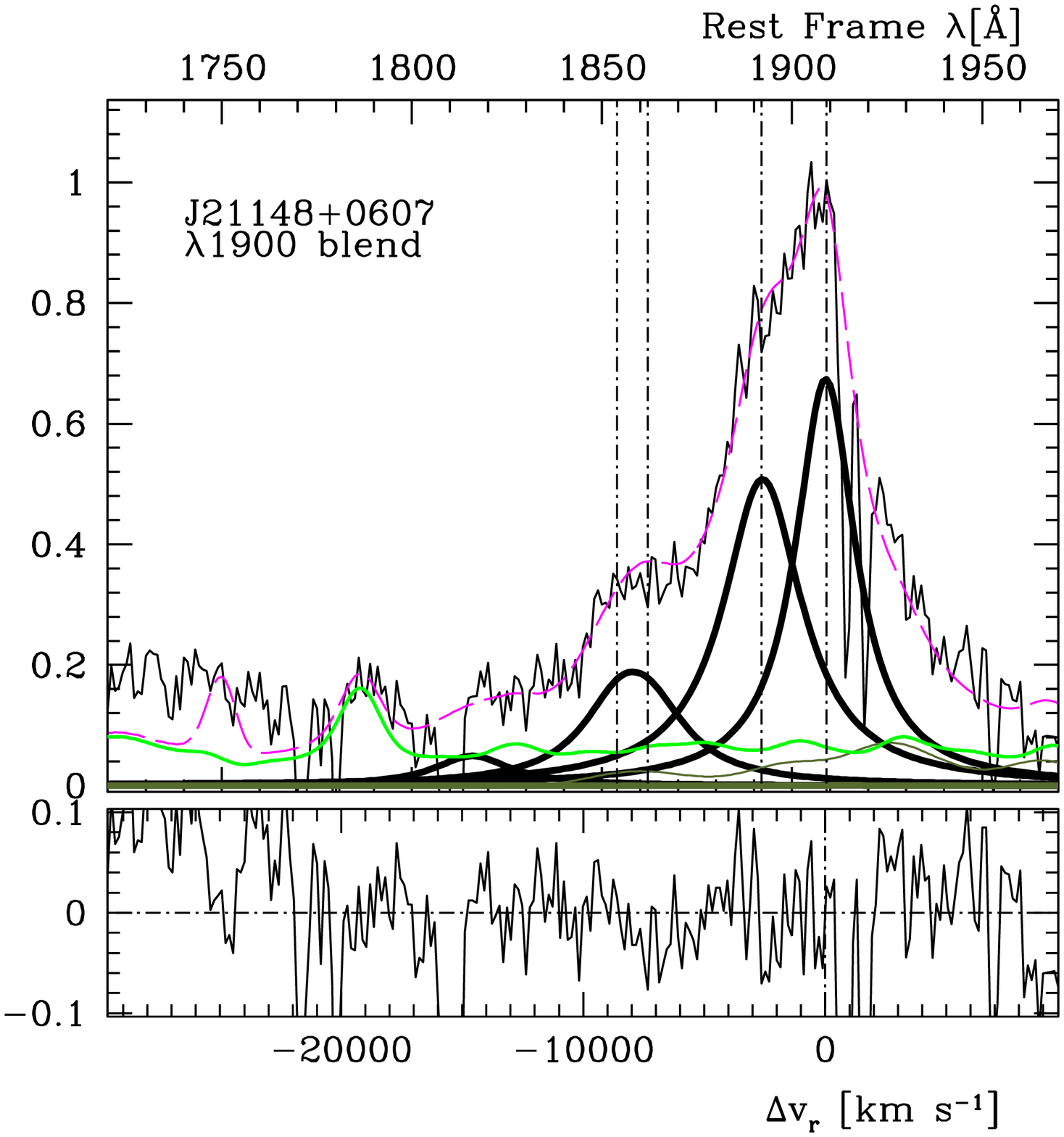}
\includegraphics[width=0.225\columnwidth]{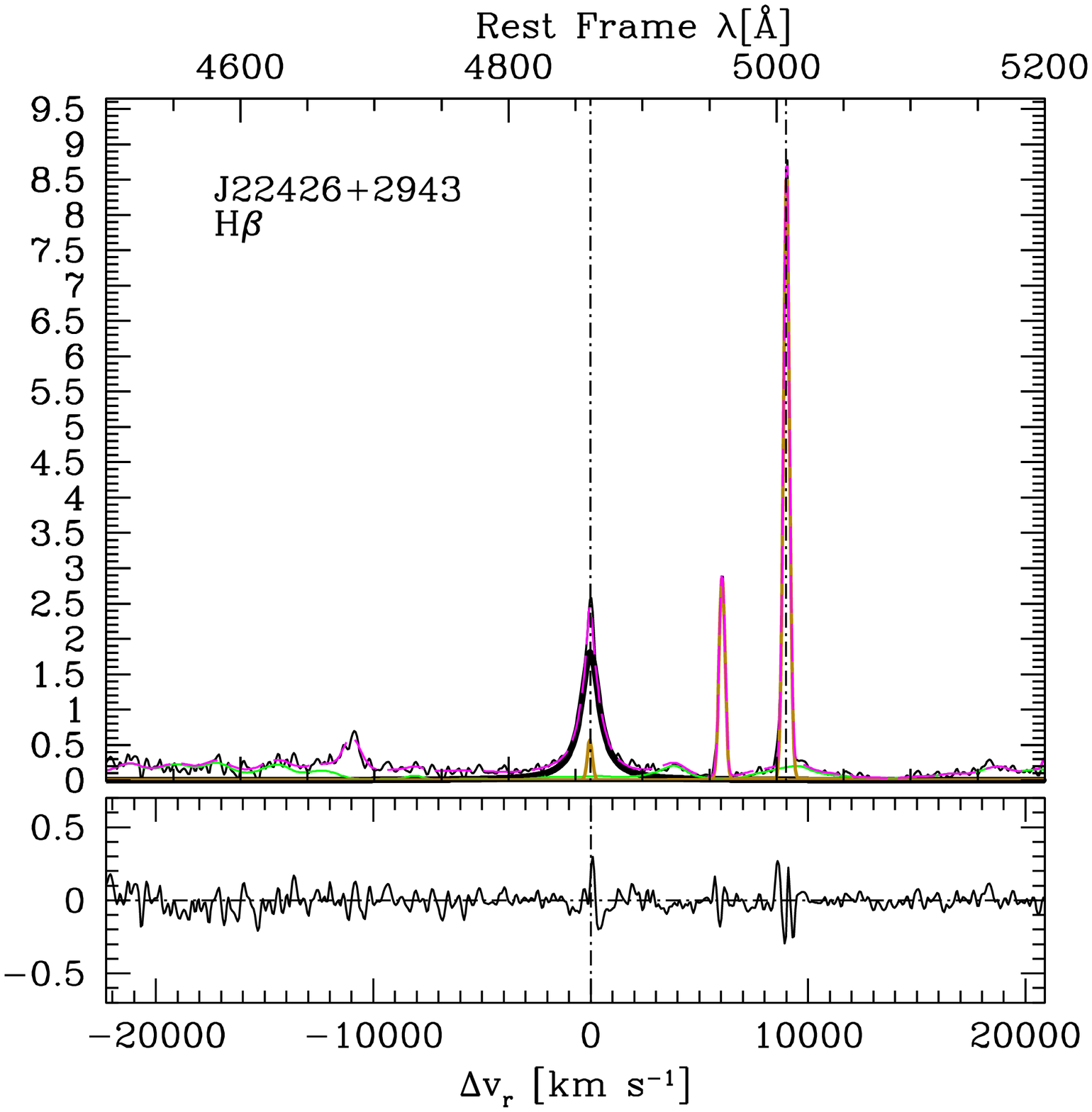}\includegraphics[width=0.225\columnwidth]{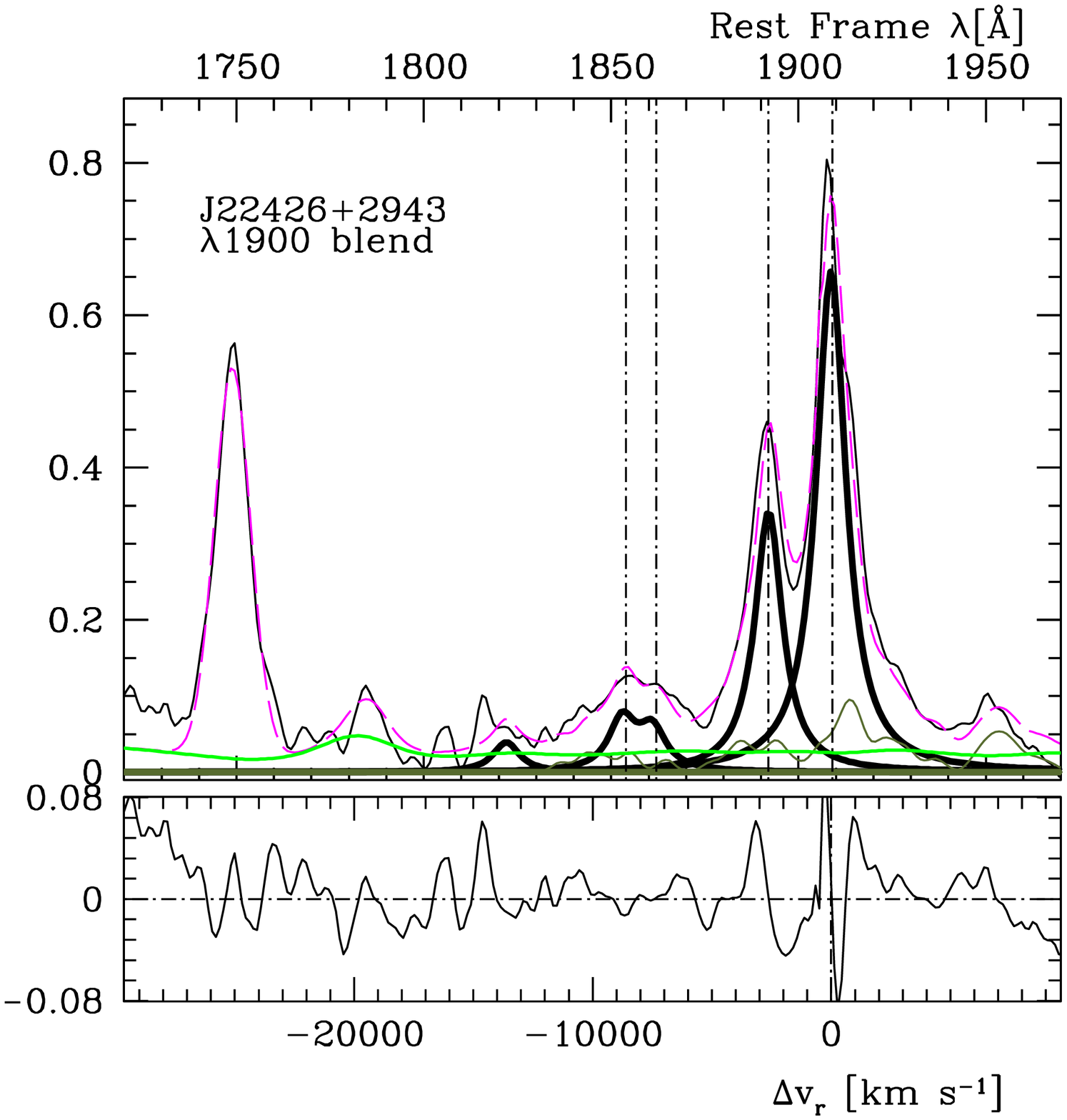}\\
\caption{Analysis of \hb\ and of the 1900 \AA\ blend\ for Pop. A sources (cont.). \label{fig:fosa3}}
\end{figure}

\begin{figure}
\includegraphics[width=0.225\columnwidth]{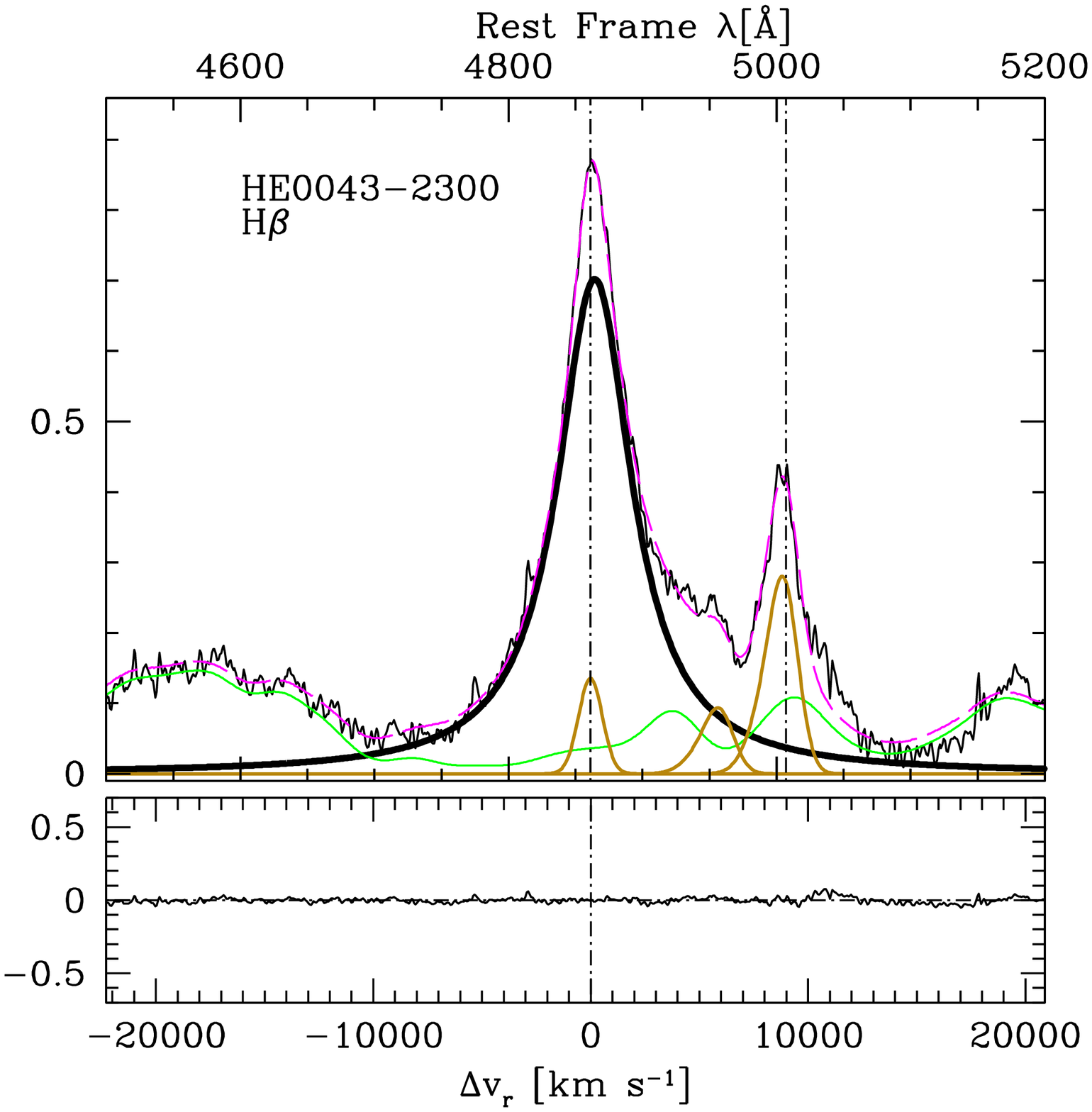} \includegraphics[width=0.225\columnwidth]{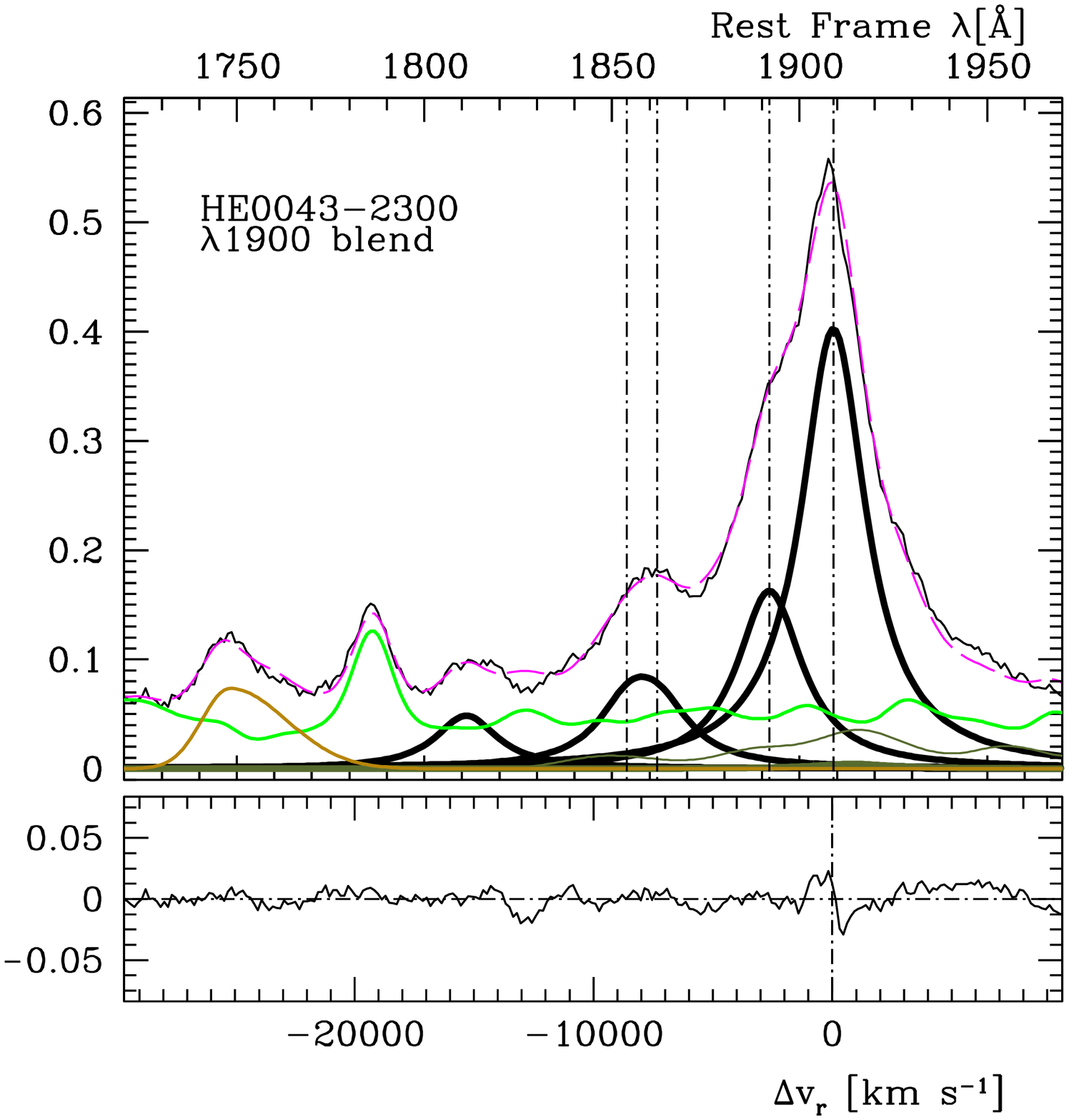}
\includegraphics[width=0.225\columnwidth]{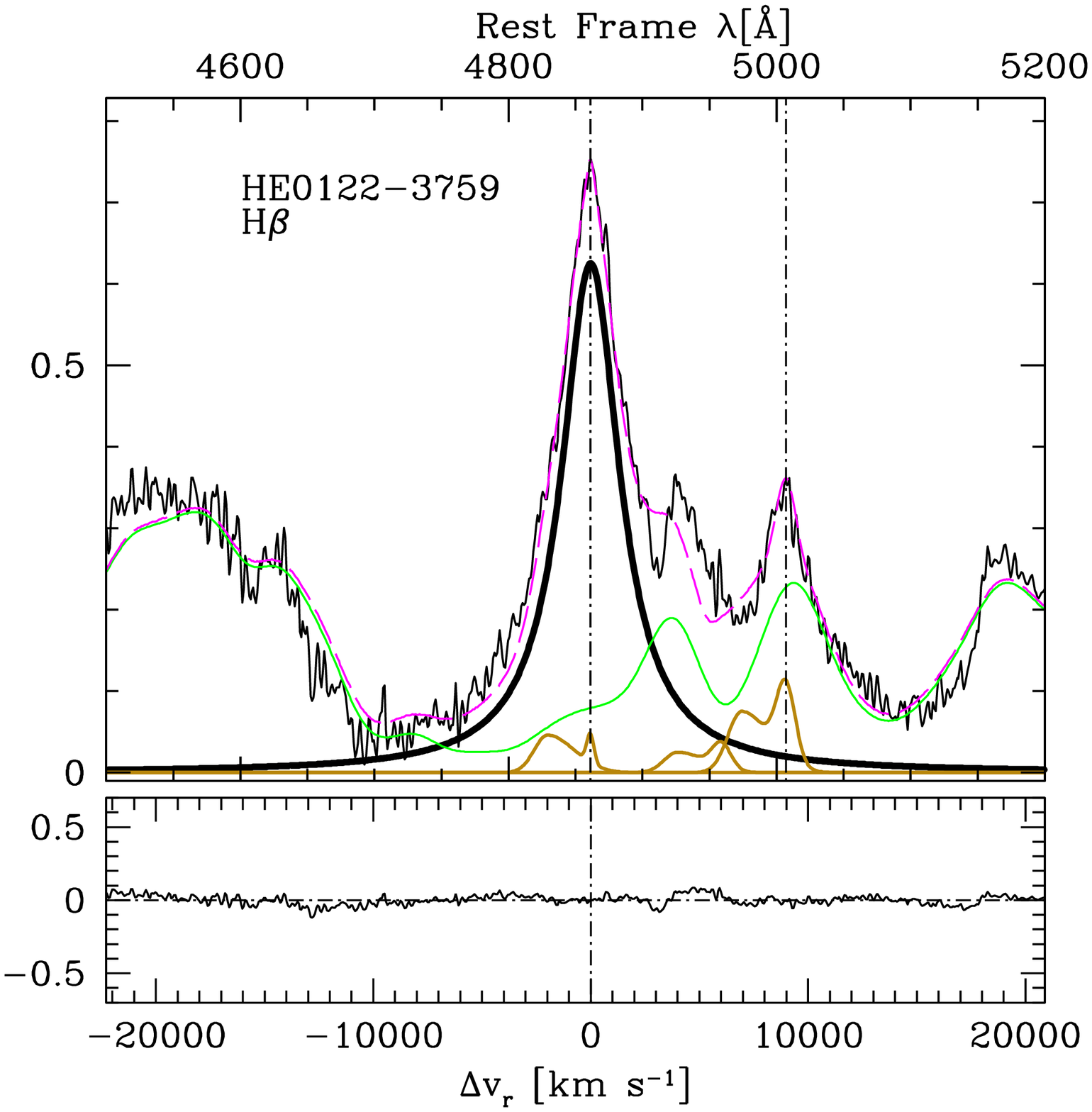} \includegraphics[width=0.225\columnwidth]{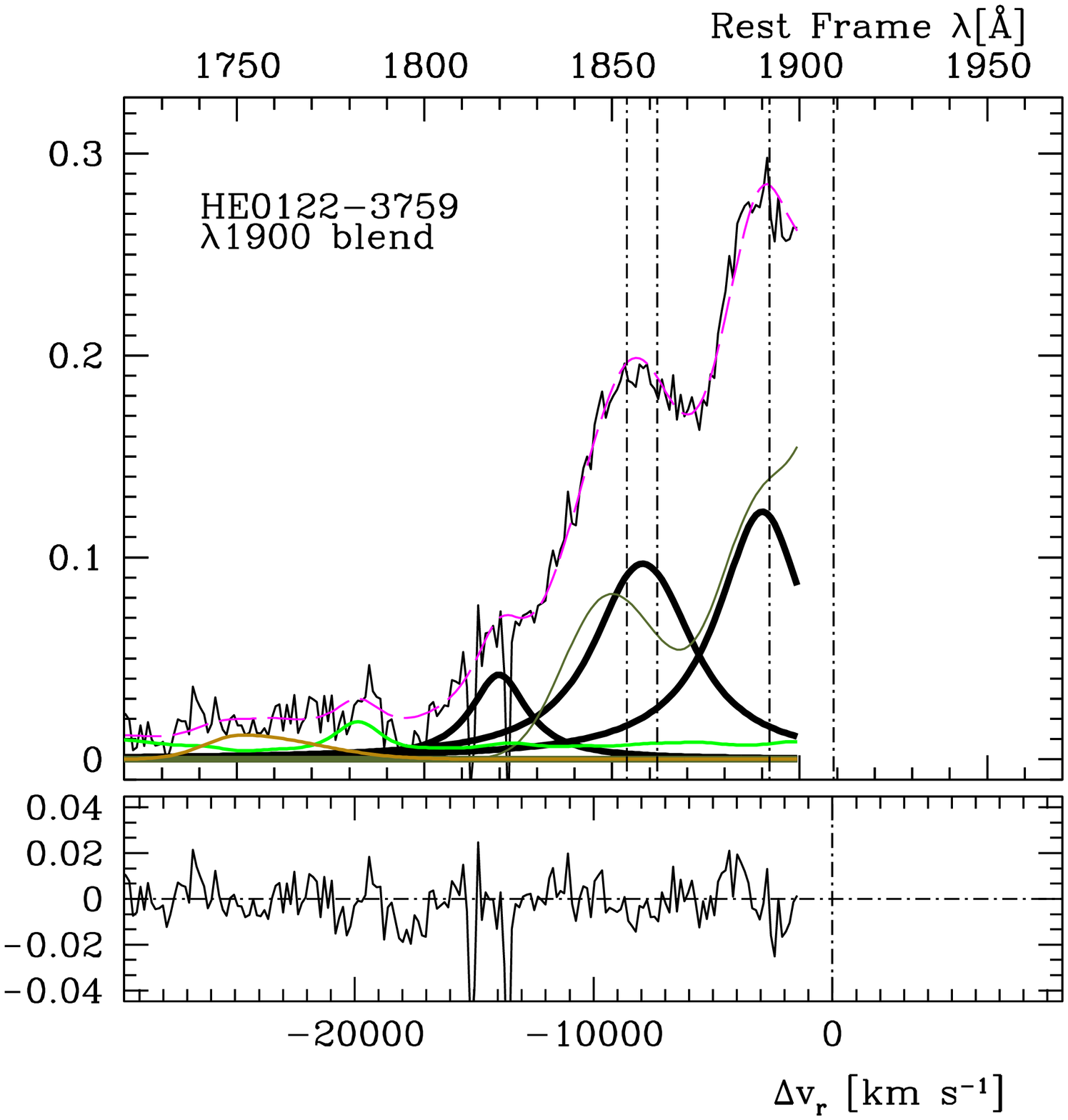}\\ 
\includegraphics[width=0.225\columnwidth]{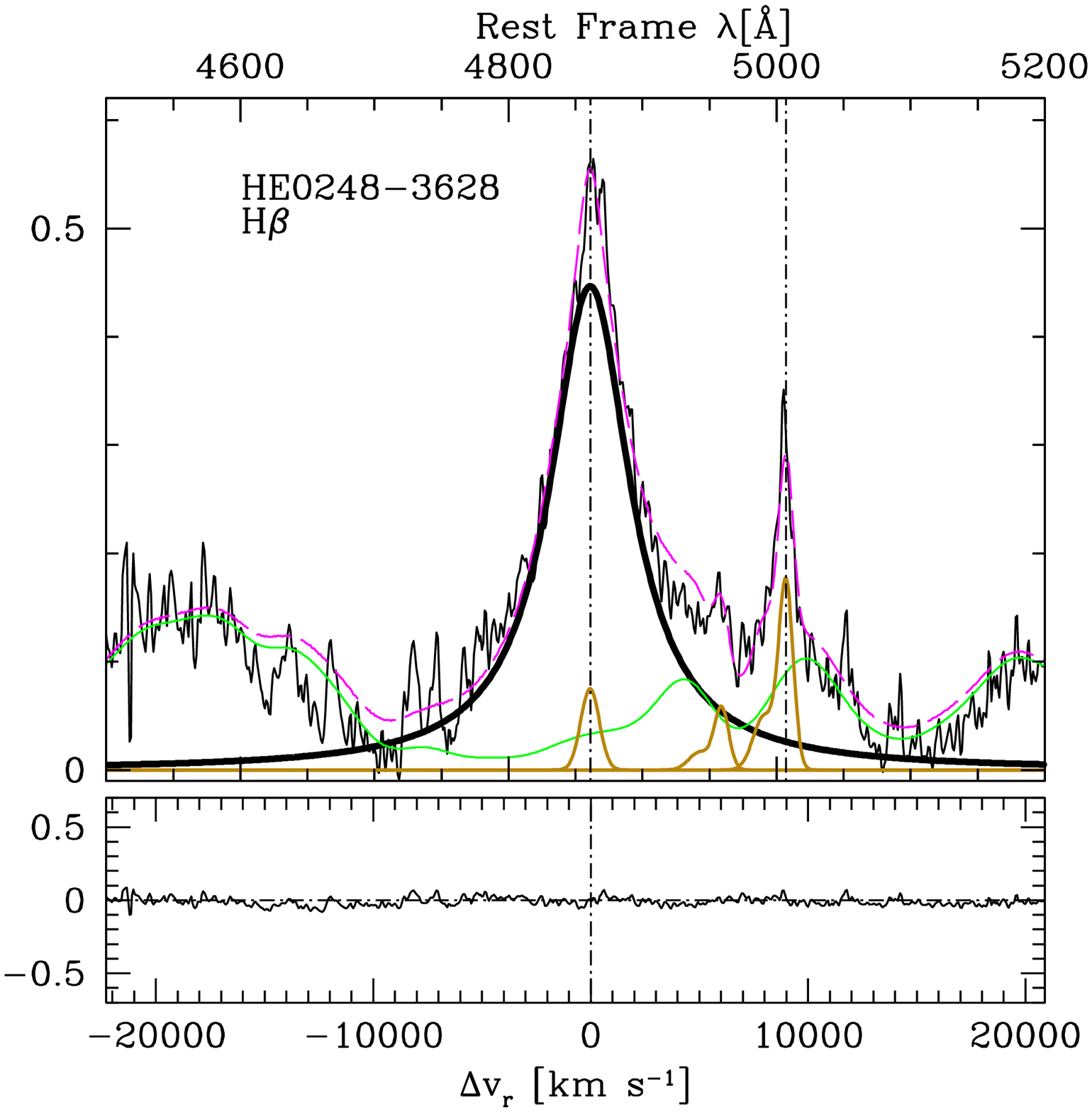} \includegraphics[width=0.225\columnwidth]{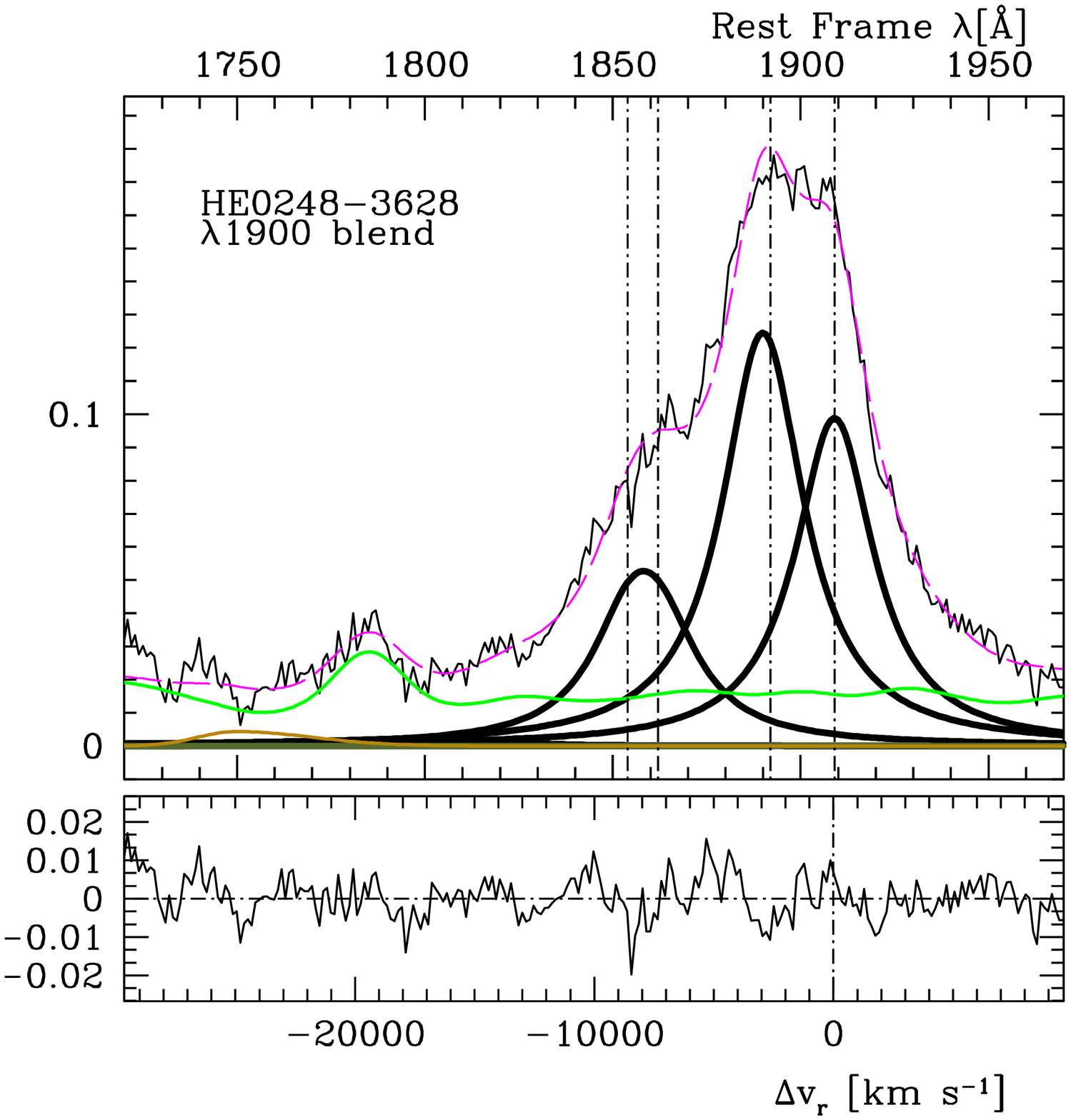} 
\includegraphics[width=0.225\columnwidth]{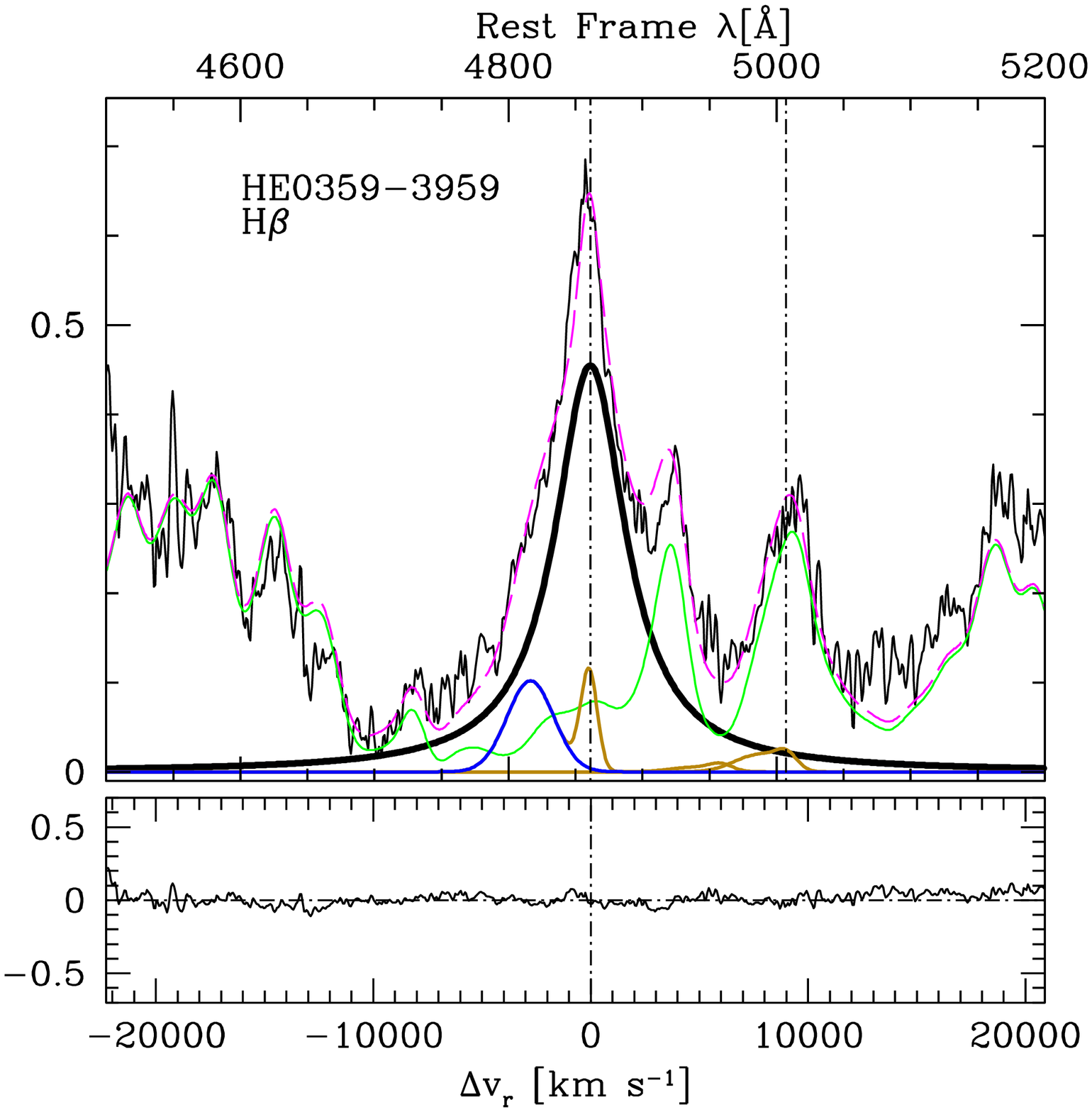} \includegraphics[width=0.225\columnwidth]{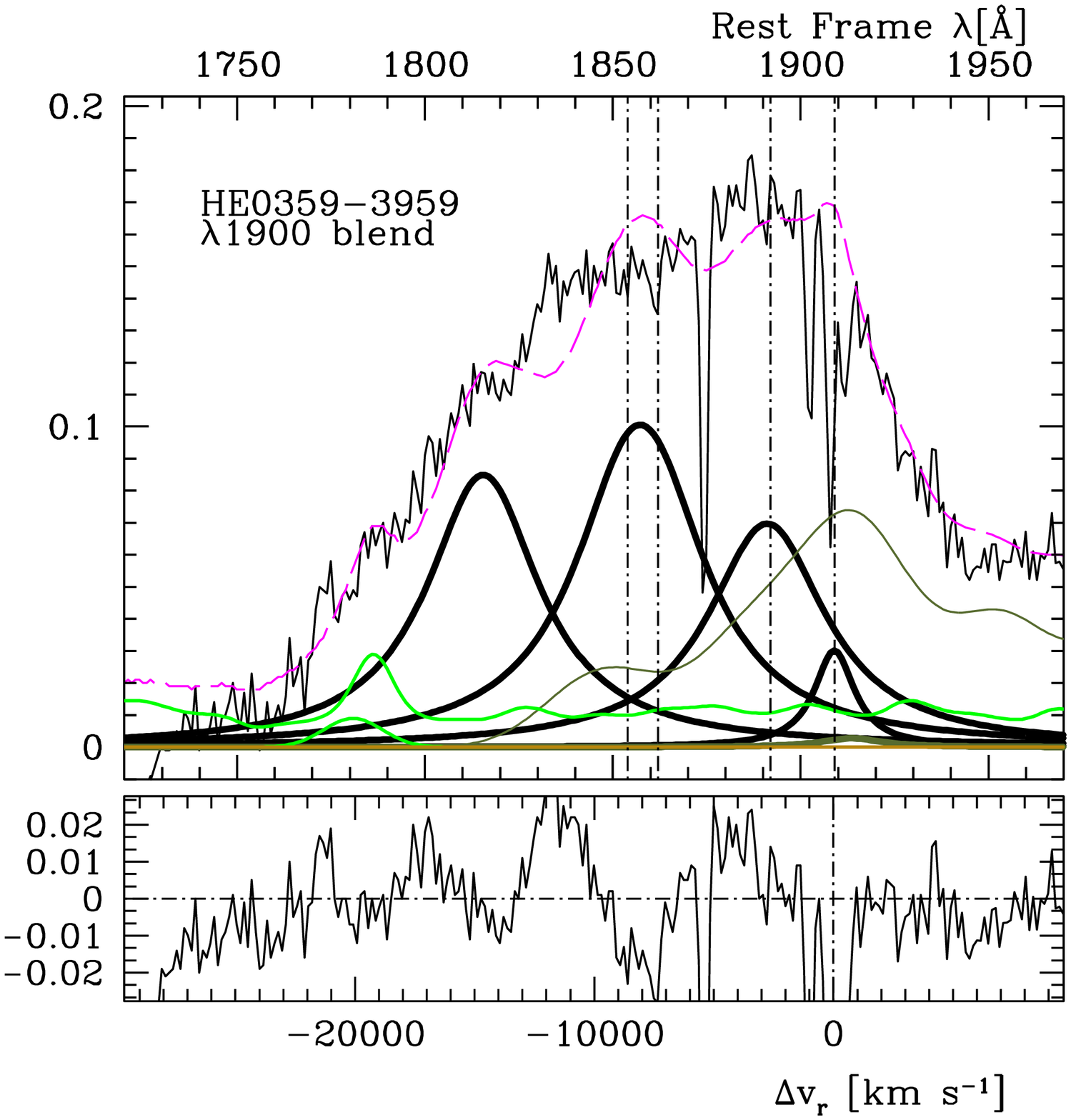}\\ 
\includegraphics[width=0.225\columnwidth]{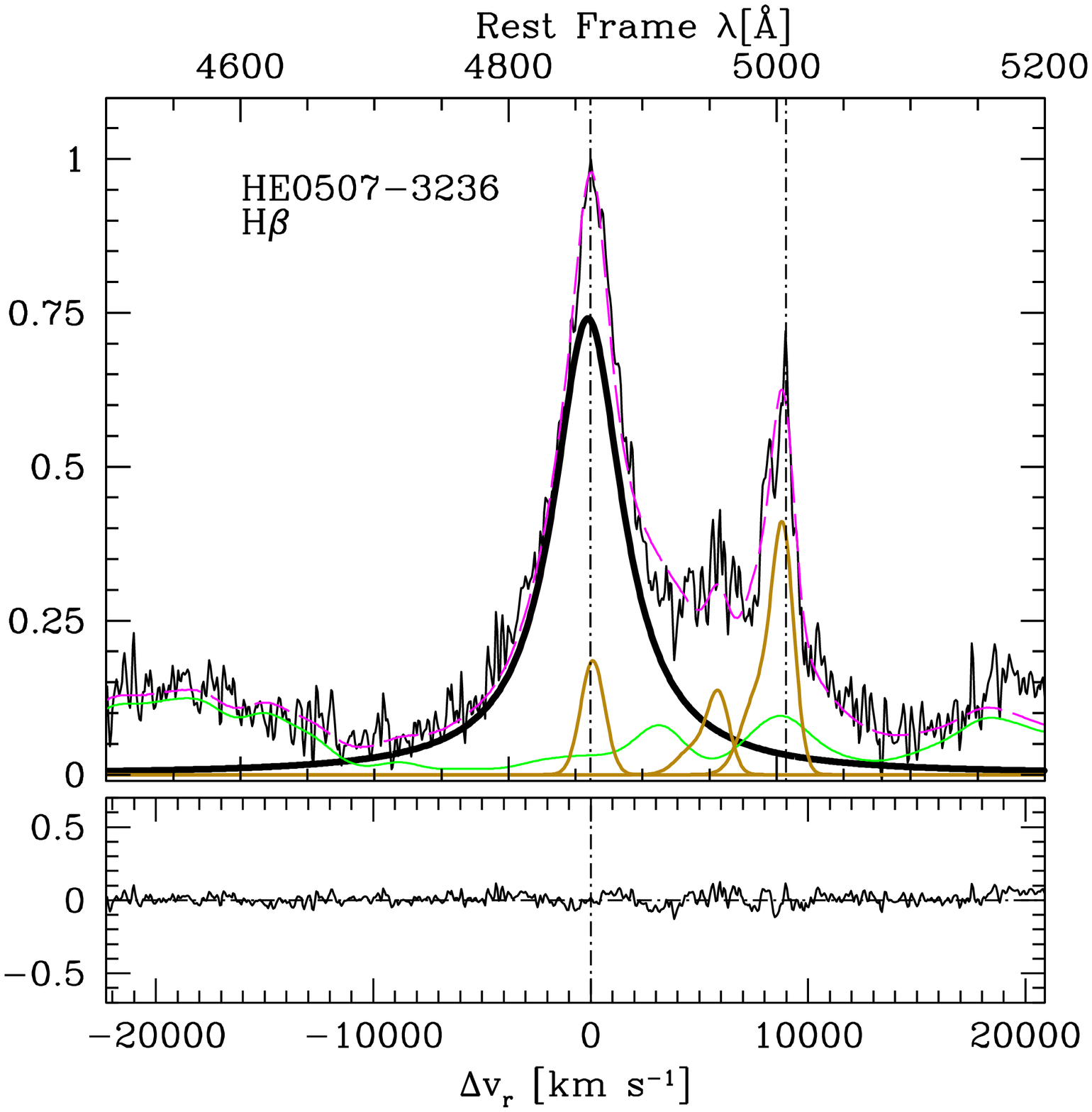} \includegraphics[width=0.225\columnwidth]{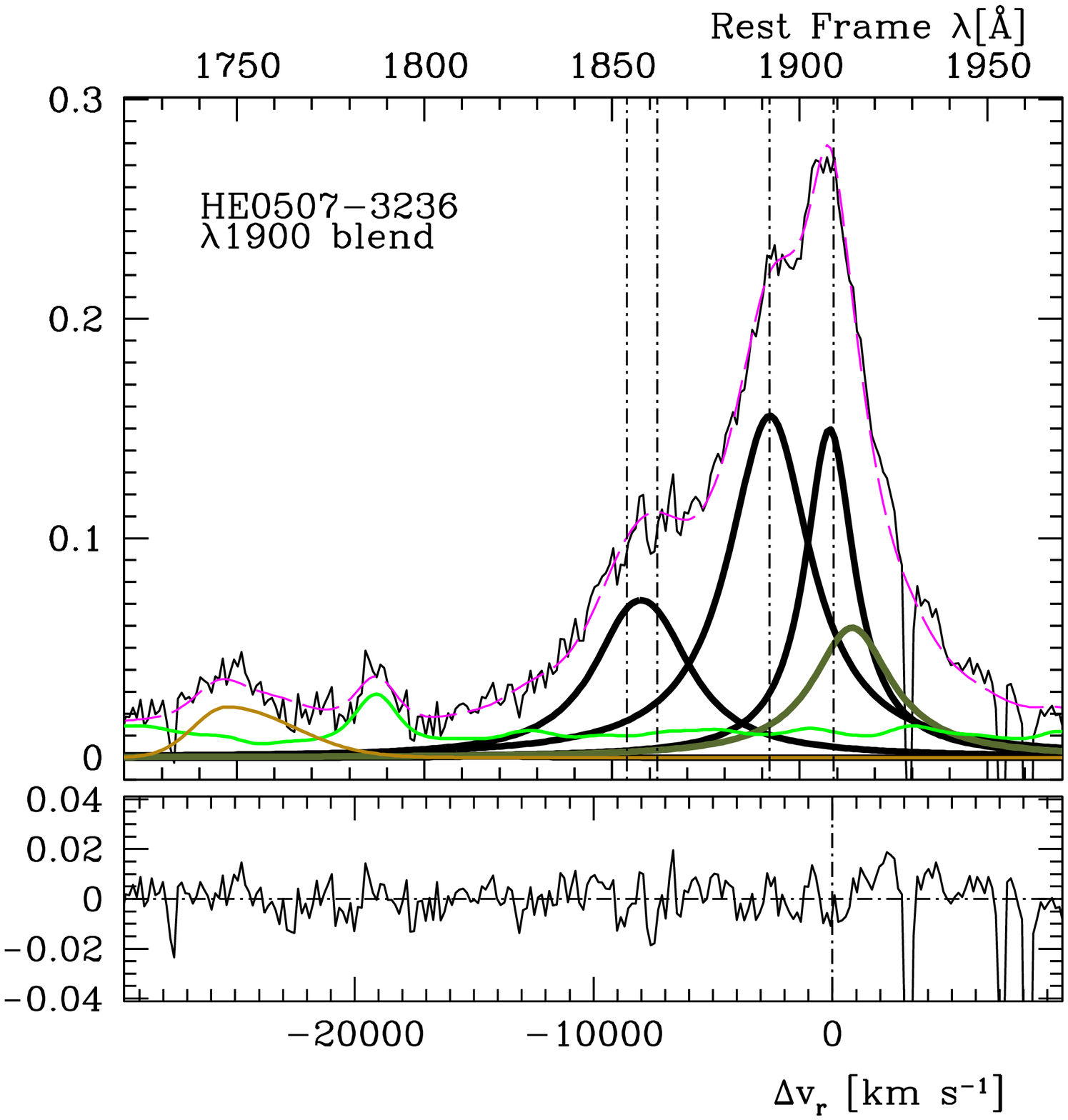}
\includegraphics[width=0.225\columnwidth]{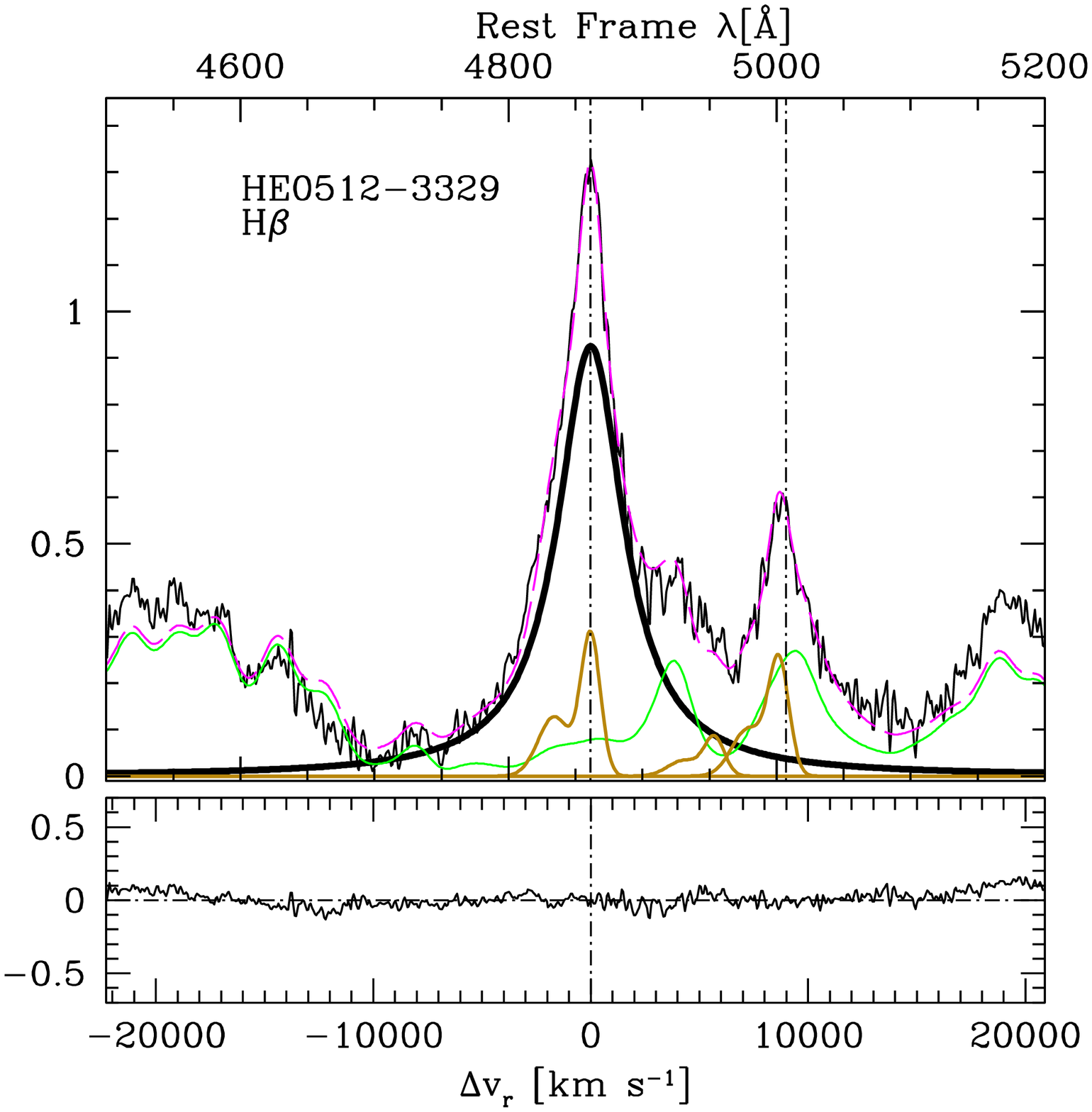} \includegraphics[width=0.225\columnwidth]{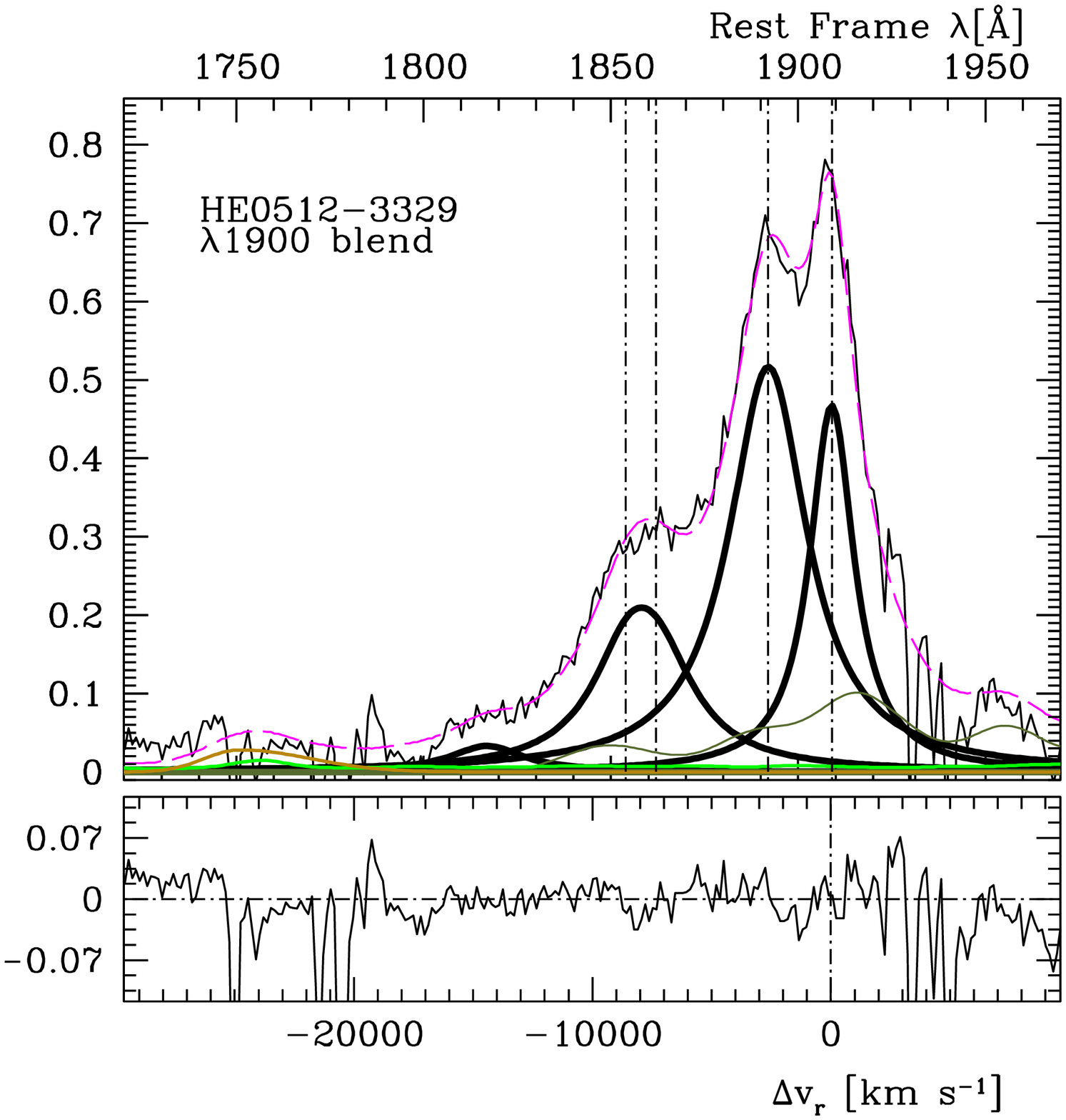}\\
\includegraphics[width=0.225\columnwidth]{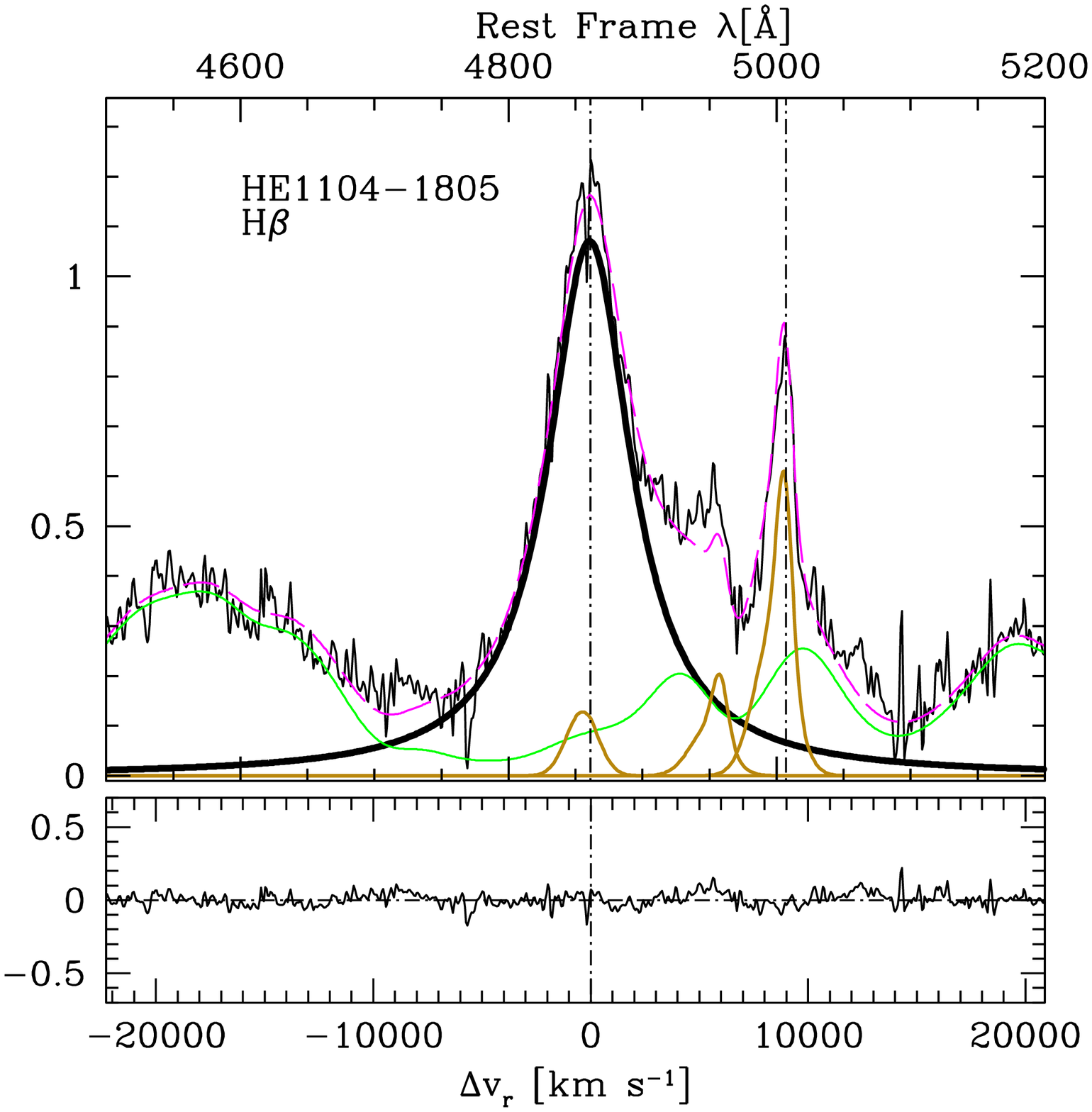} \includegraphics[width=0.225\columnwidth]{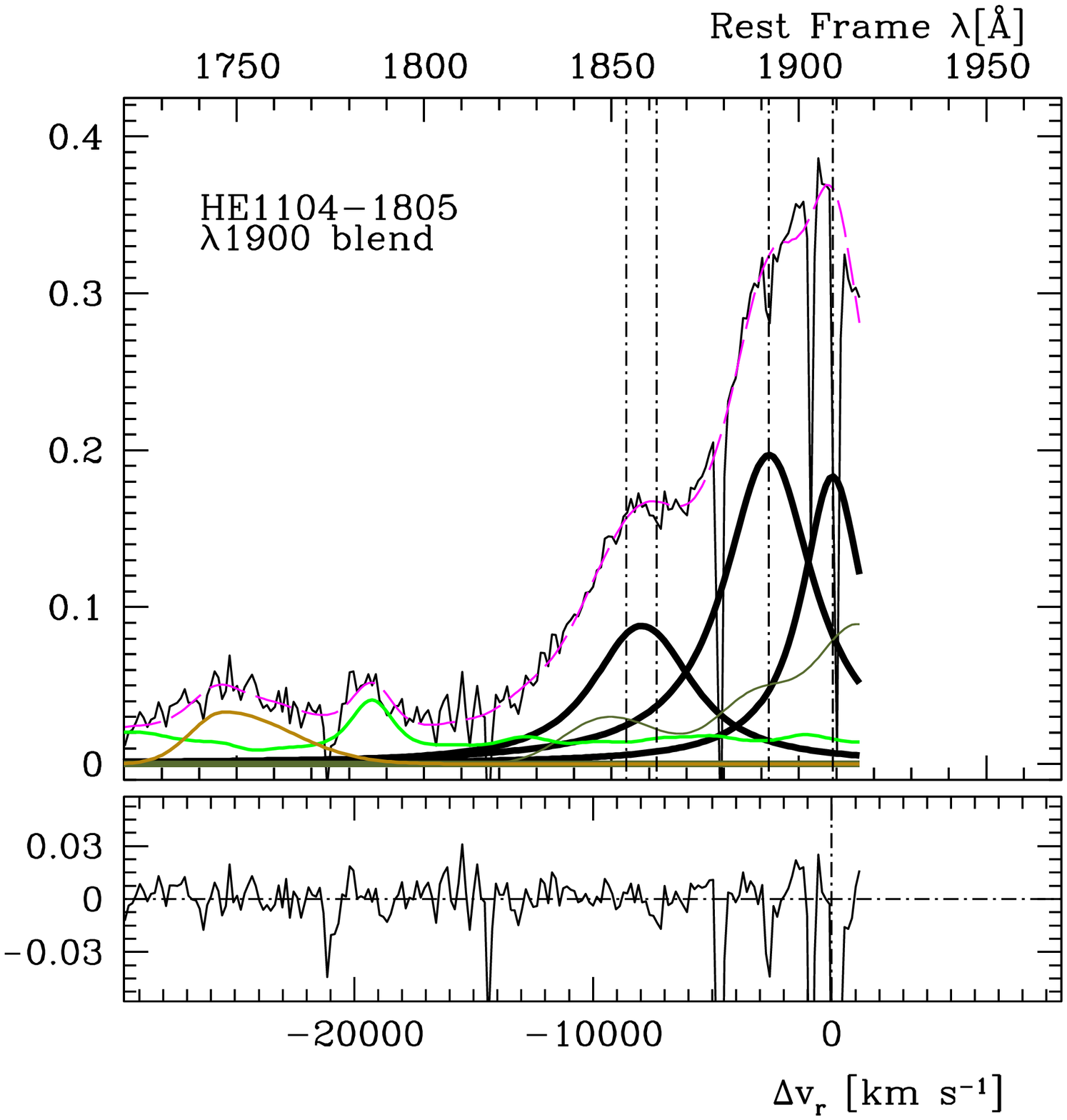}
\includegraphics[width=0.225\columnwidth]{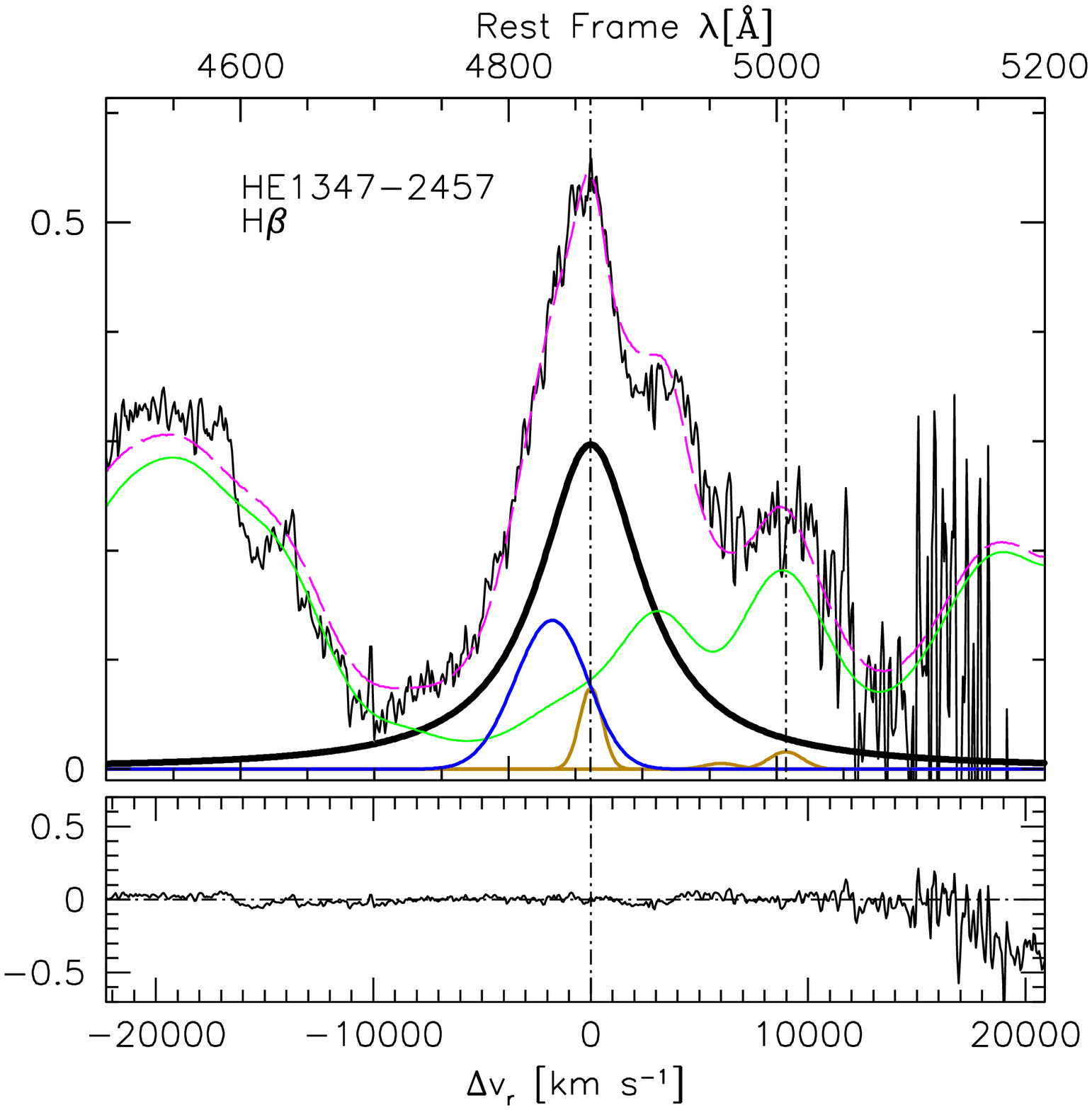} \includegraphics[width=0.225\columnwidth]{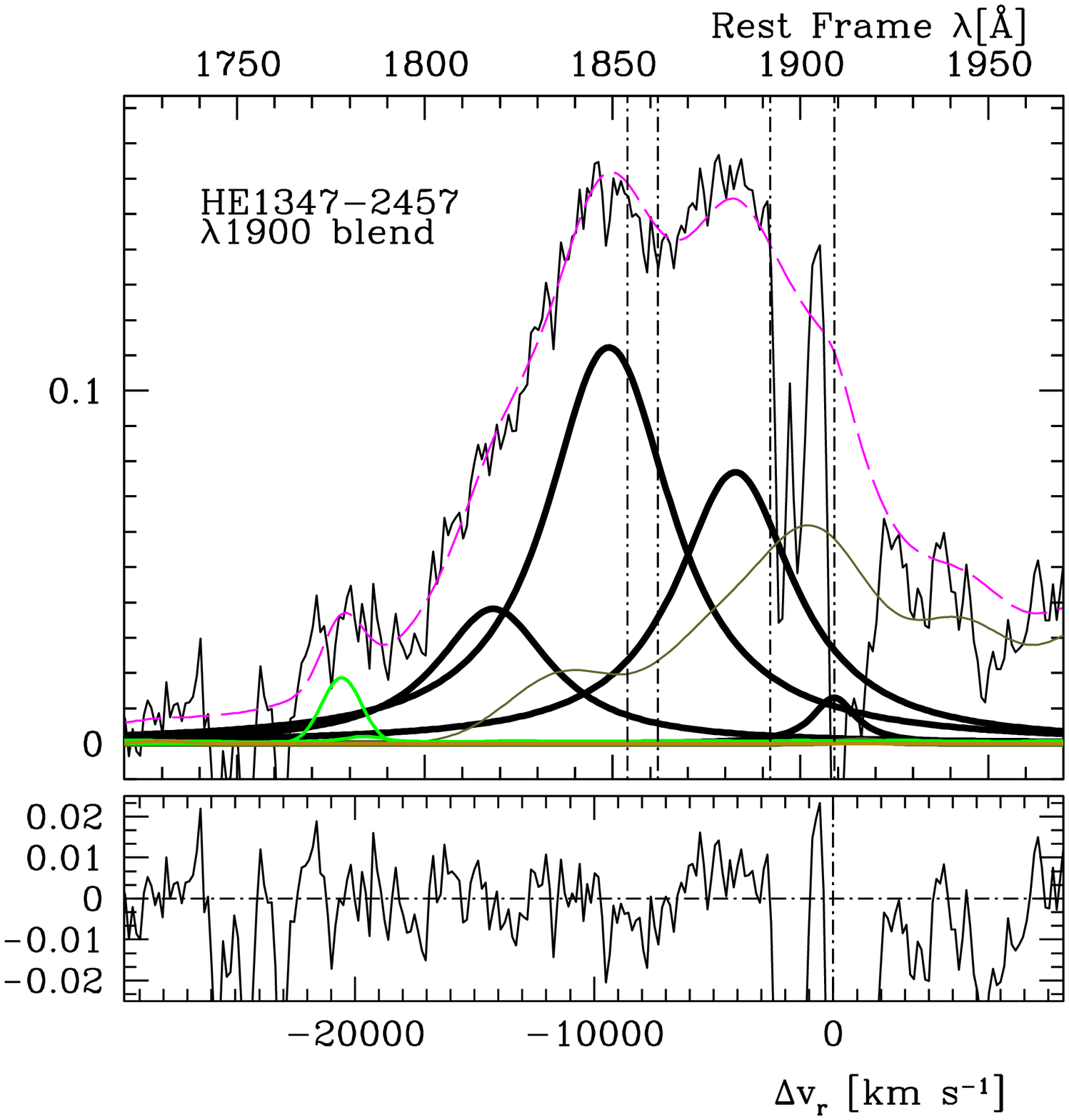}\\  
\includegraphics[width=0.225\columnwidth]{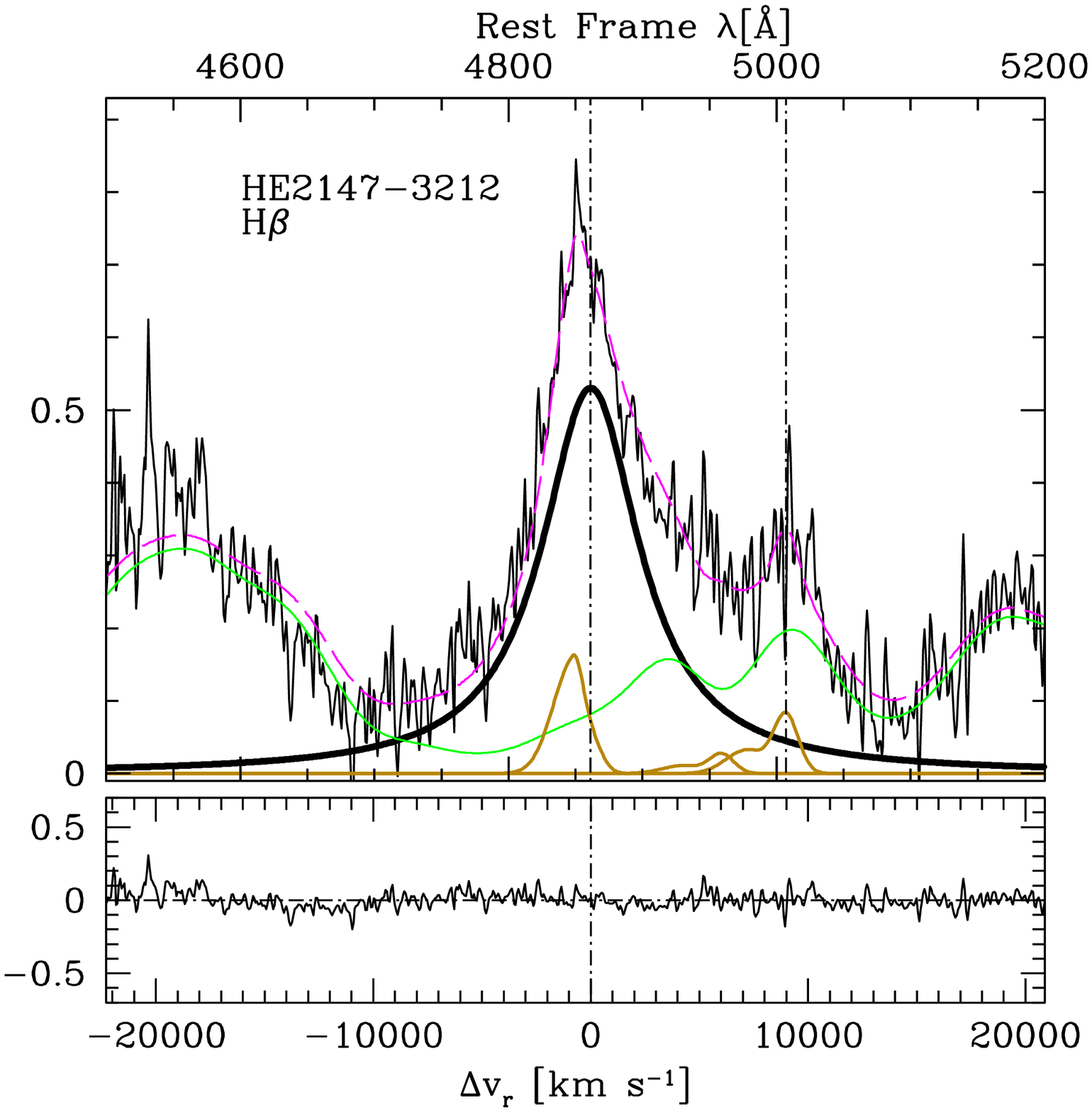} \includegraphics[width=0.225\columnwidth]{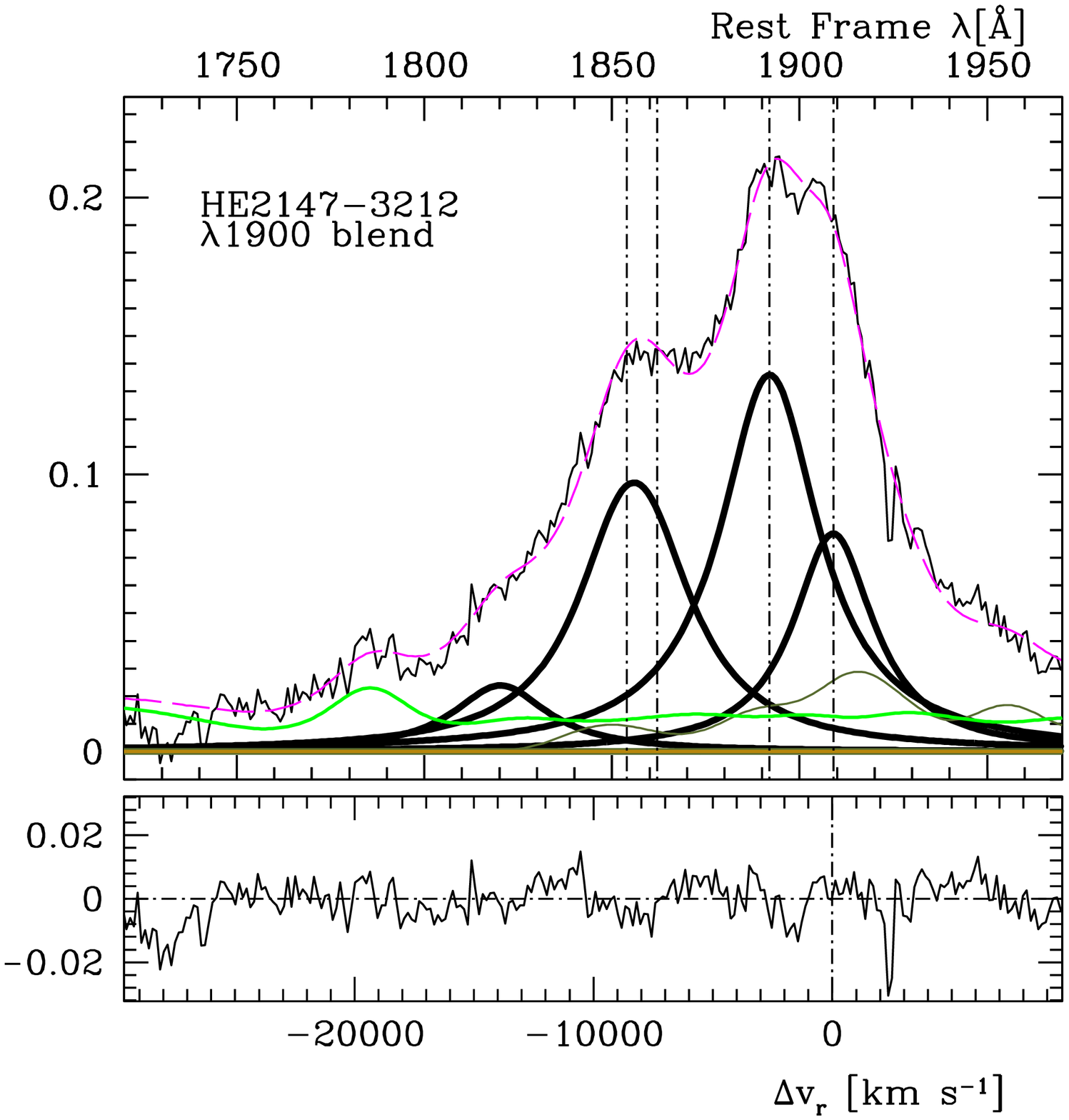}  
\includegraphics[width=0.225\columnwidth]{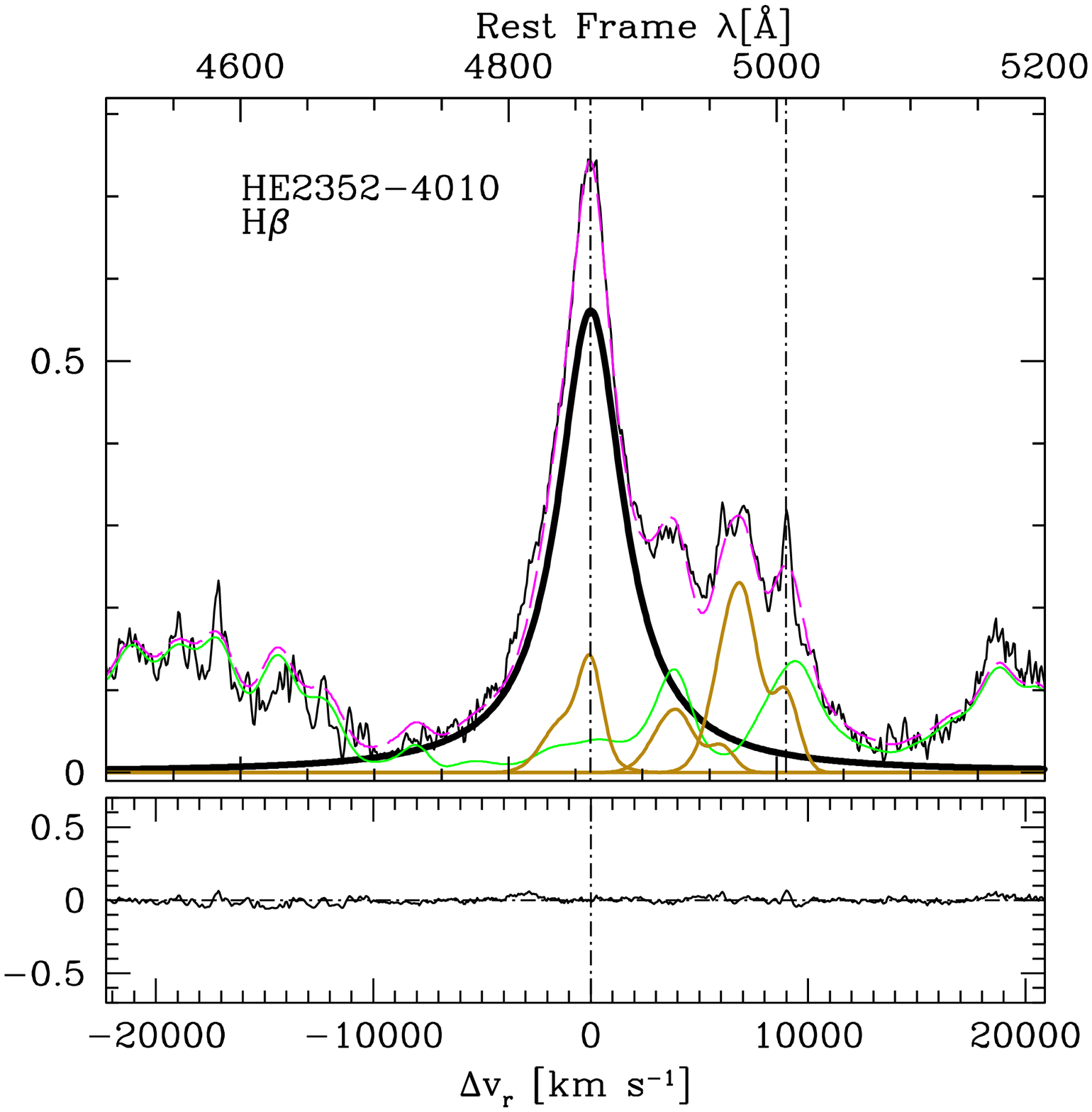} \includegraphics[width=0.225\columnwidth]{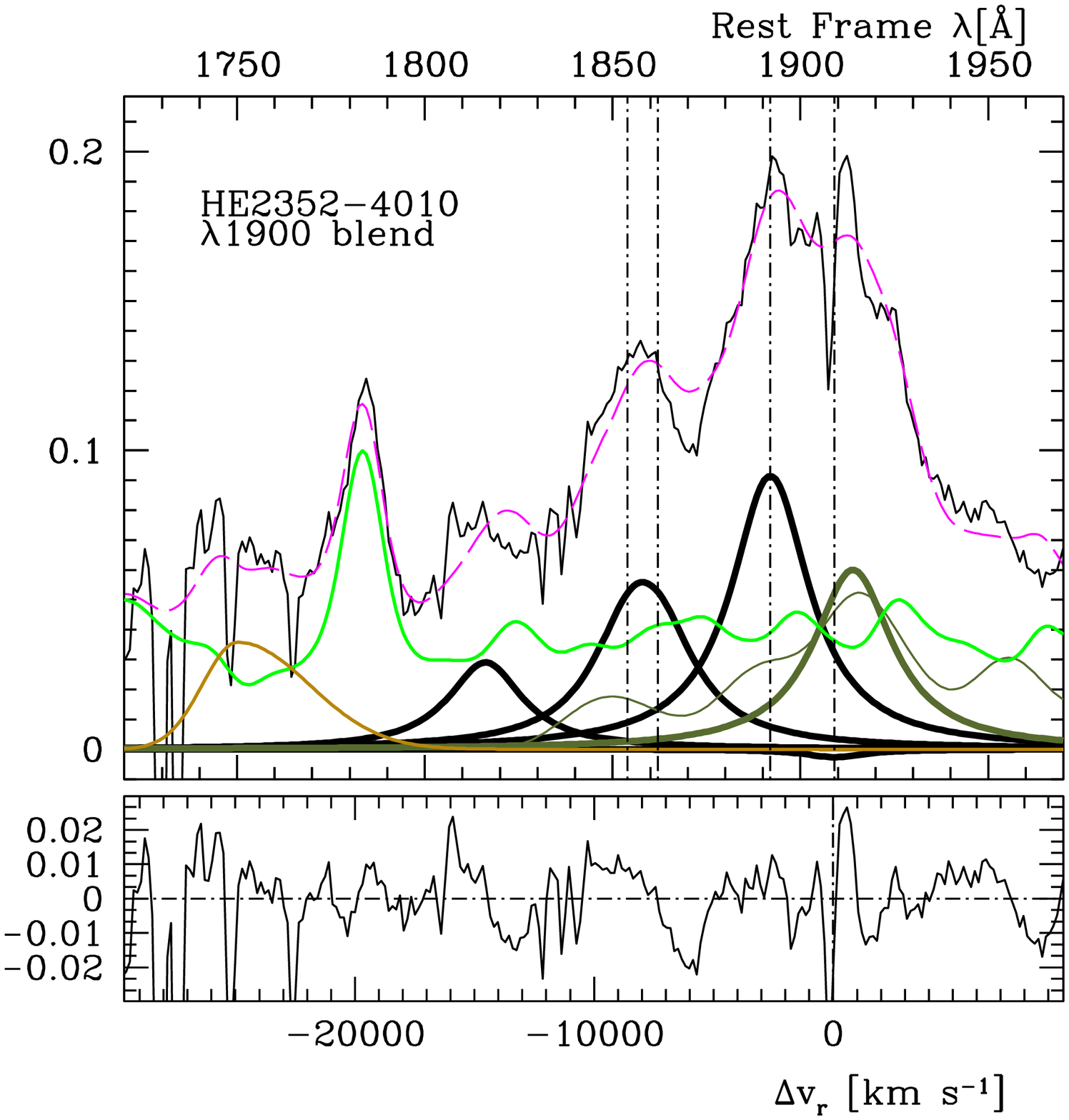}\\ 
\caption{Analysis of the 1900 \AA\ blend, for 10 Pop. A sources of the HE sub-sample. Meaning of color coding is the same as in Fig. \ref{fig:fosa1}.  \label{fig:hea}}
\end{figure}

\begin{figure}
\includegraphics[width=0.225\columnwidth]{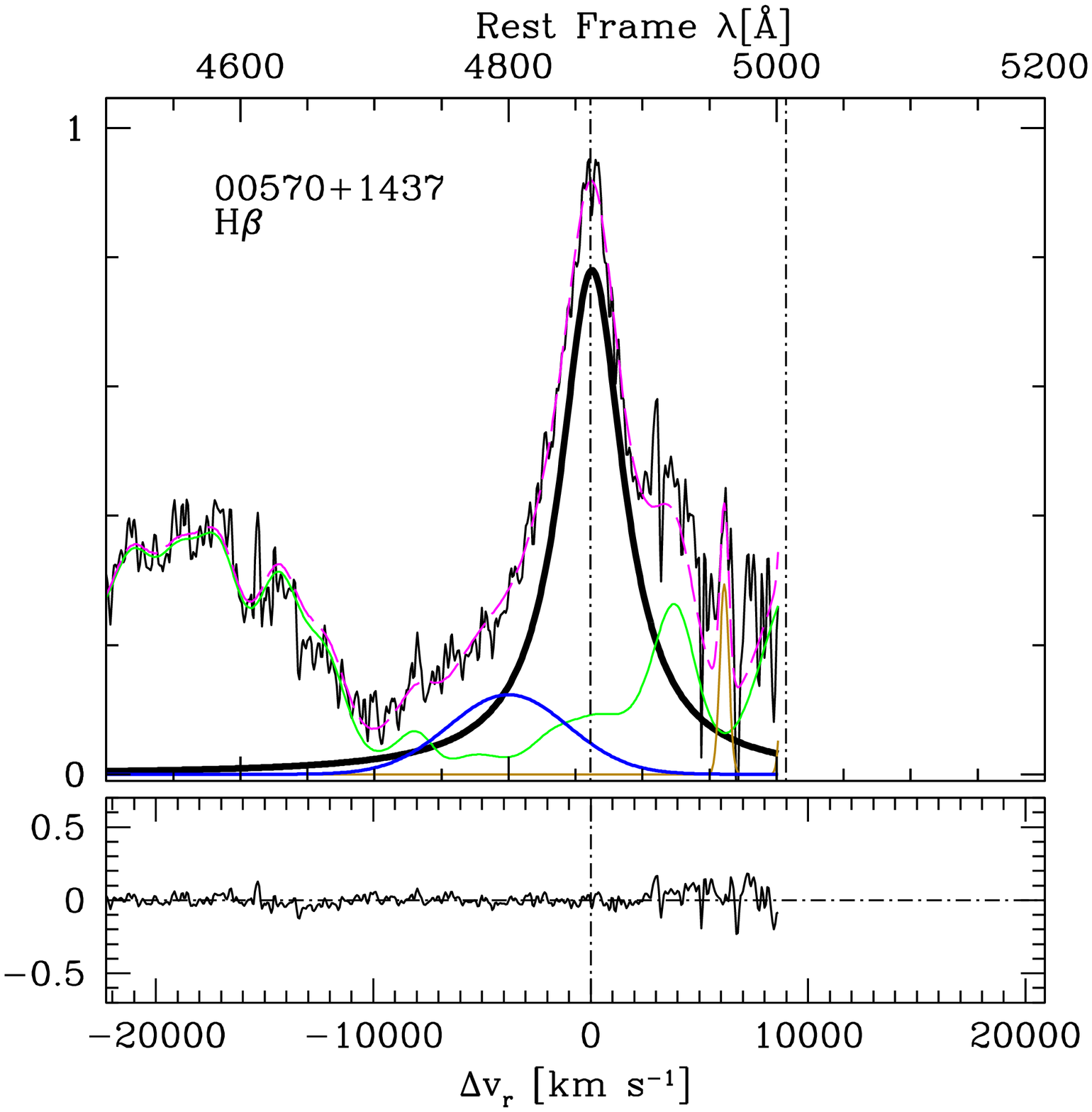}\includegraphics[width=0.225\columnwidth]{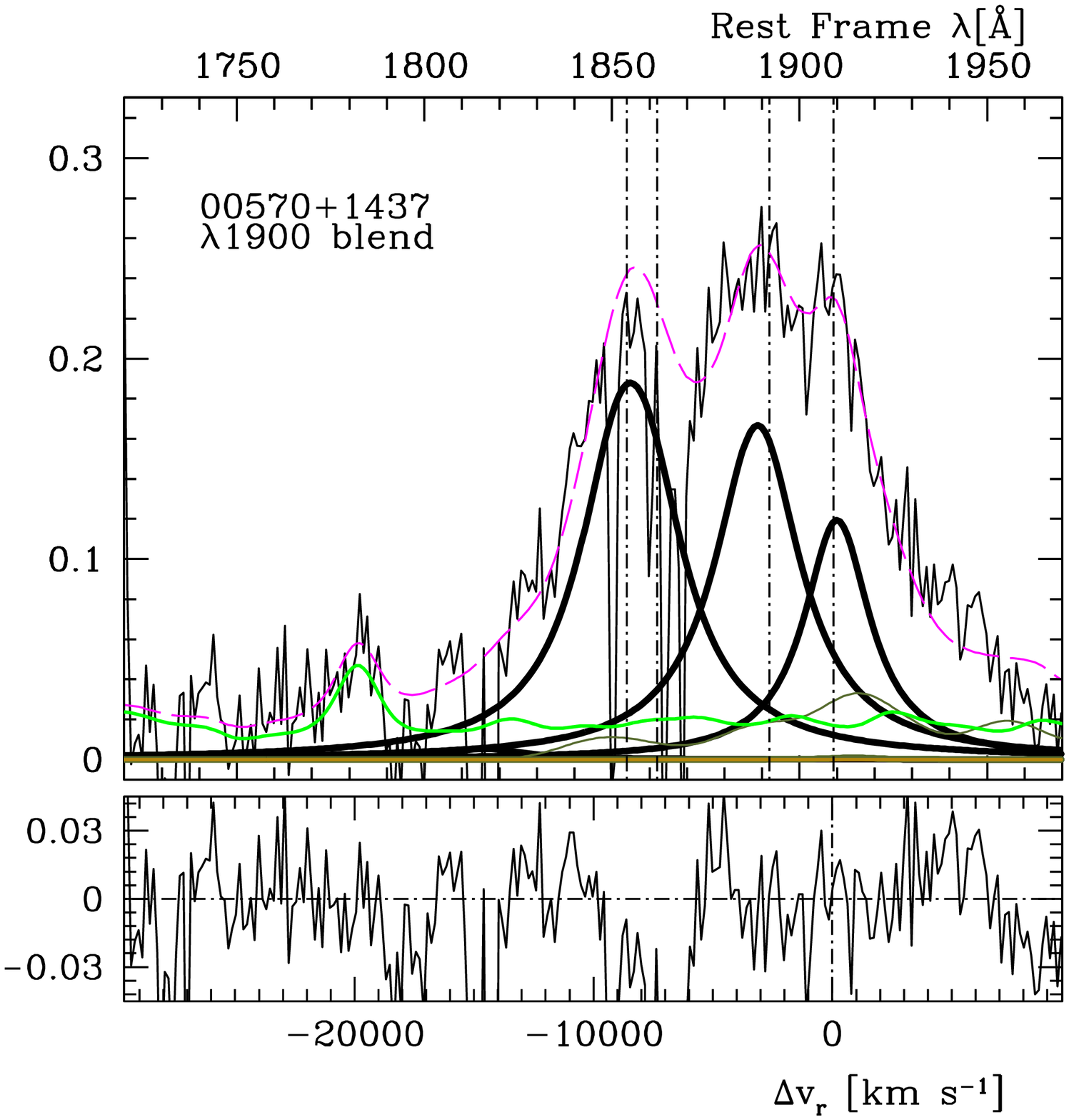}
\includegraphics[width=0.225\columnwidth]{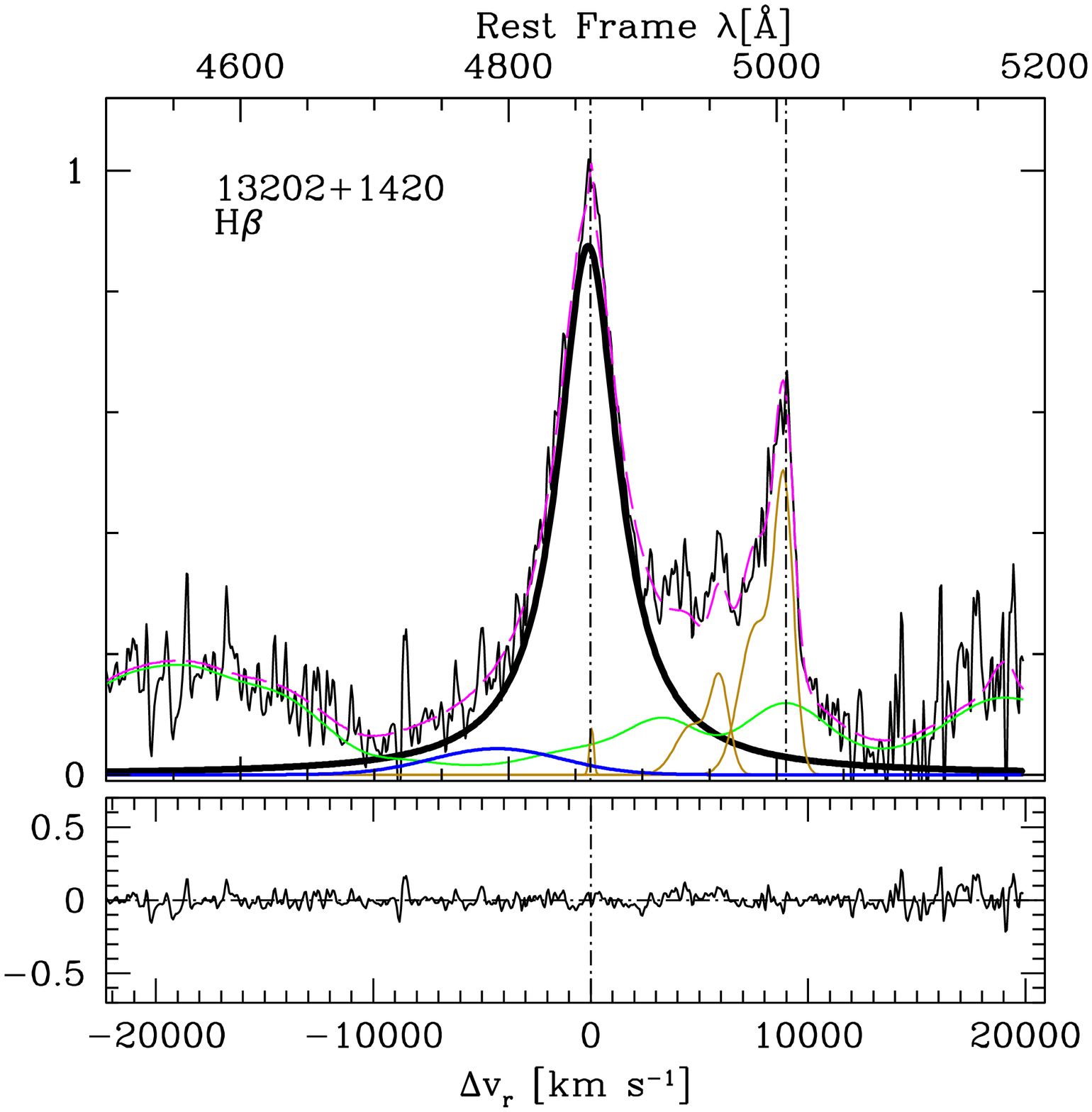}\includegraphics[width=0.225\columnwidth]{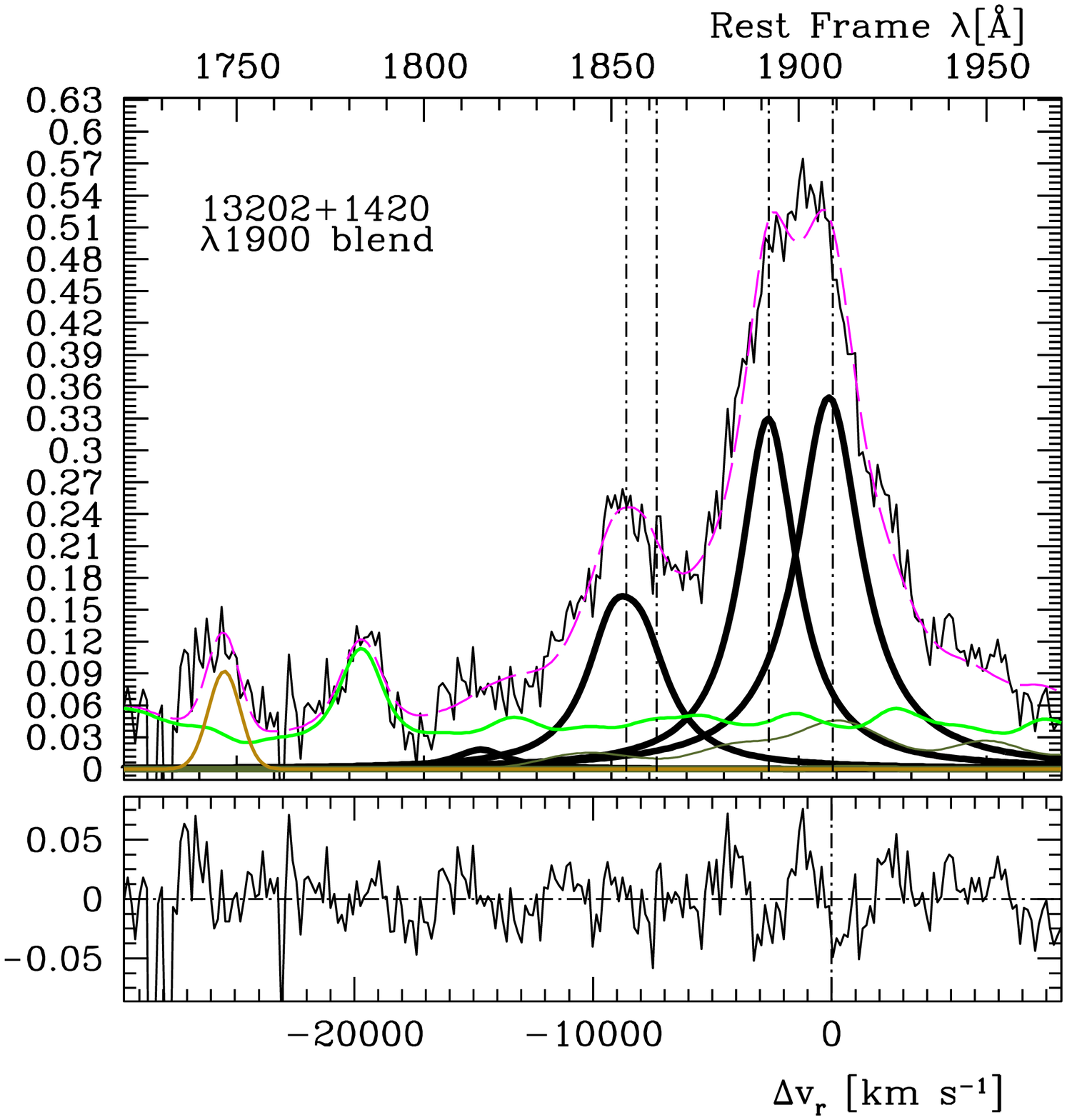}\\
\includegraphics[width=0.225\columnwidth]{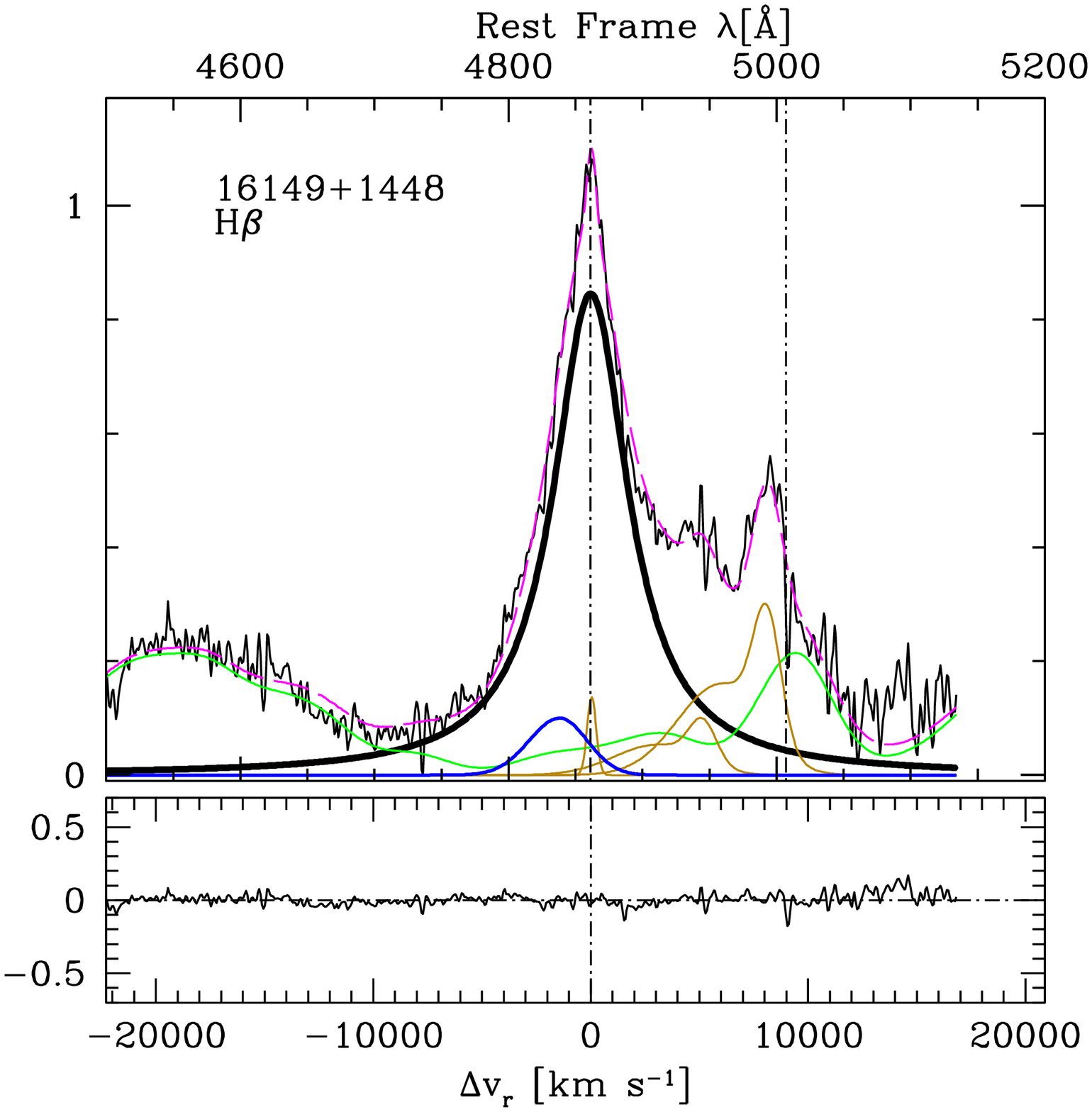}\includegraphics[width=0.225\columnwidth]{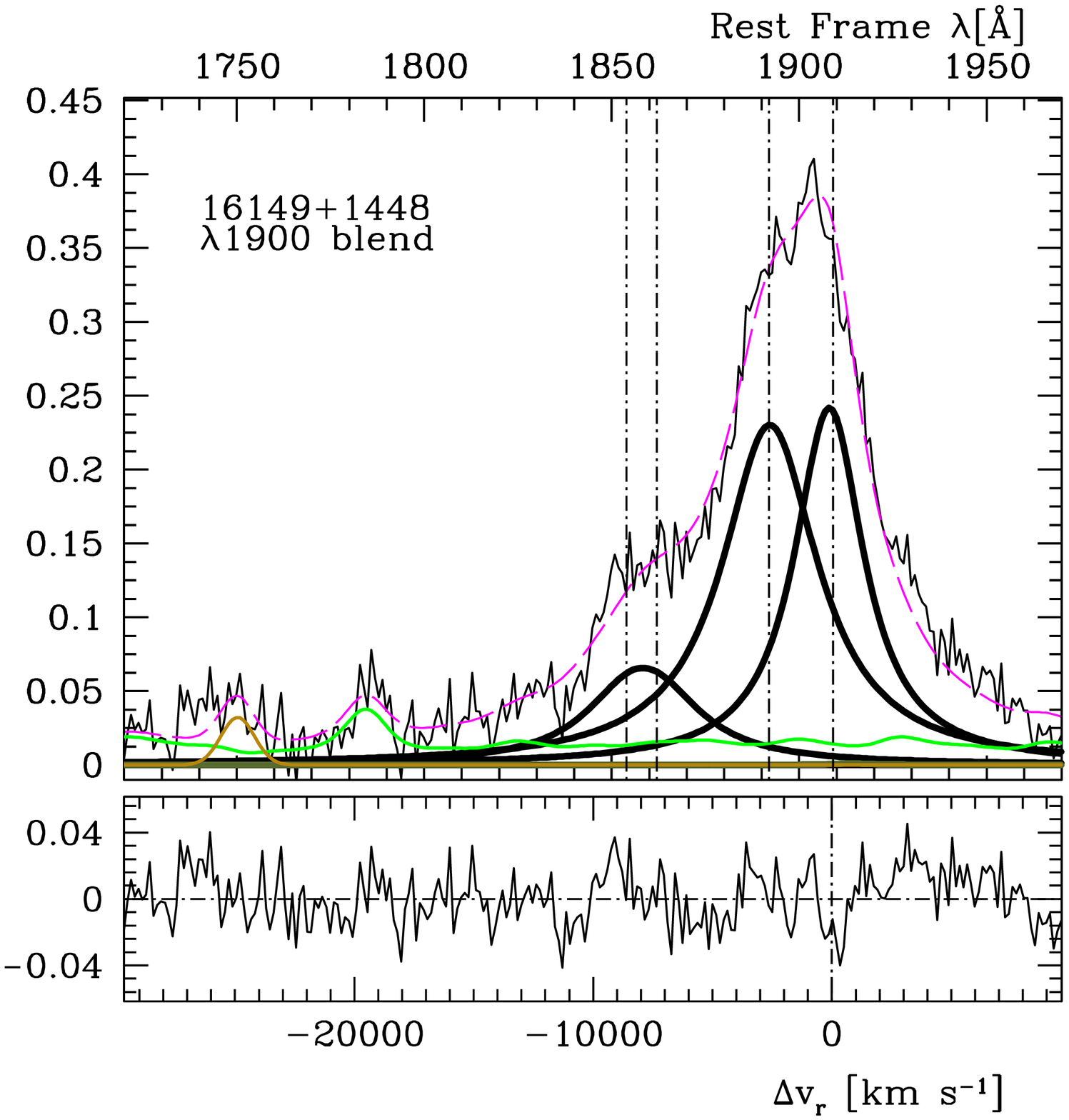}\\ 
\caption{Analysis of \hb\ and of the 1900 \AA\ blend\ for Pop. A  sources of the ISAAC sample. Meaning of color coding is the same of the previous Figures in the Appendix.  \label{fig:isaaca}}
\end{figure}

\begin{figure}
\includegraphics[width=0.225\columnwidth]{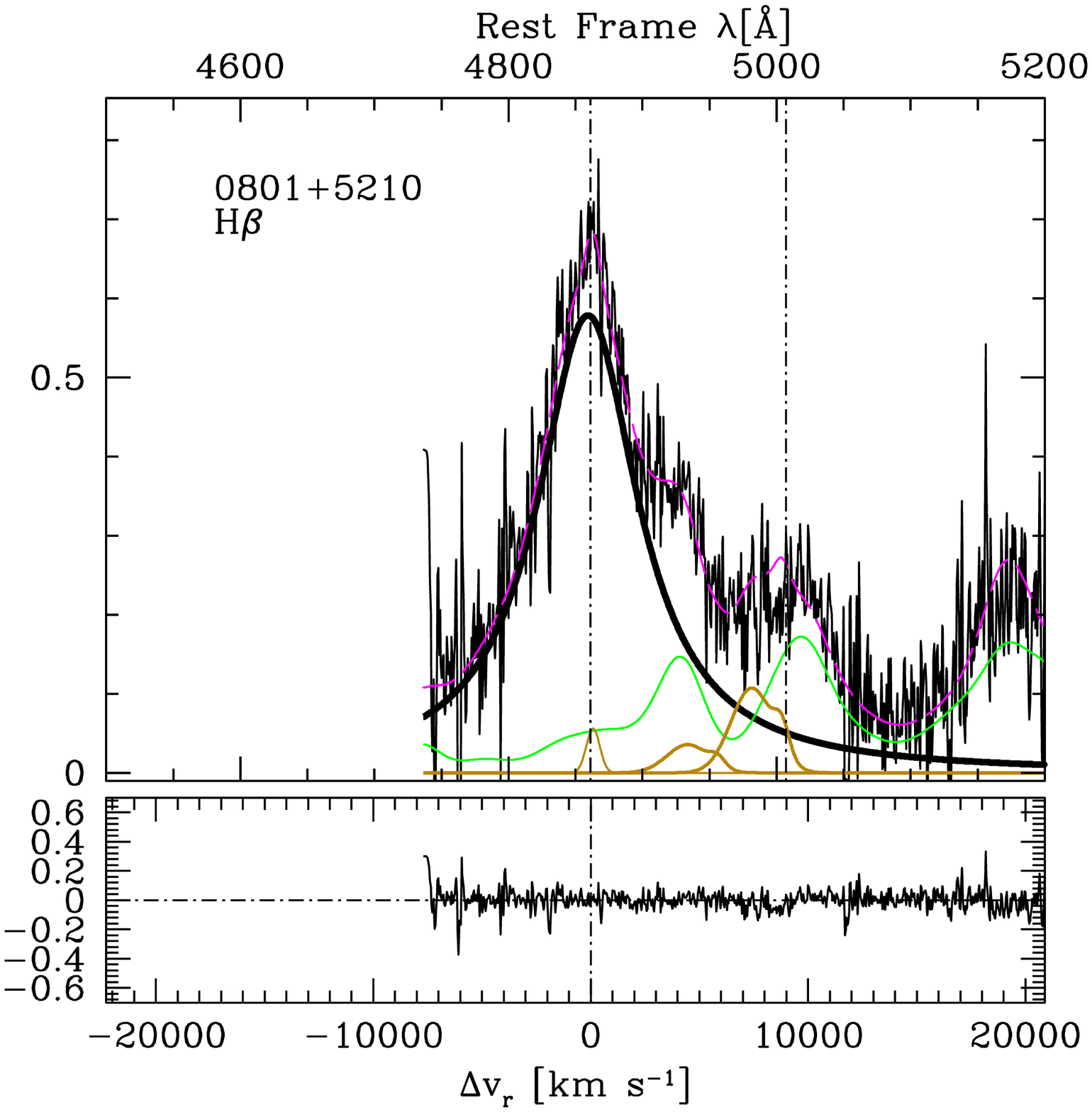}\includegraphics[width=0.225\columnwidth]{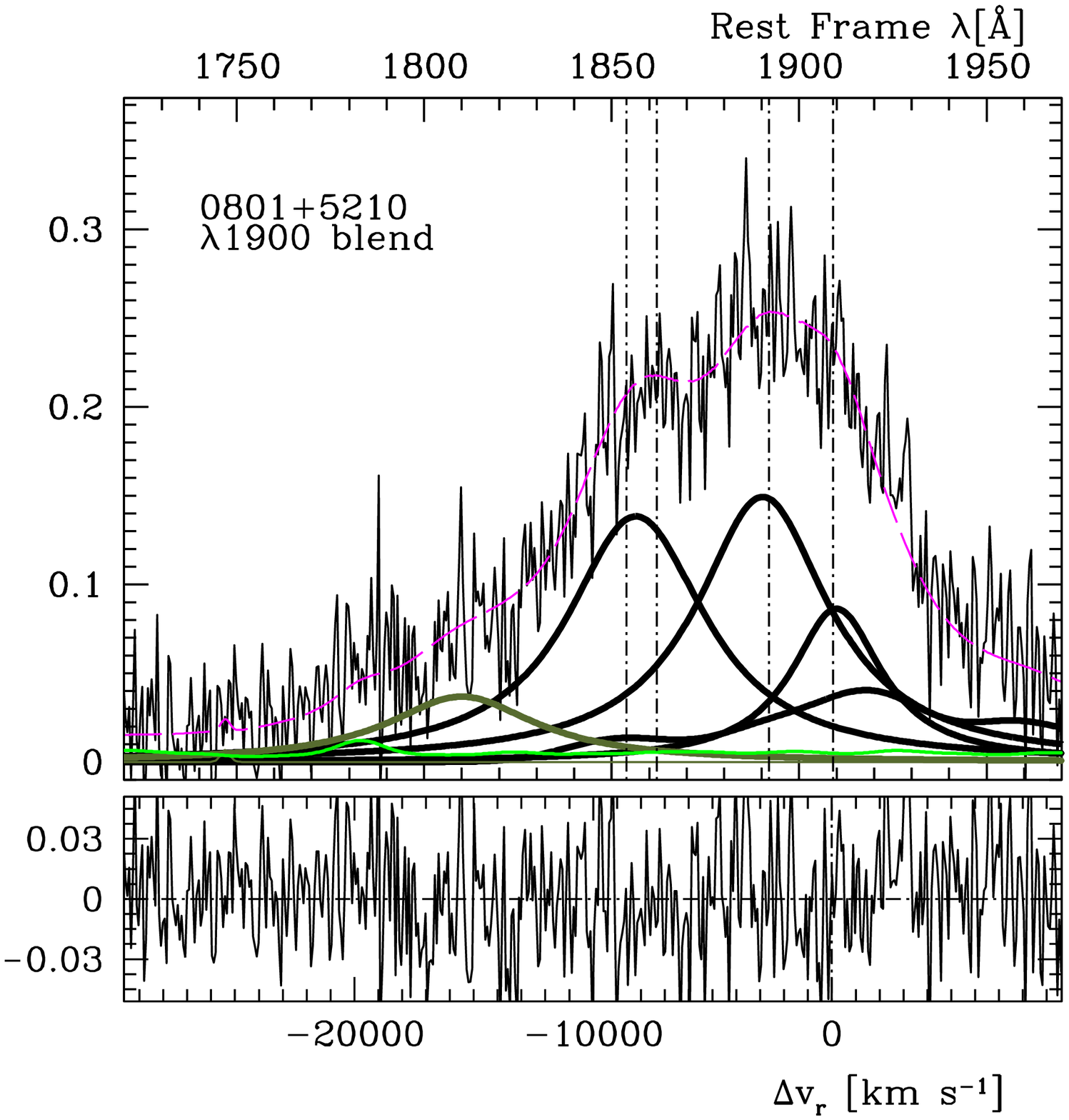}
\includegraphics[width=0.225\columnwidth]{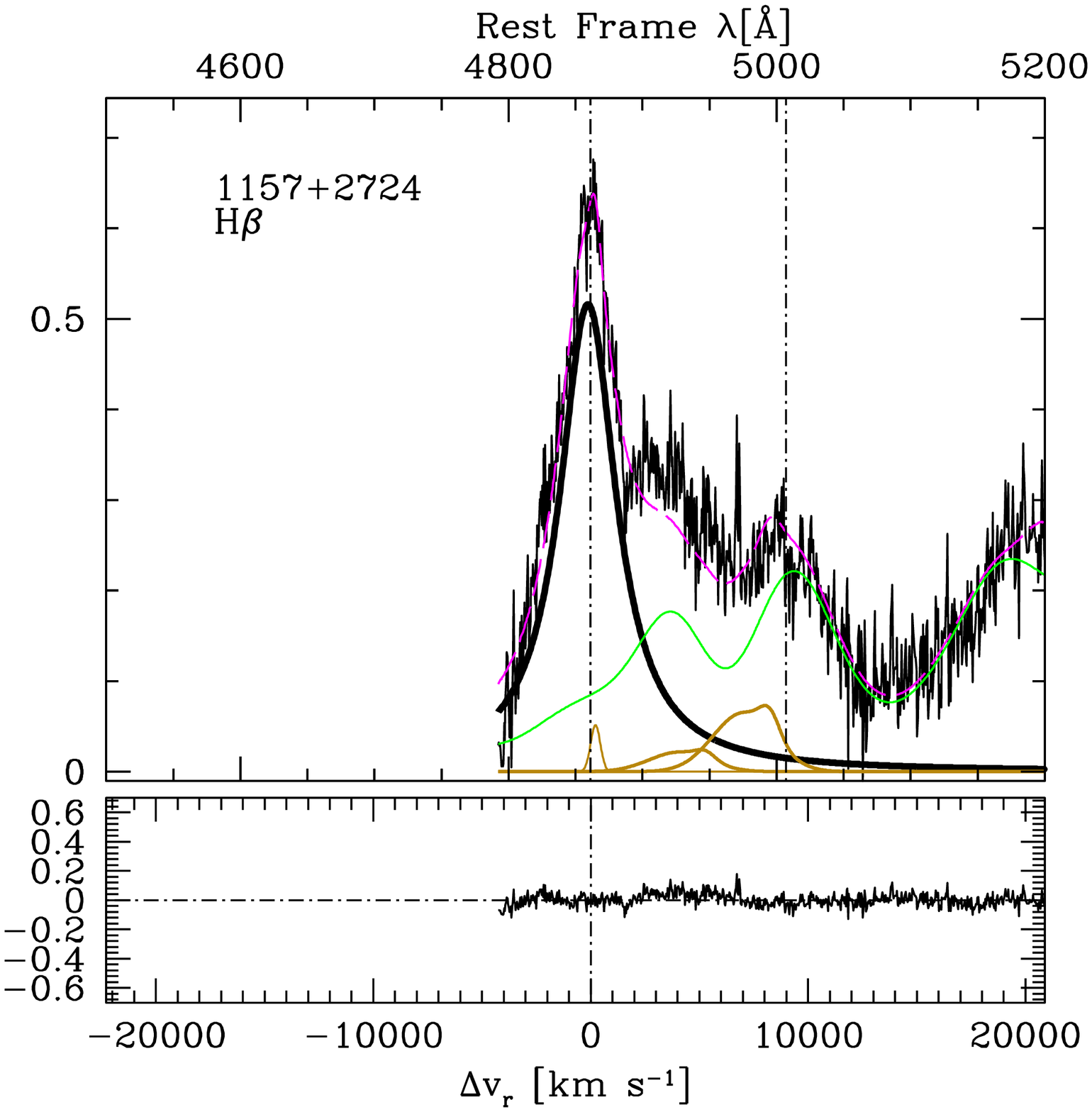}\includegraphics[width=0.225\columnwidth]{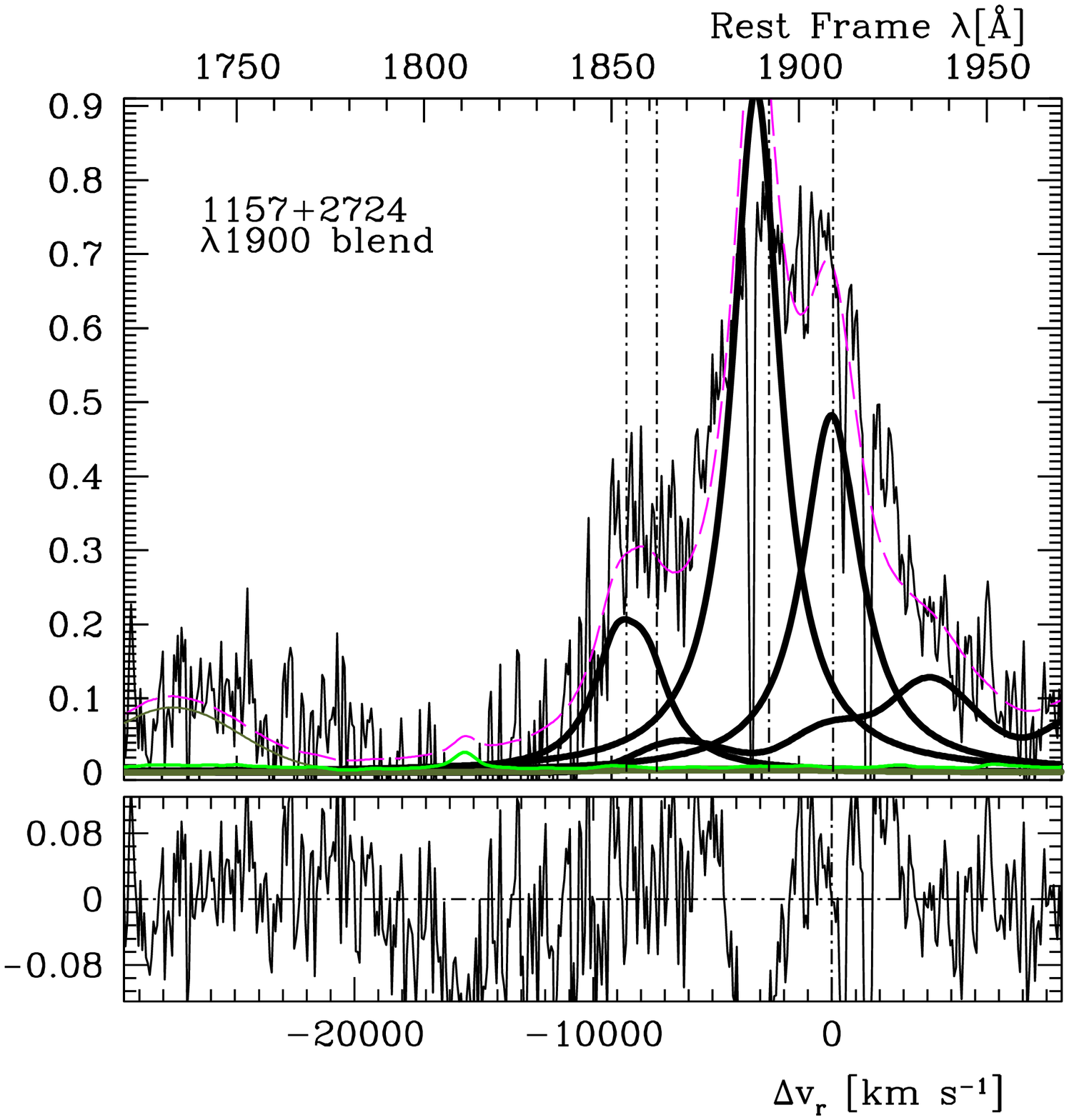}\\
\includegraphics[width=0.225\columnwidth]{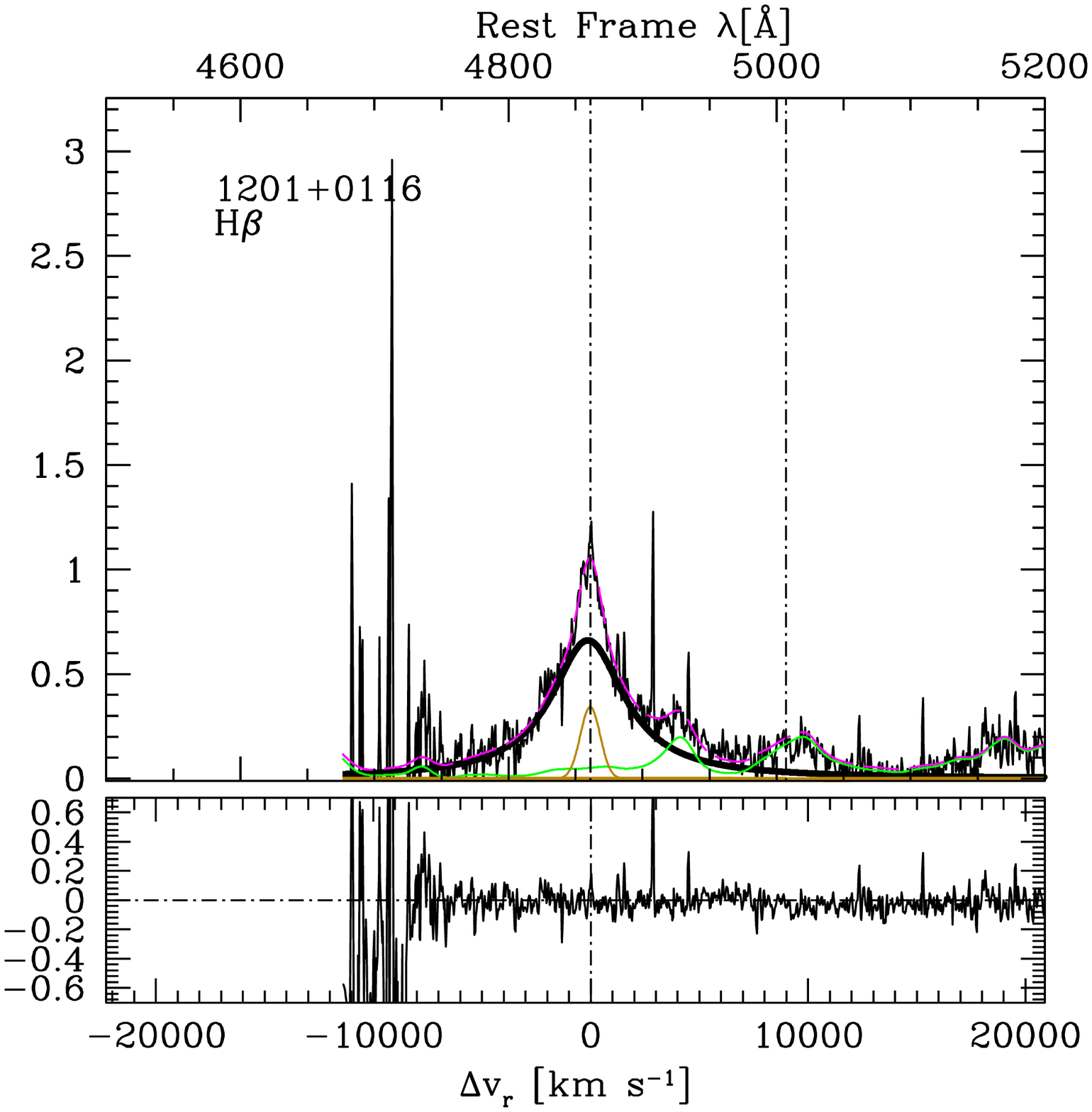}\includegraphics[width=0.225\columnwidth]{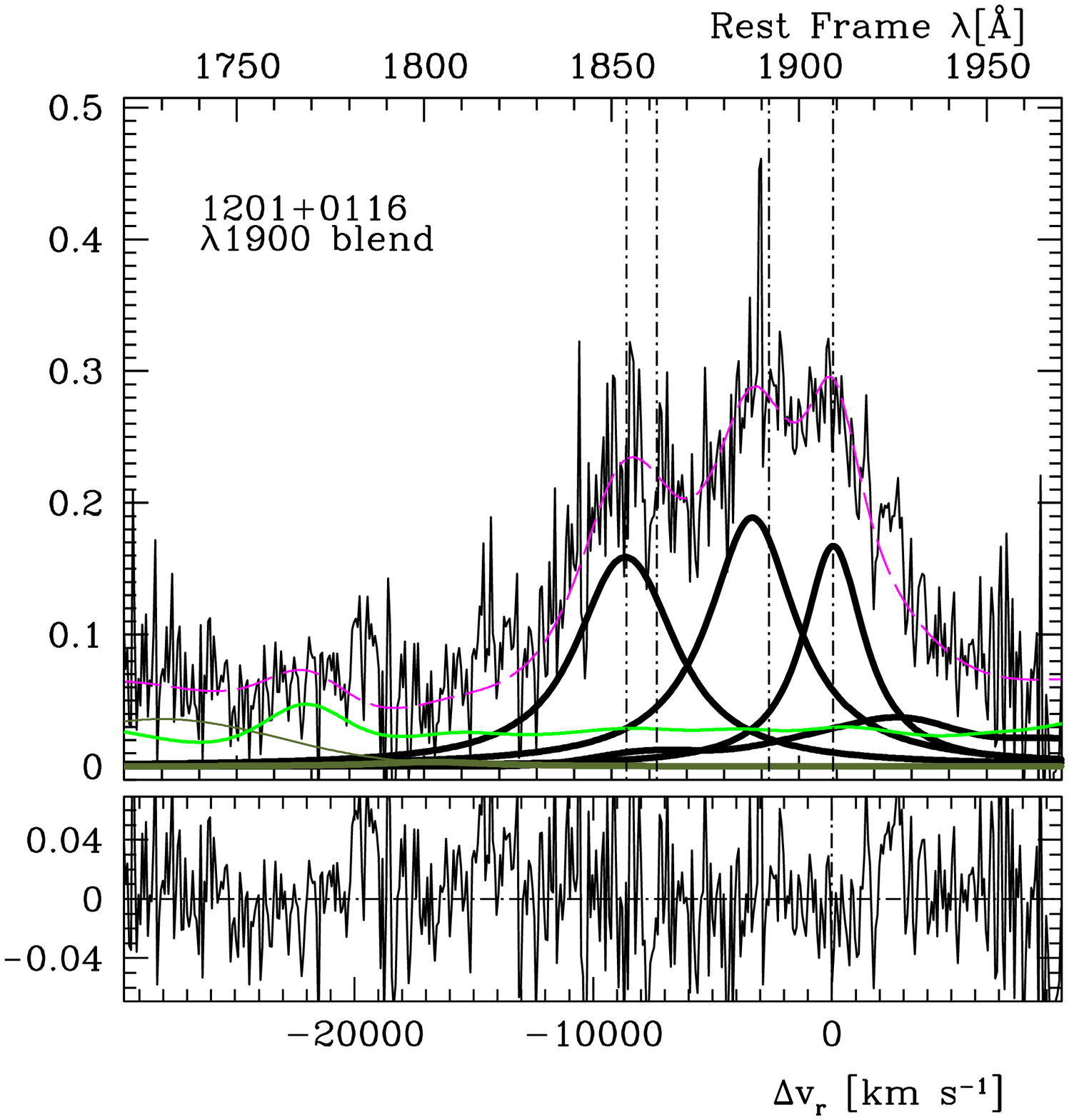} 
\includegraphics[width=0.225\columnwidth]{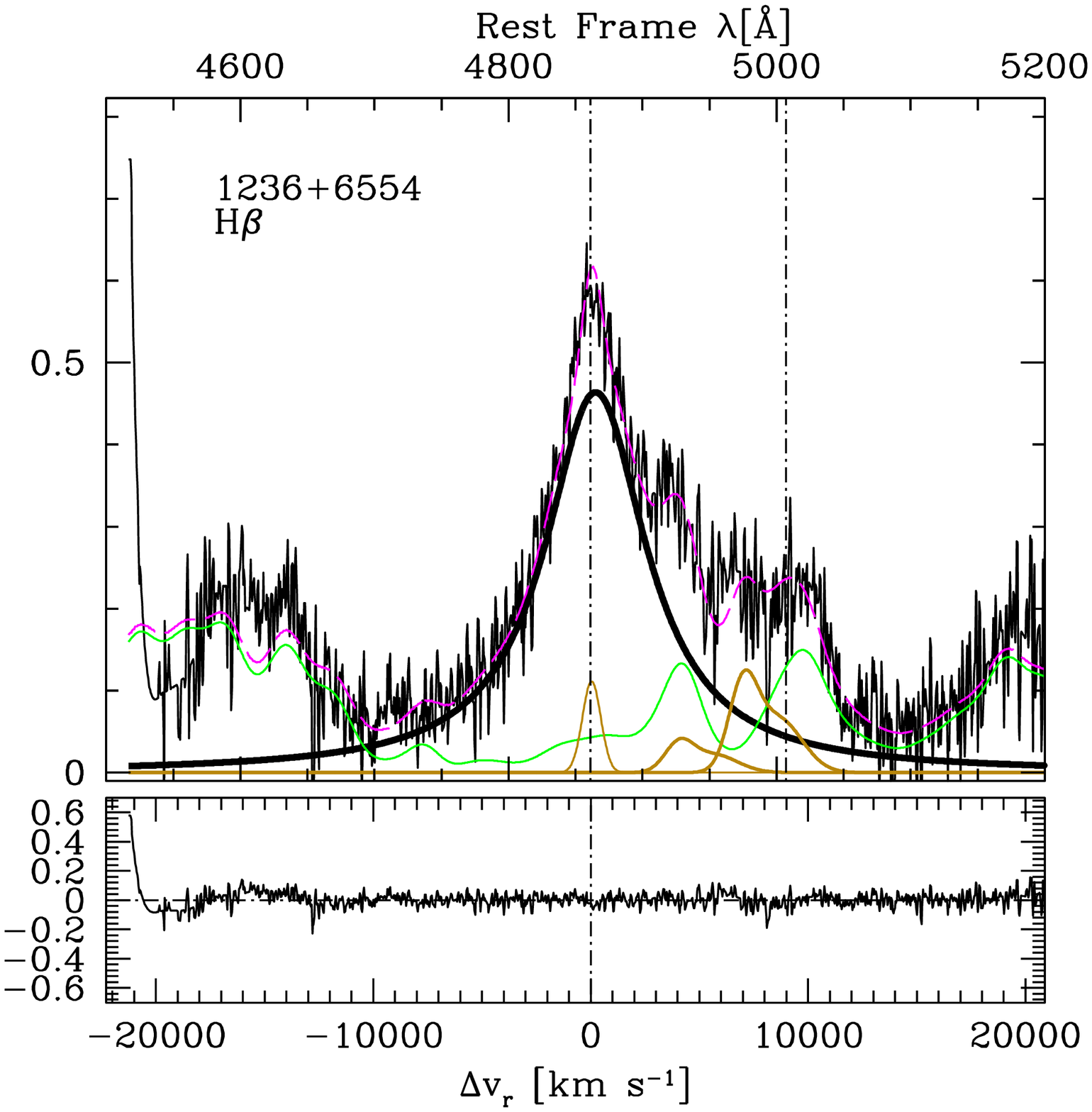}\includegraphics[width=0.225\columnwidth]{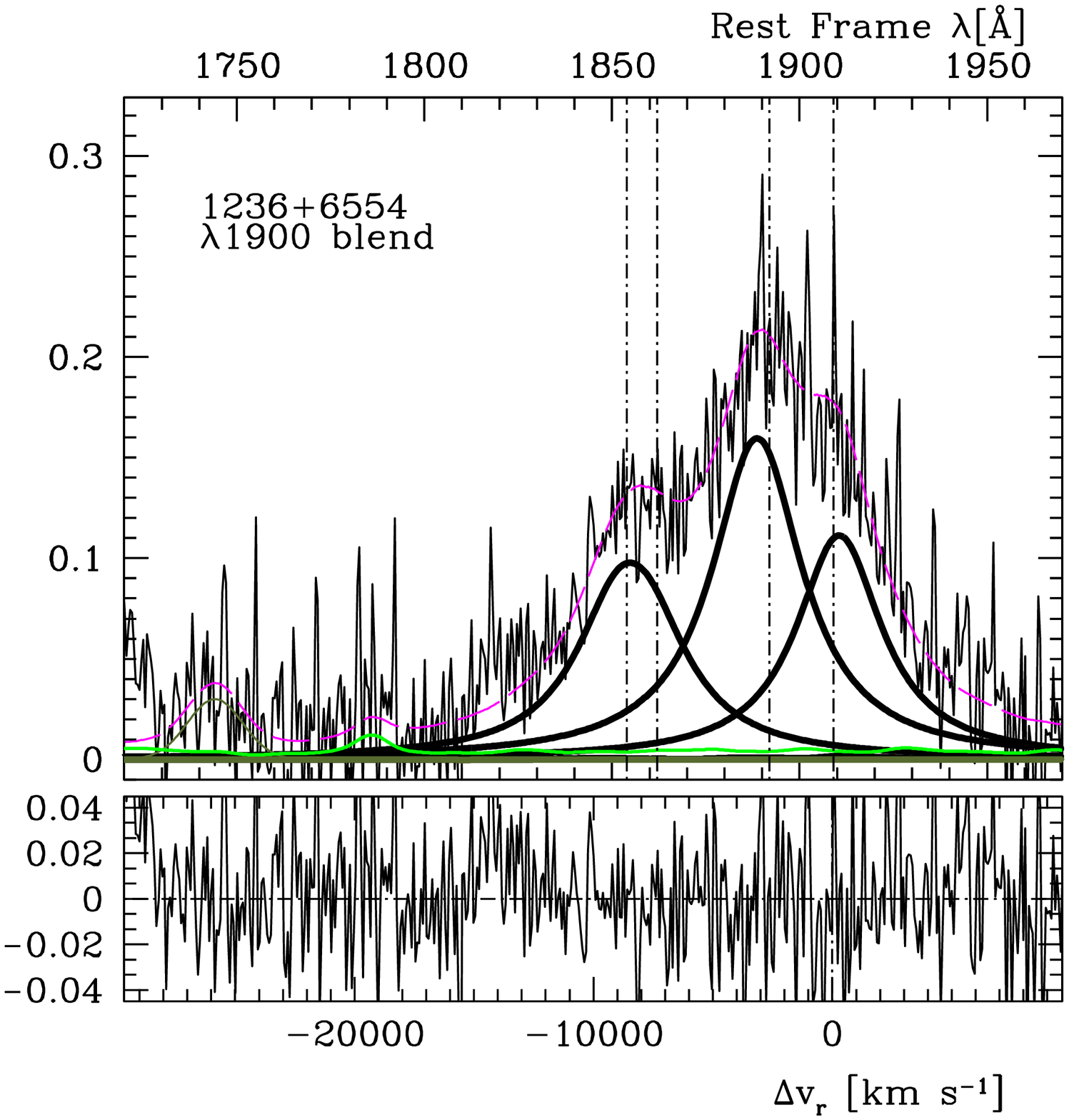}\\
\includegraphics[width=0.225\columnwidth]{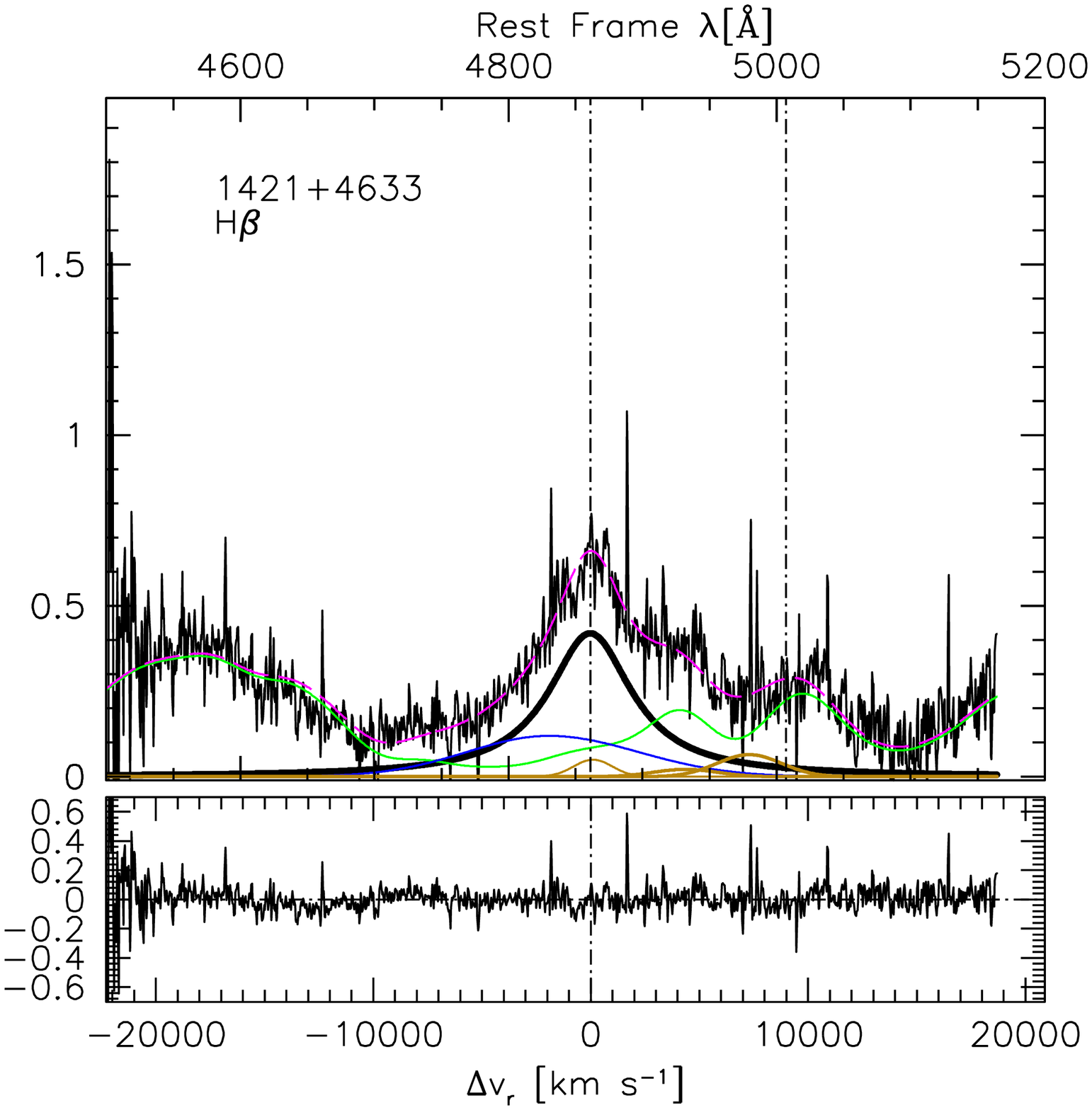}\includegraphics[width=0.225\columnwidth]{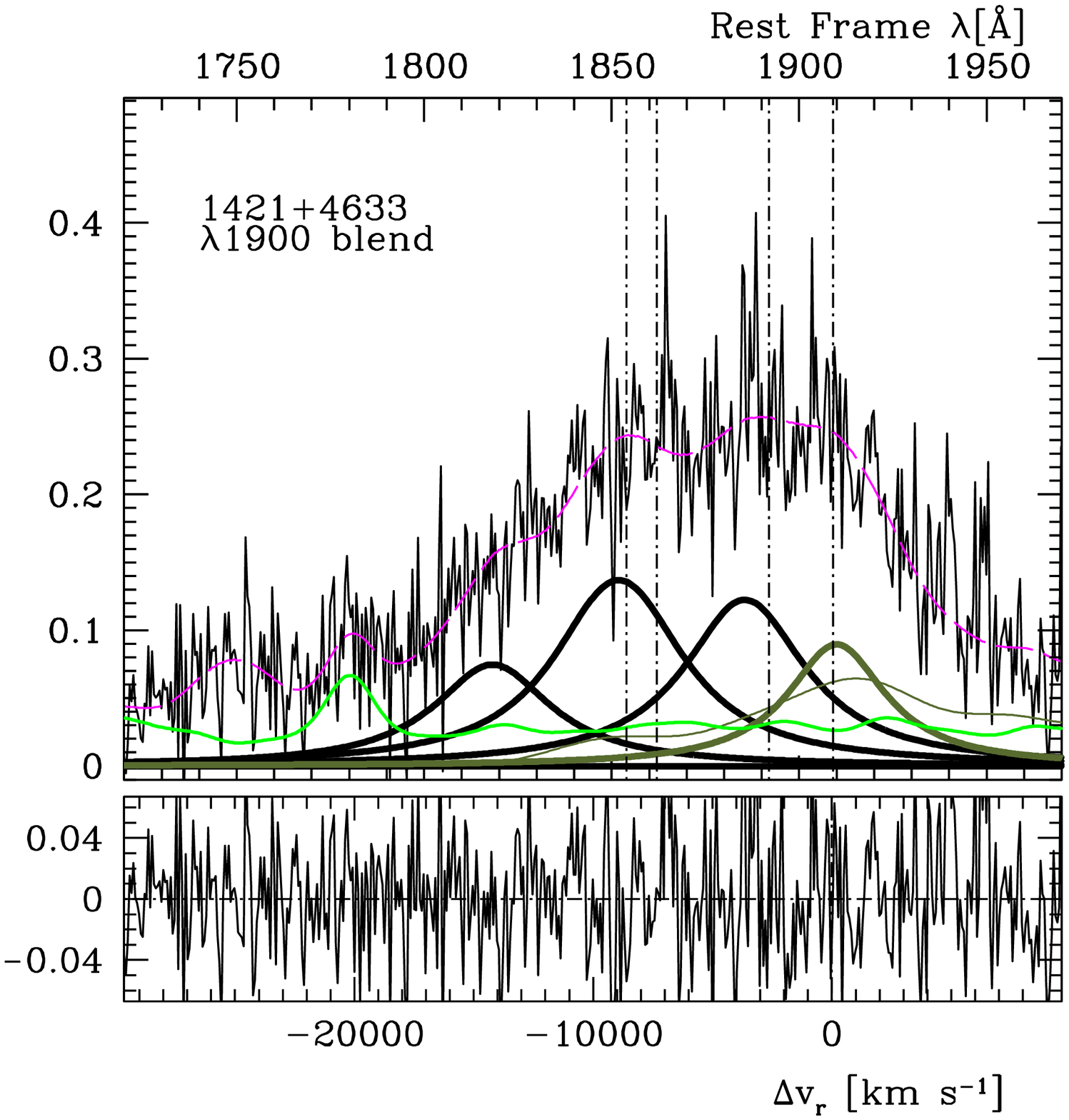}
\includegraphics[width=0.225\columnwidth]{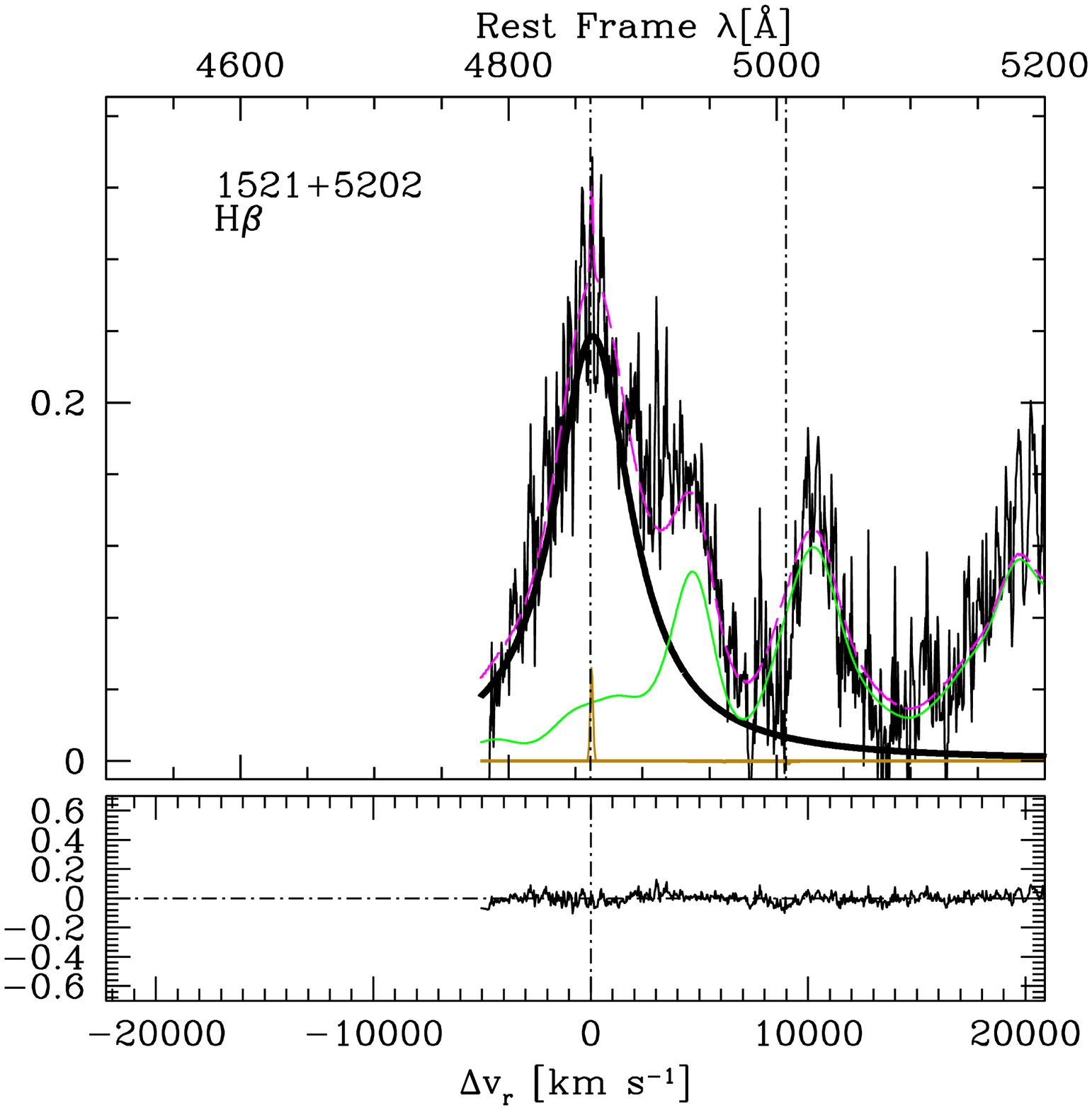}\includegraphics[width=0.225\columnwidth]{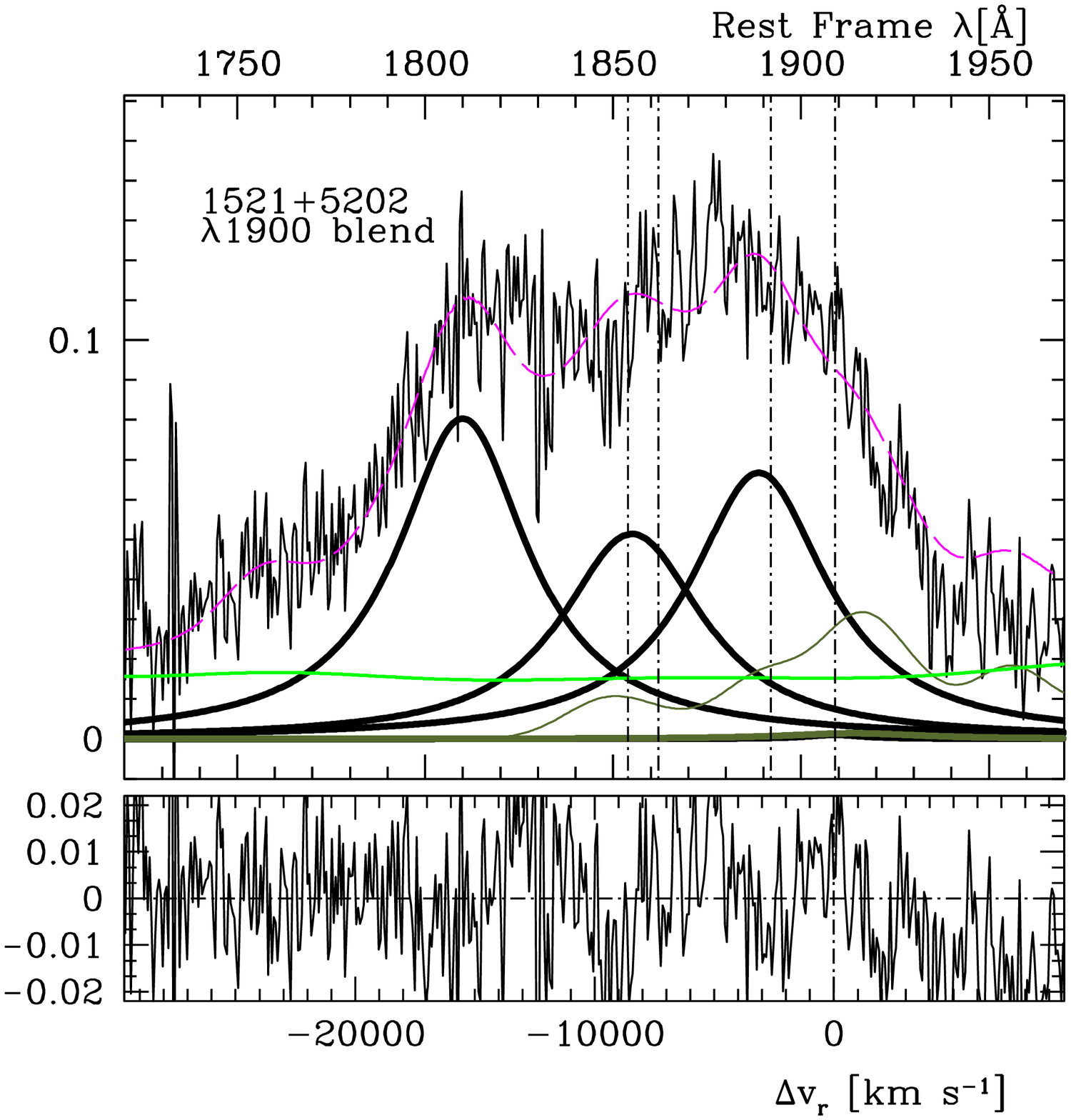}\\
\includegraphics[width=0.225\columnwidth]{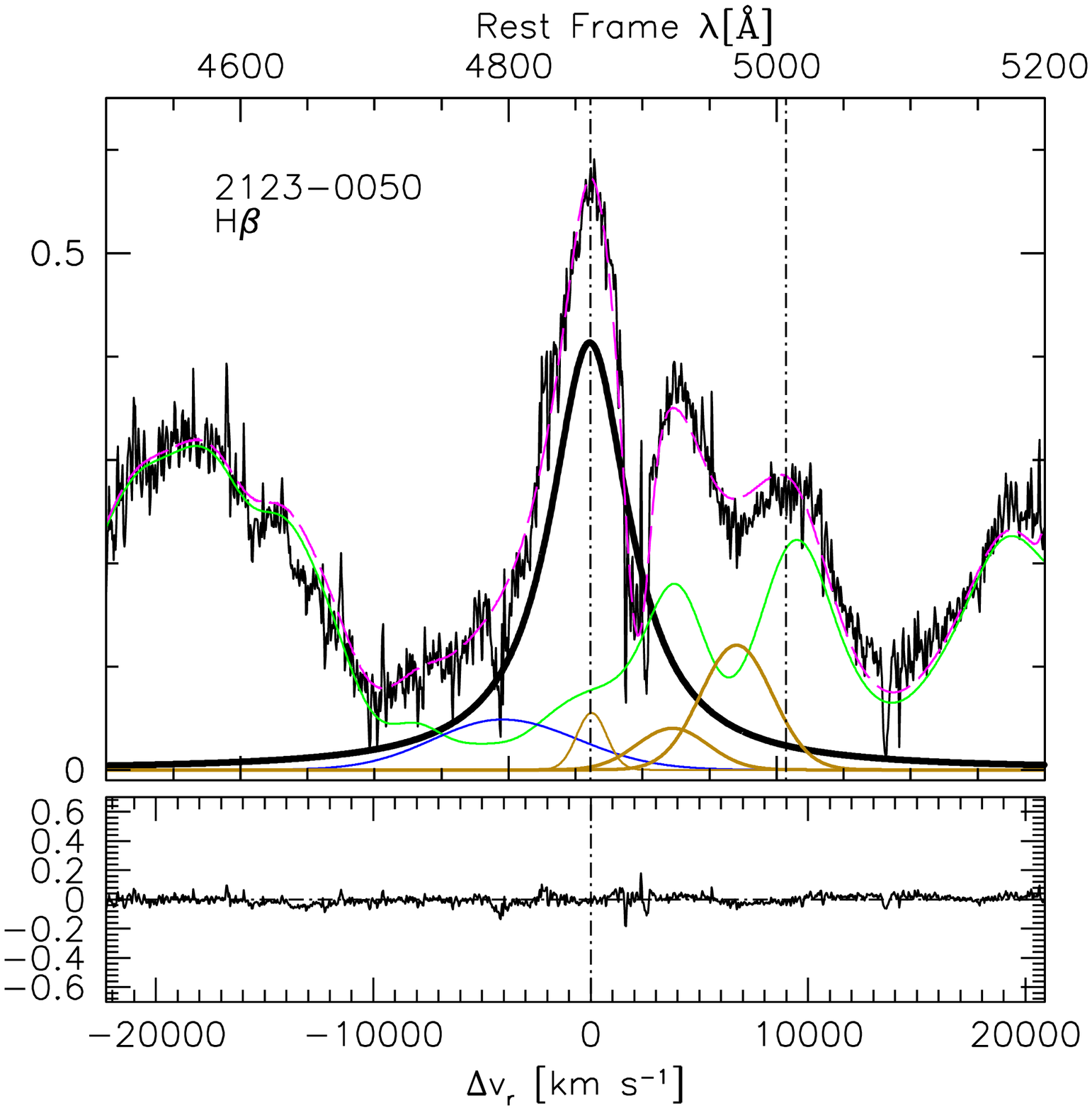}\includegraphics[width=0.225\columnwidth]{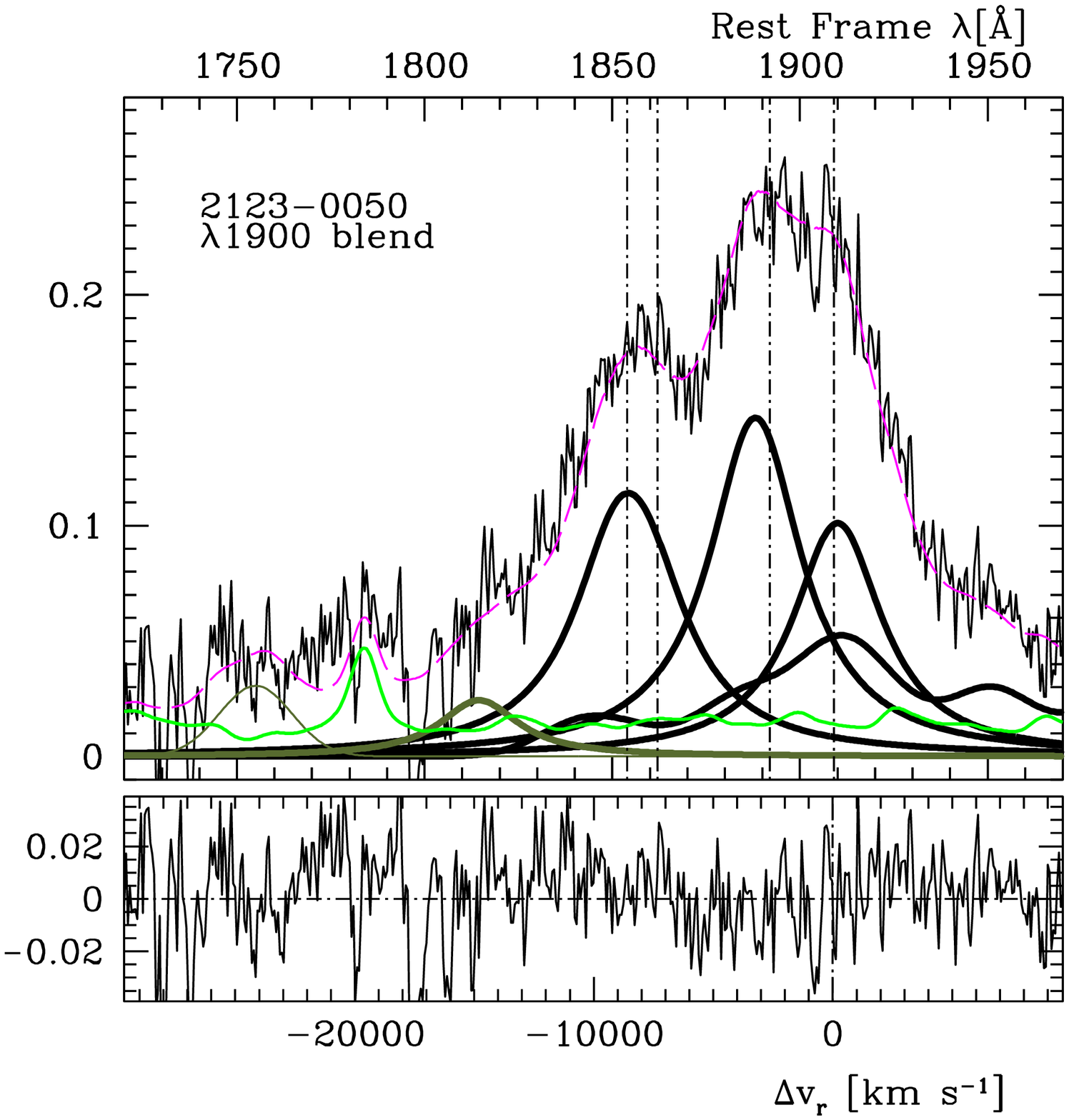} 
\caption{Analysis of \hb\ and of the 1900 \AA\ blend\ for Pop. A sources of the WISSH sample. Meaning of color coding is the same of the previous Figures in the Appendix.  \label{fig:wissha}}
\end{figure} 


\clearpage
\section{Notes on individual objects}
\label{app:notes}
\paragraph{J01342-4258}  Extreme of extreme Population A. Strong feature at 2080 \AA, extreme \feiii\ emission. In the UV spectrum a strong shelf of emission extends from the red end of the 1900 \AA\ blend to  beyond 1950 \AA. Inverted radio spectrum not accounted for by the classical synchrotron scenario.
\paragraph{J02019-1132} The CSS source 3C 57 shows a spectrum of Pop. A in the optical range. Analyzed by \citet{sulenticetal15}.
\paragraph{HE 0248--3628} Candidate high-frequency peaking object which could be associated with an incerted or self-abosbed spectrum in 5 -- 20 GHz frequency domain \citep{massardietal16}. We speculate that \object{HE 0248--3628} and \object{J01342-4258} could be both objects whose radio emission is not due to a relativistic jet but to thermal sources \citep{gancietal19}. 
\paragraph{J09199+5153} Luminous quasar, considered with ``unusually strong optical \feii\  emission"  \citep{sulenticetal90}. The \rfe\ $\approx 0.8$\ confirms that optical \feii\ emission is prominent, but not extraordinarily so. The  UV spectrum is definitely not xA, and is consistent with the A2 classification based on the optical spectrum.  
\paragraph{J07086-4933} Bad spectrum contaminated by heavy absorptions; \aliii\ lower limit. 
\paragraph{HE 0043-2300} Apart from 3C 57, the only source truly ``jetted'' radio loud. 
\paragraph{HE 0359-3959} High-luminosity analogous of J01342-4258; extreme \civ\ blueshift and extremely low ionization in the virialized BLR \citep{martinez-aldamaetal17}.
\paragraph{J1157+2724} This WISSH source has a significant difference in the redshift estimated for the present work and the one published by \citet{vietrietal18} which is estimated  from the narrow \hb\ component, 2.2133 vs	2.2170. The difference is significant. The larger redshift of \citet{vietrietal18} would imply larger shifts of \aliii.
\vfill\pagebreak

\eject\newpage\pagebreak

\clearpage
\bibliography{biblioletter2}{}
\bibliographystyle{aasjournal}

\end{document}